\documentclass[aps,prb,twocolumn,superscriptaddress]{revtex4-2}

\usepackage{amsmath}

\usepackage{amsfonts}
\usepackage{amsthm}
\usepackage{amssymb} 
\usepackage[colorlinks=true,citecolor=blue,linkcolor=red]{hyperref}
\usepackage{graphicx}
\usepackage{bbold}					
\usepackage[dvipsnames]{xcolor}
\usepackage{xcolor} 
\usepackage{bm} 

\usepackage{times}
\usepackage{bm}

\usepackage{dcolumn}
\usepackage{mathtools}

\usepackage{array,booktabs,ragged2e}	
\usepackage{soul}						
\usepackage{cancel}						
\usepackage{tabularx}
\usepackage{multirow}
\usepackage{hhline}
\usepackage{adjustbox}
\usepackage{sidecap}
\usepackage{threeparttable}
\usepackage{colortbl} 
\usepackage{tabularx}
\usepackage[utf8]{inputenc}
\usepackage[T1]{fontenc}

\usepackage{amsmath, amssymb, bm}

\newcommand{\eq}[1]{\begin{equation}#1\end{equation}}
\DeclareMathOperator{\Tr}{Tr}   

\DeclareMathOperator{\im}{Im}  			   		
\def\abs#1{\left|{#1}\right|}          			
\def\bra#1{\left<{#1}\right|}					
\def\ket#1{\left|{#1}\right>}					
\def\braket#1#2{\left<{#1}|{#2}\right>}			

\newcommand{\wh}{w_{\mathrm{qh}}}
\newcommand{\qh}{\mathrm{qh}}

\newcolumntype{P}[1]{>{\raggedleft\arraybackslash}p{#1}}
\newcolumntype{R}[1]{>{\centering\arraybackslash}p{#1}}

\usepackage{bm,color,amsmath,amssymb,mathrsfs,latexsym,graphicx,psfrag,mathtools}
\usepackage{color}
\usepackage[dvipsnames]{xcolor}

\usepackage{empheq}

\newcommand{\bsl}[1]{{\boldsymbol{#1}}}

\renewcommand{\mod}{\mathrm{mod}}

\newcommand{\ii}{\mathrm{i}}

\newcommand{\dsZ}{\mathbb{Z}}

\newcommand{\dsR}{\mathbb{R}}

\newcommand{\dsC}{\mathbb{C}}

\renewcommand{\Im}{\mathop{\mathrm{Im}}}

\newcommand{\refcite}[1]{Ref.\,\cite{#1}}

\newcommand{\mat}[1]{\left(\begin{matrix}#1\end{matrix}\right)}

\newcommand{\eqa}[1]{\begin{align}\begin{split} #1 \end{split}\end{align}}

\let\oldAA\AA
\renewcommand{\AA}{\text{\normalfont\oldAA}}

\newcommand{\ie}{{\emph{i.e.}}}

\newcommand{\N}{\mathcal{N}}

\newcommand{\Ch}{\text{Ch}}

\newtheorem{definition}{Definition}

\newtheorem{theorem}{Theorem}
\newtheorem{corollary}{Corollary}

\usepackage{comment}

\usepackage{cleveref}
\crefname{appendix}{App.}{Apps.}
\Crefname{appendix}{Appendix}{Appendices}
\crefname{equation}{Eq.}{Eqs.}
\Crefname{equation}{Equation}{Equations}
\Crefname{figure}{Figure}{Figures}
\crefname{figure}{Fig.}{Figs.}
\crefname{theorem}{Thm.}{Thms.}
\Crefname{theorem}{Theorem}{Theorems}
\crefname{table}{Tab.}{Tabs.}
\Crefname{section}{Section}{Sections}
\crefname{section}{Sec.}{Secs.}
\creflabelformat{appendix}{#2#1#3}

\newcommand{\norm}[1]{\left\lVert#1\right\rVert}

\begin{document}

\title{Analytical Interacting Tight-binding Model for Fractional Chern Insulators}

\author{Lexu Zhao}
\thanks{These authors contributed equally to this work.}
\affiliation{Department of Physics and Quantum Theory Project, University of Florida, Gainesville, Florida 32611, USA}

\author{Jayden S. Chang}
\thanks{These authors contributed equally to this work.}
\affiliation{Department of Physics and Quantum Theory Project, University of Florida, Gainesville, Florida 32611, USA}

\author{Awwab A. Azam}
\thanks{These authors contributed equally to this work.}
\affiliation{Department of Electrical and Computer Engineering, University of Florida, Gainesville, USA}

\author{Jiabin Yu}
\email{yujiabin@ufl.edu}
\affiliation{Department of Physics and Quantum Theory Project, University of Florida, Gainesville, Florida 32611, USA}

\begin{abstract}
We construct a two-dimensional 2-band tight-binding model with two-body interactions and without magnetic lattice translation, in which both the hopping and interaction are exponentially local (with Gaussian decay) in real space.
The interaction is built from pair annihilation operators whose coefficients are hopping amplitudes divided by positive powers of relative distance.
Without any band projection, the ground states at $1/3$ filling are exactly fractional Chern insulator (FCI) states (and thus have nonzero gap) for any finite lattice size with any periodic/twist boundary conditions.
The FCI ground states are analogous to the discretized version of the Laughlin-like states in the continuum Chern-ideal band, though our FCI states do not require the onsite spinor structure that is necessary in the continuum.
We further show analytically that, after adding one unit cell while keeping electron number fixed, the anyonic one-quasihole states are zero-energy ground states of the full Hamiltonian for any finite system size and any quasihole position, analogous to Laughlin quasiholes under the Haldane $V_1$ pseudopotential.
Numerical evidence is provided for the possible nonzero gap of the FCI states at $1/3$ filling in the thermodynamic limit, when interaction strength is much smaller than the single-particle band gap.
\end{abstract}

\maketitle

\section{Introduction}
Fractional Chern insulators (FCIs) realize fractional quantum Hall physics in partially filled Bloch bands, without requiring an external magnetic field \cite{Tang11,neupert2011,Sun2011,sheng2011,regnault2011}. Recent observations of the fractional quantum anomalous Hall effect in twisted bilayer MoTe$_2$~\cite{cai2023signatures,zeng2023integer,park2023observation,xu2023observation,Ji2024LocalProbetMoTe2,redekop2024direct,Kang2024_tMoTe2_2.13,park2025ferromagnetism,Xu_2025_Interplay_SecondMoireBand_tMoTe2,an2025observation,jia2025anomalous,anderson2024trion,kang2025timereversal,Wang2025HiddenStates,Holtzmann2026Optical,Huber2026Optical,Cai2026Optical,Park2026,Chang2026,xu2026signatures,Li2025Universal,Pan2026Optical,li2026signature,Sun2026TwistAngle}
and rhombohedral graphene moir\'e superlattices~\cite{Lu2024fractional,xie2025tunable,Lu2025Extended,waters2025chern,aronson2024displacement,choi2025superconductivity,zheng2025switchable,li2025stacking,xie2025unconventionalorbitalmagnetismgraphenebased,huo2025doesmoiremattercritical,Zhang2026Rhombohedral,Li2026Fractional,butler202613fractionalgaplessinteger,
hadjri2026quantumphasetransitionsfractional}
have drawn extensive interest in understanding these strongly correlated lattice states
\cite{
Li2021Spontaneous,Crepel2023Anomalous,Wang2024FCI,Reddy2023Fractional,
qiu2023interaction,dong2023composite,goldman2023zero,morales2024magic,liu2024gate,
xu2024maximally,Reddy2023toward,song2024phase,wu2024time,Yu2024FCI,
abouelkomsan_band_2024,li2024electrically,Dong2024Theory,Zhou2024Fractional,
jia2024moire,Dong2024anomalous,mao2024transfer,zhang2024polarization,Jonah2024moire,
Guo2024FCI,Kwan2025moire,wang2023topology,li2024contrasting,Reddy2024non,
lu2024interaction,xu2025multiple,ahn2024non,Wang2024higher,Chen2025,
shen2024stabilizing,wang2024phase,lu2024fractional2,Wolf2025intra,kwan2024when,
Wu2024quantum,Huang2024self,Yu2025moire,huang2025fractional,wei2025edge,zaklama2025structure,zhang2024universa1,shen2025magnetorotons,Long2026Spectra,
qiu2025topological,CHEN2026fractional,Li2025multiband,
he2025fractionalcherninsulatorscompeting,gmkp-bxf2,Liu2025characterization,
shrz-b692,Wu_2026_Nonabelian_tMoTe2,zf45-1gvb,
desrochers2026energeticsfractionalanomaloushall,
regnault2026moirecapacitoreffectstabilization,
maymann2026skyrmionfractionalcherninsulator,li2026largescaletheoreticalinvestigation}.
Most theoretical studies investigate the FCI ground states and excitations numerically, using methods such as exact diagonalization (ED) and density matrix renormalization group (DMRG). Such methods have the advantage of addressing the Hamiltonians of realistic systems that are often too complicated to solve analytically.

Besides these numerical approaches to understanding FCIs, there has been a long history of analytical studies of the fractional quantum Hall effect and FCIs. This effort dates back to Haldane's classic work~\cite{Haldane1983FQH} (as well as later works~\cite{Haldane_1985_LJ_function,Trugman1985ExactFQH}) on the fractional quantum Hall effect, which introduced the pseudopotential Hamiltonian with the Laughlin wavefunction~\cite{Laughlin1983FQHE} as an exact ground state.
Later, this analytical approach was generalized to FCIs within the framework of Chern-ideal bands 
\cite{
Parameswaran2012fci,
Roy2014QGChernBound,
claassen2015position,
Lee_2017,
ledwith2020fractional,
Jie2021IdealBands,
PhysRevLett.128.176403,
PhysRevLett.128.176404,
northe2022interplay,
Parker2023IdealBands,
Dong2023many,
Wang2023Origin,
Valentin2023IdealBands,
Queiroz_2024_TMD_Chiral,
Roy2024FCINonLL,
liu2024theorygeneralizedlandaulevels,
Sun2025IdealChern,
ji2024quantum,
Oreg2025UnifyingFrameworkFCI,
Okuma2025,
Liu_2025_idealTHF,cano2026idealquantumgeometryfractional,li2026quantumgeometricdesignlatticegeneralized,zhao2026idealbandstightbindingmodels,wan2026tunablemultibandgeometryfractional,spasicmlacak2026designinghamiltoniansexactlyflat,liu2026evaluatelaughlinspectralgap}, where the generalized $V_1$ pseudopotential can serve as an exact parent Hamiltonian. Such analytical approaches have several advantages over numerical studies. First, they set the stage for studying many-body states in the thermodynamic limit---indeed, a nonzero thermodynamic many-body gap (or incompressibility) for the Laughlin state under the $V_1$ pseudopotential was recently proven on the sphere
\cite{OpenAI2026LaughlinGap,gao2026improvinglowerboundlaughlin,
scheer2026hierarchyexactlowerbounds}.
Second, they allow us to form a clearer picture of the properties of these states, such as the formation of anyonic excitations 
\cite{
Halperin1984stat,
Arovas1984fractional,
PhysRevLett.91.036801,
PhysRevB.70.125316,
Hansson_2009,
Greiter_2016,
Nielsen_2015,
Cheng2016translational,
Nielsen_2018,
Bultinck2018filling,
Shi2025doping,
schleith2025anyondispersionnonuniformmagnetic,
Yan_2026,
li2026boundstatesanyonsgeometric,
iyer2026dispersionanyonblochbands,
xiao2026anyonchernbandsopposite,
sarkar2026latticeinducedsplittingconnectivitytopology}
and the composite fermion picture \cite{Jain1989CF,Pu2017Composite,Lopez1991,Halperin1993,Rezayi1994Fermi,pu2026compositefermionsidealchern,ding2026compositefermiontheoryfractional}.

Nevertheless, despite decades of analytical efforts, the analytical model for FCIs is currently only limited to continuum.
There have been analytical models for the fractional quantum Hall effect on lattices, but they all involve magnetic lattice translation (or external magnetic field) 
\cite{Kapit_2010,Ataki_i_2013,Kapit2013Three,Liu2013non,Scaffidi2014,Behrmann_2016,PhysRevA.101.013629,Xu_2020,Shen2026exact,cruise2026nonabelianfci,ji2026generalizedmodelfractionalquantum,ji2026modelfractionalquantumhall}.
An analytical lattice model for FCIs should not require an external magnetic field or the magnetic lattice translation symmetry. Constructing such a model remains an open problem.

In this work, we address this open question by proposing a tight-binding model with Gaussian-decaying hopping amplitudes and Gaussian-decaying two-body interactions.
We show that the lower band of the tight-binding model is an isolated Chern-ideal band with Chern number $+1$.
After projecting the Hamiltonian to the ideal band, the model has exactly three FCI ground states at filling $\nu=1/3$ for any finite lattice with periodic or twisted boundary conditions.
Adding additional unit cell while keeping the electron number unchanged, we also find that the one-quasihole states are exact zero-energy states of the projected Hamiltonian, and the localization of the anyonic excitation is checked numerically.
Combined with the frustration-free nature of the full unprojected Hamiltonian at fillings for the FCI and quasi-hole states, the three FCI ground states and the one-quasihole states are also the ground states of the full unprojected Hamiltonian at the their corresponding fillings.
Numerically for the Hamiltonian projected to the lower band, we find that the charge-neutral gap above the FCI ground states changes by only $\sim3$\% across various meshes with $24$, $30$ and $36$ unit cells, indicating the potential nonzero thermodynamic gap at filling $1/3$ when interaction is much smaller than the single-particle band gap.

\section{Interacting Lattice Model} We will first describe the spinless model we build and then
explain how it is constructed and why its projection has the
FCI states as exact ground states.
The Hamiltonian $H$ is defined on a 2D lattice with two
inequivalent sublattices per unit cell, labeled $A$ and $B$.
We use $c_{\bsl r}$ to denote the annihilation operator at
$\bsl r=\bsl R+\bsl\delta_\alpha$, where $\bsl R$ is a lattice vector with 
$\bsl a_1,\bsl a_2$ being the primitive vectors, and
$\bsl\delta_\alpha$ is the position of sublattice $\alpha=A,B$
within the $\bsl R=0$ unit cell.
In the main text, we will choose $\bsl{a}_1=(0,-1)$, $\bsl{a}_2=(1,0)$, $\bsl{\delta}_{A}=(0,0)$ and $\bsl{\delta}_{B}=(1/2,0)$ for illustrating purpose; the formulation for the general lattice structure (with $\im(a_1/a_2)>0$ for $a_j=a_{j,x}-\ii a_{j,y}$) can be found in \cref{app:E-real-space-parent}.
The model contains a single-particle part $H_0$ and an interaction
part $H_{\mathrm{int}}$, \ie, $H=H_0+H_{\mathrm{int}}$.
The single-particle part is
\eq{
H_0=\sum_{\bsl R,\bsl R'}\sum_{\alpha,\beta=A,B}
t_{\alpha\beta}(\bsl R-\bsl R')
c^\dagger_{\bsl R+\bsl\delta_\alpha}
c_{\bsl R'+\bsl\delta_\beta},
\label{eq:main-real-space-h0}
}
where the hopping amplitudes have Gaussian decay and are given by
\eqa{
t_{\alpha\beta}(\mathbf \Delta)
= \frac{\eta_{\bar\alpha}^{*}\eta_{\bar\beta}}{
\sqrt2}e^{\ii \phi_{\alpha\beta}(\mathbf \Delta)}
 \exp\left[
-\frac{\pi}{2\Omega}
|\mathbf \Delta+\bsl{\delta}_\alpha-\bsl{\delta}_\beta|^2
\right] 
\ ,
\label{eq:main-hopping-matrix}
}
where $\bar{A}=B$, $\bar{B}=A$, and $\Omega$ is the unit-cell area.
$e^{\ii \phi_{\alpha\beta}(\mathbf \Delta)}=
(-1)^{m_1m_2}e^{\ii\pi (m_1+2)(\delta_{\bar{\alpha}}+\delta_{\bar{\beta}})
}$ is a phase factor with
$\bsl \Delta=m_1\bsl a_1+m_2\bsl a_2$ and $\delta_\alpha=\delta_{\alpha,x}-\ii\delta_{\alpha,y}$.
$\eta_A,\eta_B$ are nonzero complex factors that are up to our choice. 
The Gaussian-decaying hopping amplitudes appeared in the Eliot-Mueller model~
\cite{Kapit_2010}, but the hopping in \cref{eq:main-hopping-matrix} only depends on the relative lattice vectors and thus preserves the ordinary lattice translation symmetry instead of the magnetic translation symmetry.

The interaction is generated by pair annihilation operators that are built from dividing the hopping amplitudes
by powers of the complex separation between the sites.
Specifically, the pair annihilation operators are defined as
\eq{
B_{\bsl r}^{(p)}=
\sum_{\bsl r'\ne\bsl r}
\frac{t_{\bsl r\bsl r'}}{(z_{\bsl r'}-z_{\bsl r})^p}c_{\bsl r'}c_{\bsl r},\qquad p=1,2,
\label{eq:main-pair-operators}
}
where $t_{\bsl r\bsl r'}=t_{\alpha\beta}(\bsl R-\bsl R')$
for $\bsl r=\bsl R+\bsl\delta_\alpha$ and
$\bsl r'=\bsl R'+\bsl\delta_\beta$, and
$z_{\bsl r}=r_x-\ii r_y$.
Since $t_{\bsl r\bsl r'}$ has a Gaussian dacay as
$|\bsl r-\bsl r'|$ increases, so does the coefficient in
$B_{\bsl r}^{(p)}$ for any positive $p$.
Then, the interaction Hamiltonian is
\eq{
H_{\mathrm{int}}=u\sum_{\bsl r}\sum_{p=1}^{2}
\Omega^{p-1}B_{\bsl r}^{(p)\dagger}B_{\bsl r}^{(p)},
\label{eq:main-real-space-hu}
}
where $u>0$ controls the interaction strength and makes the
interaction positive semidefinite.
At this point, we have not discussed the origin of the model. In the next section, we first discuss how we constructed $H_0$.

\section{Single-Particle Chern-Ideal Band.}

We constructed $H_0$ in order to realize the lower band as an isolated ideal band with Chern number $\Ch=1$.
We do so by Fourier transforming the Chern-ideal bands in \refcite{zhao2026idealbandstightbindingmodels} to the real space.
So we start in the momentum space using $c_{\bsl k,\alpha}=N^{-1/2}\sum_{\bsl R}
e^{-\ii\bsl k\cdot(\bsl R+\bsl\delta_\alpha)}
c_{\bsl R+\bsl\delta_\alpha}$,
where $N$ is the number of unit cells and we adopt the physical embedding where the Fourier transform includes the sublattice position $\bsl\delta_\alpha$.

\refcite{zhao2026idealbandstightbindingmodels} showed that a $\Ch=1$ ideal band on a lattice with two atoms per unit cell can be constructed from a Bloch vector of the form $U_{\bsl k}=v(\bsl k)/\sqrt{S(\bsl k)}$, where $S(\bsl{k})=v^\dagger(\bsl{k}) v(\bsl{k})>0$ and $v(\bsl{k})=(v_A(\bsl{k}),v_B(\bsl{k}))^T$.
The key requirement is that  $v_B(\bsl{k})/v_A(\bsl{k})$ is a meromorphic function of 
$k=k_x-\ii k_y$ that has one simple zero and one simple pole in the BZ and that satisfies the physical embedding. 
The physical embedding requires $v_B(\bsl k+\bsl G)/v_A(\bsl k+\bsl G)
=e^{-\ii\bsl G\cdot(\bsl\delta_B-\bsl\delta_A)}
v_B(\bsl k)/v_A(\bsl k)$
for every reciprocal-lattice vector $\bsl G$.
Then, naturally we can include $e^{-\ii\bsl{k}\cdot\bsl{\delta}_{\alpha}}$ in the expression of $v_{\alpha}(\bsl{k})$.
To achieve meromorphy, we use the ratio of the Jacobian theta functions, as first used in \refcite{Haldane_1985_LJ_function}. 
As a result, we choose
\eqa{
v_\alpha(\bsl k)
&=\eta_\alpha
\exp\left[\frac{\pi}{2\Omega}
\left(\delta_\alpha^2\frac{\bar a_2}{a_2}
-|\delta_\alpha|^2\right)\right] e^{-\ii\bsl k\cdot\bsl\delta_\alpha} g_{\bsl{k}}(\delta_\alpha)\ ,
\label{eq:main-v}
}
where $g_{\bsl{k}}(z) = \vartheta\genfrac{[}{]}{0pt}{}
{k_2}
{-k_1}
\left(\frac{z}{a_2}\middle|\frac{a_1}{a_2}\right)$ with $k_i = \bsl a_i\cdot\bsl k/(2\pi)$ and $\vartheta\genfrac{[}{]}{0pt}{}
{a}
{b}
\left(z|\tau\right)$ is the Jacobi theta function with characteristics $a,b\in\dsC$, complex argument $z\in\dsC$, and
modular parameter $\tau\in\dsC$ with
$\im(\tau)>0$ \cite{Kharchev_2015}.
For fixed $\bsl k$, the theta function has one simple zero per primitive cell. It cannot vanish at both inequivalent sublattice positions,
so $S(\bsl k)>0$.
More precisely, the Jacobi theta-function series converges uniformly in $\bsl k$ for $\im(a_1/a_2)>0$, so $S(\bsl k)$ is finite and
continuous throughout the BZ. Since the BZ is compact and
$S(\bsl k)>0$ everywhere, $S(\bsl k)$ has a strictly positive minimum.
The chosen form also guarantees that $v_B(\bsl{k})/v_A(\bsl{k})$ is a meromorphic in
$k=k_x-\ii k_y$ with one simple zero and one simple pole in the BZ ( \cref{sec:C-band-containment}).

With the chosen form of $v(\bsl{k})$, the construction of the $H_0$ is straightforward.
Since $S(\bsl{k})=v^\dagger(\bsl{k}) v(\bsl{k})$ is nonzero and finite everywhere, then a Bloch Hamiltonian that naturally realizes the lower band as the Chern-ideal band reads
\eq{
H_0=\sum_{\bsl{k}} c^\dagger_{\bsl{k}}h_0(\bsl k)c_{\bsl{k}} \text{ with } h_0(\bsl k)=S(\bsl k)I_2-v(\bsl k)v^\dagger(\bsl k),
\label{eq:main-h0}
}
where $c^\dagger_{\bsl{k}}=(c^\dagger_{\bsl{k},A},c^\dagger_{\bsl{k},B})$ and $I_2$ is the $2\times2$ identity matrix. The upper band has dispersion $S(\bsl{k}) > 0$, and the lower band is an exactly-flat Chern-ideal band at zero energy with $\Ch=1$.
Fourier transforming \cref{eq:main-h0}, we obtain  \cref{eq:main-hopping-matrix}. 
The Gaussian-decaying tight-binding model for Chern-ideal bands was constructed in \refcite{Li_2026_Lattice_GLL} on honeycomb structure. 
Nevertheless, our construction provides such models for any lattice with two inequivalent sublattice per unit cell.
(See details in \cref{sec:SP-hamiltonian}.)

\section{Many-Body Ground States as FCI States}
Before explaining the detailed form of $H_{\text{int}}$, we first explain why we choose $H_{\text{int}}$ to be positive semidefinite and construct it from pair annihilation operators.
Since our constructed $H_0$ has a strictly positive upper band and a exactly flat lower band at zero, any states that have nonzero occupation in the upper band will have a positive average value for $H_0$.
Therefore, as long as $H_{\text{int}}$ is positive semi-definite, every zero-energy state of the total $H$ must only have occupation in the lower band.
The pair annihilation form of $H_{\text{int}}$ further implies that for any $\ket{\psi}$ in the lower-band Fock space, $H\ket{\psi}=0\Leftrightarrow H_{\text{int}}\ket{\psi}=0$ is equivalent to $B_{\bsl{r}}^{(p)}\ket{\psi}=0$ for any $\bsl{r}$ and any $p$. 
In other words, when such $\ket{\psi}$ exists, $H$ is frustration-free.
As $B_{\bsl{r}}^{(p)}$ is a pair-annihilation, it is further equivalent to $\widetilde{B}_{\bsl{r}}^{(p)}\ket{\psi}=0$ for all allowed $\bsl{r}$ and $p$, where $\widetilde{B}_{\bsl{r}}^{(p)}$ is the projection of $B_{\bsl{r}}^{(p)}$ to the lower band.
Combined with the fact that the projection of $H_{\text{int}}$ is built from $\widetilde{B}_{\bsl{r}}^{(p)}$, we know 
\eq{
\label{eq:no_need_for_projection}
\ker{H} = \ker{\widetilde{H}}
}
for any electron number $N_e$, where $\widetilde{H}$ is the projection of the total Hamiltonian $H$ in the lower band.

With \cref{eq:no_need_for_projection} established, we want to choose the form of the pair-annihilation operator $B_{\bsl{r}}^{(p)}$ such that the FCI states become the exact ground states of $\widetilde{H}$ at filling $1/3$ at any finite size with any twist boundary condition.
To be specific, let us consider $N_e\geq2$ electrons on a lattice with $N=N_1N_2$ unit cells.
We choose the twisted boundary condition  $c_{\bsl r+\bsl L_j}=e^{\ii\phi_j}c_{\bsl r}$ with $\bsl L_j=N_j\bsl a_j$, meaning that the lattice is effectively on torus and $\phi_j=0$ is the periodic boundary condition.
With this condition, the site $\bsl r$ in \cref{eq:main-pair-operators} runs over only the $2N$ physical sites, while the  site $\bsl r'$ in \cref{eq:main-pair-operators} still runs over the infinite lattice with the corresponding twist phases included. (See \cref{app:E_2} for details.)

To project the Hamiltonian to the lower band, we first project the pair annihilation operator $B^{(p)}_{\bsl{r}}$:
\eq{
B^{(p)}_{\bsl{r}} \rightarrow \widetilde{B}^{(p)}_{\bsl{r}} = 
\sum_{\bsl{k},\bsl{k}'}\sum_{\bsl r'\ne\bsl r}
\frac{t_{\bsl r\bsl r'}}{(z_{\bsl r'}-z_{\bsl r})^p} \psi_{\bsl{k}'}(\bsl{r}')\psi_{\bsl{k}}(\bsl{r})\gamma_{\bsl{k}'}\gamma_{\bsl{k}}\ ,
}
where $\psi_{\bsl{k}}(\bsl{R}+\bsl{\delta}_\alpha)=e^{\ii \bsl{k}\cdot(\bsl{R}+\bsl{\delta}_\alpha)}v_\alpha(\bsl{k})/\sqrt{N S(\bsl{k})}$ and $\gamma_{\bsl{k}}$ are the Bloch wavefunction and annihilation operator for the lower band, respectively.
Then, the equivalent condition on $\ket{\psi}$ being ground state of the projected Hamiltonian $\widetilde{H}=\widetilde{H}_{\text{int}}$ is $\widetilde{H}_{\text{int}}\ket{\psi}=0$, implying $\widetilde{B}^{(p)}_{\bsl{r}} \ket{\psi} = 0$ for any $\bsl{r}$ and $p=1,2$.
We now illustrate that this condition is a lattice analogous of the zero-mode condition of the Haldane $V_1$ pseudo potential.
First, the zero energy of the lower band and the theta-function nature of $g_{\bsl{k}}$ lead to a powerful observation that $\sum_{\bsl r'\ne\bsl r}\frac{t_{\bsl r\bsl r'}}{(z_{\bsl r'}-z_{\bsl r})^p}\psi_{\bsl{k}'}(\bsl{r}')$ results in a linear combination of $\partial_{z_{\bsl{r}}}^n g_{\bsl{k}'}(z_{\bsl{r}})$ with $n=0,1,\ldots,p$ for $p=1,2$ ( \cref{sec:E-simple-projection}).
Therefore, $B_{\mathbf R+\boldsymbol\delta_\alpha}^{(p)}$ involve $g_{\bsl{k}}$ up to its $p$th-order derivatives.
Combined with the anti-commutation relation of $\gamma_{\bsl{k}}$, we obtain
\eqa{
& \widetilde B_{\mathbf R+\boldsymbol\delta_\alpha}^{(1)}
\propto\sum_{\bsl{K}}e^{\ii\bsl{K}\cdot\mathbf R}
Q_{\bsl{K}}(\delta_\alpha)\\
& \widetilde B_{\mathbf R+\boldsymbol\delta_\alpha}^{(2)}
\propto\sum_{\bsl{K}}e^{\ii\bsl{K}\cdot\mathbf R}Q_{\bsl{K}}'(\delta_\alpha)
}
where  $Q_{\bsl{K}}'(\zeta)=\partial_{\zeta}Q_{\bsl{K}}(\zeta)$ and  
\eq{
Q_{\bsl{K}}(\zeta)
=
\sum_{\mathbf k,\mathbf k'}^{\mathbf k+\mathbf k' = \bsl{K}}
\frac{\partial_{w}\left[g_{\mathbf k}(\zeta+w/2)g_{\mathbf k'}(\zeta-w/2)\right]_{w=0}}
{N\sqrt{S(\mathbf k)S(\mathbf k')}}
\gamma_{\mathbf k'}\gamma_{\mathbf k}\ .
}
It means that the zero-mode condition is equivalent to $Q_{\bsl{K}}(\delta_\alpha)\ket{\psi}=0$ and $Q_{\bsl{K}}'(\delta_\alpha)\ket{\psi}=0$ for $\alpha=A,B$ and all $\bsl{K}$, which suggests that $Q_{\bsl{K}}(\zeta)\ket{\psi}$ has two zeros---each of them has multiplicity at least two---in the unit cell unless it is identically zero for all $\zeta\in\dsC$.
However, the boundary condition of the theta function indicates that
a nonzero degree-two theta function can have only two zeros per unit cell, counting multiplicity. Therefore, the zero-mode condition now becomes that $Q_{\bsl{K}}(\zeta) \ket{\psi} = 0$ for any pair momentum $\bsl{K}$ and $\zeta\in \dsC$.

The condition $Q_{\bsl{K}}(\zeta) \ket{\psi} = 0$ is the lattice analogue of the ground state condition of the Haldane $V_1$ pseudopotential on torus.
Recall that the Haldane $V_1$ pseudopotential penalizes the term linear in the relative coordinate of any pair of electrons, and thus its zero modes on a torus must have a holomorphic many-body factor that vanishes at least cubically whenever two particles coincide, making the Laughlin states naturally the zero modes at filling $1/3$. 
Here, $Q_{\bsl{K}}(\zeta)$ does the same thing.
In our construction, a general many-body state within the Fock space of the Chern-ideal band must have an antisymmetric holomorphic factor from products of $g_{\bsl{k}}(z)$.
When $Q_{\bsl{K}}(\zeta)$ acts on $\ket{\psi}$, it effectively takes the first derivative with respect to the relative coordinate of each pair of electrons with pair momentum $\bsl{K}$  at their collision.
Therefore, the condition $Q_{\bsl{K}}(\zeta)\ket{\psi} = 0$ for all $\bsl{K}$ and all $\zeta\in\dsC$ eliminates all terms linear in the relative coordinate of any pair of electron, just like Haldane $V_1$ potential.
As a result, the many-body wavefunction $\ket{\psi}$ must have a holomorphic factor that vanishes at least cubically in the relative coordinate when two electrons collide.
Then, combined with the twisted boundary conditions, we obtain exactly three ground states at filling $1/3$, labeled by $\sigma=1,2,3$, that have the following many-body wavefunctions
\eq{
\Psi^{3,\sigma}(\{\bsl r_i\})=
F^\sigma(\{z_i\})
\prod_{i=1}^{N_e}
e^{\pi\mathcal B(\bsl r_i)/(2\Omega)}\chi(\bsl r_i),
\label{eq:main-laughlin-amplitude}
}
where $N_e=N/3$ is the electron number, and
\eq{
F^\sigma(\{z_i\})
=e^{\ii K_\sigma Z}
\prod_{\gamma=1}^{3}
\vartheta_1\left(\frac{Z-Z_\sigma^\gamma}{L_2}\middle|\tau\right)
\prod_{i<j}
\vartheta_1\left(\frac{z_i-z_j}{L_2}\middle|\tau\right)^3
\label{eq:main-holomorphic-factor}
}
is the holomorphic factor that comes from the products
of $g_{\bsl k}(z_i)$, where $z_i=z_{\bsl r_i}$,
$Z=\sum_i z_i$, $L_j=N_ja_j$, $\tau=L_1/L_2$.
Here, $\vartheta_1(\omega|\tau)=
-\vartheta\genfrac{[}{]}{0pt}{}{1/2}{1/2}(\omega|\tau)$
with $\omega\in\dsC$ is the odd Jacobi theta function
\cite{Whittaker_Watson_1996}. 
$K_\sigma$ and $Z_\sigma^\gamma$ are the center-of-mass parameters that satisfy $K_\sigma=K_1+2\pi(\sigma-1)/L_2$ and $Z_\sigma^\gamma=(Z_1+\gamma-1)L_2/3-(\sigma-1)L_1/3$, where $K_1$ and $Z_1$ are fixed by the twisted boundary conditions $\phi_1$ and $\phi_2$ ( \cref{sec:B-continuum-review}).
The single-particle factor $e^{\pi\mathcal B(\bsl{r})/(2\Omega)}\chi(\bsl{r})$ is determined by $\eta_\alpha$ via $\eta_\alpha=\chi(\bsl\delta_\alpha)e^{-\pi\mathcal K(\bsl\delta_\alpha)/(2\Omega)}$, $\mathcal B(\bsl r)=(\bar a_2/a_2)z_{\bsl r}^2-|z_{\bsl r}|^2-\mathcal K(\bsl r)$, $\mathcal K(\bsl r+\bsl a_j)=\mathcal K(\bsl r)$ and $\chi(\bsl r+\bsl a_j)=\chi(\bsl r)e^{\ii\pi\bsl a_j\times\bsl r/\Omega}$, for $j=1,2$.
The three constructed states are linearly independent (\cref{sec:C-linear-independence}), and are the only ground states of $\widetilde{H}$ at any finite sizes and twist boundary conditions (\cref{app:E-real-space-parent}).
Combined with \cref{eq:no_need_for_projection},  the three states are therefore the only ground states of $H$ at filling $1/3$ and must have a nonzero energy gap above them for any finite system size and arbitrary twist boundary conditions, since the Hilbert space at fixed electron number is finite-dimensional.

The holomorphic factor of \cref{eq:main-laughlin-amplitude} is just discretized version of the continuum ideal-band
fractional Chern many-body wavefunction in
\refcite{iyer2026dispersionanyonblochbands}
(\cref{sec:B-continuum-review}).
The difference is that $\chi(\bsl r)$ in the single-particle wavefunction here is a scalar as it is defined on discrete sites, while that in the continuum case must be a spinor owing to the translation property.
However, the connection to the fractional Chern states in the continuum does not directly establish the fractional Chern nature of the wavefunction in \cref{eq:main-laughlin-amplitude}.
To do so, we directly evaluate the total many-body Chern number of the three states with respect to the angles $\phi_1$ and $\phi_2$ for the twist boundary condition.
As a result, we obtained the total many-body Chern number as $1$, verifying the FCI nature as it implies a $e^2/(3h)$ Hall conductance of each state with $-e$ the electron charge. (See \cref{sec:C-many-body-chern} for details.)

\section{Quasihole Excitations}
With the FCI wavefunctions for $N=3N_e$, the quasihole wavefunction can be naturally constructed following the continuum constructions of \refcite{Haldane_1985_LJ_function,iyer2026dispersionanyonblochbands}.
Explicitly, the construction is for one more unit cell, \ie, $N=3N_e+1$, and is done by changing the holomorphic factor $F^\sigma$ in \cref{eq:main-laughlin-amplitude} based on the complex position of the quasihole, $w_{qh}$.
Explicitly, we have
\eq{
\Psi_{\qh}^{3,\sigma}(\{\bsl r_i\};\wh)
=F_{\qh}^{\sigma}(\{z_i\};\wh)
\prod_{i=1}^{N_e}
e^{\pi\mathcal B(\bsl r_i)/(2\Omega)}\chi(\bsl r_i)
\label{eq:main-qh-amplitude}
}
with $\sigma=1,2,3$ and
\eqa{
&F_{\qh}^{\sigma}(\{z_i\};\wh)
=F^{\sigma}\left(\left\{z_i+\frac{w_{qh}}{3N_e}\right\}\right)\prod_{i=1}^{N_e}
\vartheta_1\left(\frac{z_i-\wh}{L_2}\middle|\tau\right)\ .
\label{eq:main-qh-holomorphic}
}
Here the quasihole position $\wh$ does not need to be on a lattice site. 
Again, \cref{eq:main-qh-amplitude} is like a discretized version of the continuum quasihole wavefunction in \refcite{iyer2026dispersionanyonblochbands}, except that $\bsl r_i\in\Lambda$ are discrete lattice sites and $\chi(\bsl r_i)$ is a scalar rather than a spinor.
Notice that, compared with \cref{eq:main-holomorphic-factor}, we added a
product which introduces one zero at $z_i=\wh$ for each electron coordinate, 
which is in the same spirit as the quasihole expression in the Laughlin state \cite{Haldane_1985_LJ_function}. To keep the same twisted boundary conditions, we shifted the arguments of $F^{\sigma}$ by $\wh/(3N_e)$ (see \cref{sec:qh-der-construction}). 
With this construction, the wavefunctions lie entirely in the lower-band Fock space.

Again, establishing a connection to the continuum wave function does not, by itself, demonstrate the quasihole nature of the state. 
The fractional charge of the excitation is straightforward: compared with the background $1/3$ filling, the excess charge now is $-e(N_e-N/3)=+e/3$ for $N=3N_e+1$.
The localized nature of the excitation is checked numerically: up to a $16\times 16$ lattice, the $e/3$ charge of the excitation is mainly ($\sim 1$\% error) localized within $3|\bsl{a}_1|$ away from the $\wh$.  (See \cref{sec:qh-der-construction}).

We also showed that these one-quasihole states have exactly zero energy under the projected Hamiltonian $\widetilde H=\widetilde H_0+\widetilde H_{\mathrm{int}}$ and thus also under $H$ owing to \cref{eq:no_need_for_projection}.
Recall that a state is a zero mode of $\widetilde H$ if and only if its holomorphic factor contains no term linear in the relative coordinate near the collision of any pair of electrons.
Since the one-quasihole wavefunction has the same cubic Jastrow factors in \cref{eq:main-qh-holomorphic} as the FCI wavefunction, the quasihole holomorphic factor vanishes at least cubically in the relative displacement when any pair of electrons collide.
This implies that the projected pair-annihilation operators annihilate the quasihole states
\eq{
\widetilde B_{\bsl r}^{(p)}
\ket{\Psi_{\qh}^{3,\sigma}}=0,
\quad \bsl r\in\Lambda,\quad p=1,2,
\label{eq:main-qh-pair-zero}
}
implying
\eq{
\widetilde H\ket{\Psi_{\qh}^{3,\sigma}(\wh)}=0
\label{eq:main-qh-zero-energy}
}
for $\sigma=1,2,3$ and any quasihole position $\wh$. (See \cref{app:zero_energy_quasihole} for details.)
Therefore, the one-quasihole wavefunction has
exactly zero energy under the full unprotected Hamiltonian $H$, similar to that of the Laughlin states under Haldane $V_1$ pseudopotential \cite{Haldane1983FQH,Trugman1985ExactFQH}.

\begin{figure}[t]
\centering
\includegraphics[width=\linewidth]{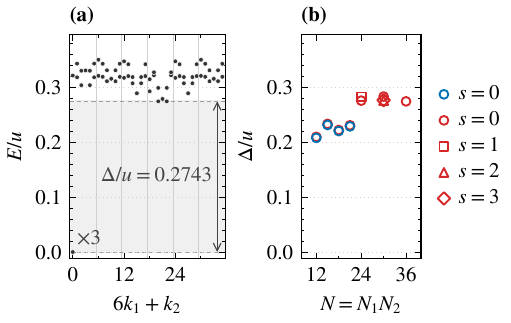}
\caption{In (a), we show the many-body spectrum of the  projected Hamiltonian $\widetilde{H}$ from the ED calculation on a $6\times 6$ $s=0$ mesh with $N_e=12$ fermions. 
Here $s$ is an integer that determines $\bsl{a}_1=(-s,-1)$.
Specifically, we plot $E/u$ ($\ie$, the energy $E$ in units of the interaction strength $u$) as a function of a linearized momentum index, which maps a given total-momentum sector labelled by the integers $(k_1, k_2)$ to a scalar $6 k_1 + k_2$. We find the numerical value of the gap to be approximately $\Delta / u = 0.2743$. The ``$\times3$'' indicates the threefold degeneracy of the ground-state sector. 
For (b), we plot the gap $\Delta/u$ as a function of the number of unit cells $N$ from the ED calculations on the full unprojected Hamiltonian $H$ (blue marker color) up to $N=21$ and the projected Hamiltonian $\widetilde{H}$ (red marker color) up to $N=36$.
Detailed information of the meshes can be found in \cref{sec:E-numerical-checks}.
}
\label{fig:ed-spectrum-6x6}
\label{fig:ed-gap-vs-N}
\end{figure}

\section{Numerical Evidence for Possible Thermodynamic Gap at  filling $1/3$}
Although we have shown that the Hamiltonian $H$ has a nonzero gap at every finite size ($\ie$, finite $N_1$ and $N_2$), it is not sufficient to demonstrate the gap in the thermodynamic limit.
We have not obtained the rigorous proof of the the thermodynamic gap, but in this section, we provide numerical evidence for its possibility, when interaction strength $u$ is $1/500$ as large as the single-particle band gap.
For the ED calculation, we choose $N_e=N/3$, $\bsl{a}_1 = (-s,-1)$ (with integer $s$), $\eta_{A}=\eta_B=1$ and $\phi_1 = \phi_2 =0$, while keeping the choice of $\bsl{a}_2$, $\bsl{\delta}_A$ and $\bsl{\delta}_B$ as those chosen for \cref{eq:main-real-space-h0}.
We focus on the many-body gap $\Delta$ between the ground states and the next-lowest energy eigenvalue as illustrated in \cref{fig:ed-gap-vs-N}(a).
As shown in \cref{fig:ed-gap-vs-N}(b), the gaps of projected and unprojected Hamiltonians align well when they are calculated on the same meshes, which is expected for the small interaction strength compared to the single-particle band gap. 
We believe such alignment holds well for $N>21$, and thus believe the gap obtained in the projected Hamiltonian is a faithful indication of the gap of the full unprojected Hamiltonian in this case of small interaction.
Particularly for the gap of the projected Hamiltonian, the difference between the minimum and maximum values of $\Delta/u$ across all six meshes for $N=24,30,36$ is $0.28305-0.27426\approx0.0088$, which is only $3.2\%$ of the mean value.
The stable values of the gap suggest the possibility for the model to remain gapped in the thermodynamic limit at filling $1/3$ when interaction is much smaller than the single-particle band gap. 
For more details about the ED calculation and results, see \cref{sec:E-numerical-checks}.

\section{Conclusions and Discussions} 

We have constructed an Gaussian-local two-band tight-binding
model with one-body and two-body terms whose lower-band projection
has an exactly FCI ground-state space at $\nu=1/3$.
The FCI has finite gap on all finite sizes, and we provide numerical evidence for the possibility that the gap remains nonzero in the thermodynamic limit when fixing the filling and when interaction strength is much smaller than the band gap. 
We also construct one-quasihole states at $N=3N_e+1$ and show that they have exactly zero energy for any quasihole position.
Establishing whether the gap
remains nonzero in the thermodynamic limit is an important
next step. 
Recent proofs of a nonzero thermodynamic gap for the continuum Laughlin state under the Haldane $V_1$ pseudopotential on the sphere
\cite{OpenAI2026LaughlinGap,gao2026improvinglowerboundlaughlin,scheer2026hierarchyexactlowerbounds} motivate exploring whether similar methods can be applied to our model.
The quasi-electron dispersion is interesting to explore: as we do not have magnetic translation symmetry, it is interesting to see whether quasi-electron excitations naturally develop dispersion.

\emph{Note added:}
During the final stage of our draft, we become aware of \refcite{Jonah2026_to_appear}, which constructed analytical parent lattice Hamiltonian for FCI states by sampling Dirac Landau levels.
We are also aware of \refcite{Wang_2026_to_appear}, which constructed a fractional Chern semimetal exhibiting a conformal criticality.

\section{Acknowledgments}
We thank Chen Fang, Jonah Herzog-Arbeitman and Zhi-Da Song for helpful discussions and coordinating the submission.
This work was supported by J.Y.’s startup funds from the University of Florida. 
A.A.A. is supported by the AI Scholars program at the University of Florida. The authors received assistance from OpenAI ChatGPT (GPT-6 Sol/Astra) and Claude Code (Claude Opus 5.5) for analytical construction, manuscript polishing, and numerical code development. The authors independently verified all outputs and retain full responsibility for this work.

\bibliography{bibfile_references_canonical,bib2.bib}

\onecolumngrid
\clearpage

\setcounter{page}{1}

\appendix

\crefalias{section}{appendix}
\crefalias{subsection}{appendix}
\crefalias{subsubsection}{appendix}

\section*{Supplemental Materials for Analytical Lattice Model for Fractional Chern Insulators}

\tableofcontents

\section{Useful Mathematical Theorems}
\label{app:A-math}

In this section, we review the relevant properties of Jacobi theta functions \cite{Haldane_1985_LJ_function,Kharchev_2015,Slavnov_2020,Bonelli_2021,kulkarni2023formfactorlocaloperators,Li_2026_Lattice_GLL}. We then construct the basis functions $g_{\bsl{k}}(z)$ and prove their linear independence and completeness under twisted boundary conditions.
This section provides the mathematical foundation for the
wavefunction and Hamiltonian constructions in
\cref{sec:B-continuum-review,app:C-lattice-states,sec:app_D,app:E-real-space-parent,sec:qh-der-construction}.

\begin{definition}
For $a,b,z,\tau\in\dsC$ with $\im(\tau)>0$, the Jacobi theta
function with characteristics $a$ and $b$ is defined by
\eq{
\vartheta
\begin{bmatrix}
a\\ b
\end{bmatrix}
(z|\tau)
\equiv
\sum_{n\in\dsZ}
e^{\ii\pi\tau(n+a)^2
+2\pi\ii(n+a)(z+b)}
. 
\label{eq:theta-with-characteristics}
}
\end{definition}
The definition implies the following transformation identities:
\begin{align}
\vartheta
\begin{bmatrix}
a\\ b
\end{bmatrix}
(z+1|\tau)
&=
e^{2\pi\ii a}
\vartheta
\begin{bmatrix}
a\\ b
\end{bmatrix}
(z|\tau),
\label{eq:theta_property_1}
\\
\vartheta
\begin{bmatrix}
a\\ b
\end{bmatrix}
(z+\tau|\tau)
&=
e^{-\ii\pi\tau-2\pi\ii(z+b)}
\vartheta
\begin{bmatrix}
a\\ b
\end{bmatrix}
(z|\tau),
\label{eq:theta_property_2}
\\
\vartheta
\begin{bmatrix}
a+c\\ b
\end{bmatrix}
(z|\tau)
&=
e^{\ii\pi\tau c^2+2\pi\ii c(z+b)}
\vartheta
\begin{bmatrix}
a\\ b
\end{bmatrix}
(z+c\tau|\tau),
\label{eq:theta_property_3}
\\
\vartheta
\begin{bmatrix}
a\\ b
\end{bmatrix}
(z+b'|\tau)
&=
\vartheta
\begin{bmatrix}
a\\ b+b'
\end{bmatrix}
(z|\tau)\ ,
\label{eq:theta_property_4}
\\
\vartheta
\begin{bmatrix}
a+1\\ b
\end{bmatrix}
(z|\tau)
&=
\vartheta
\begin{bmatrix}
a\\ b
\end{bmatrix}
(z|\tau)
\ ,
\label{eq:theta_property_5}
\end{align}
where $c,b'\in\dsC$.
The odd Jacobi theta function $\vartheta_1$ is related to the theta function by
\eq{
\vartheta_1(u|\tau)
=
-\vartheta
\begin{bmatrix}
1/2\\ 1/2
\end{bmatrix}
(u|\tau)
\label{eq:direct-theta1-series}
}
with $u\in\dsC$.
Using the transformation identities above, we can obtain
\begin{align}
\vartheta_1(u+1|\tau)
&=
-\vartheta_1(u|\tau),
\label{eq:direct-theta1-shift-one}
\\
\vartheta_1(u+\tau|\tau)
&=
-e^{-\ii\pi\tau-2\pi \ii u}
\vartheta_1(u|\tau),
\label{eq:direct-theta1-shift-tau}
\\
\vartheta_1(-u|\tau)
&=
-\vartheta_1(u|\tau),
\label{eq:theta1_-u}
\\
\vartheta_1(0|\tau)
&=0.
\label{eq:direct-theta1-oddness}
\end{align}
Conversely, a theta function can be
expressed in terms of $\vartheta_1$ as
\eqa{
\vartheta
\begin{bmatrix}
a\\ b
\end{bmatrix}
(z|\tau)
={}&
- e^{\ii\pi\tau\left(a-\frac12\right)^2
+2\pi\ii\left(a-\frac12\right)(z+b)}
\vartheta_1\left(
z+(a-\frac12)\tau+b-\frac12
\middle|\tau \right).
\label{eq:converge_relationship_theta}}
We now explicitly verify
\cref{eq:theta_property_1,eq:theta_property_2,eq:theta_property_3,eq:theta_property_4,eq:theta_property_5},
the relations for $\vartheta_1$ in \cref{eq:direct-theta1-series,eq:direct-theta1-shift-one,eq:direct-theta1-shift-tau,eq:theta1_-u,eq:direct-theta1-oddness},
and \cref{eq:converge_relationship_theta}.

To see \cref{eq:theta_property_1},
\eqa{
\vartheta\begin{bmatrix}a\\b\end{bmatrix}(z+1|\tau)
&=
\sum_{n\in\dsZ}
\exp\left[\ii\pi\tau(n+a)^2
+2\pi\ii(n+a)(z+1+b)
\right]\\
&=
\sum_{n\in\dsZ}
e^{2\pi\ii(n+a)}
\exp\left[
\ii\pi\tau(n+a)^2
+2\pi\ii(n+a)(z+b)
\right]\\
&=
e^{2\pi\ii a}
\sum_{n\in\dsZ}
\exp\left[
\ii\pi\tau(n+a)^2
+2\pi\ii(n+a)(z+b) \right]\\
&=
e^{2\pi\ii a}
\vartheta\begin{bmatrix}a\\b\end{bmatrix}(z|\tau).
\label{eq:theta-characteristic-shift-one-check}
}
To see \cref{eq:theta_property_2},
\eqa{
\vartheta\begin{bmatrix}a\\b\end{bmatrix}(z+\tau|\tau)
&=
\sum_{n\in\dsZ}\exp\left[\ii\pi\tau(n+a)^2
+2\pi\ii(n+a)(z+\tau+b)
\right]\\
&=
\sum_{n\in\dsZ}
\exp\left[\ii\pi\tau\bigl((n+a)^2+2(n+a)\bigr)
+2\pi\ii(n+a)(z+b)
\right]\\
&=
\sum_{n\in\dsZ}
\exp\left[
\ii\pi\tau\bigl((n+a+1)^2-1\bigr)
+2\pi\ii\bigl((n+a+1)-1\bigr)(z+b)\right]\\
&=
e^{-\ii\pi\tau-2\pi\ii(z+b)}
\sum_{n\in\dsZ}
\exp\left[
\ii\pi\tau(n+a+1)^2
+2\pi\ii(n+a+1)(z+b)\right]\\
&=
e^{-\ii\pi\tau-2\pi\ii(z+b)}
\sum_{k\in\dsZ}\exp\left[
\ii\pi\tau(k+a)^2
+2\pi\ii(k+a)(z+b)
\right]
\qquad (k=n+1)
\\
&= e^{-\ii\pi\tau-2\pi\ii(z+b)}
\vartheta\begin{bmatrix}a\\b\end{bmatrix}(z|\tau).
\label{eq:theta-characteristic-shift-tau-check}
}
To see \cref{eq:theta_property_3},
\eqa{
\vartheta\begin{bmatrix}a+c\\b\end{bmatrix}(z|\tau)
&=
\sum_{n\in\dsZ}
\exp\left[
\ii\pi\tau(n+a+c)^2
+2\pi\ii(n+a+c)(z+b)
\right]
\\
&=
\sum_{n\in\dsZ}
\exp\Big[
\ii\pi\tau\bigl((n+a)^2+2c(n+a)+c^2\bigr)
+2\pi\ii(n+a)(z+b)
+2\pi\ii c(z+b)
\Big]
\\
&=
e^{\ii\pi\tau c^2+2\pi\ii c(z+b)}
\sum_{n\in\dsZ}
\exp\left[
\ii\pi\tau(n+a)^2
+2\pi\ii c\tau(n+a)
+2\pi\ii(n+a)(z+b)
\right]
\\
&=
e^{\ii\pi\tau c^2+2\pi\ii c(z+b)}
\sum_{n\in\dsZ}
\exp\left[
\ii\pi\tau(n+a)^2
+2\pi\ii(n+a)(z+c\tau+b)
\right]
\\
&=
e^{\ii\pi\tau c^2+2\pi\ii c(z+b)}
\vartheta\begin{bmatrix}a\\b\end{bmatrix}(z+c\tau|\tau).
\label{eq:theta-characteristic-shift-a-check}
}
To see \cref{eq:theta_property_4},
\eq{
\vartheta\begin{bmatrix}a\\b\end{bmatrix}(z+b'|\tau)=
\sum_{n\in\dsZ}
\exp\left[
\ii\pi\tau(n+a)^2
+2\pi\ii(n+a)\bigl((z+b')+b\bigr)
\right] =
\vartheta\begin{bmatrix}a\\b+b'\end{bmatrix}(z|\tau).
\label{eq:theta-characteristic-shift-b-check}
}
To see \cref{eq:theta_property_5},
\eqa{
\vartheta
\begin{bmatrix}
a+1\\ b
\end{bmatrix}
(z|\tau)
=
\sum_{n\in\dsZ}
e^{\ii\pi\tau(n+1+a)^2
+2\pi\ii(n+1+a)(z+b)}
=
\sum_{m\in\dsZ}
e^{\ii\pi\tau(m+a)^2
+2\pi\ii(m+a)(z+b)}
=
\vartheta
\begin{bmatrix}
a\\ b
\end{bmatrix}
(z|\tau)
}
To see \cref{eq:direct-theta1-series}, using the definition of $\vartheta_1(u|\tau)$ directly,
\eqa{
\vartheta_1(u|\tau)
&\equiv
2\sum_{n=0}^{\infty}
(-1)^n e^{\ii\pi\tau(n+\frac12)^2}
\sin\left((2n+1)\pi u\right) \quad = \quad
2\sum_{n=0}^{\infty}
(-1)^n e^{\ii\pi\tau(n+\frac12)^2}
\frac{
e^{2\pi\ii(n+\frac12)u}
-
e^{-2\pi\ii(n+\frac12)u}
}{2\ii}
\\
&=
-\ii\sum_{n=0}^{\infty}
(-1)^n
e^{\ii\pi\tau(n+\frac12)^2
   +2\pi\ii(n+\frac12)u}
+
\ii\sum_{n=0}^{\infty}
(-1)^n
e^{\ii\pi\tau(n+\frac12)^2
   -2\pi\ii(n+\frac12)u}
\\
&=
-\ii\sum_{n=0}^{\infty}
(-1)^n
e^{\ii\pi\tau(n+\frac12)^2
   +2\pi\ii(n+\frac12)u}
-
\ii\sum_{n=0}^{\infty}
(-1)^{-n-1}
e^{\ii\pi\tau(-n-\frac12)^2
   +2\pi\ii(-n-\frac12)u}
\\
&=
-\ii\sum_{n=0}^{\infty}
(-1)^n
e^{\ii\pi\tau(n+\frac12)^2
   +2\pi\ii(n+\frac12)u}
-
\ii\sum_{k=-\infty}^{-1}
(-1)^k
e^{\ii\pi\tau(k+\frac12)^2
   +2\pi\ii(k+\frac12)u}
\\
&=
-\ii\sum_{n\in\dsZ}
(-1)^n
e^{\ii\pi\tau(n+\frac12)^2
   +2\pi\ii(n+\frac12)u} =
-\sum_{n\in\dsZ}
e^{\ii\pi(n+\frac12)}
e^{\ii\pi\tau(n+\frac12)^2
   +2\pi\ii(n+\frac12)u}
\qquad
(e^{\ii\pi(n+\frac12)}=\ii(-1)^n)
\\
&=
-\sum_{n\in\dsZ}
e^{\ii\pi\tau(n+\frac12)^2
+2\pi\ii(n+\frac12)(u+\frac12)} =
-\vartheta
\begin{bmatrix}\frac12\\\frac12\end{bmatrix}
(u|\tau).
}
To see \cref{eq:direct-theta1-shift-one},
\eqa{
\vartheta_1(u+1|\tau)
&= -\ii\sum_{n\in\dsZ}
(-1)^n
e^{\ii\pi\tau(n+\frac12)^2
   +2\pi\ii(n+\frac12)(u+1)} =
-\ii\sum_{n\in\dsZ}
(-1)^n e^{2\pi\ii(n+\frac12)}
e^{\ii\pi\tau\left(n+\frac12\right)^2
+2\pi\ii\left(n+\frac12\right)u}
\\
&=
-\ii\sum_{n\in\dsZ}
(-1)^n(-1)
e^{\ii\pi\tau\left(n+\frac12\right)^2
+2\pi\ii\left(n+\frac12\right)u} =
-\vartheta_1(u|\tau).
}
To see \cref{eq:direct-theta1-shift-tau},
\eqa{
\vartheta_1(u+\tau|\tau)
&=
-\ii\sum_{n\in\dsZ}
(-1)^n
e^{
\ii\pi\tau\left(n+\frac12\right)^2
+2\pi\ii\left(n+\frac12\right)(u+\tau)
} =
-\ii\sum_{n\in\dsZ}
(-1)^n
e^{
\ii\pi\tau
\left(
\left(n+\frac12\right)^2
+2\left(n+\frac12\right)
\right)
+2\pi\ii\left(n+\frac12\right)u
}
\\
&=
-\ii\sum_{n\in\dsZ}
(-1)^n
e^{
\ii\pi\tau
\left(
\left(n+\frac32\right)^2-1
\right)
+2\pi\ii
\left(
\left(n+\frac32\right)-1
\right)u
} =
-\ii e^{-\ii\pi\tau-2\pi\ii u}
\sum_{n\in\dsZ}
(-1)^n
e^{
\ii\pi\tau\left(n+\frac32\right)^2
+2\pi\ii\left(n+\frac32\right)u
}
\\
&=
-\ii e^{-\ii\pi\tau-2\pi\ii u}
\sum_{k\in\dsZ}
(-1)^{k-1}
e^{
\ii\pi\tau\left(k+\frac12\right)^2
+2\pi\ii\left(k+\frac12\right)u
}
\qquad (k=n+1)
\\
&=
\ii e^{-\ii\pi\tau-2\pi\ii u}
\sum_{k\in\dsZ}
(-1)^k
e^{
\ii\pi\tau\left(k+\frac12\right)^2
+2\pi\ii\left(k+\frac12\right)u
} =
-e^{-\ii\pi\tau-2\pi\ii u}
\vartheta_1(u|\tau).
}
To see \cref{eq:theta1_-u},
\eqa{
\vartheta_1(-u|\tau)
&=
-\ii\sum_{n\in\dsZ}
(-1)^n e^{\ii\pi\tau\left(n+\frac12\right)^2
-2\pi\ii\left(n+\frac12\right)u} =
-\ii\sum_{k\in\dsZ}
(-1)^{-k-1} e^{\ii\pi\tau\left(-k-\frac12\right)^2
-2\pi\ii\left(-k-\frac12\right)u}
\quad (n=-k-1)
\\
&=
-\ii\sum_{k\in\dsZ}
\left[-(-1)^k\right]
e^{\ii\pi\tau\left(k+\frac12\right)^2
+2\pi\ii\left(k+\frac12\right)u} =
\ii\sum_{k\in\dsZ}
(-1)^k e^{\ii\pi\tau\left(k+\frac12\right)^2
+2\pi\ii\left(k+\frac12\right)u} =
-\vartheta_1(u|\tau).
}
To see \cref{eq:direct-theta1-oddness},
\eqa{
\vartheta_1(0|\tau)
&=
-\ii\sum_{n\in\dsZ}
(-1)^n e^{\ii\pi\tau(n+\frac12)^2} =
-\ii\sum_{n=0}^{\infty}
(-1)^n e^{\ii\pi\tau(n+\frac12)^2}
-
\ii\sum_{k=-\infty}^{-1}
(-1)^k e^{\ii\pi\tau(k+\frac12)^2}
\\
&=
-\ii\sum_{n=0}^{\infty}
(-1)^n e^{\ii\pi\tau(n+\frac12)^2}
-
\ii\sum_{n=0}^{\infty}
(-1)^{-n-1}e^{\ii\pi\tau(-n-\frac12)^2}
\qquad (k=-n-1)
\\
&=
-\ii\sum_{n=0}^{\infty}
\left[
(-1)^n+(-1)^{-n-1}
\right]
e^{\ii\pi\tau(n+\frac12)^2}=0.
}
To see \cref{eq:converge_relationship_theta},
\eqa{
\vartheta\begin{bmatrix}a\\b\end{bmatrix}(z|\tau)
&=
\sum_{n\in\dsZ}
e^{\ii\pi\tau(n+a)^2+2\pi\ii(n+a)(z+b)} =
\sum_{n\in\dsZ}
e^{
\ii\pi\tau
\left(
\left(n+\frac12\right)+\left(a-\frac12\right)
\right)^2
+2\pi\ii\left(\left(n+\frac12\right)+\left(a-\frac12\right)\right)(z+b)
}
\\
&=
\sum_{n\in\dsZ}
e^{
\ii\pi\tau\left(n+\frac12\right)^2
+2\pi\ii\tau
\left(n+\frac12\right)\left(a-\frac12\right)
+\ii\pi\tau\left(a-\frac12\right)^2
+2\pi\ii\left(n+\frac12\right)(z+b)
+2\pi\ii\left(a-\frac12\right)(z+b)
}
\\
&=
e^{
\ii\pi\tau\left(a-\frac12\right)^2
+2\pi\ii\left(a-\frac12\right)(z+b)
}
\sum_{n\in\dsZ}
e^{
\ii\pi\tau\left(n+\frac12\right)^2
+2\pi\ii\left(n+\frac12\right)
\left(
z+\left(a-\frac12\right)\tau+b-\frac12+\frac12
\right)
}
\\
&=
e^{
\ii\pi\tau\left(a-\frac12\right)^2
+2\pi\ii\left(a-\frac12\right)(z+b)
}
\vartheta
\begin{bmatrix}\frac12\\\frac12\end{bmatrix}
\left(
z+(a-\frac12)\tau+b-\frac12\middle|\tau
\right)
\\
&=
- e^{ \ii\pi\tau\left(a-\frac12\right)^2
+2\pi\ii\left(a-\frac12\right)(z+b) }
\vartheta_1\left(
z+(a-\frac12)\tau+b-\frac12
\middle|\tau \right).
}

\begin{theorem}
\label{thm:theta_entire}
For any $a,b,z,\tau\in\dsC$ with $\Im(\tau)>0$, the Jacobi theta function $\vartheta\begin{bmatrix}a\\b\end{bmatrix}(z|\tau)$ is holomorphic in $z$ and is not identically zero.
\end{theorem}
\begin{proof}
Write $\tau=\tau_R+\ii\tau_I$ with $\tau_I>0$ and
$a=a_R+\ii a_I$. From the definition in \cref{eq:theta-with-characteristics},
\eq{
\vartheta\begin{bmatrix}a\\b\end{bmatrix}(z|\tau)
=\sum_{n\in\dsZ}
e^{\ii\pi\tau(n+a)^2+2\pi\ii(n+a)(z+b)}.
}
Each summand is holomorphic in $z$. Moreover,
\eq{
\left|e^{\ii\pi\tau(n+a)^2}\right|
=e^{-\pi\tau_I(n+a_R)^2-2\pi\tau_Ra_I(n+a_R)+\pi\tau_Ia_I^2}.
}
On every compact set in $z$, the remaining factor
$e^{2\pi\ii(n+a)(z+b)}$ grows at most exponentially in $|n|$.
The Gaussian decay therefore ensures locally uniform convergence of
the series, so the theta function is holomorphic in $z$.
To show that it is not identically zero, multiply the series by
$e^{-2\pi\ii az}$ and integrate over real $z=x\in[0,1]$:
\eq{
\int_0^1 dx\,e^{-2\pi\ii ax}
\vartheta\begin{bmatrix}a\\b\end{bmatrix}(x|\tau) =\sum_{n\in\dsZ}
e^{\ii\pi\tau(n+a)^2+2\pi\ii(n+a)b}
\int_0^1 dx\,e^{2\pi\ii nx}=e^{\ii\pi\tau a^2+2\pi\ii ab}\neq0.
}
Hence the theta function cannot vanish identically.
\end{proof}

$\vartheta_1$ has exactly one simple zero modulo $\mathbb{Z}+\tau\mathbb{Z}$, as established by the following theorem \cite{Whittaker_Watson_1996}:

\begin{theorem}
\label{thm:boudnary_and_zero}
Let $\tau\in\dsC$ with $\im(\tau)>0$, and let
$f:\dsC\to\dsC$ be a holomorphic function that is not identically zero satisfying
\eq{
f(z+1)=-f(z),
\qquad
f(z+\tau)
=
-e^{-\ii\pi\tau-2\pi\ii z}f(z).
}
Then $f$ has exactly one zero modulo
$\dsZ+\tau\dsZ$, and this zero is simple.
\end{theorem}

\begin{proof}
Because $f$ is a
nonzero holomorphic function, its zeros are isolated.
Choose $a\in\mathbb C$ such that the boundary of the fundamental
parallelogram
$\mathcal P_a
    =
    \left\{
        a+s+t\tau
        \ \middle|\
        0\leq s,t\leq1
    \right\}$
contains no zero of $f$. Such a choice is always possible because the
zeros of $f$ are isolated. Define the meromorphic function
\begin{equation}
    g(z)=\frac{\partial_z f(z)}{f(z)}.
    \label{eq:logarithmic-derivative}
\end{equation}
Owing to the boundary condition, we have
\eq{g(z+1)=\frac{f'(z+1)}{f(z+1)}=\frac{-f'(z)}{-f(z)}=g(z).\label{eq:g-shift-one}}
and
\eq{g(z+\tau)=\frac{f'(z+\tau)}{f(z+\tau)}=\frac{-e^{-\ii\pi\tau-2\pi \ii z}[f'(z)-2\pi \ii f(z)]}{-e^{-\ii\pi\tau-2\pi \ii z}f(z)}=\frac{f'(z)-2\pi \ii f(z)}{f(z)}=\frac{f'(z)}{f(z)}-2\pi \ii=g(z)-2\pi \ii.\label{eq:g-shift-tau}}
With the boundary condition of $g(z)$, we can now simplify the contour integral of $g$ around
$\partial\mathcal P_a$:
\eqa{
    \oint_{\partial\mathcal P_a}g(z) dz
    = &
    \int_a^{a+1}g(z)\,dz
    +\int_{a+1}^{a+1+\tau}g(z) dz
    +
    \int_{a+1+\tau}^{a+\tau}g(z) dz
    +\int_{a+\tau}^{a}g(z) dz \\
    = &
    \int_a^{a+1}g(z)dz
    +
    \int_{a+1+\tau}^{a+\tau}g(z) dz \quad = \quad
    \int_a^{a+1}g(z)dz
    + \int_{a+1}^{a}\left[ g(z)-2\pi \ii \right]dz \\ 
    =& \int_a^{a+1} g(u) du+\int_{a+1}^{a} g(u) du-2\pi i\int_{a+1}^{a}du
    =
    2\pi \ii \ ,
    \\
    \label{eq:four-boundary-integrals}
}
which means $\oint_{\partial\mathcal P_a}
\frac{f'(z)}{f(z)}dz = 2\pi\ii.$ Since $f$ is holomorphic, it has no poles.  By the argument principle, $\frac{1}{2\pi\ii}
\oint_{\partial\mathcal P_a}\frac{f'(z)}{f(z)} dz=1$
there is exactly one zero of $f$ in $\mathcal P_a$, counted with
multiplicity.  Hence, this zero is simple.
\end{proof}

In the following, we introduce the functions $g_{\bsl k}(z)$ and show its properties under lattice translations. We then show that the $N_1N_2$ allowed functions $g_{\bsl k}$ are linearly independent and form a complete basis, so that every holomorphic function satisfying the corresponding twisted boundary conditions has a unique expansion in terms of $g_{\bsl k}$.

\begin{definition}
\label{def:A-holomorphic-basis}
Given two 2D primitive lattice vectors $\bsl a_1,\bsl a_2$ with
$a_j=a_{j,x}-\ii a_{j,y}$ and $\im(a_1/a_2)>0$, we define
\eq{
g_{\bsl k}(z)
=
\vartheta\genfrac{[}{]}{0pt}{}
{\frac{\bsl a_2\cdot\bsl k}{2\pi}}
{-\frac{\bsl a_1\cdot\bsl k}{2\pi}}
\left(
\frac{z}{a_2}\,\middle|\,\frac{a_1}{a_2}
\right)\label{eq:A-gk-definition}
}
for any $\bsl k\in\dsR^2$ and $z\in\dsC$.
When $z$ represents a real-space position $\bsl r=(x,y)$, we use
$z=x-\ii y$.
\end{definition}

\begin{corollary}
For any $\bsl k\in\dsR^2$, $g_{\bsl k}(z)$ is holomorphic and is not
identically zero.
\end{corollary}
\begin{proof}
By \cref{def:A-holomorphic-basis}, $g_\bsl k(z)$ is expressed as a Jacobi theta function with characteristics $\alpha = {\frac{\bsl a_2\cdot\bsl k}{2\pi}}$ and $\beta = -\frac{\bsl a_1\cdot\bsl k}{2\pi}$, evaluated at argument $u = \frac{z}{a_2}$ with parameter $\tau = \frac{a_1}{a_2}$. Since $\text{Im}(\tau) > 0$ and $a_2 \neq 0$, this theta function is holomorphic function and not identically zero by \cref{thm:theta_entire}.    
\end{proof}

\begin{theorem}
\eq{
g_{\bsl k}(z+a_1)
=
e^{\ii\bsl a_1\cdot\bsl k}
e^{-\ii\pi a_1/a_2-2\pi\ii z/a_2}
g_{\bsl k}(z),
\label{eq:gk-a1-translation}
}
\eq{
g_{\bsl k}(z+a_2)
=
e^{\ii\bsl a_2\cdot\bsl k}
g_{\bsl k}(z).
\label{eq:gk-a2-translation}
}
\end{theorem}
\begin{proof}
Using \cref{eq:A-gk-definition,eq:theta_property_1},
\eq{
g_{\bsl k}(z+a_2) =
\vartheta\genfrac{[}{]}{0pt}{}
{\frac{\bsl a_2\cdot\bsl k}{2\pi}}
{-\frac{\bsl a_1\cdot\bsl k}{2\pi}}
\left(
\frac{z}{a_2}+1\,\middle|\,\frac{a_1}{a_2}
\right) =
e^{2\pi\ii\frac{\bsl a_2\cdot\bsl k}{2\pi}}
\vartheta\genfrac{[}{]}{0pt}{}
{\frac{\bsl a_2\cdot\bsl k}{2\pi}}
{-\frac{\bsl a_1\cdot\bsl k}{2\pi}}
\left(
\frac{z}{a_2}\,\middle|\,\frac{a_1}{a_2}
\right)
=
e^{\ii\bsl a_2\cdot\bsl k}g_{\bsl k}(z).
\label{eq:gk-a2-translation-check}
}
Using \cref{eq:theta_property_2},
\eqa{
g_{\bsl k}(z+a_1)
&=
\vartheta\genfrac{[}{]}{0pt}{}
{\frac{\bsl a_2\cdot\bsl k}{2\pi}}
{-\frac{\bsl a_1\cdot\bsl k}{2\pi}}
\left(
\frac{z}{a_2}+\frac{a_1}{a_2}\middle|\frac{a_1}{a_2}
\right) =
e^{-\ii\pi\frac{a_1}{a_2}
-2\pi\ii\left(
\frac{z}{a_2}-\frac{\bsl a_1\cdot\bsl k}{2\pi}
\right)}
\vartheta\genfrac{[}{]}{0pt}{}
{\frac{\bsl a_2\cdot\bsl k}{2\pi}}
{-\frac{\bsl a_1\cdot\bsl k}{2\pi}}
\left(
\frac{z}{a_2}\,\middle|\,\frac{a_1}{a_2}
\right)
\\
&=
e^{-\ii\pi\frac{a_1}{a_2}
-\frac{2\pi\ii z}{a_2}
+\ii\bsl a_1\cdot\bsl k}
g_{\bsl k}(z)
=
e^{\ii\bsl a_1\cdot\bsl k}
e^{-\ii\pi a_1/a_2-2\pi\ii z/a_2}
g_{\bsl k}(z).
\label{eq:gk-a1-translation-check}
}
\end{proof}

\begin{theorem}
For any positive integers $N_1$ and $N_2$,
\eqa{
g_{\bsl k}(z+N_1a_1)
&=
e^{\ii\bsl k\cdot N_1\bsl a_1}
e^{-\ii\pi N_1^2a_1/a_2}
e^{-2\pi\ii N_1z/a_2}
g_{\bsl k}(z),
\\
g_{\bsl k}(z+N_2a_2)
&=
e^{\ii\bsl k\cdot N_2\bsl a_2}
g_{\bsl k}(z).
\label{eq:g_boundary}
}
\end{theorem}
\begin{proof}
The result follows by repeated application of
\cref{eq:gk-a1-translation,eq:gk-a2-translation}.
For the second direction,
\eq{
g_{\bsl k}(z+N_2a_2)
=
\left(e^{\ii\bsl a_2\cdot\bsl k}\right)^{N_2}
g_{\bsl k}(z)
=
e^{\ii\bsl k\cdot N_2\bsl a_2}
g_{\bsl k}(z).
}
For the first direction, \cref{eq:gk-a1-translation} must be
applied at the successively shifted arguments
$z,z+a_1,\ldots,z+(N_1-1)a_1$. Thus
\eqa{
g_{\bsl k}(z+N_1a_1)
&=
\prod_{r=0}^{N_1-1}
\left[
e^{\ii\bsl a_1\cdot\bsl k}
e^{-\ii\pi a_1/a_2}
e^{-2\pi\ii(z+ra_1)/a_2}
\right]
g_{\bsl k}(z)
\\
&=
e^{\ii N_1\bsl a_1\cdot\bsl k}
e^{-\ii\pi N_1a_1/a_2}
e^{-2\pi\ii N_1z/a_2}
e^{-2\pi\ii(a_1/a_2)\sum_{r=0}^{N_1-1}r}
g_{\bsl k}(z)
\\
&=
e^{\ii N_1\bsl a_1\cdot\bsl k}
e^{-\ii\pi N_1a_1/a_2}
e^{-2\pi\ii N_1z/a_2}
e^{-\ii\pi N_1(N_1-1)a_1/a_2}
g_{\bsl k}(z)
\\
&=
e^{\ii\bsl k\cdot N_1\bsl a_1}
e^{-\ii\pi N_1^2a_1/a_2}
e^{-2\pi\ii N_1z/a_2}
g_{\bsl k}(z),
}
where we used $\sum_{r=0}^{N_1-1}r=N_1(N_1-1)/2$.
\end{proof}

\begin{theorem}
Let $\phi_1,\phi_2\in\dsR$ and let the reciprocal vectors $\bsl{b}_1,\bsl{b}_2$ satisfy
$\bsl b_i\cdot\bsl a_j=2\pi\delta_{ij}$.
For any positive integers $N_1$ and $N_2$, the functions
$g_{\bsl k_{\kappa_1,\kappa_2}^{\phi_1,\phi_2}}$ are
$N_1N_2$ linearly independent functions, where
\eq{
\bsl k_{\kappa_1,\kappa_2}^{\phi_1,\phi_2}
=
\left(
\frac{\kappa_1}{N_1}+\frac{\phi_1}{2\pi N_1}
\right)\bsl b_1
+
\left(
\frac{\kappa_2}{N_2}+\frac{\phi_2}{2\pi N_2}
\right)\bsl b_2,\label{eq:A-allowed-momenta}
}
and $\kappa_j=0,1,\ldots,N_j-1$.
\end{theorem}
\begin{proof}
Introduce the two linear operators acting on holomorphic functions $h$,
\eqa{
(\mathcal M_1h)(z)
&=
e^{\ii\pi a_1/a_2+2\pi\ii z/a_2}h(z+a_1),
\\
(\mathcal M_2h)(z)
&=h(z+a_2).
}
Using \cref{eq:gk-a1-translation,eq:gk-a2-translation,eq:A-allowed-momenta}, we obtain
\eq{\mathcal M_1 g_{\bsl k_{\kappa_1,\kappa_2}^{\phi_1,\phi_2}}= e^{\ii(2\pi\kappa_1+\phi_1)/N_1}
g_{\bsl k_{\kappa_1,\kappa_2}^{\phi_1,\phi_2}},\label{eq:gk-M1-eigenvalue}}
\eq{
\mathcal M_2 g_{\bsl k_{\kappa_1,\kappa_2}^{\phi_1,\phi_2}} = e^{\ii(2\pi\kappa_2+\phi_2)/N_2}
g_{\bsl k_{\kappa_1,\kappa_2}^{\phi_1,\phi_2}}.\label{eq:gk-M2-eigenvalue}
}
The operators commute because
\eq{
(\mathcal M_1\mathcal M_2h)(z) =
e^{\ii\pi a_1/a_2+2\pi\ii z/a_2}h(z+a_1+a_2) =
e^{\ii\pi a_1/a_2+2\pi\ii(z+a_2)/a_2}h(z+a_2+a_1) =
(\mathcal M_2\mathcal M_1h)(z),
}
where $e^{2\pi\ii}=1$. We know that eigenvectors of a linear operator with different eigenvalues are linearly independent.
Here, according to \cref{eq:gk-M1-eigenvalue} and \cref{eq:gk-M2-eigenvalue}, the functions
$g_{\bsl k_{\kappa_1,\kappa_2}^{\phi_1,\phi_2}}$
have distinct simultaneous eigenvalue pairs under the operators $\mathcal M_1$ and $\mathcal M_2$.
Since none of the functions $g_{\bsl k_{\kappa_1,\kappa_2}^{\phi_1,\phi_2}}$ is identically zero, they are
linearly independent.
\end{proof}

\begin{theorem}\label{thm:f-gk-completeness}
Let $\phi_1,\phi_2\in\dsR$.
For any positive integers $N_1$ and $N_2$, suppose that an holomorphic
function $f(z)$ satisfies
\eq{f(z+N_1a_1)=
e^{\ii\phi_1}
e^{-\ii\pi N_1^2a_1/a_2}
e^{-2\pi\ii N_1z/a_2}
f(z),\label{eq:f-full-a1-boundary}}
\eq{
f(z+N_2a_2)
=e^{\ii\phi_2}f(z).
\label{eq:f-full-a2-boundary}}
Then there exist constants
$A_{\kappa_1,\kappa_2}\in\dsC$, independent of $z$, such that
\eq{
f(z)
=
\sum_{\kappa_1=0}^{N_1-1}
\sum_{\kappa_2=0}^{N_2-1}
A_{\kappa_1,\kappa_2}
g_{\bsl k_{\kappa_1,\kappa_2}^{\phi_1,\phi_2}}(z).
\label{eq:f-gk-expansion}
}
Moreover, this expansion is unique.
\end{theorem}

\begin{proof}
We will show that the vector space of holomorphic functions satisfying \cref{eq:f-full-a1-boundary} and
\cref{eq:f-full-a2-boundary} has dimension at most $N_1N_2$. Set $x=\frac{z}{N_2a_2}$, and define a new function
\eq{
\widetilde f(x)
=
e^{-\ii\phi_2x}f(N_2a_2x).
\label{eq:f-periodic-gauge}
}
\cref{eq:f-full-a2-boundary} gives that it is periodic:
\eqa{
\tilde{f}(x+1) = e^{-\ii\phi_2(x+1)} f(N_2a_2x+N_2a_2) = e^{-\ii\phi_2(x+1)}e^{\ii\phi_2}f(N_2a_2x) = \tilde{f}(x).}
Therefore, $\widetilde f$ has a Fourier expansion, convergent
locally uniformly in $x$,
\eq{
\widetilde f(x)
=
\sum_{n\in\dsZ}c_ne^{2\pi\ii nx}.
\label{eq:f-Fourier-expansion}
}
\cref{eq:f-full-a1-boundary} implies, when $z$ is translated by $N_1 a_1$,
\eq{
\widetilde f\left(x+\frac{N_1a_1}{N_2a_2}\right)
=
e^{\ii\phi_1}
e^{-\ii\phi_2N_1a_1/(N_2a_2)}
e^{-\ii\pi N_1^2a_1/a_2}
e^{-2\pi\ii N_1N_2x}
\widetilde f(x).
\label{eq:f-tilde-second-boundary}
}
Substituting \cref{eq:f-Fourier-expansion} into both sides of \cref{eq:f-tilde-second-boundary}
gives
\eqa{
\sum_{n\in\dsZ}
c_n e^{2\pi\ii nN_1a_1/(N_2a_2)}e^{2\pi\ii nx}
&=
e^{\ii\phi_1}
e^{-\ii\phi_2N_1a_1/(N_2a_2)}
e^{-\ii\pi N_1^2a_1/a_2}
\sum_{n\in\dsZ}c_ne^{2\pi\ii(n-N_1N_2)x}.
}
Reindexing the sum on the right-hand side gives
$\sum_{n\in\mathbb Z} c_n e^{2\pi\ii(n-N_1N_2)x} =
\sum_{n\in\mathbb Z} c_{n+N_1N_2}e^{2\pi\ii nx}$.
Comparing the coefficients of $e^{2\pi\ii nx}$ on both sides,
we obtain
\eq{c_n e^{2\pi\ii nN_1a_1/(N_2a_2)} =
e^{\ii\phi_1}
e^{-\ii\phi_2N_1a_1/(N_2a_2)}
e^{-\ii\pi N_1^2a_1/a_2}
c_{n+N_1N_2}.}
Rearranging yields
\eq{
c_{n+N_1N_2} =
e^{-\ii\phi_1}
e^{\ii\phi_2N_1a_1/(N_2a_2)}
e^{\ii\pi N_1^2a_1/a_2}
e^{2\pi\ii nN_1a_1/(N_2a_2)}c_n.
\label{eq:f-Fourier-coefficient-recurrence}
}
\cref{eq:f-Fourier-coefficient-recurrence} shows that all Fourier coefficients can be
uniquely determined from just $N_1N_2$ coefficients:
$c_0,c_1,\ldots,c_{N_1N_2-1}$.
Therefore, the space of holomorphic functions $f(z)$
satisfying the boundary conditions
has dimension at most $N_1N_2$.
The $N_1N_2$ functions
$g_{\bsl k_{\kappa_1,\kappa_2}^{\phi_1,\phi_2}}(z)$
belong to this space as they satisfy the same boundary conditions by \cref{eq:g_boundary}, since
$\bsl k_{\kappa_1,\kappa_2}^{\phi_1,\phi_2}\cdot N_j\bsl a_j
=2\pi\kappa_j+\phi_j$, and they are linearly independent.
Therefore, they form a basis. This proves both the existence and
the uniqueness of the expansion in \cref{eq:f-gk-expansion}.
\end{proof}

\begin{theorem}
\label{thm:sum_t_r_p}
Let $\bar{\Lambda}\subset\dsR^2$ be an infinite lattice
with primitive translation vectors $\mathbf a_1,\mathbf a_2$
and finitely many sites per unit cell.
For each lattice site, write
$z_{\mathbf s}=s_x-\mathrm i s_y$.
Fix $\mathbf r\in\bar{\Lambda}$, and let
$M_{\mathbf r\mathbf s}\in\mathbb C$ be independent of
$\mathbf k$. Suppose
\begin{equation}
\sum_{\mathbf s\in\bar{\Lambda}}
M_{\mathbf r\mathbf s}g_{\mathbf k}(z_{\mathbf s})=0
\qquad
\text{for every }\mathbf k\in\dsR^2,
\end{equation}
and suppose that the series
$
\sum_{\mathbf s\in\bar{\Lambda}}
\left|
M_{\mathbf r\mathbf s}g_{\mathbf k}(z_{\mathbf s})
\right|
$
converges uniformly for $\bsl{k}\in\text{BZ}$.
Then, for any nonnegative integer $p$, we have
\begin{equation}
\sum_{\mathbf s\in\bar\Lambda\setminus\{\mathbf r\}}
\frac{M_{\mathbf r\mathbf s}}
{(z_{\mathbf s}-z_{\mathbf r})^p}
g_{\mathbf k}(z_{\mathbf s})
=
-\frac{M_{\mathbf r\mathbf r}}{p!}
g_{\mathbf k}^{(p)}(z_{\mathbf r})+
\sum_{j=0}^{p-1}
\frac{g_{\mathbf k}^{(j)}(z_{\mathbf r})}{j!}
\sum_{\mathbf s\in\bar\Lambda\setminus\{\mathbf r\}}
\frac{M_{\mathbf r\mathbf s}}
{(z_{\mathbf s}-z_{\mathbf r})^{p-j}}
\qquad
\text{for every }\mathbf k\in\dsR^2\ ,
\end{equation}
where $g_{\mathbf k}^{(j)}=\partial_z^j g_{\mathbf k}$, with $g_{\mathbf k}^{(0)}=g_{\mathbf k}$, and all the lattice sums  converge absolutely.
\end{theorem}

\begin{proof}
Define $k_i = \bsl{a}_i\cdot\bsl{k}/(2\pi)$
and $\tau_0=a_1/a_2$.
From the theta-function definition in
\cref{eq:A-gk-definition}, Fourier orthogonality gives
\eq{
\int_0^1 dk_1\,
e^{2\pi\ii(n+k_2)k_1}g_{\bsl k}(z)
=
e^{\ii\pi\tau_0(n+k_2)^2}
e^{2\pi\ii(n+k_2)z/a_2},
\qquad n\in\dsZ.
\label{eq:A-fourier-coefficient-proof}
}
Multiplying the assumed identity
$\sum_{\bsl s}M_{\bsl r\bsl s}g_{\bsl k}(z_{\bsl s})=0$
by $e^{2\pi\ii(n+k_2)k_1}$ and integrating
over $k_1$ yields
\eq{
\sum_{\bsl s\in\bar\Lambda}
M_{\bsl r\bsl s}e^{\lambda z_{\bsl s}}=0,
\qquad
\lambda=\frac{2\pi\ii\xi}{a_2},
\quad \xi\in\dsR.
\label{eq:A-exponential-identity}
}
Here we used uniform absolute convergence to
interchange the sum and integral, and the fact
that every $\xi\in\dsR$ can be written as
$n+k_2$, with $n\in\dsZ$ and $0\le k_2<1$. The Fourier identity and the assumed uniform
absolute convergence also imply that
$\sum_{\bsl s}|M_{\bsl r\bsl s}e^{\lambda z_{\bsl s}}|$
converges uniformly for $\xi$ in bounded intervals.
In particular, setting $\xi=0$ gives
$\sum_{\bsl s}|M_{\bsl r\bsl s}|<\infty$. 
Since distinct lattice sites have a minimum separation
$d>0$, for any nonnegative integer $q$, we have
\eqa{
\sum_{\bsl s\ne\bsl r}
\left|
\frac{M_{\bsl r\bsl s}g_{\bsl k}(z_{\bsl s})}
{(z_{\bsl s}-z_{\bsl r})^q}
\right|
&\leq
\frac{1}{d^q}
\sum_{\bsl s}
|M_{\bsl r\bsl s}g_{\bsl k}(z_{\bsl s})|
<\infty,\\
\sum_{\bsl s\ne\bsl r}
\left|
\frac{M_{\bsl r\bsl s}}
{(z_{\bsl s}-z_{\bsl r})^q}
\right|
&\leq
\frac{1}{d^q}
\sum_{\bsl s}
|M_{\bsl r\bsl s}|
<\infty.
}
Thus, all lattice sums in the theorem converge
absolutely.
For any $p\ge0$, define
\eq{
F_p(\lambda)=
\sum_{\substack{\bsl s\in\bar\Lambda\\
\bsl s\ne\bsl r}}
\frac{M_{\bsl r\bsl s}}
{(z_{\bsl s}-z_{\bsl r})^p}
e^{\lambda(z_{\bsl s}-z_{\bsl r})}.
}
The identity above gives
\eq{
F_0(\lambda)
=e^{-\lambda z_{\bsl r}}
\sum_{\bsl s\in\bar\Lambda}
M_{\bsl r\bsl s}e^{\lambda z_{\bsl s}}
-M_{\bsl r\bsl r}
=-M_{\bsl r\bsl r}.
}
For $p\ge1$, differentiating term by term
with respect to $\lambda$ gives
\eq{
F_p'(\lambda)=\sum_{\bsl s\ne\bsl r}
\frac{M_{\bsl r\bsl s}}
{(z_{\bsl s}-z_{\bsl r})^p}
(z_{\bsl s}-z_{\bsl r})
e^{\lambda(z_{\bsl s}-z_{\bsl r})}=F_{p-1}(\lambda).
}
Termwise differentiation is justified by
the convergence established above and the positive separation of different lattice sites. Repeated  therefore gives
\eq{
F_p^{(j)}(\lambda)=F_{p-j}(\lambda)
\quad (0\le j\le p),
\qquad
F_p^{(p)}(\lambda)=-M_{\bsl r\bsl r}.
}
Since the $p$th derivative is constant,
$F_p$ is a polynomial of degree at most $p$.
Its exact Taylor expansion at $\lambda=0$ is
\eq{
F_p(\lambda)=
\sum_{j=0}^{p-1}
\frac{\lambda^j}{j!}F_{p-j}(0)
-\frac{M_{\bsl r\bsl r}}{p!}\lambda^p.
}
Substituting the definition of $F_p$ into the
previous equation and multiplying both sides by
$e^{\lambda z_{\bsl r}}$, we obtain
\eq{
\sum_{\bsl s\ne\bsl r}
\frac{M_{\bsl r\bsl s}e^{\lambda z_{\bsl s}}}
{(z_{\bsl s}-z_{\bsl r})^p}=
-\frac{M_{\bsl r\bsl r}}{p!}
\lambda^p e^{\lambda z_{\bsl r}}
+\sum_{j=0}^{p-1}
\frac{\lambda^j e^{\lambda z_{\bsl r}}}{j!}
\sum_{\bsl s\ne\bsl r}
\frac{M_{\bsl r\bsl s}}
{(z_{\bsl s}-z_{\bsl r})^{p-j}}.
\label{eq:A-exponential-derivative-identity}
}
We now return to $g_{\bsl k}$.
Recall from the theta-function definition that
\eq{
g_{\bsl k}(z)=\sum_{m\in\dsZ}c_m e^{\lambda_m z},
\qquad
g_{\bsl k}^{(j)}(z)
=\sum_{m\in\dsZ}c_m\lambda_m^j e^{\lambda_m z},
}
where $c_m=e^{\ii\pi\tau_0(m+k_2)^2
-2\pi\ii(m+k_2)k_1}$, $\lambda_m=\frac{2\pi\ii(m+k_2)}{a_2}$. Since \cref{eq:A-exponential-derivative-identity}
holds for every $\lambda_m$, we multiply it
by $c_m$ and sum over $m\in\dsZ$.
This gives
\eqa{
\sum_{\bsl s\ne\bsl r}
\frac{M_{\bsl r\bsl s}g_{\bsl k}(z_{\bsl s})}
{(z_{\bsl s}-z_{\bsl r})^p}
&=\sum_{m\in\dsZ}c_m
\sum_{\bsl s\ne\bsl r}
\frac{M_{\bsl r\bsl s}e^{\lambda_m z_{\bsl s}}}
{(z_{\bsl s}-z_{\bsl r})^p}=\sum_{m\in\dsZ}c_m
\left[
-\frac{M_{\bsl r\bsl r}}{p!}
\lambda_m^p e^{\lambda_m z_{\bsl r}}
+\sum_{j=0}^{p-1}
\frac{\lambda_m^j e^{\lambda_m z_{\bsl r}}}{j!}
\sum_{\bsl s\ne\bsl r}
\frac{M_{\bsl r\bsl s}}
{(z_{\bsl s}-z_{\bsl r})^{p-j}}
\right]\\
&=-\frac{M_{\bsl r\bsl r}}{p!}
g_{\bsl k}^{(p)}(z_{\bsl r})
+\sum_{j=0}^{p-1}
\frac{g_{\bsl k}^{(j)}(z_{\bsl r})}{j!}
\sum_{\bsl s\ne\bsl r}
\frac{M_{\bsl r\bsl s}}
{(z_{\bsl s}-z_{\bsl r})^{p-j}}.
}
This is exactly the claimed identity for $p\ge1$. For $p=0$, the result follows directly
from the initial assumption of the theorem in \cref{eq:A-exponential-identity}.

\end{proof}

\section{Review of Continuum Wavefunctions}
\label{sec:B-continuum-review}

In this section, we briefly review the continuum many-body wavefunctions for fractional Chern states in the absence
of an external magnetic field, following
Ref.~\cite{iyer2026dispersionanyonblochbands}.
These wavefunctions motivate the lattice construction
in \cref{app:C-lattice-states}.

Consider a 2D system with periodic/twist boundary condition in real space specified by two vectors $\bsl L_1,\bsl L_2$.
For a real-space position $\bsl r=(x,y)$, we use the complex
coordinate $z=x-\ii y$, with 
$L_j=L_{j,x}-\ii L_{j,y}$.
The primitive lattice vectors are $\bsl a_j=\bsl L_j/N_j$,
where $N_j$ is the number of unit cells along direction $j$,
with complex representations $a_j=a_{j,x}-\ii a_{j,y}$.
We define the torus parameter $\tau=L_1/L_2$ and take
$\im(a_1/a_2)>0$, so the unit-cell area is
$\Omega=\bsl a_1\times\bsl a_2=\im(a_1\bar a_2)>0$,
and $\im\tau>0$.
In this continuum review, we choose $L_2>0$ real. We consider the filling $\nu=1/m$,
where $m$ is a positive odd integer, which means the total number of unit cells $N$
and the number of electrons $N_e$ satisfy
\eq{
 N=mN_e=N_1N_2.
 \label{eq:B-filling}
}
The continuum many-body wavefunction consists of an antisymmetric holomorphic factor and a product of single-particle factors. For sector $\sigma=1,\ldots,m$, the holomorphic factor is
\eq{
F^\sigma(\{z_i\})
=
A_\sigma e^{\ii K_\sigma Z}
\prod_{\gamma=1}^{m}
\vartheta_1\left(
\frac{Z-Z_\sigma^\gamma}{L_2}\Big|\tau
\right)
\prod_{i<j}^{N_e}
\vartheta_1\left(
\frac{z_i-z_j}{L_2}\Big|\tau
\right)^m,
\qquad Z=\sum_i z_i\ ,
\label{eq:ideal-fci-holomorphic-factor}
}
where $A_\sigma$ fixes the overall normalization of the sector. In addition,
\eq{
K_\sigma=K_1+\frac{2\pi}{L_2}(\sigma-1),
\quad
Z_\sigma^\gamma=
\left(\frac{Z_1}{m}+\frac{\gamma-1}{m}\right)L_2
-\frac{\sigma-1}{m}L_1.
\label{eq:K-Z-sector-definition}
}
Here $K_1$ is the sector-one ($\sigma=1$) parameter and $Z_1$ is its dimensionless
center-of-mass offset. 

Combining the holomorphic factor with the
single-particle factors of a Chern-ideal band with
$\Ch=1$, the full many-body wavefunction reads
\eq{
\Psi^\sigma_{\alpha_1\ldots\alpha_{N_e}}
(\bsl r_1,\ldots,\bsl r_{N_e}) = F^\sigma(\{z_i\}) 
\prod_{i=1}^{N_e}
e^{\pi\mathcal B(\bsl r_i)/(2\Omega)}
\chi_{\alpha_i}(\bsl r_i).
\label{eq:ideal-fci-ground-state}
}
Here $\alpha_i$ labels the component of the same ideal-band spinor $\bsl\chi(\bsl r_i)$ associated with the $i$th particle. In the chosen complex coordinates, the scalar factor is defined by $\mathcal B(\bsl r)=z^2-|z|^2-\mathcal K(\bsl r)$, where $\mathcal K(\bsl r)$ is a real K\"ahler potential periodic under primitive lattice translations, $\mathcal K(\bsl r+\bsl a_j)=\mathcal K(\bsl r)$. We choose its additive constant such that $\int_{\mathrm{UC}}d^2r\,\mathcal K(\bsl r)=0$.
The spinor is normalized and satisfies $\bsl\chi^\dagger(\bsl r)\bsl\chi(\bsl r)=1$ and $\bsl\chi(\bsl r+\bsl a_j)
=e^{\frac{\ii\pi}{\Omega}\bsl a_j\times\bsl r}
\bsl\chi(\bsl r), j=1,2$,
where $\bsl u\times\bsl v=u_xv_y-u_yv_x$.

For a fixed particle coordinate $z_i$, the pair factors $\prod_{i<j}^{N_e}\vartheta_1((z_i-z_j)/L_2\mid\tau)^m$
contribute $m(N_e-1)$ zeros and the center-of-mass factor $\prod_{\gamma=1}^{m}\vartheta_1((Z-Z_\sigma^\gamma)/L_2\mid\tau)$ contributes
$m$ zeros, giving $mN_e=N$ zeros in total. For twisted-boundary phases $\phi_1,\phi_2\in\dsR$,
the theta-function quasiperiodicity constrains $K_1$ and $Z_1$ through
\eqa{
(-1)^{N}e^{\ii K_1L_2}&=e^{\ii\phi_2},\\
(-1)^{N+m-1}e^{\ii K_1L_1}e^{2\pi \ii Z_1}&=e^{\ii\phi_1}.
\label{eq:fci-boundary-constraint}
}
With the above convention, the many-body wavefunction obeys
\eq{
\Psi^\sigma_{\alpha_1\ldots\alpha_{N_e}}
(\ldots,\bsl r_i+\bsl L_j,\ldots)
=e^{\ii\phi_j}
\Psi^\sigma_{\alpha_1\ldots\alpha_{N_e}}
(\ldots,\bsl r_i,\ldots),
\qquad j=1,2.
\label{eq:fci-twisted-boundary-condition}
}
Moreover, in the lattice construction, primitive lattice translations either multiply a many-body wavefunction by an overall phase or map it to a state in a different sector $\sigma$, again up to a phase. Thus, the subspace spanned by these states is invariant under lattice translations, consistent with the absence of an external magnetic field in our construction. These translation properties are explicitly established in \cref{sec:C-boundaries-translations}.

\section{Many-Body Wavefunctions for Fractional Chern States on Lattices}
\label{app:C-lattice-states}

In this section, we construct many-body wavefunctions for fractional Chern states on lattices at filling $1/m$, where $m$ is a positive odd integer.
We show that these wavefunctions can be constructed by discretizing the continuum many-body wavefunctions in \cref{eq:ideal-fci-ground-state}, with the simplification that the spinor $\bsl\chi$ can be replaced by a scalar function $\chi$ on the lattice.
We explicitly verify the twisted boundary conditions and lattice translation properties, calculate the many-body Chern number, and show that the resulting states lie in the Fock space of a Chern-ideal band at filling $1/m$.

\subsection{Discretization of the Continuum Wavefunction on Lattices}
\label{sec:C-setup}

We discretize the continuum many-body wavefunction in
\cref{eq:ideal-fci-ground-state} using the geometry and filling
conventions of \cref{sec:B-continuum-review}.
Unlike the continuum review, we allow $a_2$ to be complex.
We denote the primitive reciprocal lattice vectors by $\bsl b_1$ and
$\bsl b_2$, where $b_i=b_{i,x}-\ii b_{i,y}$ and
$\bsl b_i\cdot\bsl a_j=2\pi\delta_{ij}$.
The standard reciprocal vectors are $\bsl{b}_1
= \frac{2\pi}{\Omega}(a_{2y},-a_{2x})$ and $\bsl{b}_2 = \frac{2\pi}{\Omega}(-a_{1y},a_{1x})$. Now convert to our complex convention:
\eqa{
b_1
&= \frac{2\pi}{\Omega}(a_{2y}+\ii a_{2x})
= \frac{2\pi \ii}{\Omega}(a_{2x}-\ii a_{2y})
= \frac{2\pi \ii a_2}{\Omega},
\\
b_2
&= \frac{2\pi}{\Omega}(-a_{1y}-\ii a_{1x})
= -\frac{2\pi \ii a_1}{\Omega}.
\label{eq:complex-reciprocal-vectors}
}
Consequently, $b_1/b_2=-a_2/a_1$, and hence
$\Im(b_1/b_2)=\Im(a_1/a_2)/|a_1/a_2|^2>0$.

We consider two atoms per unit cell, labeled by $\eta=A,B$.
The set of physical sites in the infinite lattice is
\eq{\overline{\Lambda}
    =
    \left\{
      n_1\bm a_1+n_2\bm a_2+\bm\delta_\eta
      \ \middle|\
      n_j\in\dsZ,\ \eta=A,B
    \right\},
    \label{eq:infinite-lattice-site-set}}
and a fundamental set of sites on the finite torus is
\eq{\Lambda
    =
    \left\{
      n_1\bm a_1+n_2\bm a_2+\bm\delta_\eta
      \ \middle|\
      n_j=0,\ldots,N_j-1,\ \eta=A,B
    \right\},
    \label{eq:finite-torus-site-set}}
where $\bsl\delta_\eta$ denotes the position of sublattice $\eta$
within one unit cell. The two sublattice positions are inequivalent
under Bravais-lattice translations, i.e.,
\eq{
 \bsl\delta_B-\bsl\delta_A\notin\dsZ\bsl a_1+\dsZ\bsl a_2.
 \label{eq:C-orbital-inequivalence}
}
We use $\delta_\eta=\delta_{\eta,x}-\ii\delta_{\eta,y}$
for the complex representative of $\bsl\delta_\eta$. We denote the single-particle site basis by $\ket{\bsl r}$,
with $\braket{\bsl r}{\bsl r'}=\delta_{\bsl r,\bsl r'}$
for $\bsl r,\bsl r'\in\Lambda$, and use $c^\dagger_{\bsl r}$
for the corresponding creation operator.
Under twisted boundary conditions, the site states satisfy
$\ket{\bsl r+\bsl L_j}=e^{-\ii\phi_j}\ket{\bsl r}$,
where $\phi_j\in\dsR$, $j=1,2$, are the boundary phases.

We construct the lattice many-body amplitudes by evaluating the continuum wavefunction at the lattice sites
$\bsl r_i\in\overline\Lambda$:
\eq{\Psi^{m,\sigma}(\{\bm r_i\})
= F^\sigma(\{z_i\})
\prod_{i=1}^{N_e}
e^{\pi\mathcal B(\bm r_i)/(2\Omega)}
\chi(\bm r_i),
\label{eq:C-discretized-amplitude}}
where $\sigma=1,\ldots,m$ labels the different sectors,
and $F^\sigma$ is specified in
\cref{eq:ideal-fci-holomorphic-factor,eq:K-Z-sector-definition},
with $A_\sigma=1$. Since we now allow complex $a_2$, we generalize the single-particle exponent $\mathcal B$ in \cref{sec:B-continuum-review} to
\eq{
\mathcal B(\bsl r)
=
z^2\frac{\bar a_2}{a_2}
-|z|^2-\mathcal K(\bsl r),
\qquad
\mathcal K(\bsl r+\bsl a_j)
=
\mathcal K(\bsl r).
\label{eq:C-B-definition}
}
Unlike the continuum case, $\chi(\bsl r)$ can be chosen as a scalar $U(1)$ function on the discrete lattice, satisfying
\eq{\chi(\bsl r+\bsl a_j)
=
\chi(\bsl r)
\exp\left[
\frac{\ii\pi}{\Omega}
\bsl a_j\times\bsl r
\right],
\qquad j=1,2\ .
\label{eq:C-chi-translation}}
This scalar form is possible because $\chi$ is defined only on discrete lattice sites. The translation rule in \cref{eq:C-chi-translation} is consistent, since translating by $\bsl a_1$ and $\bsl a_2$ in either order gives the same result.

Under particle exchange, the single-particle factors
$\prod_{i=1}^{N_e}e^{\pi\mathcal B(\bsl r_i)/(2\Omega)}
\chi(\bsl r_i)$ are symmetric, while the Jastrow factor
$\prod_{i<j}^{N_e}
\vartheta_1((z_i-z_j)/L_2\mid\tau)^m$
contributes $(-1)^m$:
\eq{
\Psi^{m,\sigma}(\ldots,\bsl r_j,\ldots,\bsl r_i,\ldots)
=(-1)^m\Psi^{m,\sigma}(\ldots,\bsl r_i,\ldots,\bsl r_j,\ldots).
\label{eq:C-fermionic-exchange}
}
Thus odd $m$ gives fermionic states. 
The zeros discussed in \cref{sec:B-continuum-review} now need not lie on physical lattice sites.

Although the wavefunction is defined on the infinite lattice $\overline{\Lambda}$, the physical state requires only its values in the fundamental region $\Lambda$, provided the twisted boundary conditions are satisfied. Thus, up to an overall normalization, the many-body state reads
\eq{\ket{\Psi^{m,\sigma}}
= \frac{1}{\sqrt{N_e!}}
\sum_{\bsl r_1\in\Lambda}\cdots
\sum_{\bsl r_{N_e}\in\Lambda}
\Psi^{m,\sigma}(\{\bm r_i\})
c^\dagger_{\bsl r_1}\cdots
c^\dagger_{\bsl r_{N_e}}
\ket{0}\ .
\label{eq:C-discretized-many-body-state}}

\subsection{Boundary Conditions and Lattice Translations}
\label{sec:C-boundaries-translations}

We now verify the twisted boundary conditions and lattice translation properties of the discrete wavefunction in \cref{eq:C-discretized-amplitude}, following the continuum approach of Ref. \cite{iyer2026dispersionanyonblochbands}.

\subsubsection{Twisted Boundary Conditions}
\label{sec:TBC}
We first check how the single-particle site factor $e^{\pi\mathcal B(\bsl r)/(2\Omega)}\chi(\bsl r)$ in \cref{eq:C-discretized-amplitude} transforms under lattice translations.
Since $\mathcal K$ is periodic, \cref{eq:C-B-definition} gives
\eqa{
\mathcal B(\bsl r+\bsl a_j)-\mathcal B(\bsl r)
&=(z+a_j)^2\frac{\bar a_2}{a_2}
-|z+a_j|^2-\mathcal K(\bsl r+\bsl a_j)
-\left[z^2\frac{\bar a_2}{a_2}-|z|^2-\mathcal K(\bsl r)\right]\\
&=\frac{\bar a_2}{a_2}(2a_jz+a_j^2)
-z\bar a_j-\bar z a_j-|a_j|^2.
\label{eq:B-primitive-shift}
}
Combining this with \cref{eq:C-chi-translation}, and using
$\bsl a_j\times\bsl r=(a_j\bar z-\bar a_jz)/(2\ii)$, we obtain
\eqa{
e^{\pi\mathcal B(\bsl r+\bsl a_j)/(2\Omega)}
\chi(\bsl r+\bsl a_j)
&=e^{\pi\mathcal B(\bsl r)/(2\Omega)}\chi(\bsl r)
e^{\frac{\pi}{2\Omega}
\left\{\mathcal B(\bsl r+\bsl a_j)-\mathcal B(\bsl r)
+a_j\bar z-\bar a_jz\right\}}\\
&=e^{\pi\mathcal B(\bsl r)/(2\Omega)}\chi(\bsl r)
e^{\frac{\pi}{2\Omega}
\left\{\frac{\bar a_2}{a_2}(2a_jz+a_j^2)
-2z\bar a_j-|a_j|^2\right\}}\\
&=\exp\left[\frac{\pi}{2\Omega}
\left(\frac{\bar a_2}{a_2}a_j-\bar a_j\right)(2z+a_j)\right]
e^{\pi\mathcal B(\bsl r)/(2\Omega)}\chi(\bsl r).
\label{eq:combined-band-factor-translation}
}
It is valid for
arbitrary complex $a_2$. In particular, the factor in the exponential part of \cref{eq:combined-band-factor-translation} is
\eqa{
\frac{\bar a_2}{a_2}a_2-\bar a_2&=0,\\
\frac{\bar a_2}{a_2}a_1-\bar a_1
&=\frac{a_1\bar a_2-a_2\bar a_1}{a_2}
=\frac{2\ii\Omega}{a_2}.
\label{eq:C-site-factor-geometric-identities}
}
Thus the site factor is unchanged along $\bsl a_2$, while along
$\bsl a_1$ it acquires $e^{2\pi\ii z/a_2+\ii\pi a_1/a_2}$.
Repeated translation through the full period gives the following for each direction
\eqa{
e^{\pi\mathcal B(\bsl r+\bsl L_1)/(2\Omega)}\chi(\bsl r+\bsl L_1)
&=\prod_{n=0}^{N_1-1}
 e^{2\pi\ii(z+na_1)/a_2+\ii\pi a_1/a_2}
 e^{\pi\mathcal B(\bsl r)/(2\Omega)}\chi(\bsl r)\\
&=e^{2\pi\ii N_1z/a_2+\ii\pi N_1^2a_1/a_2}
 e^{\pi\mathcal B(\bsl r)/(2\Omega)}\chi(\bsl r)\\
&=e^{2\pi\ii Nz/L_2+\ii\pi N\tau}
 e^{\pi\mathcal B(\bsl r)/(2\Omega)}\chi(\bsl r),\\
e^{\pi\mathcal B(\bsl r+\bsl L_2)/(2\Omega)}\chi(\bsl r+\bsl L_2)
&=e^{\pi\mathcal B(\bsl r)/(2\Omega)}\chi(\bsl r),
\label{eq:C-site-factor-full-periods}
}
where we used $\sum_{n=0}^{N_1-1}n=N_1(N_1-1)/2$,
$N=N_1N_2$, and $\tau=N_1a_1/(N_2a_2)$.

With the boundary condition of the site factors, we now verify the boundary conditions for the full wavefunction.
For $z_i\mapsto z_i+L_2$ with a fixed $i$, we have $Z\mapsto Z+L_2$, and each factor of $F^\sigma$ in \cref{eq:ideal-fci-holomorphic-factor} transforms as below:
\eqa{
e^{\ii K_\sigma Z}
&\longmapsto e^{\ii K_\sigma L_2}e^{\ii K_\sigma Z},\\
\prod_{\gamma=1}^{m}\vartheta_1\left(
\frac{Z-Z_\sigma^\gamma}{L_2}\Big|\tau\right)
&\longmapsto (-1)^m
\prod_{\gamma=1}^{m}\vartheta_1\left(
\frac{Z-Z_\sigma^\gamma}{L_2}\Big|\tau\right),\\
\prod_{j\neq i}\vartheta_1\left(
\frac{z_i-z_j}{L_2}\Big|\tau\right)^m
&\longmapsto (-1)^{m(N_e-1)}
\prod_{j\neq i}\vartheta_1\left(
\frac{z_i-z_j}{L_2}\Big|\tau\right)^m.
}
The site factor is unchanged owing to \cref{eq:C-site-factor-full-periods}.
Therefore, using $e^{\ii(K_\sigma-K_1)L_2}=e^{2\pi \ii(\sigma-1)}=1$,
\eq{
\Psi^{m,\sigma}(\ldots,\bsl r_i+\bsl L_2,\ldots)
=(-1)^{mN_e}e^{\ii K_\sigma L_2}\Psi^{m,\sigma}(\{\bsl r_j\})
=(-1)^N e^{\ii K_1L_2}\Psi^{m,\sigma}(\{\bsl r_j\}).
\label{eq:fci-L2-boundary-factor}
}

For $z_i\mapsto z_i+L_1$, we have $Z\mapsto Z+L_1$. Using
\cref{eq:direct-theta1-shift-tau,eq:C-site-factor-full-periods}, each factor of $\Psi^{m,\sigma}$ in \cref{eq:C-discretized-amplitude} transforms as
\eqa{
e^{\ii K_\sigma Z}
&\longmapsto e^{\ii K_\sigma L_1}e^{\ii K_\sigma Z},\\
\prod_{\gamma=1}^{m}\vartheta_1\left(
\frac{Z-Z_\sigma^\gamma}{L_2}\Big|\tau\right)
&\longmapsto (-1)^m e^{-\ii\pi m\tau}
 e^{-2\pi \ii mZ/L_2}
 e^{\frac{2\pi \ii}{L_2}\sum_{\gamma=1}^{m}Z_\sigma^\gamma}
\prod_{\gamma=1}^{m}\vartheta_1\left(
\frac{Z-Z_\sigma^\gamma}{L_2}\Big|\tau\right),\\
\prod_{j\neq i}\vartheta_1\left(
\frac{z_i-z_j}{L_2}\Big|\tau\right)^m
&\longmapsto (-1)^{m(N_e-1)} e^{-\ii\pi m\tau(N_e-1)}
 e^{-\frac{2\pi \ii m}{L_2}(N_ez_i-Z)}
\prod_{j\neq i}\vartheta_1\left(
\frac{z_i-z_j}{L_2}\Big|\tau\right)^m,\\
e^{\pi\mathcal B(\bsl r_i)/(2\Omega)}\chi(\bsl r_i)
&\longmapsto e^{2\pi \ii Nz_i/L_2+\ii\pi N\tau}
 e^{\pi\mathcal B(\bsl r_i)/(2\Omega)}\chi(\bsl r_i).
}
Multiplying these factors, the position-dependent exponentials cancel
because $N=mN_e$:
\eqa{
\Psi^{m,\sigma}(\ldots,\bsl r_i+\bsl L_1,\ldots)
&=(-1)^{mN_e}e^{\ii K_\sigma L_1}
 e^{-\ii\pi m\tau-\ii\pi m\tau(N_e-1)+\ii\pi N\tau}\\
&\qquad\times e^{-2\pi \ii mZ/L_2}
 e^{-\frac{2\pi \ii m}{L_2}(N_ez_i-Z)}e^{2\pi \ii Nz_i/L_2}
 e^{\frac{2\pi \ii}{L_2}\sum_\gamma Z_\sigma^\gamma}
 \Psi^{m,\sigma}(\{\bsl r_j\})\\
&=(-1)^N e^{\ii K_\sigma L_1}
 e^{\frac{2\pi \ii}{L_2}\sum_\gamma Z_\sigma^\gamma}
 \Psi^{m,\sigma}(\{\bsl r_j\}).
\label{eq:Psi_trans}
}
The center-of-mass zero positions $Z_\sigma^\gamma$ satisfy the following by definition in \cref{eq:K-Z-sector-definition}
\eqa{
\sum_{\gamma=1}^{m}Z_\sigma^\gamma
&=\sum_{\gamma=1}^{m}\left[
\left(\frac{Z_1}{m}+\frac{\gamma-1}{m}\right)L_2
-\frac{\sigma-1}{m}L_1\right]\\
&=Z_1L_2+\frac{L_2}{m}\sum_{\gamma=1}^{m}(\gamma-1)
-(\sigma-1)L_1\\
&=Z_1L_2+\frac{m-1}{2}L_2-(\sigma-1)L_1.
\label{eq:sum_Z}
}
Substituting \cref{eq:sum_Z} and \cref{eq:K-Z-sector-definition} into
\cref{eq:Psi_trans},
\eqa{
\Psi^{m,\sigma}(\ldots,\bsl r_i+\bsl L_1,\ldots)
&=(-1)^N
 \left[e^{\ii K_1L_1}e^{2\pi \ii(\sigma-1)L_1/L_2}\right]
 \left[e^{2\pi \ii Z_1}e^{\ii\pi(m-1)}
 e^{-2\pi \ii(\sigma-1)L_1/L_2}\right]
 \Psi^{m,\sigma}(\{\bsl r_j\})\\
&=(-1)^{N+m-1}e^{\ii K_1L_1}e^{2\pi \ii Z_1}
 \Psi^{m,\sigma}(\{\bsl r_j\})\\
&=(-1)^N e^{\ii K_1L_1}e^{2\pi \ii Z_1}
 \Psi^{m,\sigma}(\{\bsl r_j\}),
\label{eq:fci-L1-boundary-factor}
}
where the final equality uses odd $m$. Both full-period multipliers
are independent of $\sigma$.

Having obtained the full-torus translation factors in each direction \cref{eq:fci-L2-boundary-factor} and \cref{eq:fci-L1-boundary-factor}, we now require them to equal the boundary phases $e^{\ii\phi_j}$:
\eq{\Psi^{m,\sigma}
(\ldots,\bsl r_\ell+\bsl L_j,\ldots)
=
e^{\ii\phi_j}
\Psi^{m,\sigma}
(\ldots,\bsl r_\ell,\ldots),
\qquad
j=1,2.
\label{eq:C-twisted-boundary}}
The multipliers in
\cref{eq:fci-L2-boundary-factor,eq:fci-L1-boundary-factor}
reproduce \cref{eq:fci-boundary-constraint} for odd $m$.
In order to satisfy \cref{eq:fci-boundary-constraint}, we may choose $K_1(\phi_1,\phi_2)$ and $Z_1(\phi_1,\phi_2)$ as below
\eqa{
K_1(\phi_1,\phi_2)
&= K_1^{(0)} +\frac{\phi_2}{L_2},
\\
Z_1(\phi_1,\phi_2)
&= Z_1^{(0)}
+\frac{\phi_1-\tau\phi_2}{2\pi},
\label{eq:chern-general-K1-Z1-lift}}
where $K_1^{(0)}$ and $Z_1^{(0)}$ are independent of
$\phi_1,\phi_2$ and satisfy
\eqa{
(-1)^N e^{\ii K_1^{(0)}L_2}
&=1,\\
(-1)^N
e^{\ii K_1^{(0)}L_1}
e^{2\pi\ii Z_1^{(0)}}
&=1.
\label{eq:C-zero-twist-K1-Z1}}
To see this choice of \cref{eq:chern-general-K1-Z1-lift} and \cref{eq:C-zero-twist-K1-Z1} satisfies the conditions in \cref{eq:fci-boundary-constraint}: for the $L_2$ condition,
\eq{
(-1)^N e^{\ii K_1L_2} =
(-1)^N e^{\ii\left(K_1^{(0)}+\frac{\phi_2}{L_2}\right)L_2} =
(-1)^N e^{\ii K_1^{(0)}L_2}e^{\ii\phi_2} =
e^{\ii\phi_2},
}
and for the $L_1$ condition,
\eqa{
(-1)^N e^{\ii K_1L_1}e^{2\pi \ii Z_1}
&=
(-1)^N
e^{\ii\left(K_1^{(0)}+\frac{\phi_2}{L_2}\right)L_1}
e^{2\pi \ii\left(
Z_1^{(0)}+\frac{\phi_1-\tau\phi_2}{2\pi}
\right)} =
(-1)^N e^{\ii K_1^{(0)}L_1}e^{2\pi \ii Z_1^{(0)}}
e^{\ii\phi_2L_1/L_2}
e^{\ii(\phi_1-\tau\phi_2)}
\\
&=
e^{\ii\tau\phi_2}e^{\ii\phi_1-\ii\tau\phi_2} =
e^{\ii\phi_1}.
}

Thus, the lattice wavefunctions satisfy the required twisted boundary conditions along both torus periods. This also justifies the form of the many-body state in \cref{eq:C-discretized-many-body-state}, which includes only $\bsl r_i\in\Lambda$.

\subsubsection{Lattice Translations}

We now determine how simultaneous translation of all
electrons by each primitive lattice vector $\bsl a_j$
acts on the $m$ many-body wavefunctions.

Acting directly on the explicit wavefunction in \cref{eq:C-discretized-amplitude}, define the simultaneous translation of all particles by one unit cell:
$\{\bsl r_i\}\longmapsto\{\bsl r_i+\bsl a_j\}$ with $\bsl a_j=\frac{\bsl L_j}{N_j}$.
Under this operation,
$Z\longmapsto Z+N_ea_j$, and $z_i-z_k\longmapsto z_i-z_k.$
Thus the holomorphic Jastrow factor is unchanged.
For compactness, denote the center-of-mass theta product $\prod_{\gamma=1}^{m}\vartheta_1((Z-Z_\sigma^\gamma)/L_2\mid\tau)$ by
$\mathcal P_\sigma(Z)$. Substitution of
\cref{eq:K-Z-sector-definition} gives
\eq{
\mathcal P_\sigma(Z)
=
\prod_{\gamma=1}^{m}\vartheta_1(u_{\sigma\gamma}|\tau),
\quad
u_{\sigma\gamma}
=
\frac{Z}{L_2}
-\frac{Z_1+\gamma-1}{m}
+\frac{\sigma-1}{m}\tau.
\label{eq:com-theta-product}
}
Multiplying \cref{eq:combined-band-factor-translation} over all
particles gives
\eqa{
\prod_{i=1}^{N_e}\left[
 e^{\pi\mathcal B(\bsl r_i+\bsl a_j)/(2\Omega)}
 \chi(\bsl r_i+\bsl a_j)\right]
&=\prod_{i=1}^{N_e}
 \exp\left[\frac{\pi}{2\Omega}
 \left(\frac{\bar a_2}{a_2}a_j-\bar a_j\right)(2z_i+a_j)\right]
 \prod_{i=1}^{N_e}\left[e^{\pi\mathcal B(\bsl r_i)/(2\Omega)}\chi(\bsl r_i)\right]\\
&=\exp\left[\frac{\pi}{2\Omega}
 \left(\frac{\bar a_2}{a_2}a_j-\bar a_j\right)
 \sum_{i=1}^{N_e}(2z_i+a_j)\right]
 \prod_{i=1}^{N_e}\left[e^{\pi\mathcal B(\bsl r_i)/(2\Omega)}\chi(\bsl r_i)\right]\\
&=\exp\left[\frac{\pi}{2\Omega}
 \left(\frac{\bar a_2}{a_2}a_j-\bar a_j\right)(2Z+N_ea_j)\right]
 \prod_{i=1}^{N_e}\left[e^{\pi\mathcal B(\bsl r_i)/(2\Omega)}\chi(\bsl r_i)\right].
\label{eq:many-body-band-factor-translation}
}
Here $\sum_i(2z_i+a_j)=2Z+N_ea_j$.
By \cref{eq:C-site-factor-geometric-identities}, this multiplier
is one for $j=2$, without choosing $a_2$ real.

Since $N=mN_e=N_1N_2$ and $a_2=\frac{L_2}{N_2}$, we have
\eq{
N_ea_2
=
\frac{N_1N_2}{m}\frac{L_2}{N_2}
=
\frac{N_1}{m}L_2.
\label{eq:com-shift-a2}
}
Since all $N_e$ electrons move by $a_2$, $Z=\sum_i z_i$ changes by
$Z\rightarrow Z+N_ea_2=Z+\frac{N_1}{m}L_2.$
Under $Z\mapsto Z+L_2/m$, from \cref{eq:com-theta-product} we know
$u_{\sigma\gamma}\mapsto u_{\sigma\gamma}+1/m$.
Now for $\gamma=2,\ldots,m$, we can simply relabel it as
\eq{
u_{\sigma\gamma}+\frac{1}{m}
=
\frac{Z}{L_2}
-\frac{(Z_1+\gamma-1)-1}{m}
+\frac{\sigma-1}{m}\tau
=
u_{\sigma,\gamma-1},
}
For $\gamma=1$, we can check to see
$u_{\sigma1}+1/m=u_{\sigma m}+1$.
Then after $Z\rightarrow Z+L_2/m$, the factor transforms as
\eqa{
\mathcal P_\sigma\left(Z+\frac{L_2}{m}\right)
&=
\prod_{\gamma=1}^{m}
\vartheta_1\left(
u_{\sigma\gamma}+\frac{1}{m}\Big|\tau
\right) =
\vartheta_1\left(u_{\sigma m}+1\Big|\tau\right)
\prod_{\gamma=2}^{m}
\vartheta_1\left(u_{\sigma,\gamma-1}\Big|\tau\right)
\\
&=
-\vartheta_1(u_{\sigma m}|\tau)
\prod_{\gamma=1}^{m-1}
\vartheta_1(u_{\sigma\gamma}|\tau)
=
-\mathcal P_\sigma(Z),
\label{eq:com-product-elementary-a2-shift}
}
using \cref{eq:direct-theta1-shift-one}.
Iterating this relation $N_1$ times and including the exponential
$e^{\ii K_\sigma Z}$ yields explicitly
\eq{
\Psi^{m,\sigma}
[\{\bsl r_i+\bsl a_2\}]
=
\lambda_{2,\sigma}
\Psi^{m,\sigma}[\{\bsl r_i\}],
\qquad
\lambda_{2,\sigma}
=
(-1)^{N_1}e^{\ii K_\sigma N_1L_2/m}.
\label{eq:explicit-a2-translation}
}
Thus the state $\Psi^{m,\sigma}$ is an eigenstate of the
coordinate-shift operator that shifts
every electron by $\bsl a_2$, with eigenvalue $\lambda_{2,\sigma}$
(it does not mix the $\sigma$ sectors).

For the $a_1$ direction, similar to \cref{eq:com-shift-a2},
\eq{
N_ea_1=\frac{N_2}{m}L_1.
\label{eq:com-shift-a1}
}
So translating every electron by $\bsl a_1$ lets
$Z\rightarrow Z+\frac{N_2}{m}L_1$.
Using \cref{eq:C-site-factor-geometric-identities} and
\cref{eq:com-shift-a1}, the site-factor multiplier becomes
\eqa{
\prod_{i=1}^{N_e}\left[
 e^{\pi\mathcal B(\bsl r_i+\bsl a_1)/(2\Omega)}\chi(\bsl r_i+\bsl a_1)\right]
&=\exp\left[\frac{\pi}{2\Omega}
 \frac{2\ii\Omega}{a_2}(2Z+N_ea_1)\right]
 \prod_{i=1}^{N_e}\left[e^{\pi\mathcal B(\bsl r_i)/(2\Omega)}\chi(\bsl r_i)\right]\\
&=\exp\left[\frac{\ii\pi N_2}{L_2}
 \left(2Z+\frac{N_2}{m}L_1\right)\right]
 \prod_{i=1}^{N_e}\left[e^{\pi\mathcal B(\bsl r_i)/(2\Omega)}\chi(\bsl r_i)\right]\\
&=e^{2\pi\ii N_2Z/L_2+\ii\pi\tau N_2^2/m}
 \prod_{i=1}^{N_e}\left[e^{\pi\mathcal B(\bsl r_i)/(2\Omega)}\chi(\bsl r_i)\right].
\label{eq:many-body-band-factor-a1}
}
Notice that
$Z\rightarrow Z+\frac{N_2}{m}L_1$
means
$u_{\sigma\gamma}\rightarrow
u_{\sigma\gamma}+\frac{N_2}{m}\tau$,
which is equivalent to $\sigma\rightarrow\sigma+N_2$.
And the center-of-mass zeros obey, by definition in \cref{eq:K-Z-sector-definition},
\eq{
Z_{\sigma+N_2}^{\gamma}
=
Z_\sigma^\gamma-\frac{N_2}{m}L_1,
\label{eq:com-zero-a1-shift}
}
for integer $\sigma$. Using \cref{eq:com-theta-product} and \cref{eq:com-zero-a1-shift}, we obtain
\eq{
\mathcal P_\sigma\left(Z+\frac{N_2}{m}L_1\right)
=
\prod_{\gamma}
\vartheta_1\left(
\frac{Z+\frac{N_2}{m}L_1-Z_\sigma^\gamma}{L_2}\Big|\tau
\right)
=
\prod_{\gamma}
\vartheta_1\left(
\frac{Z-Z_{\sigma+N_2}^\gamma}{L_2}\Big|\tau
\right)
=
\mathcal P_{\sigma+N_2}(Z).
\label{eq:COMidentity}
}
Similarly, the definition of $K_\sigma$ in \cref{eq:K-Z-sector-definition} gives
\eq{
K_{\sigma+N_2}
=
K_\sigma+\frac{2\pi N_2}{L_2}.
\label{eq:K_sigma+N2}
}
We now apply the translation $\{\bsl r_i\}\to\{\bsl r_i+\bsl a_1\}$ to the full wavefunction in \cref{eq:C-discretized-amplitude} and \cref{eq:ideal-fci-holomorphic-factor}. Using the unchanged Jastrow factor, the site-factor transformation in \cref{eq:many-body-band-factor-a1}, and the center-of-mass identities in \cref{eq:COMidentity} and \cref{eq:K_sigma+N2}, we obtain
\eqa{
\Psi^{m,\sigma}
\left[\{\bsl r_i+\bsl a_1\}\right]
&=
e^{\,\ii K_\sigma\left(Z+\frac{N_2}{m}L_1\right)}
\mathcal P_\sigma\left(
Z+\frac{N_2}{m}L_1
\right)
\prod_{i<j}
\vartheta_1\left(
\frac{z_i-z_j}{L_2}\Big|\tau
\right)^m
e^{2\pi \ii N_2Z/L_2+\ii\pi\tau N_2^2/m}
\prod_{i=1}^{N_e}
\left[
e^{\pi\mathcal B(\bsl r_i)/(2\Omega)}
\chi(\bsl r_i)
\right]
\\
&=
e^{\,\ii K_\sigma N_2L_1/m+\ii\pi\tau N_2^2/m}
e^{\,\ii K_\sigma Z}
e^{\,2\pi \ii N_2Z/L_2}
\mathcal P_{\sigma+N_2}(Z)
\prod_{i<j}
\vartheta_1\left(
\frac{z_i-z_j}{L_2}\Big|\tau
\right)^m
\prod_{i=1}^{N_e}
\left[
e^{\pi\mathcal B(\bsl r_i)/(2\Omega)}
\chi(\bsl r_i)
\right]
\\
&=
e^{\,\ii K_\sigma N_2L_1/m+\ii\pi\tau N_2^2/m}
e^{\,\ii\left(K_\sigma+\frac{2\pi N_2}{L_2}\right)Z}
\mathcal P_{\sigma+N_2}(Z)
\prod_{i<j}
\vartheta_1\left(
\frac{z_i-z_j}{L_2}\Big|\tau
\right)^m
\prod_{i=1}^{N_e}
\left[
e^{\pi\mathcal B(\bsl r_i)/(2\Omega)}
\chi(\bsl r_i)
\right]
\\
&=
e^{\,\ii K_\sigma N_2L_1/m+\ii\pi\tau N_2^2/m}
e^{\,\ii K_{\sigma+N_2}Z}
\mathcal P_{\sigma+N_2}(Z)
\prod_{i<j}
\vartheta_1\left(
\frac{z_i-z_j}{L_2}\Big|\tau
\right)^m
\prod_{i=1}^{N_e}
\left[
e^{\pi\mathcal B(\bsl r_i)/(2\Omega)}
\chi(\bsl r_i)
\right]
\\
&=
e^{\,\ii K_\sigma N_2L_1/m+\ii\pi\tau N_2^2/m}
\Psi^{m,\sigma+N_2}
\left[\{\bsl r_i\}\right].
\label{eq:explicit-a1-translation-extended}
}

Equation \cref{eq:explicit-a1-translation-extended} contains the sector label $\sigma+N_2$, which may lie outside the physical range $1,\ldots,m$. To identify the corresponding physical sector, we reduce this label modulo $m$. Since $\operatorname{mod}(n,m)$ gives a remainder in $\{0,\ldots,m-1\}$, we first subtract $1$ from the sector label before taking the modulo, and then add $1$ to return to the physical range $1,\ldots,m$. Explicitly,
\eq{\sigma+N_2=1+\operatorname{mod}(\sigma-1+N_2,m)+pm,
\qquad p=\left\lfloor\frac{\sigma-1+N_2}{m}\right\rfloor\in\dsZ.\label{eq:reduced-sector-index}}
The floor function $\lfloor x\rfloor$ denotes the greatest integer less than or equal to $x$. Here $1+\operatorname{mod}(\sigma-1+N_2,m)$ is the
corresponding physical sector label, while $pm$ is the
difference between the original and reduced labels.
Thus, $p$ counts how many multiples of $m$ are removed
in the reduction. 

Here $\Psi^{m,\sigma+N_2}$ is obtained by evaluating
\cref{eq:C-discretized-amplitude} at the integer label
$\sigma+N_2$, even when it lies outside $1,\ldots,m$.
We now relate it to the corresponding physical sector
wavefunction and determine the proportionality factor.
For an integer shift $\sigma\to\sigma+pm$, the theta arguments in
\cref{eq:com-theta-product} satisfy
\eq{
u_{\sigma+pm,\gamma}=u_{\sigma\gamma}+p\tau.
}
Repeated application of \cref{eq:direct-theta1-shift-tau} gives,
for any integer $p$,
\eq{
\vartheta_1(u+p\tau|\tau)
=(-1)^p e^{-\ii\pi p^2\tau-2\pi\ii p u}
\vartheta_1(u|\tau).
\label{eq:C-theta-integer-shift}
}
Furthermore, summing the arguments in \cref{eq:com-theta-product} gives
\eq{
\sum_{\gamma=1}^{m}u_{\sigma\gamma}
=\frac{mZ}{L_2}-Z_1-\frac{m-1}{2}+(\sigma-1)\tau.
}
Applying \cref{eq:C-theta-integer-shift} to all $m$ theta
functions in \cref{eq:com-theta-product}, we obtain
\eqa{\mathcal P_{\sigma+pm}(Z)&=\prod_{\gamma=1}^{m}
\vartheta_1(u_{\sigma+pm,\gamma}|\tau)=\prod_{\gamma=1}^{m}\vartheta_1(u_{\sigma\gamma}+p\tau|\tau)=\prod_{\gamma=1}^{m}\left[(-1)^p e^{-\ii\pi p^2\tau-2\pi\ii p u_{\sigma\gamma}}\vartheta_1(u_{\sigma\gamma}|\tau)\right]\\
&=(-1)^{pm}
e^{-\ii\pi mp^2\tau}
e^{-2\pi\ii p\sum_{\gamma=1}^{m}u_{\sigma\gamma}}
\mathcal P_\sigma(Z)\\
&=(-1)^{pm}
e^{-\ii\pi mp^2\tau}
e^{-2\pi\ii p\left[
\frac{mZ}{L_2}-Z_1-\frac{m-1}{2}+(\sigma-1)\tau
\right]}\mathcal P_\sigma(Z)\\
&=(-1)^{pm}e^{\ii\pi p(m-1)}
e^{-\ii\pi mp^2\tau+2\pi\ii pZ_1
-2\pi\ii p(\sigma-1)\tau}
e^{-2\pi\ii pmZ/L_2}\mathcal P_\sigma(Z)\\
&=(-1)^p
e^{-\ii\pi mp^2\tau+2\pi\ii pZ_1
-2\pi\ii p(\sigma-1)\tau}
e^{-2\pi\ii pmZ/L_2}\mathcal P_\sigma(Z).
}
Here the last equality uses $(-1)^{pm}e^{\ii\pi p(m-1)}=(-1)^p$. In the full wavefunction, the factor $e^{-2\pi\ii pmZ/L_2}$
cancels the extra factor $e^{2\pi\ii pmZ/L_2}$ from
$K_{\sigma+pm}=K_\sigma+2\pi pm/L_2$:
\eq{
e^{\ii K_{\sigma+pm}Z}e^{-2\pi\ii pmZ/L_2} =e^{\ii K_\sigma Z}
e^{2\pi\ii pmZ/L_2}e^{-2\pi\ii pmZ/L_2}=e^{\ii K_\sigma Z}.
}
The Jastrow and site factors are independent of $\sigma$, hence
\eq{
\Psi^{m,\sigma+pm}[\{\bsl r_i\}]
=(-1)^p e^{-\ii\pi mp^2\tau+2\pi\ii pZ_1
 -2\pi\ii p(\sigma-1)\tau}
\Psi^{m,\sigma}[\{\bsl r_i\}].
\label{eq:C-sector-integer-shift}
}
Using \cref{eq:reduced-sector-index} to rewrite the
integer sector label and then applying
\cref{eq:C-sector-integer-shift}, we obtain
\eqa{
\Psi^{m,\sigma+N_2}[\{\bsl r_i\}]
&=\Psi^{m,\,1+\operatorname{mod}(\sigma-1+N_2,m)+pm}
[\{\bsl r_i\}]\\
&=(-1)^p e^{-\ii\pi mp^2\tau+2\pi\ii pZ_1
-2\pi\ii p\tau \big(1+\operatorname{mod}(\sigma-1+N_2,m)-1\big)}
\Psi^{m,\,1+\operatorname{mod}(\sigma-1+N_2,m)}
[\{\bsl r_i\}]\\
&=(-1)^p e^{-\ii\pi mp^2\tau+2\pi\ii pZ_1
-2\pi\ii p\tau\operatorname{mod}(\sigma-1+N_2,m)}
\Psi^{m,\,1+\operatorname{mod}(\sigma-1+N_2,m)}
[\{\bsl r_i\}].
\label{eq:C-sector-reduced-wavefunction}
}

Substituting \cref{eq:C-sector-reduced-wavefunction} into
\cref{eq:explicit-a1-translation-extended} and collecting
the exponential factors, we obtain
\eqa{
\Psi^{m,\sigma}[\{\bsl r_i+\bsl a_1\}]
&=e^{\frac{\ii K_\sigma N_2L_1}{m}
+\frac{\ii\pi\tau N_2^2}{m}}
(-1)^p e^{-\ii\pi mp^2\tau+2\pi\ii pZ_1
-2\pi\ii p\tau\operatorname{mod}(\sigma-1+N_2,m)}
\Psi^{m,\,1+\operatorname{mod}(\sigma-1+N_2,m)}
[\{\bsl r_i\}]\\
&=(-1)^p e^{\frac{\ii K_\sigma N_2L_1}{m}
+\frac{\ii\pi\tau N_2^2}{m}
-\ii\pi mp^2\tau +2\pi\ii pZ_1 -2\pi\ii p\tau\operatorname{mod}(\sigma-1+N_2,m)}
\Psi^{m,\,1+\operatorname{mod}(\sigma-1+N_2,m)}
[\{\bsl r_i\}]\\
&=\lambda_{1,\sigma}
\Psi^{m,\,1+\operatorname{mod}(\sigma-1+N_2,m)}
[\{\bsl r_i\}].
\label{eq:explicit-a1-translation}
}
Here we defined the coefficient $\lambda_{1,\sigma}$ which can be simplified
using $K_\sigma=K_1+2\pi(\sigma-1)/L_2$ and
$\tau=L_1/L_2$:
\eqa{
\lambda_{1,\sigma}
&=(-1)^p e^{\frac{\ii K_\sigma N_2L_1}{m}
+\frac{\ii\pi\tau N_2^2}{m} -\ii\pi mp^2\tau +2\pi\ii pZ_1 -2\pi\ii p\tau\operatorname{mod}(\sigma-1+N_2,m)}\\
&=(-1)^p e^{\frac{\ii K_1N_2L_1}{m}
+\frac{2\pi\ii\tau(\sigma-1)N_2}{m}
+\frac{\ii\pi\tau N_2^2}{m}
-\ii\pi mp^2\tau
+2\pi\ii pZ_1
-2\pi\ii p\tau\operatorname{mod}(\sigma-1+N_2,m)}\\
&=(-1)^p e^{\frac{\ii K_1N_2L_1}{m}
+2\pi\ii pZ_1+\frac{\ii\pi\tau}{m}\Big(2(\sigma-1)N_2+N_2^2-m^2p^2-2mp\operatorname{mod}(\sigma-1+N_2,m)
\Big)}\\
&=(-1)^p e^{\frac{\ii K_1N_2L_1}{m}
+2\pi\ii pZ_1
+\frac{\ii\pi\tau}{m}
\Big(\operatorname{mod}(\sigma-1+N_2,m)^2 -(\sigma-1)^2
\Big)}.
\label{eq:C-lambda1-simplified}
}
In the last equality, we use \cref{eq:reduced-sector-index}
to obtain
\eqa{
&2(\sigma-1)N_2+N_2^2-m^2p^2
-2mp\operatorname{mod}(\sigma-1+N_2,m)\\
&=(\sigma-1+N_2)^2-(\sigma-1)^2
-m^2p^2-2mp\operatorname{mod}(\sigma-1+N_2,m)\\
&=\Big[\operatorname{mod}(\sigma-1+N_2,m)+pm\Big]^2
-(\sigma-1)^2-m^2p^2
-2mp\operatorname{mod}(\sigma-1+N_2,m)\\
&=\operatorname{mod}(\sigma-1+N_2,m)^2
-(\sigma-1)^2.
\label{eq:C-translation-exponent-simplification}
}

\cref{eq:explicit-a1-translation} shows that
translation along $\bsl a_1$ maps each sector $\sigma$
to $1+\operatorname{mod}(\sigma-1+N_2,m)$, multiplied
by a nonzero, coordinate-independent factor
$\lambda_{1,\sigma}$. As $\sigma$ runs from $1$ to $m$,
the resulting sector labels also cover $1,\ldots,m$
exactly once. Thus, translation permutes the $m$
sectors and preserves their span. If $m$ divides
$N_2$, each sector maps to itself.

\subsubsection{Many-Body Crystal Momentum}
We now use the primitive-translation laws to identify the
many-body crystal momentum. Let $T_{\bsl a_j}$ act on amplitudes by
$(T_{\bsl a_j}\Psi)(\{\bsl r_i\})=\Psi(\{\bsl r_i+\bsl a_j\})$.
The two operators commute, $T_{\bsl a_1}T_{\bsl a_2}=T_{\bsl a_2}T_{\bsl a_1}$, directly from their
definition, since the coordinate shifts by $\bsl a_1$ and
$\bsl a_2$ commute.

The translation results derived in the preceding subsection show that $T_{\bsl a_1}$ and $T_{\bsl a_2}$ preserve the space spanned by the $m$ manybody states, although they may change the sector label $\sigma$. Since the two translation operators are unitary and commute, we can choose a new basis of this space consisting of simultaneous translation eigenstates. 
Specifically, we construct linear combinations
\eq{
\Psi^{(q)}
=
\sum_{\sigma=1}^{m}
c_\sigma^{(q)}\Psi^{m,\sigma},
\qquad q=1,\ldots,m,
}
such that
\eq{
T_{\bsl a_j}\Psi^{(q)}
=
\Lambda_j^{(q)}\Psi^{(q)},
\qquad j=1,2.
\label{eq:newbasisEigen}
}
Since $T_{\bsl a_j}$ is unitary, $|\Lambda_j^{(q)}|=1$.
For any lattice vector
$\bsl R=n_1\bsl a_1+n_2\bsl a_2$,
with $n_1,n_2\in\dsZ$, let $T_{\bsl R}$
denote the simultaneous coordinate shift of all
electrons by $\bsl R$.
Successive translations satisfy
$T_{\bsl R}T_{\bsl R'}=T_{\bsl R+\bsl R'}$.
Therefore, $T_{\bsl R}=T_{\bsl a_1}^{n_1}T_{\bsl a_2}^{n_2}$.
Using \cref{eq:newbasisEigen}, we obtain
\eqa{
T_{\bsl R}\Psi^{(q)}=T_{\bsl a_1}^{n_1}T_{\bsl a_2}^{n_2}\Psi^{(q)}=(\Lambda_1^{(q)})^{n_1}
(\Lambda_2^{(q)})^{n_2}\Psi^{(q)}.
}
Thus, $\Psi^{(q)}$ is an eigenstate of
translation by any lattice vector $\bsl R$.
Since $|\Lambda_j^{(q)}|=1$, we can define
the many-body crystal momentum
$\bsl K_{\mathrm{MB}}^{(q)}$ by
$\Lambda_j^{(q)}
=e^{\ii\bsl K_{\mathrm{MB}}^{(q)}\cdot\bsl a_j}$,
for $j=1,2$. Consequently,
\eq{
T_{\bsl R}\Psi^{(q)}
=e^{\ii\bsl K_{\mathrm{MB}}^{(q)}\cdot\bsl R}
\Psi^{(q)}.
}
Setting $\bsl R=\bsl a_j$ and using the definition
of $T_{\bsl a_j}$, we obtain
\eq{
\Psi^{(q)}
[\{\bsl r_i+\bsl a_j\}]
=
e^{\ii\bsl K_{\mathrm{MB}}^{(q)}\cdot\bsl a_j}
\Psi^{(q)}
[\{\bsl r_i\}],
\qquad j=1,2.
\label{eq:many-body-lattice-translation-rule}
}
This is the many-body analogue of Bloch's translation law: translating all electrons by one primitive lattice vector multiplies the wavefunction by a phase.

We now calculate these momenta explicitly.
By \cref{eq:explicit-a1-translation}, $T_{\bsl a_1}$
changes the sector label by $N_2$ modulo $m$.
Let $\ell$ be the smallest positive integer
such that $\ell N_2$ is a multiple of $m$.
After $\ell$ translations, each sector returns
to itself. Multiplying the successive coefficients
in \cref{eq:C-lambda1-simplified},
\eq{
T_{\bsl a_1}^\ell\Psi^{m,\sigma}
=(-1)^{\ell N_2/m}
e^{\frac{\ii\ell N_2}{m}(K_1L_1+2\pi Z_1)}
\Psi^{m,\sigma}.
}
Because this multiplier is independent of
$\sigma$, the same relation holds for every
$\Psi^{(q)}$. 

Comparing with
\cref{eq:newbasisEigen},
we obtain
\eq{
(\Lambda_1^{(q)})^\ell=(-1)^{\ell N_2/m}
e^{\frac{\ii\ell N_2}{m}(K_1L_1+2\pi Z_1)}=e^{\frac{\ii\ell N_2}{m}
(K_1L_1+2\pi Z_1+\pi)}.
}
Taking the $\ell$-th roots gives the possible
$T_{\bsl a_1}$ eigenvalues,
\eq{
\Lambda_1^{(q)}
=e^{\frac{\ii N_2}{m}
(K_1L_1+2\pi Z_1+\pi)
+\frac{2\pi\ii t}{\ell}},
\qquad t=0,\ldots,\ell-1.
}
We next determine the $T_{\bsl a_2}$ eigenvalues.
By \cref{eq:explicit-a2-translation},
each original sector is already an eigenstate
of $T_{\bsl a_2}$ with eigenvalue $\lambda_{2,\sigma}$.
Since $T_{\bsl a_1}$ and $T_{\bsl a_2}$ commute,
\eq{
T_{\bsl a_2}(T_{\bsl a_1}\Psi^{m,\sigma})
=T_{\bsl a_1}(T_{\bsl a_2}\Psi^{m,\sigma})
=\lambda_{2,\sigma}T_{\bsl a_1}\Psi^{m,\sigma}.
}
Thus, sectors connected by $T_{\bsl a_1}$ have the
same $T_{\bsl a_2}$ eigenvalue. Their linear combinations
that diagonalize $T_{\bsl a_1}$ are therefore also
eigenstates of $T_{\bsl a_2}$, with
\eq{
\Lambda_2^{(q)}
=\lambda_{2,\sigma}
=(-1)^{N_1}e^{\ii K_\sigma N_1L_2/m}.
}
The $m$ sectors form $m/\ell$ disjoint cycles
under $T_{\bsl a_1}$, each containing $\ell$ sectors.
We choose $\sigma=1,\ldots,m/\ell$ as one
representative from each cycle. Since $T_{\bsl a_1}$
cyclically permutes the sectors with nonzero
coefficients, all $\ell$ roots correspond to
eigenstates within each cycle. We label the
resulting $m$ eigenstates by
$q=(\sigma-1)\ell+t+1$, where
$t=0,\ldots,\ell-1$.
Using
$\Lambda_j^{(q)}
=e^{\ii\bsl K_{\mathrm{MB}}^{(q)}\cdot\bsl a_j}$,
the two translation eigenvalues give
\eqa{
\bsl K_{\mathrm{MB}}^{(q)}\cdot\bsl a_1
&=\frac{N_2}{m}(K_1L_1+2\pi Z_1+\pi)
+\frac{2\pi t}{\ell},\\
\bsl K_{\mathrm{MB}}^{(q)}\cdot\bsl a_2
&=\pi N_1+\frac{K_\sigma N_1L_2}{m},
}
with both equalities understood modulo $2\pi$.
Since $\bsl b_i\cdot\bsl a_j=2\pi\delta_{ij}$,
we obtain
\eq{
\bsl K_{\mathrm{MB}}^{(q)}
=\left[
\frac{N_2}{2\pi m}(K_1L_1+2\pi Z_1+\pi)
+\frac{t}{\ell}
\right]\bsl b_1+
\left[
\frac{N_1}{2}
+\frac{N_1K_\sigma L_2}{2\pi m}
\right]\bsl b_2.
\label{eq:C-explicit-MB-momentum}
}
The momentum is defined modulo
reciprocal-lattice vectors.

We finally check consistency with the twisted
boundary conditions in \cref{eq:C-twisted-boundary}.
Since $\bsl L_j=N_j\bsl a_j$, the momentum
eigenvalue gives a factor
$e^{\ii\bsl K_{\mathrm{MB}}^{(q)}\cdot\bsl L_j}$
after $N_j$ primitive translations.
Translating all $N_e$ electrons by $\bsl L_j$
instead gives $e^{\ii N_e\phi_j}$.
Thus,
\eq{
e^{\ii\bsl K_{\mathrm{MB}}^{(q)}\cdot\bsl L_j}
=e^{\ii N_e\phi_j}.
\label{eq:many-body-momentum-boundary-consistency}
}

For later use, we express the translation law in the active Fock-space convention.
Since $\hat T_{\bsl a_j}$ acts as $T_{\bsl a_j}^{-1}$
on wavefunction amplitudes,
\eq{
\hat T_{\bsl a_j}\ket{\Psi^{(q)}}
=e^{-\ii\bsl K_{\mathrm{MB}}^{(q)}\cdot\bsl a_j}
\ket{\Psi^{(q)}},
\qquad j=1,2.
\label{eq:C-active-many-body-translation}
}

\subsection{Linear Independence of the Lattice States}
\label{sec:C-linear-independence}

We now prove that the $m$ constructed many-body states
remain nonzero and linearly independent after lattice
discretizations. We work at fixed twists with $N_e\geq1$ and the filling in
\cref{eq:B-filling}. The site and nonvanishing
site factors are those of \cref{sec:C-setup}.

To isolate the dependence on the sector $\sigma$, write the center-of-mass factor as
\eq{
 H_\sigma(Z)=e^{\ii K_\sigma Z}\mathcal P_\sigma(Z),
 \label{eq:discretized-com-factor}
}
using \cref{eq:com-theta-product}. This is the only
sector-dependent factor in \cref{eq:C-discretized-amplitude}.
To prove that the $m$ constructed states are
linearly independent, we need to show that
\eq{
\sum_{\sigma=1}^{m}
c_\sigma\ket{\Psi^{m,\sigma}}=0
\quad\Rightarrow\quad
c_\sigma=0
\quad\text{for all }\sigma.
}
Suppose, for contradiction, that there exist
coefficients $c_\sigma\in\dsC$, not all zero, such that
\eq{
\sum_{\sigma=1}^{m}
c_\sigma
\Psi^{m,\sigma}(\bsl r_1,\ldots,\bsl r_{N_e}) = 0
\label{eq:assumed-discretized-linear-relation}
}
for every lattice occupation configuration, i.e., every choice of $N_e$ distinct occupied sites $\bsl r_i\in\Lambda$.
In \cref{eq:C-discretized-amplitude,eq:ideal-fci-holomorphic-factor}, the Jastrow
and site factors are common to all sectors and
nonzero when the occupied sites are distinct, thus our assumption \cref{eq:assumed-discretized-linear-relation} becomes
\eq{
\sum_{\sigma=1}^{m}c_\sigma H_\sigma(Z)=0
\label{eq:com-linear-relation-at-discretized-Z}
}
at every value of $Z$ obtained by placing the
electrons on distinct lattice sites. With the same coefficients, define
\eq{
H(Z)
=
\sum_{\sigma=1}^{m}c_\sigma H_\sigma(Z).
\label{eq:combined-center-of-mass-section}
}
So far, our assumption shows that $H(Z)=0$ at the values obtained from lattice occupation
configurations. To prove linear independence,
we will first show that this forces $H(Z)$
to vanish identically for every complex $Z\in\dsC$.
We will then show that this is possible only
if all $c_\sigma=0$, contradicting our assumption.

We now show that a nonzero $H(Z)$ cannot vanish at every value of $Z$ obtained from lattice configurations. This will establish that our assumed linear relation (\cref{eq:com-linear-relation-at-discretized-Z}) forces $H(Z)\equiv0$ for every complex $Z$. To see that, we compare
the number of these values with the number of zeros
a nonzero $H$ can have. The explicit formula in \cref{eq:discretized-com-factor}
defines each $H_\sigma(Z)$ as an holomorphic function
that is not identically zero.

The center-of-mass transformations already used in
\cref{sec:C-boundaries-translations}, together with
\cref{eq:K-Z-sector-definition,eq:sum_Z}, give:
In the $L_2$ direction,
\eq{
H_\sigma(Z+L_2)
=
e^{\ii K_\sigma(Z+L_2)}
\prod_{\gamma=1}^{m}
\vartheta_1\left(
\frac{Z-Z_\sigma^\gamma}{L_2}+1
\middle|\tau
\right)
=
e^{\ii K_\sigma L_2}(-1)^m H_\sigma(Z)
=
(-1)^m e^{\ii K_1L_2}H_\sigma(Z),
\label{eq:C-com-torus-boundary2}
}
where
$e^{\ii K_\sigma L_2}
=e^{\ii K_1L_2}e^{2\pi \ii(\sigma-1)}
=e^{\ii K_1L_2}$.
In the $L_1$ direction,
\eqa{
H_\sigma(Z+L_1)
&=
e^{\ii K_\sigma(Z+L_1)}
\prod_{\gamma=1}^{m}
\vartheta_1\left(
\frac{Z-Z_\sigma^\gamma}{L_2}+\tau
\,\middle|\,
\tau
\right)
\\
&=
e^{\ii K_\sigma L_1}
(-1)^m
e^{-\ii\pi m\tau}
e^{-2\pi \ii\sum_\gamma (Z-Z_\sigma^\gamma)/L_2}
H_\sigma(Z)
\\
&=
e^{\ii K_\sigma L_1}
(-1)^m
e^{-\ii\pi m\tau}
e^{-2\pi \ii mZ/L_2}
e^{\frac{2\pi \ii}{L_2}\sum_\gamma Z_\sigma^\gamma}
H_\sigma(Z)
\\
&=
e^{\ii K_\sigma L_1}
(-1)^m
e^{-\ii\pi m\tau}
e^{-2\pi \ii mZ/L_2}
e^{2\pi \ii Z_1}
e^{\ii\pi(m-1)}
e^{-2\pi \ii(\sigma-1)\tau}
H_\sigma(Z)
\\
&=
-
e^{\ii K_1L_1}
e^{2\pi \ii Z_1}
e^{-\ii\pi m\tau}
e^{-2\pi \ii mZ/L_2}
H_\sigma(Z).
\label{eq:C-com-torus-boundary1}
}

Since the translation factors in
\cref{eq:C-com-torus-boundary2,eq:C-com-torus-boundary1}
are independent of $\sigma$, every linear combination
of the $H_\sigma$ satisfies the same boundary conditions, thus they apply to $H$ defined in
\cref{eq:combined-center-of-mass-section}.

We now show that a nonzero $H(Z)$ has exactly $m$ zeros on the torus.
We count its zeros using the argument principle.
Taking logarithmic derivatives of the two boundary
conditions gives
\eqa{
\frac{H'(Z+L_2)}{H(Z+L_2)}
&=
\frac{H'(Z)}{H(Z)},
\\
\frac{H'(Z+L_1)}{H(Z+L_1)}
&=
\frac{H'(Z)}{H(Z)}
-\frac{2\pi\ii m}{L_2}.
\label{eq:C-com-logarithmic-derivative}
}
Choose a fundamental parallelogram $\mathcal P$
whose boundary contains no zero of $H$, with vertices
$Z_0$, $Z_0+L_2$, $Z_0+L_2+L_1$, and $Z_0+L_1$,
in counterclockwise order.
This orientation is positive in the complex $Z$-plane
because $\im(L_1/L_2)>0$.
Using \cref{eq:C-com-logarithmic-derivative}, the
integrals along the two edges parallel to $L_1$ cancel,
and the argument principle gives
\eq{
\frac{1}{2\pi\ii}
\oint_{\partial\mathcal P}
\frac{H'(Z)}{H(Z)}dZ
=
\frac{1}{2\pi\ii}
\int_{Z_0}^{Z_0+L_2}
\left[
\frac{H'(Z)}{H(Z)}
-
\frac{H'(Z+L_1)}{H(Z+L_1)}
\right]dZ
=
\frac{1}{2\pi\ii}
\int_{Z_0}^{Z_0+L_2}
\frac{2\pi\ii m}{L_2}\,dZ
=
m.
\label{eq:C-com-zero-count}
}
Since $H$ is holomorphic, it has no poles.
Thus, any linear combination that is not identically
zero has exactly $m$ zeros on the torus, counted
with multiplicity.

We now show that our lattice produces more than $m$ distinct values of $Z$ that leads to more than $m$ zeros.
Fix $N_e-1$ electrons at distinct sites and let the
remaining electron occupy any unoccupied site.
There are
\eq{
2N-(N_e-1)
=
(2m-1)N_e+1
>m
\label{eq:C-number-of-available-COM-values}
}
possible positions for this electron.
Since the other coordinates are fixed, each different
position gives a different value of
$Z=\sum_i z_i$, modulo $L_1\dsZ+L_2\dsZ$. The assumed relation
\cref{eq:com-linear-relation-at-discretized-Z}
requires $H(Z)=0$ at all these values, which is more than $m$
distinct values.

This is impossible if $H$ is not identically zero on $\dsC$, by \cref{eq:C-com-zero-count}.
Therefore, $H$ must be identically zero on $\dsC$:
\eq{
H(Z)
=
\sum_{\sigma=1}^{m}c_\sigma H_\sigma(Z)
\equiv0,
\qquad \forall Z\in\dsC.
\label{eq:C-com-global-linear-relation}
}
So we have turned the assumed relation on the
finite lattice in \cref{eq:com-linear-relation-at-discretized-Z} into an identity valid for every
complex $Z$ in \cref{eq:C-com-global-linear-relation}.

We have shown in \cref{eq:C-com-global-linear-relation}
that $H(Z)=0$ for every complex $Z$, and we now show that this requires $c_\sigma=0$
for every $\sigma$.
First, we show that the different $H_\sigma(Z)$
acquire different factors under the same shift
$Z\to Z+L_2/m$. From the definition of the center-of-mass zeros in
\cref{eq:K-Z-sector-definition}, we have
\eq{
Z_\sigma^{\gamma+1}-Z_\sigma^\gamma
=
\frac{L_2}{m},
\qquad \gamma=1,\ldots,m-1.
\label{eq:spacing_of_COM_zeros}
}
Using the spacing in \cref{eq:spacing_of_COM_zeros}
and the theta-function identity in
\cref{eq:direct-theta1-shift-one}, we obtain
\eqa{
H_\sigma\left(Z+\frac{L_2}{m}\right)
&=
e^{\ii K_\sigma(Z+L_2/m)}
\prod_{\gamma=1}^{m}
\vartheta_1\left(
\frac{Z+L_2/m-Z_\sigma^\gamma}{L_2}
\Big|\tau
\right)
\\
&=
e^{\ii K_\sigma(Z+L_2/m)}
\vartheta_1\left(
\frac{Z-Z_\sigma^m}{L_2}+1
\Big|\tau
\right)
\prod_{\gamma=1}^{m-1}
\vartheta_1\left(
\frac{Z-Z_\sigma^\gamma}{L_2}
\Big|\tau
\right)
\\
&=
-e^{\ii K_\sigma L_2/m}
e^{\ii K_\sigma Z}
\prod_{\gamma=1}^{m}
\vartheta_1\left(
\frac{Z-Z_\sigma^\gamma}{L_2}
\Big|\tau
\right) =
-e^{\ii K_\sigma L_2/m}H_\sigma(Z)
\\
&=
\mu_\sigma H_\sigma(Z),
\label{eq:Hsigma-fractional-shift}
}
where
$\mu_\sigma=-e^{\ii K_\sigma L_2/m}
=-e^{\ii K_1L_2/m}e^{2\pi\ii(\sigma-1)/m}$. These factors are distinct for different $\sigma=1,\ldots,m$.

Now recall that $H(Z)=0$ for every complex $Z$
by \cref{eq:C-com-global-linear-relation}.
Therefore, it must also hold at
$Z+qL_2/m$ for any integer $q$, and then using $H(Z)=\sum_{\sigma=1}^{m}c_\sigma H_\sigma(Z)$ from \cref{eq:combined-center-of-mass-section}, and using \cref{eq:Hsigma-fractional-shift} repeatedly, we get
\eq{
0 =H\left(Z+\frac{qL_2}{m}\right) =\sum_{\sigma=1}^{m}
c_\sigma H_\sigma\left(Z+\frac{qL_2}{m}\right) =\sum_{\sigma=1}^{m}
c_\sigma\mu_\sigma^q H_\sigma(Z),
\quad q=0,\ldots,m-1.
}
These are $m$ linear equations for the $m$
quantities $c_\sigma H_\sigma(Z)$.
Writing them in matrix form gives
\eq{
\begin{pmatrix}
1&1&\cdots&1\\
\mu_1&\mu_2&\cdots&\mu_m\\
\vdots&\vdots&&\vdots\\
\mu_1^{m-1}&\mu_2^{m-1}&\cdots&\mu_m^{m-1}
\end{pmatrix}
\begin{pmatrix}
c_1H_1(Z)\\
c_2H_2(Z)\\
\vdots\\
c_mH_m(Z)
\end{pmatrix}
=
0.
\label{eq:com-Vandermonde-system}
}
Since all $\mu_\sigma$ are distinct, this
Vandermonde matrix has a nonzero determinant,
$\prod_{1\leq\sigma<\sigma'\leq m}
(\mu_{\sigma'}-\mu_\sigma)\neq0$,
with determinant $1$ when $m=1$.
Therefore, the matrix is invertible, which implies
$c_\sigma H_\sigma(Z)=0$ for every $\sigma$
and every complex $Z$.
However, each $H_\sigma(Z)$ is not identically zero
by its explicit definition in
\cref{eq:discretized-com-factor}.
Hence,
\eq{
c_\sigma=0,
\qquad
\sigma=1,\ldots,m.
}
This contradicts our assumption that the coefficients are not all zero. This proves linear independence.

Linear independence also implies that each discretized state is nonzero.

\subsection{Many-body Chern Number}
\label{sec:C-many-body-chern}

We now compute analytically the many-body Chern number \cite{Niu1985PRB,Qi2006PRB,Lin2023chern} of the full
$m$-dimensional manifold spanned by $\ket{\Psi^{m,\sigma}}$ with
$\sigma=1,\ldots,m$.
First we note that the states $\ket{\Psi^{m,\sigma}}$
are not necessarily orthonormal, though they are linearly independent. So we need to prepare an orthonormal basis.
To do so, define the $G$ matrix to find their overlaps
\eq{
\left[G(\phi_1,\phi_2)\right]_{\sigma_1\sigma_2}
=
\braket{\Psi^{m,\sigma_1}}{\Psi^{m,\sigma_2}}.
\label{eq:many-body-gram-matrix}
}
Explicitly, using the finite-torus state convention in
\cref{eq:C-discretized-many-body-state},
\eq{
\left[G(\phi_1,\phi_2)\right]_{\sigma_1\sigma_2}
=
\sum_{\bsl r_1\in\Lambda}\cdots
\sum_{\bsl r_{N_e}\in\Lambda}
\left[\Psi^{m,\sigma_1}(\{\bsl r_i\})\right]^*
\Psi^{m,\sigma_2}(\{\bsl r_i\}),
\label{eq:many-body-gram-matrix-finite}
}
where both amplitudes are evaluated at the same phases for twisted boundary conditions $(\phi_1,\phi_2)$.
Thus $G(\phi_1,\phi_2)$ is finite and depends smoothly on
$\phi_1,\phi_2$.
The linear independence further implies that
$G(\phi_1,\phi_2)$ is positive definite, so its positive inverse
square root $G^{-1/2}(\phi_1,\phi_2)$ exists.
Combined with the smoothness of $G$, this implies that
$G^{-1/2}$ also depends smoothly on $(\phi_1,\phi_2)$. Note that multiplying our states by this matrix $G^{-1/2}$  will make their overlap matrix equal to the identity.

Before computing the many-body Chern number, we first remove the twisted-boundary phases from the many-body states.
To do this, we introduce the total position operator:
\eq{
\hat{\bsl X}
=
\sum_{\bsl r\in\Lambda}
\bsl r\,c^\dagger_{\bsl r}c_{\bsl r}.
\label{eq:many-body-total-position}
}
On a configuration
$c^\dagger_{\bsl r_1}\cdots
c^\dagger_{\bsl r_{N_e}}\ket{0}$, it has eigenvalue
$\bsl X=\sum_{j=1}^{N_e}\bsl r_j.$ We choose the twist momentum
\eq{
\bsl k_{\phi}(\phi_1,\phi_2)
=
\frac{\phi_1}{2\pi N_1}\bsl b_1
+
\frac{\phi_2}{2\pi N_2}\bsl b_2,
\label{eq:many-body-twist-momentum}
}
so that it obeys $\bsl k_{\phi}\cdot\bsl L_i
=
\phi_i$, with $i=1,2.$
Consequently, $e^{\ii\bsl k_{\phi}\cdot(\bsl X+\bsl L_i)}
= e^{\ii\phi_i} e^{\ii\bsl k_{\phi}\cdot\bsl X}.$
Thus the factor $e^{\ii\bsl k_{\phi}\cdot\bsl X}$
carries precisely the twisted-boundary phases. Removing it gives
amplitudes that are periodic under full-torus translations. And in complex notation,
\eq{
\bsl k_{\phi}\cdot\bsl X
=
\frac{1}{2}
\left(
k_{\phi}\bar Z
+
\bar k_{\phi}Z
\right),
}

Collecting the $m$ states:
\eq{
\ket{\Psi^m}
=
\left(
\ket{\Psi^{m,1}},
\ldots,
\ket{\Psi^{m,m}}
\right)
}
and using $G^{-1/2}$ to orthonormalize the states and $e^{-\ii\bsl k_{\phi}\cdot\hat{\bsl X}}$ to remove the boundary-twist phases, we define
\eq{
\ket{U(\phi_1,\phi_2)}
=
e^{-\ii\bsl k_{\phi}\cdot\hat{\bsl X}}
\ket{\Psi^m}G^{-1/2},
\qquad
\braket{U}{U}=I_m.
\label{eq:many-body-orthonormal-frame}
}
We have thus obtained an orthonormal basis with periodic spatial boundary conditions. 
Here $\ket{U}$ denotes the matrix of orthonormal column states,
and $\bra{U}$ denotes its Hermitian adjoint.

We now use this basis to define the Berry connection:
\eq{
\mathcal A_\mu
=
\ii\,\Tr\left[
\bra{U(\phi_1,\phi_2)}
\partial_{\phi_\mu}
\ket{U(\phi_1,\phi_2)}
\right],
\qquad \mu=1,2.
\label{eq:many-body-berry-connection-smooth-part}
}
Using the coordinate order $(\phi_1,\phi_2)$, we define the Berry curvature as
\eq{
\mathcal F_{12}
=
\partial_{\phi_1}\mathcal A_2
-
\partial_{\phi_2}\mathcal A_1.
\label{eq:many-body-covariant-curvature}
}
The many-body Chern number is therefore
\eq{
C_{\mathrm{MB}} =
\frac{1}{2\pi}
\int_0^{2\pi}d\phi_1
\int_0^{2\pi}d\phi_2\,
\mathcal F_{12} =
\frac{1}{2\pi}
\int_0^{2\pi}d\phi_2\,
\left[
\mathcal A_2(2\pi,\phi_2)-\mathcal A_2(0,\phi_2)
\right] +
\frac{1}{2\pi}
\int_0^{2\pi}d\phi_1
\left[
\mathcal A_1(\phi_1,0)-\mathcal A_1(\phi_1,2\pi)
\right] \ ,
\label{eq:many-body-covariant-chern-definition}
}
where we have used the fact that $\mathcal A_\mu$
is defined in a smooth gauge.

We now evaluate \cref{eq:many-body-covariant-chern-definition}
by calculating the two differences
$\mathcal A_2(2\pi,\phi_2)-\mathcal A_2(0,\phi_2)$ and
$\mathcal A_1(\phi_1,0)-\mathcal A_1(\phi_1,2\pi)$.
We first relate the states at the opposite boundaries,
and then substitute these relations into the Berry connection.

To determine how the orthonormal states $|U\rangle$ transform under a $2\pi$ change of either boundary phase, we use \cref{eq:many-body-orthonormal-frame} and examine how its three factors, $|\Psi^m\rangle$, $G^{-1/2}$, and $e^{-\ii\bsl k_{\phi}\cdot\hat{\bsl X}}$, transform.

We first determine how the original many-body states transform when either boundary twist increases by $2\pi$.
The Jastrow and site factors in
\cref{eq:ideal-fci-holomorphic-factor,eq:C-discretized-amplitude}
are independent of the twists.
Thus, we only need to calculate how the center-of-mass factor changes, using $K_1(\phi_1,\phi_2)$, $Z_1(\phi_1,\phi_2)$ (\cref{eq:chern-general-K1-Z1-lift}). 
For boundaries $\phi_1=0$ and $\phi_1=2\pi$,
\cref{eq:chern-general-K1-Z1-lift,eq:K-Z-sector-definition} give
\eqa{
K_\sigma(2\pi,\phi_2)
&=K_\sigma(0,\phi_2),
\\
Z_1(2\pi,\phi_2)
&=Z_1(0,\phi_2)+1,
\\
Z_\sigma^\gamma(2\pi,\phi_2)
&=Z_\sigma^\gamma(0,\phi_2)+\frac{L_2}{m}.
}
Using the spacing of the zeros and
\cref{eq:direct-theta1-shift-one}, and the definition of $H_\sigma$ from \cref{eq:discretized-com-factor} we can obtain
\eqa{
H_\sigma(Z;2\pi,\phi_2)
&=e^{\ii K_\sigma(0,\phi_2)Z}
\prod_{\gamma=1}^{m}
\vartheta_1\left(
\frac{Z-Z_\sigma^\gamma(0,\phi_2)}{L_2}-\frac1m
\Big|\tau\right)\\
&=-e^{\ii K_\sigma(0,\phi_2)Z}
\prod_{\gamma=1}^{m}
\vartheta_1\left(
\frac{Z-Z_\sigma^\gamma(0,\phi_2)}{L_2}
\Big|\tau\right) =-H_\sigma(Z;0,\phi_2).
}
The remaining wavefunction factors are unchanged, so
$\ket{\Psi^{m,\sigma}(2\pi,\phi_2)}
=-\ket{\Psi^{m,\sigma}(0,\phi_2)}$.
Collecting the $m$ states gives
\eq{
\ket{\Psi^m(2\pi,\phi_2)}
=\ket{\Psi^m(0,\phi_2)}S_1,
\qquad S_1=-I_m.
\label{eq:many-body-S1}
}

For the boundaries $\phi_2=0$ and $\phi_2=2\pi$,
\cref{eq:chern-general-K1-Z1-lift,eq:K-Z-sector-definition} gives
\eqa{
K_\sigma(\phi_1,2\pi)
&=K_\sigma(\phi_1,0)+\frac{2\pi}{L_2},
\\
Z_1(\phi_1,2\pi)
&=Z_1(\phi_1,0)-\tau,
\\
Z_\sigma^\gamma(\phi_1,2\pi)
&
=Z_\sigma^\gamma(\phi_1,0)-\frac{L_1}{m}.
}
For $\sigma=1,\ldots,m-1$, these are precisely the
parameters of sector $\sigma+1$ at $(\phi_1,0)$ by \cref{eq:K-Z-sector-definition}:
\eq{
K_\sigma(\phi_1,2\pi)=K_{\sigma+1}(\phi_1,0),
\qquad
Z_\sigma^\gamma(\phi_1,2\pi)
=Z_{\sigma+1}^\gamma(\phi_1,0),
}
and using \cref{eq:discretized-com-factor} gives
\eq{
H_\sigma(Z;\phi_1,2\pi)=H_{\sigma+1}(Z;\phi_1,0),
\qquad \sigma=1,\ldots,m-1.
\label{eq:C_138}
}
For the last sector $\sigma=m$, by \cref{eq:chern-general-K1-Z1-lift,eq:K-Z-sector-definition},
\eq{
K_m(\phi_1,2\pi)=K_1(\phi_1,0)+\frac{2\pi m}{L_2},
\qquad
Z_m^\gamma(\phi_1,2\pi)=Z_1^\gamma(\phi_1,0)-L_1,
}
and by \cref{eq:direct-theta1-shift-tau} and \cref{eq:discretized-com-factor}, we obtain
\eqa{
H_m(Z;\phi_1,2\pi)
&=e^{\ii[K_1(\phi_1,0)+2\pi m/L_2]Z}
\prod_{\gamma=1}^{m}
\vartheta_1\left(
\frac{Z-Z_1^\gamma(\phi_1,0)}{L_2}+\tau
\Big|\tau\right)
\\
&=(-1)^m e^{-\ii\pi m\tau}
 e^{2\pi\ii mZ/L_2}e^{-2\pi\ii mZ/L_2}
 e^{\frac{2\pi\ii}{L_2}\sum_{\gamma=1}^{m}
 Z_1^\gamma(\phi_1,0)}H_1(Z;\phi_1,0)
\\
&=(-1)^m e^{-\ii\pi m\tau}
 e^{2\pi\ii[Z_1(\phi_1,0)+(m-1)/2]}
 H_1(Z;\phi_1,0)
\\
&=-e^{-\ii\pi m\tau+2\pi\ii Z_1(\phi_1,0)}
H_1(Z;\phi_1,0).
\label{eq:C_140}
}
Here we used
$\sum_\gamma Z_1^\gamma(\phi_1,0)
=[Z_1(\phi_1,0)+(m-1)/2]L_2$
from \cref{eq:sum_Z}. Consequently, the full states obey (combining \cref{eq:C_138} and \cref{eq:C_140})
\eqa{
\ket{\Psi^{m,\sigma}(\phi_1,2\pi)}
&=\ket{\Psi^{m,\sigma+1}(\phi_1,0)},
\qquad \sigma=1,\ldots,m-1,
\\
\ket{\Psi^{m,m}(\phi_1,2\pi)}
&=q(\phi_1)\ket{\Psi^{m,1}(\phi_1,0)},
\label{eq:C_141}
}
where
\eqa{
q(\phi_1) =-e^{-\ii\pi m\tau+2\pi\ii Z_1(\phi_1,0)} =-e^{-\ii\pi m\tau+2\pi\ii[Z_1^{(0)}+\phi_1/(2\pi)]} =-e^{-\ii\pi m\tau+2\pi\ii Z_1^{(0)}}e^{\ii\phi_1}.
\label{eq:many-body-wraparound-factor}
}
For $m>1$, collecting \cref{eq:C_141} gives
\eq{
\ket{\Psi^m(\phi_1,2\pi)}=\left(
\ket{\Psi^{m,2}(\phi_1,0)},\ldots,
\ket{\Psi^{m,m}(\phi_1,0)},
q(\phi_1)\ket{\Psi^{m,1}(\phi_1,0)}
\right) =\ket{\Psi^m(\phi_1,0)}S_2(\phi_1),
}
where
\eq{
S_2(\phi_1)
=\mat{
0&0&\cdots&0&q(\phi_1)\\
1&0&\cdots&0&0\\
0&1&\cdots&0&0\\
\vdots&&\ddots&&\vdots\\
0&0&\cdots&1&0
}.
\label{eq:many-body-S2}
}
For $m=1$, there is only one sector,
so $S_2(\phi_1)=\mat{q(\phi_1)}$.
We can also calculate the determinant of $S_1$ and $S_2$ for later use: for $S_1=-I_m$,
\eq{
\det S_1=\det(-I_m)=(-1)^m.
}
For $m>1$, expanding $\det S_2$ along its last column gives
\eqa{
\det S_2(\phi_1) =(-1)^{1+m}q(\phi_1)\det I_{m-1}=(-1)^{m-1}q(\phi_1)=(-1)^m e^{-\ii\pi m\tau+2\pi\ii Z_1^{(0)}}
e^{\ii\phi_1}.
\label{eq:many-body-S2-determinant}
}
The same final formula holds directly for $m=1$.

We have obtained \cref{eq:many-body-S1,eq:C_141}, which
show how the original $m$ manybody states transform
when either boundary phase increases by $2\pi$.

The Berry connection uses $\ket{U}$ rather than
$\ket{\Psi^m}$ (see \cref{eq:many-body-orthonormal-frame}). We therefore substitute the state transformations
above into the definitions of $G$ and $\ket{U}$.
From \cref{eq:many-body-gram-matrix},
\eqa{
G(2\pi,\phi_2)
&=\langle\Psi^m(2\pi,\phi_2)|\Psi^m(2\pi,\phi_2)\rangle =S_1^\dagger
\langle\Psi^m(0,\phi_2)|\Psi^m(0,\phi_2)\rangle S_1=S_1^\dagger G(0,\phi_2)S_1
=G(0,\phi_2),
\\
G(\phi_1,2\pi)
&=\langle\Psi^m(\phi_1,2\pi)|\Psi^m(\phi_1,2\pi)\rangle=S_2(\phi_1)^\dagger
\langle\Psi^m(\phi_1,0)|\Psi^m(\phi_1,0)\rangle S_2(\phi_1)=S_2(\phi_1)^\dagger G(\phi_1,0)S_2(\phi_1).
\label{eq:many-body-gram-edge-relations}
}
Also, \cref{eq:many-body-twist-momentum} gives
\eqa{
\bsl k_{\phi}(2\pi,\phi_2)
&=\frac{\bsl b_1}{N_1}+\frac{\phi_2}{2\pi N_2}\bsl b_2
=\bsl k_{\phi}(0,\phi_2)+\frac{\bsl b_1}{N_1},
\\
\bsl k_{\phi}(\phi_1,2\pi)
&=\frac{\phi_1}{2\pi N_1}\bsl b_1+\frac{\bsl b_2}{N_2}
=\bsl k_{\phi}(\phi_1,0)+\frac{\bsl b_2}{N_2}.
\label{eq:howG_transform}
}
Substituting \cref{eq:many-body-gram-edge-relations} and \cref{eq:howG_transform}
 into \cref{eq:many-body-orthonormal-frame},
\eqa{
\ket{U(2\pi,\phi_2)}
&=e^{-\ii\bsl k_{\phi}(2\pi,\phi_2)\cdot\hat{\bsl X}}
\ket{\Psi^m(2\pi,\phi_2)}G^{-1/2}(2\pi,\phi_2)=e^{-\ii\frac{\bsl b_1}{N_1}\cdot\hat{\bsl X}}
 e^{-\ii\bsl k_{\phi}(0,\phi_2)\cdot\hat{\bsl X}}
\ket{\Psi^m(0,\phi_2)}S_1G^{-1/2}(2\pi,\phi_2)
\\
&=e^{-\ii\frac{\bsl b_1}{N_1}\cdot\hat{\bsl X}}
\ket{U(0,\phi_2)}G^{1/2}(0,\phi_2)
S_1G^{-1/2}(2\pi,\phi_2)=e^{-\ii\frac{\bsl b_1}{N_1}\cdot\hat{\bsl X}}
\ket{U(0,\phi_2)}V_1(\phi_2).
\label{eq:many-body-U-edge1}
}
The third equality uses
$e^{-\ii\bsl k_{\phi}\cdot\hat{\bsl X}}\ket{\Psi^m}
=\ket{U}G^{1/2}$.
For the other pair of boundaries,
\eqa{
\ket{U(\phi_1,2\pi)}
&=e^{-\ii\bsl k_{\phi}(\phi_1,2\pi)\cdot\hat{\bsl X}}
\ket{\Psi^m(\phi_1,2\pi)}G^{-1/2}(\phi_1,2\pi)=e^{-\ii\frac{\bsl b_2}{N_2}\cdot\hat{\bsl X}}
 e^{-\ii\bsl k_{\phi}(\phi_1,0)\cdot\hat{\bsl X}}
\ket{\Psi^m(\phi_1,0)}S_2(\phi_1)G^{-1/2}(\phi_1,2\pi)
\\
&=e^{-\ii\frac{\bsl b_2}{N_2}\cdot\hat{\bsl X}}
\ket{U(\phi_1,0)}G^{1/2}(\phi_1,0)
S_2(\phi_1)G^{-1/2}(\phi_1,2\pi)=e^{-\ii\frac{\bsl b_2}{N_2}\cdot\hat{\bsl X}}
\ket{U(\phi_1,0)}V_2(\phi_1),
\label{eq:many-body-U-edge2}
}
where we defined
\eqa{
V_1(\phi_2)
&=G^{1/2}(0,\phi_2)S_1G^{-1/2}(2\pi,\phi_2),
\\
V_2(\phi_1)
&=G^{1/2}(\phi_1,0)S_2(\phi_1)G^{-1/2}(\phi_1,2\pi).
\label{eq:many-body-V-definitions}
}
We thus showed how the orthonormal states $|U\rangle$ transform between opposite boundaries in $(\phi_1,\phi_2)$, through the matrices $V_1$ and $V_2$ and the corresponding phase factors. These matrices are unitary: for $V_1$, \cref{eq:many-body-gram-edge-relations} gives
\eq{
V_1(\phi_2)
=G^{1/2}(0,\phi_2)(-I_m)G^{-1/2}(0,\phi_2)
=-I_m,
\qquad V_1^\dagger V_1=I_m.
}
For $V_2$, the positive square roots of $G$ are Hermitian,
so
\eqa{
V_2(\phi_1)^\dagger V_2(\phi_1)
&=G^{-1/2}(\phi_1,2\pi)S_2(\phi_1)^\dagger
G^{1/2}(\phi_1,0)G^{1/2}(\phi_1,0)
S_2(\phi_1)G^{-1/2}(\phi_1,2\pi)
\\
&=G^{-1/2}(\phi_1,2\pi)
\left[S_2(\phi_1)^\dagger G(\phi_1,0)S_2(\phi_1)\right]
G^{-1/2}(\phi_1,2\pi)
\\
&=G^{-1/2}(\phi_1,2\pi)G(\phi_1,2\pi)
G^{-1/2}(\phi_1,2\pi)=I_m.
}
Since $V_1$ and $V_2$ are square matrices, we also have
$V_iV_i^\dagger=I_m$. 

Now we find the berry connections at opposite boundaries. We now substitute
\cref{eq:many-body-U-edge1,eq:many-body-U-edge2}
into the Berry-connection definition
\cref{eq:many-body-berry-connection-smooth-part}.
The operators
$e^{-\ii(\bsl b_i/N_i)\cdot\hat{\bsl X}}$
are independent of the twists, so the derivatives below
act only on $\ket{U}$ and $V_i$.
For the first pair of boundaries, the product rule gives
\eqa{
\mathcal A_2(2\pi,\phi_2)
&=\ii\Tr\left[
\bra{U(2\pi,\phi_2)}\partial_{\phi_2}\ket{U(2\pi,\phi_2)}
\right]
\\
&=\ii\Tr\left[
V_1(\phi_2)^\dagger\bra{U(0,\phi_2)}
e^{+\ii\frac{\bsl b_1}{N_1}\cdot\hat{\bsl X}}
\partial_{\phi_2}\left(
e^{-\ii\frac{\bsl b_1}{N_1}\cdot\hat{\bsl X}}
\ket{U(0,\phi_2)}V_1(\phi_2)
\right)\right]
\\
&=\ii\Tr\left[
V_1(\phi_2)^\dagger\bra{U(0,\phi_2)}
\partial_{\phi_2}\ket{U(0,\phi_2)}V_1(\phi_2)
\right]+\ii\Tr\left[
V_1(\phi_2)^\dagger
\bra{U(0,\phi_2)}\ket{U(0,\phi_2)}
\partial_{\phi_2}V_1(\phi_2)
\right]
\\
&=\ii\Tr\left[
\bra{U(0,\phi_2)}\partial_{\phi_2}\ket{U(0,\phi_2)}
V_1(\phi_2)V_1(\phi_2)^\dagger
\right]
+\ii\Tr\left[V_1(\phi_2)^\dagger
\partial_{\phi_2}V_1(\phi_2)\right]
\\
&=\mathcal A_2(0,\phi_2)
+\ii\Tr\left[V_1(\phi_2)^\dagger
\partial_{\phi_2}V_1(\phi_2)\right].
}
We used $\langle U|U\rangle=I_m$
and $V_1V_1^\dagger=I_m$.
For the second pair of boundaries, the same substitution gives
\eqa{
\mathcal A_1(\phi_1,2\pi)
&=\ii\Tr\left[
\bra{U(\phi_1,2\pi)}\partial_{\phi_1}\ket{U(\phi_1,2\pi)}
\right]
\\
&=\ii\Tr\left[
V_2(\phi_1)^\dagger\bra{U(\phi_1,0)}
e^{+\ii\frac{\bsl b_2}{N_2}\cdot\hat{\bsl X}}
\partial_{\phi_1}\left(
e^{-\ii\frac{\bsl b_2}{N_2}\cdot\hat{\bsl X}}
\ket{U(\phi_1,0)}V_2(\phi_1)
\right)\right]
\\
&=\ii\Tr\left[
V_2(\phi_1)^\dagger\bra{U(\phi_1,0)}
\partial_{\phi_1}\ket{U(\phi_1,0)}V_2(\phi_1)
\right]+\ii\Tr\left[
V_2(\phi_1)^\dagger
\bra{U(\phi_1,0)}\ket{U(\phi_1,0)}
\partial_{\phi_1}V_2(\phi_1)
\right]
\\
&=\ii\Tr\left[
\bra{U(\phi_1,0)}\partial_{\phi_1}\ket{U(\phi_1,0)}
V_2(\phi_1)V_2(\phi_1)^\dagger
\right]
+\ii\Tr\left[V_2(\phi_1)^\dagger
\partial_{\phi_1}V_2(\phi_1)\right]
\\
&=\mathcal A_1(\phi_1,0)
+\ii\Tr\left[V_2(\phi_1)^\dagger
\partial_{\phi_1}V_2(\phi_1)\right].
}
Thus, the two differences needed in
\cref{eq:many-body-covariant-chern-definition} are
\eqa{
\mathcal A_2(2\pi,\phi_2)-\mathcal A_2(0,\phi_2)
&=\ii\Tr\left[V_1(\phi_2)^\dagger
\partial_{\phi_2}V_1(\phi_2)\right],
\\
\mathcal A_1(\phi_1,0)-\mathcal A_1(\phi_1,2\pi)
&=-\ii\Tr\left[V_2(\phi_1)^\dagger
\partial_{\phi_1}V_2(\phi_1)\right].
\label{eq:many-body-berry-edge-differences}
}

We now evaluate the Chern number.
For a differentiable invertible matrix, Jacobi's determinant
identity \cite{Abadir_Magnus_2005} gives the logarithmic derivative of its determinant.
Since $V_i^{-1}=V_i^\dagger$, we have
\eqa{
\partial_{\phi_2}\log\det V_1(\phi_2)
&=\frac{\partial_{\phi_2}\det V_1(\phi_2)}{\det V_1(\phi_2)}
=\Tr\left[V_1(\phi_2)^{-1}\partial_{\phi_2}V_1(\phi_2)\right]
=\Tr\left[V_1(\phi_2)^\dagger\partial_{\phi_2}V_1(\phi_2)\right],
\\
\partial_{\phi_1}\log\det V_2(\phi_1)
&=\frac{\partial_{\phi_1}\det V_2(\phi_1)}{\det V_2(\phi_1)}
=\Tr\left[V_2(\phi_1)^{-1}\partial_{\phi_1}V_2(\phi_1)\right]
=\Tr\left[V_2(\phi_1)^\dagger\partial_{\phi_1}V_2(\phi_1)\right].
\label{eq:Jacobi_det_identity}
}
Here the logarithmic derivative means
$(\det V_i)^{-1}\partial(\det V_i)$. Substituting \cref{eq:many-body-berry-edge-differences} into
\cref{eq:many-body-covariant-chern-definition} and using \cref{eq:Jacobi_det_identity} gives
\eqa{
C_{\mathrm{MB}}
&=\frac{\ii}{2\pi}\int_0^{2\pi}d\phi_2\,
\Tr\left[V_1(\phi_2)^\dagger\partial_{\phi_2}V_1(\phi_2)\right]
-\frac{\ii}{2\pi}\int_0^{2\pi}d\phi_1\,
\Tr\left[V_2(\phi_1)^\dagger\partial_{\phi_1}V_2(\phi_1)\right]
\\
&=\frac{\ii}{2\pi}\int_0^{2\pi}d\phi_2\,
\partial_{\phi_2}\log\det V_1(\phi_2)
-\frac{\ii}{2\pi}\int_0^{2\pi}d\phi_1\,
\partial_{\phi_1}\log\det V_2(\phi_1).
\label{eq:many-body-chern-from-transition-matrices}
}
It remains to evaluate the two scalar determinants.
First, $V_1=-I_m$ gives
\eq{
\det V_1(\phi_2)=(-1)^m.
}
For $V_2$, taking determinants of
\cref{eq:many-body-gram-edge-relations} gives
\eq{
\det G(\phi_1,2\pi) =\det\left[S_2(\phi_1)^\dagger
G(\phi_1,0)S_2(\phi_1)\right]
=[\det S_2(\phi_1)]^*\det G(\phi_1,0)\det S_2(\phi_1)
=|\det S_2(\phi_1)|^2\det G(\phi_1,0).
}
Because $G$ is positive definite, its square roots have positive
determinants. Hence, \cref{eq:many-body-V-definitions} gives
\eqa{
\det V_2(\phi_1)
&=\det G^{1/2}(\phi_1,0)\,
\det S_2(\phi_1)\,
\det G^{-1/2}(\phi_1,2\pi)=\frac{\sqrt{\det G(\phi_1,0)}\,\det S_2(\phi_1)}
{\sqrt{\det G(\phi_1,2\pi)}}
\\
&=\frac{\sqrt{\det G(\phi_1,0)}\,\det S_2(\phi_1)}
{|\det S_2(\phi_1)|\sqrt{\det G(\phi_1,0)}}=\frac{\det S_2(\phi_1)}{|\det S_2(\phi_1)|}=\frac{(-1)^m e^{-\ii\pi m\tau+2\pi\ii Z_1^{(0)}}}
{|e^{-\ii\pi m\tau+2\pi\ii Z_1^{(0)}}|}
e^{\ii\phi_1},
}
where the last equality uses
\cref{eq:many-body-S2-determinant} and
$|e^{\ii\phi_1}|=1$.
All factors except $e^{\ii\phi_1}$ are independent of the twists and by \cref{eq:Jacobi_det_identity},
\eqa{
\partial_{\phi_2}\log\det V_1(\phi_2)
&=\frac{\partial_{\phi_2}(-1)^m}{(-1)^m}=0,
\\
\partial_{\phi_1}\log\det V_2(\phi_1)
&=\frac{\partial_{\phi_1}e^{\ii\phi_1}}{e^{\ii\phi_1}}
=\frac{\ii e^{\ii\phi_1}}{e^{\ii\phi_1}}=\ii.
}
Finally, substitution into
\cref{eq:many-body-chern-from-transition-matrices} gives
\eqa{
C_{\mathrm{MB}}
&=\frac{\ii}{2\pi}\int_0^{2\pi}0\,d\phi_2
-\frac{\ii}{2\pi}\int_0^{2\pi}\ii\,d\phi_1=-\frac{\ii}{2\pi}(2\pi\ii)
=1.
}
Thus, the $m$-dimensional manifold spanned by the constructed lattice many-body states $\{|\Psi^{m,\sigma}\rangle\}_{\sigma=1}^{m}$ has total many-body Chern number $C_{\mathrm{MB}}=1$.

\subsection{The Fractional Chern State is Contained in the Fock Space of an Ideal Chern Band}
\label{sec:C-band-containment}

We now expand the lattice states in the ideal-band basis. The full-period calculation in
\cref{sec:C-boundaries-translations} gives the holomorphic boundary
conditions
\eqa{
F^\sigma(\ldots,z_j+L_2,\ldots)
&=
e^{\ii\phi_2}F^\sigma(\ldots,z_j,\ldots),
\\
F^\sigma(\ldots,z_j+L_1,\ldots)
&=
e^{\ii\phi_1}
e^{-\ii\pi N\tau}
e^{-2\pi\ii Nz_j/L_2}
F^\sigma(\ldots,z_j,\ldots).
\label{eq:F-twisted-holomorphic-boundary}
}
These boundary conditions agree with
\cref{eq:g_boundary} for the functions in
\cref{eq:A-gk-definition} at the allowed momenta in
\cref{eq:A-allowed-momenta}.
Using \cref{eq:f-gk-expansion} successively for each particle
coordinate, we can expand $F^\sigma$ as
\eq{
F^\sigma(z_1,\ldots,z_{N_e})
=
\sum_{\bsl k_1,\ldots,\bsl k_{N_e}
\in B_{\phi_1,\phi_2}^{N_1,N_2}}
A_{\bsl k_1,\ldots,\bsl k_{N_e}}
g_{\bsl k_1}(z_1)\cdots
g_{\bsl k_{N_e}}(z_{N_e}).
\label{eq:F_expansion}
}
Here
\eq{
 B_{\phi_1,\phi_2}^{N_1,N_2}
 \equiv\left\{\bsl k_{\kappa_1,\kappa_2}^{\phi_1,\phi_2} \middle|\kappa_j\in\dsZ,\ 0\leq\kappa_j<N_j\right\},
 \label{eq:allowed+momenta}
}
with $\bsl k_{\kappa_1,\kappa_2}^{\phi_1,\phi_2}$ defined in
\cref{eq:A-allowed-momenta}.

Recall that we discretize \cref{eq:F_expansion}. To capture the physical meaning, we define
\eq{
\psi_{\bsl k}(\bsl r)
=
\frac{1}{\mathcal N_{\bsl k}}
e^{\pi\mathcal B(\bsl r)/(2\Omega)}
\chi(\bsl r)g_{\bsl k}(z),
\qquad
\mathcal N_{\bsl k}^2
=
\sum_{\bsl r\in\Lambda}
\left|
e^{\pi\mathcal B(\bsl r)/(2\Omega)}
\chi(\bsl r)g_{\bsl k}(z)
\right|^2.
\label{eq:C-normalized-band-orbital}
}
We choose $\mathcal N_{\bsl k}>0$.
For fixed $\bsl k$, $g_{\bsl k}(z)$ with continuous $z$ has exactly one simple zero
modulo $a_1\dsZ+a_2\dsZ$.
Since the two physical sites in a unit cell are inequivalent,
$\delta_B-\delta_A\notin a_1\dsZ+a_2\dsZ$, the values
$g_{\bsl k}(\delta_A)$ and $g_{\bsl k}(\delta_B)$ cannot both
vanish.
The remaining site factors are nonzero, so
$\mathcal N_{\bsl k}$ is strictly positive for every real
$\bsl k$. Using
\eq{
e^{\pi\mathcal B(\bsl r)/(2\Omega)}
\chi(\bsl r)g_{\bsl k}(z)
=
\mathcal N_{\bsl k}\psi_{\bsl k}(\bsl r),
}
each term in \cref{eq:F_expansion}, after including the
one-particle factors, becomes a product of the Bloch wavefunctions
$\psi_{\bsl k}$.
Therefore the full discretized many-body amplitude $
\Psi^{m,\sigma}
=
F^\sigma
\prod_i
e^{\pi \mathcal B(\bsl r_i)/(2\Omega)}
\chi(\bsl r_i)$ can be written as
\eq{
\Psi^{m,\sigma}(\{\bsl r_i\})
=
\sum_{\bsl k_1,\ldots,\bsl k_{N_e}
\in B_{\phi_1,\phi_2}^{N_1,N_2}}
C_{\bsl k_1,\ldots,\bsl k_{N_e}}
\psi_{\bsl k_1}(\bsl r_1)\cdots
\psi_{\bsl k_{N_e}}(\bsl r_{N_e}),
}
where
\eq{
C_{\bsl k_1,\ldots,\bsl k_{N_e}}
=
A_{\bsl k_1,\ldots,\bsl k_{N_e}}
\mathcal N_{\bsl k_1}\cdots
\mathcal N_{\bsl k_{N_e}}.
}
Since $m$ is odd, $F^\sigma$ is antisymmetric under
particle exchange, as shown in \cref{eq:C-fermionic-exchange}.
By the uniqueness of the expansion established in
\cref{eq:f-gk-expansion}, exchanging two particle coordinates
in \cref{eq:F_expansion} requires the coefficients
$A_{\bsl k_1,\ldots,\bsl k_{N_e}}$ to be antisymmetric
under permutations of their momentum labels.
The same holds for $C_{\bsl k_1,\ldots,\bsl k_{N_e}}$. Define the corresponding fermionic creation operator (where the Bloch wavefunction $\psi_{\bsl k}$ will later be shown to belong to an ideal Chern band):
\eq{
\gamma^\dagger_{\bsl k}
= \sum_{\bsl r\in\Lambda}
\psi_{\bsl k}(\bsl r)c^\dagger_{\bsl r}.
\label{eq:creation_operator}
}
Then, with the convention of
\cref{eq:C-discretized-many-body-state},
\eq{
\ket{\Psi^{m,\sigma}}
=
\frac{1}{\sqrt{N_e!}}
\sum_{\bsl k_1,\ldots,\bsl k_{N_e}
\in B_{\phi_1,\phi_2}^{N_1,N_2}}
C_{\bsl k_1,\ldots,\bsl k_{N_e}}
\gamma^\dagger_{\bsl k_1}\cdots
\gamma^\dagger_{\bsl k_{N_e}}\ket{0}.
\label{eq:C-band-fock-expansion}
}
Thus $\ket{\Psi^{m,\sigma}}$ is a superposition of Slater determinants constructed entirely from the one-particle wavefunctions
$\psi_{\bsl k}$. In other words, it lies in the Fock space
generated by these Bloch states.
It remains to show that the $\psi_{\bsl k}$ themselves form an
isolated Chern-ideal band. There are $N=N_1N_2$ allowed momenta, and hence $N$ one-particle
states in this band.
Since $N_e=N/m$, the filling of the band is $\frac{N_e}{N}=\frac{1}{m}$.

We now show that the single particle states we just constructed,
\eq{
\ket{\psi_{\bsl k}}
=
\gamma^\dagger_{\bsl k}\ket{0}
=
\sum_{\bsl r\in\Lambda}\psi_{\bsl k}(\bsl r)\ket{\bsl r},
}
are Bloch states that define an isolated Chern-ideal band.
The following translation formulas also apply when the translated position lies outside the finite lattice $\Lambda$.

First, under $\bsl r\mapsto\bsl r+\bsl a_2$, we have
\eq{
\psi_{\bsl k}(\bsl r+\bsl a_2)
=
e^{\ii\bsl k\cdot\bsl a_2}\psi_{\bsl k}(\bsl r).
\label{eq:continuous-bloch-translation-a2}
}
Under $\bsl r\mapsto\bsl r+\bsl a_1$, we obtain
\eq{
\psi_{\bsl k}(\bsl r+\bsl a_1) =
e^{2\pi\ii z/a_2+\ii\pi a_1/a_2}
e^{-\ii\pi a_1/a_2
-2\pi\ii\left(z/a_2-\bsl a_1\cdot\bsl k/(2\pi)\right)}
\psi_{\bsl k}(\bsl r) =
e^{\ii\bsl k\cdot\bsl a_1}\psi_{\bsl k}(\bsl r).
\label{eq:continuous-bloch-translation-a1}
}
Under $\bsl k\mapsto\bsl k+\bsl b_2$, by \cref{eq:A-gk-definition} and theta-function identities, we have
\eq{
\psi_{\bsl k+\bsl b_2}(\bsl r)=\psi_{\bsl k}(\bsl r).
\label{eq:psi_k+b2}
}
Under $\bsl k\mapsto\bsl k+\bsl b_1$, similarly, we have
\eq{
\psi_{\bsl k+\bsl b_1}(\bsl r)
=
e^{-\ii\bsl k\cdot\bsl a_2}\psi_{\bsl k}(\bsl r).
\label{eq:psi_k+b1}
}
Here $\mathcal N_{\bsl k+\bsl b_j}=\mathcal N_{\bsl k}$. Combining the real-space translation laws in \cref{eq:continuous-bloch-translation-a2,eq:continuous-bloch-translation-a1} and the reciprocal-space relations in \cref{eq:psi_k+b2,eq:psi_k+b1}, we obtain the corresponding Fock-space expressions, with $\hat T_{\bsl a_j}$ denoting lattice translation by $\bsl a_j$:
\eqa{
\hat T_{\bsl a_j}\ket{\psi_{\bsl k}}
&=
e^{-\ii\bsl k\cdot\bsl a_j}\ket{\psi_{\bsl k}},
\qquad j=1,2,
\\
\ket{\psi_{\bsl k+\bsl b_2}}
&=\ket{\psi_{\bsl k}},
\qquad
\ket{\psi_{\bsl k+\bsl b_1}}
=e^{-\ii\bsl k\cdot\bsl a_2}\ket{\psi_{\bsl k}}.
}
For distinct momenta in $B_{\phi_1,\phi_2}^{N_1,N_2}$,
orthogonality follows from the finite Bravais-lattice sum
\eq{
\sum_{n_1=0}^{N_1-1}\sum_{n_2=0}^{N_2-1}
e^{\ii(\bsl k'-\bsl k)\cdot(n_1\bsl a_1+n_2\bsl a_2)}
=
N\delta_{\bsl k,\bsl k'}.
}
Together with the normalization above, this gives
\eq{
\braket{\psi_{\bsl k}}{\psi_{\bsl k'}}
=\delta_{\bsl k,\bsl k'},
\qquad
\{\gamma_{\bsl k},\gamma^\dagger_{\bsl k'}\}
=\delta_{\bsl k,\bsl k'}.
}
Notice that writing
$\bsl k=k_1\bsl b_1+k_2\bsl b_2$, we have
$\bsl k\cdot\bsl a_2=2\pi k_2$, so \cref{eq:psi_k+b1} becomes
$\ket{\psi_{k_1+1,k_2}}
=e^{-2\pi\ii k_2}\ket{\psi_{k_1,k_2}}$,
whereas \cref{eq:psi_k+b2} is periodic in the other reciprocal direction.
Thus the transition phase winds once as $k_2$ goes from $0$ to $1$. This gives $\Ch=+1$.

We next check the periodic part of the Bloch state to verify the ideal-band condition.
As we are working on a lattice with a finite number of degrees
of freedom in a unit cell, it is equivalent to look at the
vector representation as we stick to the physical embedding.
Let $\bsl R=n_1\bsl a_1+n_2\bsl a_2$, with
$n_j=0,\ldots,N_j-1$.

Using the lattice-site decomposition in
\cref{eq:finite-torus-site-set} and the Bloch
translation laws in
\cref{eq:continuous-bloch-translation-a1,eq:continuous-bloch-translation-a2},
we have
$\psi_{\bsl k}(\bsl R+\bsl\delta_\alpha)
=e^{\ii\bsl k\cdot\bsl R}\psi_{\bsl k}(\bsl\delta_\alpha)$.
Therefore,
\eqa{
\ket{\psi_{\bsl k}}
&=\sum_{\bsl r\in\Lambda}
\psi_{\bsl k}(\bsl r)\ket{\bsl r}=\sum_{\alpha=A,B}\sum_{\bsl R}
\psi_{\bsl k}(\bsl R+\bsl\delta_\alpha)
\ket{\bsl R+\bsl\delta_\alpha}=\sum_{\alpha=A,B}
\psi_{\bsl k}(\bsl\delta_\alpha)
\sum_{\bsl R}
e^{\ii\bsl k\cdot\bsl R}
\ket{\bsl R+\bsl\delta_\alpha}\\
&=\sum_{\alpha=A,B}
e^{-\ii\bsl k\cdot\bsl\delta_\alpha}
\psi_{\bsl k}(\bsl\delta_\alpha)
\sum_{\bsl R}
e^{\ii\bsl k\cdot(\bsl R+\bsl\delta_\alpha)}
\ket{\bsl R+\bsl\delta_\alpha}=\sum_{\alpha=A,B}
\sqrt N e^{-\ii\bsl k\cdot\bsl\delta_\alpha}
\psi_{\bsl k}(\bsl\delta_\alpha)
\ket{\bsl k,\alpha},
\label{eq:psi_k_in_k_alpha_basis}
}
where we define the normalized Bloch basis
\eq{
\ket{\bsl k,\alpha}
=
\frac{1}{\sqrt N}\sum_{\bsl R}
e^{\ii\bsl k\cdot(\bsl R+\bsl\delta_\alpha)}
\ket{\bsl R+\bsl\delta_\alpha}
\label{eq:C-orbital-Bloch-basis}
}
in the physical embedding. The coefficients of the two Bloch basis states
$\ket{\bsl k,A}$ and $\ket{\bsl k,B}$ in \cref{eq:psi_k_in_k_alpha_basis}
form the two-component spinor $U_{\bsl k}$.

Substituting the Bloch state definition in
\cref{eq:C-normalized-band-orbital} and the theta-function
expression in \cref{eq:A-gk-definition}, we obtain
\eq{
U_{\bsl k}
=
\frac{\sqrt N}{\mathcal N_{\bsl k}}\mat{
e^{-\ii\bsl\delta_A\cdot\bsl k}
e^{\pi\mathcal B(\bsl\delta_A)/(2\Omega)}\chi(\bsl\delta_A)
\vartheta\genfrac{[}{]}{0pt}{}
{\frac{\bsl a_2\cdot\bsl k}{2\pi}}
{-\frac{\bsl a_1\cdot\bsl k}{2\pi}}
\left(\frac{\delta_A}{a_2}\,\middle|\,\frac{a_1}{a_2}\right)
\\
e^{-\ii\bsl\delta_B\cdot\bsl k}
e^{\pi\mathcal B(\bsl\delta_B)/(2\Omega)}\chi(\bsl\delta_B)
\vartheta\genfrac{[}{]}{0pt}{}
{\frac{\bsl a_2\cdot\bsl k}{2\pi}}
{-\frac{\bsl a_1\cdot\bsl k}{2\pi}}
\left(\frac{\delta_B}{a_2}\,\middle|\,\frac{a_1}{a_2}\right)
}.
}
Using the normalization in \cref{eq:C-normalized-band-orbital}
and the Bloch translation laws, we find
\eq{
\mathcal N_{\bsl k}^2
=
N\sum_{\alpha=A,B}
\left|
e^{\pi\mathcal B(\bsl\delta_\alpha)/(2\Omega)}
\chi(\bsl\delta_\alpha)g_{\bsl k}(\delta_\alpha)
\right|^2,
\qquad U_{\bsl k}^\dagger U_{\bsl k}=1. \label{eq:C-normalization-cell}
}
To simplify the theta-function characteristics,
we introduce the reciprocal-lattice coordinates
\eq{
\bsl k=k_1\bsl b_1+k_2\bsl b_2,
\qquad
k_1=\frac{\bsl a_1\cdot\bsl k}{2\pi},
\qquad
k_2=\frac{\bsl a_2\cdot\bsl k}{2\pi}.
\label{eq:reciprocal-coordinate-definition}
}
Its complex representative is $k=k_x-\ii k_y=k_1b_1+k_2b_2$.
Substituting these coordinates
gives the equivalent spinor expression
\eq{
U_{\bsl k}
=
\frac{\sqrt N}{\mathcal N_{\bsl k}}\mat{
e^{-\ii\bsl\delta_A\cdot\bsl k}
e^{\pi\mathcal B(\bsl\delta_A)/(2\Omega)}
\chi(\bsl\delta_A)
\vartheta\genfrac{[}{]}{0pt}{}{k_2}{-k_1}
\left(\frac{\delta_A}{a_2}\,\middle|\,\frac{a_1}{a_2}\right)
\\
e^{-\ii\bsl\delta_B\cdot\bsl k}
e^{\pi\mathcal B(\bsl\delta_B)/(2\Omega)}
\chi(\bsl\delta_B)
\vartheta\genfrac{[}{]}{0pt}{}{k_2}{-k_1}
\left(\frac{\delta_B}{a_2}\,\middle|\,\frac{a_1}{a_2}\right)
}.
\label{eq:two-component-characteristic-spinor}
}
Take the first component as the common scalar factor, $(U_{\bsl k})_A$.
Wherever $(U_{\bsl k})_A\neq0$, the spinor can be written as
\eq{
U_{\bsl k}
=
(U_{\bsl k})_A\mat{1\\f(k)},
\qquad
f(k)\equiv\frac{(U_{\bsl k})_B}{(U_{\bsl k})_A}.
\label{eq:projective-spinor-form}
}
Directly from \cref{eq:two-component-characteristic-spinor},
\eq{
f(k)=
e^{\frac{\pi[\mathcal B(\bsl\delta_B)-\mathcal B(\bsl\delta_A)]}
{2\Omega}}
\frac{\chi(\bsl\delta_B)}{\chi(\bsl\delta_A)}
e^{-\ii(\bsl\delta_B-\bsl\delta_A)\cdot\bsl k}
\frac{
\vartheta\genfrac{[}{]}{0pt}{}{k_2}{-k_1}
\left(\frac{\delta_B}{a_2}\,\middle|\,\frac{a_1}{a_2}\right)
}{
\vartheta\genfrac{[}{]}{0pt}{}{k_2}{-k_1}
\left(\frac{\delta_A}{a_2}\,\middle|\,\frac{a_1}{a_2}\right)
}.
\label{eq:f-characteristic-ratio}
}
From the relation to the odd Jacobi theta function, we have
\eq{
f(k) =
e^{\frac{\pi[\mathcal B(\bsl\delta_B)-\mathcal B(\bsl\delta_A)]}
{2\Omega}}
\frac{\chi(\bsl\delta_B)}{\chi(\bsl\delta_A)}
e^{-\ii(\bsl\delta_B-\bsl\delta_A)\cdot\bsl k
+
2\pi\ii\left(k_2-\frac12\right)
\frac{\delta_B-\delta_A}{a_2}}
\frac{
\vartheta_1\left(
\frac{\delta_B-(a_1+a_2)/2}{a_2}
+k_2\frac{a_1}{a_2}-k_1
\middle|\frac{a_1}{a_2}
\right)
}{
\vartheta_1\left(
\frac{\delta_A-(a_1+a_2)/2}{a_2}
+k_2\frac{a_1}{a_2}-k_1
\,\middle|\,\frac{a_1}{a_2}
\right)
}.
\label{eq:f-with-real-space-modulus}
}

We now express every momentum-dependent factor in terms of the
complex momentum $k=k_x-\ii k_y$.
From \cref{eq:complex-reciprocal-vectors}, $k
=
\frac{2\pi\ii}{\Omega}(k_1a_2-k_2a_1)$.
It follows that $k_2\frac{a_1}{a_2}-k_1
=
\frac{\ii\Omega}{2\pi a_2}k$.
For an arbitrary complex position $\delta=\delta_x-\ii\delta_y$,
we have
\eq{
2\pi k_2\frac{\delta}{a_2}-\bsl\delta\cdot\bsl k =
\frac{\delta}{2a_2}(a_2\bar k+\bar a_2 k)
-\frac12(\delta\bar k+\bar\delta k) =
\frac12\left(\frac{\bar a_2}{a_2}\delta-\bar\delta\right)k.
\label{eq:position-momentum-complex-identity}
}
This identity is valid for arbitrary complex $a_2$ so no assumption
$a_2=\bar a_2$ is made. Applying \cref{eq:position-momentum-complex-identity} to
$\delta=\delta_B-\delta_A$ gives
\eq{
-\ii(\bsl\delta_B-\bsl\delta_A)\cdot\bsl k
+2\pi\ii\left(k_2-\frac12\right)
\frac{\delta_B-\delta_A}{a_2} =
-\frac{\ii\pi(\delta_B-\delta_A)}{a_2}
+ \frac{\ii}{2}
\left[
\frac{\bar a_2}{a_2}(\delta_B-\delta_A)
-(\bar\delta_B-\bar\delta_A)
\right]k.
\label{eq:position-dependent-phase-holomorphic}
}
Therefore, every momentum-dependent factor in
\cref{eq:f-with-real-space-modulus} depends on $k$ and not on
$\bar k$. We next transform the theta functions from the real-space modular
parameter $a_1/a_2$ to the reciprocal-space modular parameter
$b_1/b_2$. Since $\frac{b_1}{b_2}=-\frac{a_2}{a_1}$,
the required modular transformation, for complex $u$, is
\eq{
\vartheta_1\left(u\,\middle|\,\frac{a_1}{a_2}\right) =
\ii\left(-\ii\frac{a_1}{a_2}\right)^{-1/2}
\exp\left(-\ii\pi\frac{a_2}{a_1}u^2\right)
\vartheta_1\left(
\frac{a_2}{a_1}u\middle|\frac{b_1}{b_2}\right).
\label{eq:theta-one-reciprocal-modular-transformation}
}
The overall square-root factor cancels from the ratio of theta
functions.
For $\alpha=A,B$, the transformed theta argument is
\eq{
\frac{a_2}{a_1}
\left[
\frac{\delta_\alpha-(a_1+a_2)/2}{a_2}
+\frac{\ii\Omega}{2\pi a_2}k
\right]
=
\frac{\delta_\alpha-(a_1+a_2)/2}{a_1}
+\frac{\ii\Omega}{2\pi a_1}k.
\label{eq:transformed-theta-argument-first}
}
\cref{eq:complex-reciprocal-vectors} implies
$\frac{\ii\Omega}{2\pi a_1}=\frac{1}{b_2}$.
The transformed theta argument is therefore
\eq{
\frac{k}{b_2}
+
\frac{\delta_\alpha-(a_1+a_2)/2}{a_1},
\qquad \alpha=A,B.
\label{eq:dimensionless-reciprocal-theta-argument}
}
Combining the exponential produced by the modular transformation
with \cref{eq:position-dependent-phase-holomorphic}, we obtain
\eqa{
f(k)
&=
e^{\frac{\pi[\mathcal B(\bsl\delta_B)-\mathcal B(\bsl\delta_A)]}
{2\Omega}}
\frac{\chi(\bsl\delta_B)}{\chi(\bsl\delta_A)}
e^{-\frac{\ii\pi(\delta_B-\delta_A)(\delta_A+\delta_B-a_2)}
{a_1a_2}
+ \frac{\ii}{2}
\left[
\frac{\bar a_1}{a_1}(\delta_B-\delta_A)
-(\bar\delta_B-\bar\delta_A)
\right]k}
\frac{
\vartheta_1\left(
\frac{k}{b_2}+\frac{\delta_B-(a_1+a_2)/2}{a_1}
\,\middle|\,\frac{b_1}{b_2}
\right)
}{
\vartheta_1\left(
\frac{k}{b_2}+\frac{\delta_A-(a_1+a_2)/2}{a_1}
\,\middle|\,\frac{b_1}{b_2}
\right)
}.
\label{eq:explicit-meromorphic-f}
}
\Cref{eq:explicit-meromorphic-f} shows that the coordinate $f(k)$ in \cref{eq:projective-spinor-form} is meromorphic in $k$, with poles where $(U_{\bsl k})_A=0$, and independent of $k^*$. Using the expression for $f(k)$ in \cref{eq:f-characteristic-ratio}, we now determine how $f(k)$ transforms under a reciprocal-lattice shift $k\to k+b_j$:
\eq{
f(k+b_j)
=
e^{-\ii\bsl b_j\cdot(\bsl\delta_B-\bsl\delta_A)}f(k),
\qquad j=1,2.
\label{eq:f-general-reciprocal-translation}
}
The meromorphic projective form of $f(k)$ establishes
the ideal band condition
\cite{Jian_2012,claassen2015position,Lee_2017,
Mera_2021,Parker2023IdealBands,
cano2026idealquantumgeometryfractional,
zhao2026idealbandstightbindingmodels}.
Together with the reciprocal-lattice translation
properties in
\cref{eq:psi_k+b1,eq:psi_k+b2,eq:f-general-reciprocal-translation},
which give $\Ch=+1$ as discussed, this shows that
$U_{\bsl k}$ defines an ideal Chern band.
Therefore, \cref{eq:C-band-fock-expansion} shows that $\ket{\Psi^{m,\sigma}}$ lies entirely within
the Fock space of this band at filling $1/m$.

\section{General Projected Two-Body Interactions with Exact Fractional Chern Ground States}
\label{sec:app_D}

In this section, we construct a class of positive-semidefinite
two-body Hamiltonians $H_p$ within the Fock space of the
ideal band discussed in \cref{sec:C-band-containment}. 

Since they act in this one-band Fock space, they can be viewed as the projected version of the Hamiltonian defined on lattice.  Our goal is to construct $H_p$ such that it has and only has $\ket{\Psi^{3,\sigma}}$ defined in
\cref{eq:C-discretized-many-body-state} as ground states at $\nu= N_e/N=N_e/(N_1 N_2)=1/3 $ with
$N_e\geq2$ for arbitrary  finite $(N_1,N_2)$ and arbitrary boundary twists:
\eq{
H_p\ket{\Psi}=0 \text{ with } \hat{N}_e\ket{\Psi}=\frac{N}{3}\ket{\Psi}
\quad \Leftrightarrow \quad
\ket{\Psi}=\sum_{\sigma=1}^{3}c_\sigma\ket{\Psi^{3,\sigma}},
\qquad c_\sigma\in\dsC\ ,
\label{eq:two_directions_to_proof}
}
where $\hat{N}_e$ is the number operator for electrons.

In the following, we first construct the pair-annihilation operators and briefly explain why their common zero modes are the fractional Chern states, and we use these operators to construct the projected Hamiltonian in \cref{sec:D-general-projected-interaction}.
Then, we prove the two directions of \cref{eq:two_directions_to_proof} in \cref{sec:D2,sec:D3}, respectively.

\subsection{Pair-Annihilation Operators}
\label{sec:D1}

We first construct the pair-annihilation operators $Q_{\bsl K}(\zeta)$ directly from the discrete ideal-band Bloch states in \cref{eq:C-normalized-band-orbital}, where $\chi(\bsl r)$ is defined only on lattice sites introduced in \cref{eq:C-chi-translation}.
We follow the geometric conventions of
\cref{sec:B-continuum-review,sec:C-setup}, and the construction can be thought as
the discretized versions of the projected Haldane $V_1$ pseudopotential
\cite{Haldane1983FQH,Trugman1985ExactFQH}. All one-particle momenta belong to $B_{\phi_1,\phi_2}^{N_1,N_2}$ in
\cref{eq:allowed+momenta}, which is the set of allowed crystal momenta on a finite lattice with twisted boundary conditions, and all physical-site
sums run over $\Lambda$ defined in
\cref{eq:finite-torus-site-set}.
For a pair of particle coordinates, we introduce
the center and relative coordinates
\eq{
\zeta=\frac{z_1+z_2}{2},
\qquad
w=z_1-z_2,
\qquad
z_1=\zeta+\frac{w}{2},
\qquad
z_2=\zeta-\frac{w}{2}.
\label{eq:D-relative-coordinate-convention}
}
To construct the pair-annihilation operator, we first define
\eq{
J_{\bsl k,\bsl k'}(\zeta)
=
\frac{g_{\bsl k}'(\zeta)g_{\bsl k'}(\zeta)}
{\mathcal N_{\bsl k}\mathcal N_{\bsl k'}}\ ,
\label{eq:relative-one-jet-wronskian}
}
where $g'_{\bsl k}(\zeta)$ denotes
$\left.\partial_z g_{\bsl k}(z)\right|_{z=\zeta}$, and $g_{\bsl k}$ is defined in
\cref{eq:A-gk-definition}.
Differentiatin it gives
\eq{
g_{\bsl k}'(\zeta)
=
\frac{1}{a_2}
\vartheta'\genfrac{[}{]}{0pt}{}
{\frac{\bsl a_2\cdot\bsl k}{2\pi}}
{-\frac{\bsl a_1\cdot\bsl k}{2\pi}}
\left(
\frac{\zeta}{a_2}\middle|\frac{a_1}{a_2}
\right),
\label{eq:explicit-derivative-of-g}
}
where the prime on $\vartheta$ denotes differentiation with
respect to its first argument $\zeta/a_2$. We now define the band-projected pair-annihilation operator
\eq{
Q_{\bsl K}(\zeta)
=
\sum_{\bsl k,\bsl k'\in B_{\phi_1,\phi_2}^{N_1,N_2}}
\delta^{\mathrm{BZ}}_{\bsl k+\bsl k',\bsl K}
J_{\bsl k,\bsl k'}(\zeta)
\gamma_{\bsl k'}\gamma_{\bsl k},
\label{eq:discrete-V1-pair-annihilator}
}
where $\bsl K$ is the total pair crystal momentum, and
$\delta^{\mathrm{BZ}}_{\bsl k+\bsl k',\bsl K}$ equals one if
$\bsl k+\bsl k'-\bsl K\in\dsZ\bsl b_1+\dsZ\bsl b_2$,
and zero otherwise. 
Note that $\bsl K$ takes value from \eq{
B_{2\phi_1,2\phi_2}^{N_1,N_2}
=\left\{
\left(\frac{\kappa_1}{N_1}+\frac{\phi_1}{\pi N_1}\right)\bsl b_1
+\left(\frac{\kappa_2}{N_2}+\frac{\phi_2}{\pi N_2}\right)\bsl b_2
\ \middle|\
\kappa_j\in\dsZ,\ 0\leq\kappa_j<N_j
\right\},
\label{eq:D-allowed-pair-momenta}
}
which is the mesh
with doubled twist angles $B_{2\phi_1,2\phi_2}^{N_1,N_2}$. The reason of having $2\phi_j$ is that, following \cref{eq:allowed+momenta}, each
single-particle momentum satisfies
$e^{\ii\bsl k\cdot\bsl L_j}=e^{\ii\phi_j}$, and since $\bsl K=\bsl k+\bsl k'$ modulo reciprocal-lattice vectors, the pair momentum satisfies
$e^{\ii\bsl K\cdot\bsl L_j}
=e^{\ii\bsl k\cdot\bsl L_j}
e^{\ii\bsl k'\cdot\bsl L_j}
=e^{2\ii\phi_j}$, $j=1,2$. Next, using $\gamma_{\bsl k}\gamma_{\bsl k'}
=-\gamma_{\bsl k'}\gamma_{\bsl k}$ and exchanging
$\bsl k\leftrightarrow\bsl k'$, we get
\eq{
Q_{\bsl K}(\zeta)
=
-\sum_{\bsl k,\bsl k'\in B_{\phi_1,\phi_2}^{N_1,N_2}}
\delta^{\mathrm{BZ}}_{\bsl k+\bsl k',\bsl K}
J_{\bsl k',\bsl k}(\zeta)
\gamma_{\bsl k'}\gamma_{\bsl k},
}
Average this with \cref{eq:discrete-V1-pair-annihilator}:
\eq{
Q_{\bsl K}(\zeta)
=
\frac12
\sum_{\bsl k,\bsl k'\in B_{\phi_1,\phi_2}^{N_1,N_2}}
\delta^{\mathrm{BZ}}_{\bsl k+\bsl k',\bsl K}
\left[J_{\bsl k,\bsl k'}(\zeta)-J_{\bsl k',\bsl k}(\zeta)\right]
\gamma_{\bsl k'}\gamma_{\bsl k}.
\label{eq:Q_K}
}
The coefficient
$\frac12[J_{\bsl k,\bsl k'}(\zeta)
-J_{\bsl k',\bsl k}(\zeta)]$ in \cref{eq:Q_K}
can be expressed as a relative-coordinate derivative.
At fixed $\zeta$, we have
\eq{
\left.
\partial_w
\left[
g_{\bsl k}\left(\zeta+\frac{w}{2}\right)
g_{\bsl k'}\left(\zeta-\frac{w}{2}\right)
\right]
\right|_{w=0} =
\frac12
\left[
g'_{\bsl k}(\zeta)g_{\bsl k'}(\zeta)
-
g'_{\bsl k'}(\zeta)g_{\bsl k}(\zeta)
\right].
\label{eq:D-relative-derivative-pair}
}
Substituting this identity and
\cref{eq:relative-one-jet-wronskian} into \cref{eq:Q_K}
gives
\eq{
Q_{\bsl K}(\zeta)
=
\sum_{\bsl k,\bsl k'\in B_{\phi_1,\phi_2}^{N_1,N_2}}
\delta^{\mathrm{BZ}}_{\bsl k+\bsl k',\bsl K}
\frac{
\left.\partial_w
\left[
g_{\bsl k}\left(\zeta+\frac w2\right)
g_{\bsl k'}\left(\zeta-\frac w2\right)
\right]\right|_{w=0}
}{
\mathcal N_{\bsl k}\mathcal N_{\bsl k'}
}
\gamma_{\bsl k'}\gamma_{\bsl k}.
\label{eq:D-Q-relative-derivative-form}
}
The reason behind such construction is that \cref{eq:D-Q-relative-derivative-form} mimics the Haldane $V_1$ potential~\cite{Haldane1983FQH,Trugman1985ExactFQH}, as $Q_{\bsl K}(\zeta)$'s zero modes will have a holomorphic factor that is cubic in relative position when any two particles collide.
This connection will be transparent after we show why the zero modes of $Q_{\bsl K}(\zeta)$ are the three fractional Chern states
$\ket{\Psi^{3,\sigma}}$, as discussed in the next section.

\subsection{Zero modes of $Q_{\mathbf{K}}(\zeta)$ at $N_e/N=1/3$}
\label{sec:D2}

In this part, we study the zero modes $Q_{\bsl K}(\zeta)$ at $N_e/N=1/3$, \ie, finding $\ket{\Psi}$ in the Fock space generated by $\gamma^\dagger_{\bsl{k}}$ that satisfies $ \hat{N}_e\ket{\Psi}=\frac{N}{3}\ket{\Psi}$ and 
\eq{
Q_{\bsl K}(\zeta)\ket{\Psi}=0 \ \quad \forall \bsl K\in B_{2\phi_1,2\phi_2}^{N_1,N_2}\ ,
\zeta\in\dsC\ .
\label{eq:Q_zero_mode}
}

\subsubsection{Zero modes are fractional Chern states }
\label{sec:D-zero-modes-are-FCI}

Given a generic $N_e$-electron state $\ket{\Psi}$
in the ideal-band Fock space at filling $1/3$,
\eq{
\ket{\Psi}
=
\frac{1}{\sqrt{N_e!}}
\sum_{\bsl k_1,\ldots,\bsl k_{N_e}\in B_{\phi_1,\phi_2}^{N_1,N_2}}
\left(\prod_{j=1}^{N_e}\mathcal N_{\bsl k_j}\right)
\widetilde A_{\bsl k_1,\ldots,\bsl k_{N_e}}
\gamma^\dagger_{\bsl k_1}\cdots
\gamma^\dagger_{\bsl k_{N_e}}\ket{0},
\label{eq:D-generic-band-state}
}
where $\widetilde A_{\bsl k_1,\ldots,\bsl k_{N_e}}$ are the
expansion coefficients of this arbitrary state, antisymmetric
under permutations of their momentum labels.
The normalization factors $\mathcal N_{\bsl k}$ are defined in
\cref{eq:C-normalized-band-orbital}, while the factor
$1/\sqrt{N_e!}$ follows the convention of
\cref{eq:C-discretized-many-body-state}.
To obtain the coordinate-space wavefunction, we replace the
Fock states by their Slater-determinant wavefunctions.
Using the antisymmetry of $\widetilde A$, the permutation sum
cancels the factorial normalization factors, giving
\eq{
\Psi(\bsl r_1,\ldots,\bsl r_{N_e})
=
\sum_{\bsl k_1,\ldots,\bsl k_{N_e}\in B_{\phi_1,\phi_2}^{N_1,N_2}}
\widetilde A_{\bsl k_1,\ldots,\bsl k_{N_e}}
\prod_{j=1}^{N_e}
\left[\mathcal N_{\bsl k_j}
\psi_{\bsl k_j}(\bsl r_j)\right].
}
Substituting the normalized ideal-band Bloch wavefunctions from
\cref{eq:C-normalized-band-orbital}, the normalization
factors cancel:
\eq{
\Psi(\bsl r_1,\ldots,\bsl r_{N_e})
=
\sum_{\bsl k_1,\ldots,\bsl k_{N_e}\in B_{\phi_1,\phi_2}^{N_1,N_2}}
\widetilde A_{\bsl k_1,\ldots,\bsl k_{N_e}}
g_{\bsl k_1}(z_1)\cdots g_{\bsl k_{N_e}}(z_{N_e})
\mathcal S(\bsl r_1,\ldots,\bsl r_{N_e})
=
\widetilde F(z_1,\ldots,z_{N_e})
\mathcal S(\bsl r_1,\ldots,\bsl r_{N_e}),
}
where we define the known common lattice factor
\eq{
\mathcal S(\bsl r_1,\ldots,\bsl r_{N_e})
=
\prod_{j=1}^{N_e}
e^{\pi\mathcal B(\bsl r_j)/(2\Omega)}
\chi(\bsl r_j)
\label{eq:commonlatticefactor}
}
and the unknown holomorphic factor
\eq{
\widetilde F(z_1,\ldots,z_{N_e})
=
\sum_{\bsl k_1,\ldots,\bsl k_{N_e}\in B_{\phi_1,\phi_2}^{N_1,N_2}}
\widetilde A_{\bsl k_1,\ldots,\bsl k_{N_e}}
g_{\bsl k_1}(z_1)\cdots g_{\bsl k_{N_e}}(z_{N_e}).
\label{eq:D-generic-holomorphic-factor}
}
Now we impose the zero-mode condition \cref{eq:Q_zero_mode}.
Summing over the allowed pair momenta, \cref{eq:Q_zero_mode} gives
\eq{
Q(\zeta)\ket{\Psi}=0,
\qquad
Q(\zeta)=\sum_{\bsl K\in B_{2\phi_1,2\phi_2}^{N_1,N_2}}
Q_{\bsl K}(\zeta).
\label{eq:D-total-pair-annihilator}
}
Using \cref{eq:discrete-V1-pair-annihilator,eq:relative-one-jet-wronskian},
we have
$Q(\zeta)=\sum_{\bsl k,\bsl k'\in B_{\phi_1,\phi_2}^{N_1,N_2}}
J_{\bsl k,\bsl k'}(\zeta)
\gamma_{\bsl k'}\gamma_{\bsl k}$.
Acting on the generic state in
\cref{eq:D-generic-band-state}, and using the fermionic
anticommutation relations
$\{\gamma_{\bsl k},\gamma_{\bsl p}^\dagger\}
=\delta_{\bsl k,\bsl p}$
and the antisymmetry of $\widetilde A$, we get 
\eqa{
Q(\zeta)\ket{\Psi}
&=
\frac{1}{\sqrt{N_e!}}
\sum_{\bsl k_1,\ldots,\bsl k_{N_e}\in B_{\phi_1,\phi_2}^{N_1,N_2}}
\left(\prod_{j=1}^{N_e}\mathcal N_{\bsl k_j}\right)
\widetilde A_{\bsl k_1,\ldots,\bsl k_{N_e}}
\sum_{\bsl k,\bsl k'\in B_{\phi_1,\phi_2}^{N_1,N_2}}
J_{\bsl k,\bsl k'}(\zeta)
\gamma_{\bsl k'}\gamma_{\bsl k}
\gamma^\dagger_{\bsl k_1}\cdots
\gamma^\dagger_{\bsl k_{N_e}}\ket{0}
\\
&=
\frac{1}{\sqrt{N_e!}}
\sum_{\bsl k_1,\ldots,\bsl k_{N_e}\in B_{\phi_1,\phi_2}^{N_1,N_2}}
\left(\prod_{j=1}^{N_e}\mathcal N_{\bsl k_j}\right)
\widetilde A_{\bsl k_1,\ldots,\bsl k_{N_e}}
\left[\begin{aligned}
&\left(J_{\bsl k_1,\bsl k_2}(\zeta)
-J_{\bsl k_2,\bsl k_1}(\zeta)\right)
\gamma^\dagger_{\bsl k_3}\cdots
\gamma^\dagger_{\bsl k_{N_e}}\\
&-\left(J_{\bsl k_1,\bsl k_3}(\zeta)
-J_{\bsl k_3,\bsl k_1}(\zeta)\right)
\gamma^\dagger_{\bsl k_2}
\gamma^\dagger_{\bsl k_4}\cdots
\gamma^\dagger_{\bsl k_{N_e}}
+\cdots
\end{aligned}\right]\ket{0}
\\
&=
\frac{\binom{N_e}{2}}{\sqrt{N_e!}}
\sum_{\bsl k_1,\ldots,\bsl k_{N_e}\in B_{\phi_1,\phi_2}^{N_1,N_2}}
\left(\prod_{j=1}^{N_e}\mathcal N_{\bsl k_j}\right)
\widetilde A_{\bsl k_1,\ldots,\bsl k_{N_e}}
\left[
J_{\bsl k_1,\bsl k_2}(\zeta)
-J_{\bsl k_2,\bsl k_1}(\zeta)
\right]
\gamma^\dagger_{\bsl k_3}\cdots
\gamma^\dagger_{\bsl k_{N_e}}\ket{0}
\\
&=
\frac{2\binom{N_e}{2}}{\sqrt{N_e!}}
\sum_{\bsl k_1,\ldots,\bsl k_{N_e}\in B_{\phi_1,\phi_2}^{N_1,N_2}}
\left(\prod_{j=1}^{N_e}\mathcal N_{\bsl k_j}\right)
\widetilde A_{\bsl k_1,\ldots,\bsl k_{N_e}}
J_{\bsl k_1,\bsl k_2}(\zeta)
\gamma^\dagger_{\bsl k_3}\cdots
\gamma^\dagger_{\bsl k_{N_e}}\ket{0}
\\
&=
\frac{N_e(N_e-1)}{\sqrt{N_e!}}
\sum_{\bsl k_1,\ldots,\bsl k_{N_e}\in B_{\phi_1,\phi_2}^{N_1,N_2}}
\left(\prod_{j=1}^{N_e}\mathcal N_{\bsl k_j}\right)
\widetilde A_{\bsl k_1,\ldots,\bsl k_{N_e}}
J_{\bsl k_1,\bsl k_2}(\zeta)
\gamma^\dagger_{\bsl k_3}\cdots
\gamma^\dagger_{\bsl k_{N_e}}\ket{0}
=0.
\label{eq:D-Q-action-generic_k}
}
We now group the momentum sums according to the two
annihilated particles and the $N_e-2$ remaining particles.
Rearranging \cref{eq:D-Q-action-generic_k}, we obtain
\eq{
0=Q(\zeta)\ket{\Psi}=
\frac{N_e(N_e-1)}{\sqrt{N_e!}}
\sum_{\bsl k_3,\ldots,\bsl k_{N_e}\in B_{\phi_1,\phi_2}^{N_1,N_2}}
\left(\prod_{j=3}^{N_e}\mathcal N_{\bsl k_j}\right)
\left[
\sum_{\bsl k_1,\bsl k_2\in B_{\phi_1,\phi_2}^{N_1,N_2}}
\widetilde A_{\bsl k_1,\ldots,\bsl k_{N_e}}
\mathcal N_{\bsl k_1}\mathcal N_{\bsl k_2}
J_{\bsl k_1,\bsl k_2}(\zeta)
\right]
\gamma^\dagger_{\bsl k_3}\cdots
\gamma^\dagger_{\bsl k_{N_e}}\ket{0}.
\label{eq:D-Q-action-generic}
}
In \cref{eq:D-Q-action-generic}, exchanging the remaining momenta
changes the signs of both $\widetilde A$ and the Fock state,
so the signs cancel. Since distinct momentum occupation
configurations give orthogonal Fock states,
$Q(\zeta)\ket{\Psi}=0$ requires each coefficient to vanish:
\eq{
\sum_{\bsl k_1,\bsl k_2\in B_{\phi_1,\phi_2}^{N_1,N_2}}
\widetilde A_{\bsl k_1,\ldots,\bsl k_{N_e}}
\mathcal N_{\bsl k_1}\mathcal N_{\bsl k_2}
J_{\bsl k_1,\bsl k_2}(\zeta)
=0.
\label{eq:D-pair-coefficient-constraint}
}
Using \cref{eq:relative-one-jet-wronskian},
we have
$\mathcal N_{\bsl k_1}\mathcal N_{\bsl k_2}
J_{\bsl k_1,\bsl k_2}(\zeta)
=g'_{\bsl k_1}(\zeta)g_{\bsl k_2}(\zeta)$.
Multiplying \cref{eq:D-pair-coefficient-constraint}
by $\prod_{j=3}^{N_e}g_{\bsl k_j}(z_j)$ and
summing over the remaining momenta, we obtain
\eqa{
0
&=
\sum_{\bsl k_3,\ldots,\bsl k_{N_e}\in B_{\phi_1,\phi_2}^{N_1,N_2}}\sum_{\bsl k_1,\bsl k_2\in B_{\phi_1,\phi_2}^{N_1,N_2}}
\widetilde A_{\bsl k_1,\ldots,\bsl k_{N_e}}
\mathcal N_{\bsl k_1}\mathcal N_{\bsl k_2}
J_{\bsl k_1,\bsl k_2}(\zeta)
\prod_{j=3}^{N_e}g_{\bsl k_j}(z_j)\\
&=
\sum_{\bsl k_1,\ldots,\bsl k_{N_e}
\in B_{\phi_1,\phi_2}^{N_1,N_2}}
\widetilde A_{\bsl k_1,\ldots,\bsl k_{N_e}}
g'_{\bsl k_1}(\zeta)g_{\bsl k_2}(\zeta)
\prod_{j=3}^{N_e}g_{\bsl k_j}(z_j)
\\
&=
\frac12
\sum_{\bsl k_1,\ldots,\bsl k_{N_e}
\in B_{\phi_1,\phi_2}^{N_1,N_2}}
\widetilde A_{\bsl k_1,\ldots,\bsl k_{N_e}}
\left[
g'_{\bsl k_1}(\zeta)g_{\bsl k_2}(\zeta)
-g'_{\bsl k_2}(\zeta)g_{\bsl k_1}(\zeta)
\right]
\prod_{j=3}^{N_e}g_{\bsl k_j}(z_j)
\\
&=
\left.\partial_w
\left[
\sum_{\bsl k_1,\ldots,\bsl k_{N_e}
\in B_{\phi_1,\phi_2}^{N_1,N_2}}
\widetilde A_{\bsl k_1,\ldots,\bsl k_{N_e}}
g_{\bsl k_1}\left(\zeta+\frac w2\right)
g_{\bsl k_2}\left(\zeta-\frac w2\right)
\prod_{j=3}^{N_e}g_{\bsl k_j}(z_j)
\right]\right|_{w=0}
\\
&=
\left.\partial_w
\widetilde F\left(
\zeta+\frac w2,\zeta-\frac w2,
z_3,\ldots,z_{N_e}
\right)
\right|_{w=0},
\label{eq:D-vanishing-relative-derivative}
} 
for all $\zeta,z_3,\ldots,z_{N_e}\in\dsC$. Here we used \cref{eq:D-relative-derivative-pair} and the definition of
$\widetilde F$ in \cref{eq:D-generic-holomorphic-factor}. 
Thus, for any zero mode, the holomorphic factor
$\widetilde F(\{z_i\})$ defined in
\cref{eq:D-generic-holomorphic-factor} has no term
linear in $w$ near $w=0$.
Because $\widetilde F$ is antisymmetric under exchange
of two particles, sending $w=z_1-z_2$ to $-w$ at fixed $\zeta$,
\eq{
\widetilde F\left(
\zeta-\frac{w}{2},\zeta+\frac{w}{2},z_3,\ldots,z_{N_e}
\right)
=
-\widetilde F\left(
\zeta+\frac{w}{2},\zeta-\frac{w}{2},z_3,\ldots,z_{N_e}
\right).
}
Hence, as a holomorphic function of $w$, $\widetilde F$ contains
only odd powers near $w=0$.
By \cref{eq:D-vanishing-relative-derivative}, its linear term vanishes.
Therefore, 
$\widetilde F(\zeta+w/2,\zeta-w/2,z_3,\ldots,z_{N_e})=O(w^3)$
as $w\to0$. Combined with anti-symmetry of the fermionic many-body wavefunction, every pair of electrons must have a zero of at least third
order in $\widetilde F$ when their coordinates coincide.
Therefore, one can clearly see that the zero-mode condition enforces that the terms linear in the relative displacement when two particles collide must vanish, which is exactly what Haldane $V_1$ potential does.

We now factor out these required zeros and determine what freedom
remains in the wavefunction.
The twisted boundary conditions extend these cubic zeros
to $z_i-z_j\in L_1\dsZ+L_2\dsZ$.
On the other hand, $\vartheta_1(u|\tau)$ has a simple zero whenever
$u\in\dsZ+\tau\dsZ$. Hence $\vartheta_1\left(
\frac{z_i-z_j}{L_2}\,\middle|\,\tau
\right)^3$
has a cubic zero at exactly the same points. We can therefore factor
out these universal cubic zeros and define
\eq{
\mathcal R(z_1,\ldots,z_{N_e})
=
\frac{\widetilde F(z_1,\ldots,z_{N_e})}
{\displaystyle\prod_{i<j}^{N_e}
\vartheta_1\left(
\frac{z_i-z_j}{L_2}\,\middle|\,\tau
\right)^3}.
\label{eq:def_of_R}
}
Since $\widetilde F$ vanishes to at least the same order as the
denominator at every zero of the denominator, the apparent
singularities of $\mathcal R$ are removable. Thus $\mathcal R$
is a holomorphic function of all particle coordinates. 

We now use the twisted boundary conditions to show that
$\mathcal R$ cannot depend on the relative particle
coordinates, but only on their sum $Z=\sum_j z_j$.
Although $\mathcal R$ is unknown, its boundary conditions can
be derived from its definition in \cref{eq:def_of_R}.
The numerator $\widetilde F$ satisfies the same boundary
conditions as $g_{\bsl k}$, given in
\cref{eq:D-generic-holomorphic-factor,eq:g_boundary,eq:allowed+momenta},
while the denominator transforms according to
\cref{eq:direct-theta1-shift-one,eq:direct-theta1-shift-tau}.
Taking their ratio and using $N=3N_e$, we obtain
\eqa{
\mathcal R(\ldots,z_i+L_2,\ldots)
&=
(-1)^{3(N_e-1)}e^{\ii\phi_2}\mathcal R(\{z_j\}),
\\
\mathcal R(\ldots,z_i+L_1,\ldots)
&=
(-1)^{3(N_e-1)}e^{\ii\phi_1}
e^{-3\ii\pi\tau-6\pi\ii Z/L_2}\mathcal R(\{z_j\})\ .
\label{eq:D-quotient-boundary}
}
The multipliers depend only on the total coordinate $Z=\sum_jz_j$. Opposite full-period shifts
of two particle coordinates therefore leave $\mathcal R$ unchanged.
After setting
$z_{N_e}
= Z-\sum_{j=1}^{N_e-1} z_j$
we can view $\mathcal{R}$ as
$\mathcal{R}(z_1,\ldots,z_{N_e-1};Z)$.
At fixed $Z$, $\mathcal R$ is an holomorphic, doubly periodic function
of each of $z_1,\ldots,z_{N_e-1}$, and hence is constant in
$z_1,\ldots,z_{N_e-1}$. Thus there exists a function $H$ of $Z$
alone such that
$\mathcal R(z_1,\ldots,z_{N_e})=H(Z)$,
so \cref{eq:def_of_R} gives
\eq{
\widetilde F(z_1,\ldots,z_{N_e}) =
H(Z)\prod_{i<j}^{N_e}
\vartheta_1\left(
\frac{z_i-z_j}{L_2}\,\middle|\,\tau
\right)^3.
\label{eq:D-cubic-jastrow-factorization}
}
So now we have shown that the relative-coordinate dependence of any zero-energy
state is fixed by the cubic Jastrow factor in \cref{eq:D-cubic-jastrow-factorization}. The only remaining
unknown is $H(Z)$. 

We next use
its twisted boundary conditions to prove that there are at most
three independent choices of $H(Z)$. Since $\mathcal R=H(Z)$ and shifting any $z_i$ by $L_j$
also shifts $Z$ by $L_j$, the boundary conditions
in \cref{eq:D-quotient-boundary} apply to $H(Z)$.
Substituting the constraints in \cref{eq:fci-boundary-constraint}
and using $N=3N_e$, we obtain
\eqa{
H(Z+L_2)
&=-e^{\ii K_1L_2}H(Z),
\\
H(Z+L_1)
&=-e^{\ii K_1L_1}e^{2\pi\ii Z_1}
e^{-3\ii\pi\tau-6\pi\ii Z/L_2}H(Z).
\label{eq:D-center-of-mass-boundary}
}
These are exactly the boundary conditions satisfied by
the $m=3$ center-of-mass factors $H_\sigma$ in
\cref{eq:C-com-torus-boundary2,eq:C-com-torus-boundary1}.
The first condition in \cref{eq:D-center-of-mass-boundary} implies
\eqa{
e^{-\ii K_1(Z+L_2)}H(Z+L_2)
&=
-e^{-\ii K_1Z}H(Z).
}
Thus $e^{-\ii K_1Z}H(Z)$ is antiperiodic under
$Z\to Z+L_2$. Since
$e^{2\pi\ii(n+1/2)}=-1$ for $n\in\dsZ$, its Fourier expansion therefore contains half-integer harmonics,
with coefficients $d_n\in\dsC$:
\eq{
e^{-\ii K_1Z}H(Z)
=
\sum_{n\in\dsZ}
d_n e^{2\pi\ii(n+1/2)Z/L_2}.
\label{eq:D-H-Fourier}
}
Now evaluate \cref{eq:D-H-Fourier} at $Z+L_1$, and use
$L_1=\tau L_2$:
\eq{
e^{-\ii K_1(Z+L_1)}H(Z+L_1)=
\sum_{n\in\dsZ}
d_n
e^{2\pi\ii(n+1/2)\tau}
e^{2\pi\ii(n+1/2)Z/L_2}.
\label{eq:version1}
}
On the other hand, the second boundary condition in
\cref{eq:D-center-of-mass-boundary} gives:
\eqa{
e^{-\ii K_1(Z+L_1)}H(Z+L_1)
&=
-e^{2\pi\ii Z_1}
e^{-3\ii\pi\tau}
e^{-6\pi\ii Z/L_2}
e^{-\ii K_1Z}H(Z)
\\
&=-e^{2\pi\ii Z_1}
e^{-3\ii\pi\tau}
e^{-6\pi\ii Z/L_2}
\sum_{n\in\dsZ}
d_n e^{2\pi\ii(n+1/2)Z/L_2}
\\
&=
-e^{2\pi\ii Z_1}
e^{-3\ii\pi\tau}
\sum_{n\in\dsZ}
d_n
e^{2\pi\ii(n-5/2)Z/L_2}
\\
&=
-e^{2\pi\ii Z_1}
e^{-3\ii\pi\tau}
\sum_{n\in\dsZ}
d_{n+3}
e^{2\pi\ii(n+1/2)Z/L_2},
\label{eq:version2}
}
Comparing the coefficients of
$e^{2\pi\ii(n+1/2)Z/L_2}$ in
\cref{eq:version1} and \cref{eq:version2},
\eqa{
d_n e^{2\pi\ii(n+1/2)\tau}
&=
-e^{2\pi\ii Z_1}
e^{-3\ii\pi\tau}d_{n+3},
}
and hence
\eq{
d_{n+3}
=
-e^{-2\pi\ii Z_1}
e^{2\pi\ii(n+2)\tau}d_n.
\label{eq:D-H-Fourier-recurrence}
}

Thus, all Fourier coefficients are determined by $d_0,d_1,d_2$,
so the space of holomorphic functions satisfying
\cref{eq:D-center-of-mass-boundary} has dimension at most three.
The three functions $H_\sigma$, $\sigma=1,2,3$, satisfy
these same boundary conditions by
\cref{eq:C-com-torus-boundary2,eq:C-com-torus-boundary1}
and are linearly independent by \cref{sec:C-linear-independence}.
They therefore form a basis, giving
\eq{
H(Z)=\sum_{\sigma=1}^{3}c_\sigma H_\sigma(Z),
}
and substituting it into \cref{eq:D-cubic-jastrow-factorization}
we obtain
\eq{
\widetilde F(z_1,\ldots,z_{N_e}) =
\left[
\sum_{\sigma=1}^{3}c_\sigma H_\sigma(Z)
\right]
\prod_{i<j}^{N_e}
\vartheta_1\left(
\frac{z_i-z_j}{L_2}\middle|\tau
\right)^3 =
\sum_{\sigma=1}^{3}c_\sigma
\left[
H_\sigma(Z)
\prod_{i<j}^{N_e}
\vartheta_1\left(
\frac{z_i-z_j}{L_2}\middle|\tau
\right)^3
\right].
}
By \cref{eq:discretized-com-factor} and
\cref{eq:ideal-fci-holomorphic-factor}, we have
\eq{
\widetilde F(z_1,\ldots,z_{N_e})
=
\sum_{\sigma=1}^{3}c_\sigma F^\sigma(z_1,\ldots,z_{N_e})
}
where $F^\sigma$ is evaluated at $m=3$.
Multiplying by the common lattice factor $\mathcal S$ (\cref{eq:commonlatticefactor})
and using \cref{eq:C-discretized-amplitude}, we obtain
\eqa{
\Psi(\bsl r_1,\ldots,\bsl r_{N_e}) =
\widetilde F(z_1,\ldots,z_{N_e})
\mathcal S(\bsl r_1,\ldots,\bsl r_{N_e}) =
\sum_{\sigma=1}^{3}
c_\sigma
F^\sigma(z_1,\ldots,z_{N_e})
\mathcal S(\bsl r_1,\ldots,\bsl r_{N_e}) =
\sum_{\sigma=1}^{3}
c_\sigma
\Psi^{3,\sigma}(\bsl r_1,\ldots,\bsl r_{N_e}).
}
Therefore,
\eq{
\ket{\Psi}=\sum_{\sigma=1}^{3}c_\sigma\ket{\Psi^{3,\sigma}}.
}
Thus any filling $1/3$ state that satisfy the zero-mode condition in \cref{eq:Q_zero_mode} must be a linear combination of the
three constructed fractional Chern states in \cref{eq:C-discretized-many-body-state}.

\subsubsection{Fractional Chern states in \cref{eq:C-discretized-many-body-state} are all zero modes}
\label{sec:D3}

It remains to prove whether the zero modes of span the same space as the fractional Chern states in \cref{eq:C-discretized-many-body-state}. 
We now resolve the question by showing that each of the
three constructed $m=3$ fractional Chern states in \cref{eq:C-discretized-many-body-state} satisfy the zero-mode condition in \cref{eq:Q_zero_mode}.

Using the momentum-space expansion in \cref{eq:C-band-fock-expansion}, the band creation operators in \cref{eq:creation_operator}, and the relative-coordinate derivative identity in \cref{eq:D-relative-derivative-pair} for $F^\sigma$, we obtain
\eqa{
Q(\zeta)\ket{\Psi^{3,\sigma}}
&=
\frac{N_e(N_e-1)}{\sqrt{N_e!}}
\sum_{\bsl k_1,\ldots,\bsl k_{N_e}\in B_{\phi_1,\phi_2}^{N_1,N_2}}
\left(\prod_{j=1}^{N_e}\mathcal N_{\bsl k_j}\right)
A_{\bsl k_1,\ldots,\bsl k_{N_e}}
J_{\bsl k_1,\bsl k_2}(\zeta)
\gamma^\dagger_{\bsl k_3}\cdots
\gamma^\dagger_{\bsl k_{N_e}}\ket{0}
\\
&=
\frac{N_e(N_e-1)}{\sqrt{N_e!}}
\sum_{\bsl r_3,\ldots,\bsl r_{N_e}\in\Lambda}
\left[
\prod_{j=3}^{N_e}
e^{\pi\mathcal B(\bsl r_j)/(2\Omega)}
\chi(\bsl r_j)
\right]\left[
\sum_{\bsl k_1,\ldots,\bsl k_{N_e}\in B_{\phi_1,\phi_2}^{N_1,N_2}}
A_{\bsl k_1,\ldots,\bsl k_{N_e}}
g'_{\bsl k_1}(\zeta)g_{\bsl k_2}(\zeta)
\prod_{j=3}^{N_e}g_{\bsl k_j}(z_j)
\right]
c^\dagger_{\bsl r_3}\cdots
c^\dagger_{\bsl r_{N_e}}\ket{0}
\\
&=
\frac{N_e(N_e-1)}{\sqrt{N_e!}}
\sum_{\bsl r_3,\ldots,\bsl r_{N_e}\in\Lambda}
\left[
\prod_{j=3}^{N_e}
e^{\pi\mathcal B(\bsl r_j)/(2\Omega)}
\chi(\bsl r_j)
\right]\left.
\partial_w F^\sigma\left(
\zeta+\frac w2,\zeta-\frac w2,
z_3,\ldots,z_{N_e}
\right)
\right|_{w=0}
c^\dagger_{\bsl r_3}\cdots
c^\dagger_{\bsl r_{N_e}}\ket{0}.
}
By \cref{eq:ideal-fci-holomorphic-factor}, $F^\sigma$
contains the factor
$\vartheta_1(w/L_2\mid\tau)^3$,
while all remaining factors are holomorphic near $w=0$.
Since $\vartheta_1(w/L_2\mid\tau)$ has a simple zero
at $w=0$, the relative-coordinate derivative vanishes.
Therefore,
$Q(\zeta)\ket{\Psi^{3,\sigma}}=0$
for every $\zeta$ and every $\sigma=1,2,3$.
To conclude that each $Q_{\bsl K}$ annihilates these states,
we use the lattice-translation invariance of their span,
established in \cref{app:C-lattice-states}, and the simultaneous
translation eigenbasis of \cref{eq:C-active-many-body-translation}.
Let $\ket{\Psi_{\bsl K_{\mathrm{MB}}}}$ denote any
simultaneous translation eigenstate in this span,
with many-body momentum $\bsl K_{\mathrm{MB}}$, satisfying
$\hat T_{\bsl a_j}\ket{\Psi_{\bsl K_{\mathrm{MB}}}}
=e^{-\ii\bsl K_{\mathrm{MB}}\cdot\bsl a_j}
\ket{\Psi_{\bsl K_{\mathrm{MB}}}}$.
The band operators transform as
\eq{
\hat T_{\bsl a_j}\gamma^\dagger_{\bsl k}\hat T_{\bsl a_j}^{-1}
=
e^{-\ii\bsl k\cdot\bsl a_j}\gamma^\dagger_{\bsl k},
\qquad
\hat T_{\bsl a_j}Q_{\bsl K}(\zeta)\hat T_{\bsl a_j}^{-1}
=
e^{\ii\bsl K\cdot\bsl a_j}Q_{\bsl K}(\zeta).
}
Consequently,
\eqa{
\hat T_{\bsl a_j}Q_{\bsl K}(\zeta) \ket{\Psi_{\bsl K_{\mathrm{MB}}}}
&=
(\hat T_{\bsl a_j}Q_{\bsl K}(\zeta)\hat T_{\bsl a_j}^{-1})
(\hat T_{\bsl a_j}\ket{\Psi_{\bsl K_{\mathrm{MB}}}})
\\
&=
e^{\ii\bsl K\cdot\bsl a_j}
Q_{\bsl K}(\zeta)
e^{-\ii\bsl K_{\mathrm{MB}}\cdot\bsl a_j}
\ket{\Psi_{\bsl K_{\mathrm{MB}}}}
\\
&=
e^{\ii(\bsl K-\bsl K_{\mathrm{MB}})\cdot\bsl a_j}
Q_{\bsl K}(\zeta)\ket{\Psi_{\bsl K_{\mathrm{MB}}}}.
}
For fixed many-body momentum $\bsl K_{\mathrm{MB}}$, different pair momenta $\bsl K$ (modulo reciprocal
lattice vectors) give different eigenvalue pairs under
$\hat T_{\bsl a_1}$ and $\hat T_{\bsl a_2}$. The corresponding states are therefore
orthogonal and cannot cancel in
\eq{
0=Q(\zeta)\ket{\Psi_{\bsl K_{\mathrm{MB}}}}
=
\sum_{\bsl K\in B_{2\phi_1,2\phi_2}^{N_1,N_2}}
Q_{\bsl K}(\zeta)\ket{\Psi_{\bsl K_{\mathrm{MB}}}}.
}
Thus each term in the sum vanishes. Transforming back to the original
sector basis gives
\eq{
Q_{\bsl K}(\zeta)\ket{\Psi^{3,\sigma}}=0,
\qquad \sigma=1,2,3,
}
for every $\zeta\in\dsC$ and every
$\bsl K\in B_{2\phi_1,2\phi_2}^{N_1,N_2}$.
Thus, every constructed $m=3$ fractional Chern state in \cref{eq:C-discretized-many-body-state} is annihilated by all $Q_{\bsl K}(\zeta)$.

In summary, the zero-mode space for \cref{eq:Q_zero_mode}
in the ideal-band Fock space with $N_e=N/3$ particles is exactly $\operatorname{span}\left\{
\ket{\Psi^{3,1}},\ket{\Psi^{3,2}},\ket{\Psi^{3,3}}
\right\}$.

\subsection{General Projected Interaction}
\label{sec:D-general-projected-interaction}

Now the route to the construction of the projected Hamiltonian is clear.
We want a projected Hamiltonian $H_p$ such that $H_p\ket{\psi}=0$ is equivalent to the zero-mode condition in \cref{eq:Q_zero_mode}.
To keep $H_p$ general, we will first show that $Q_{\bsl K}(\zeta)$ can be expressed as a linear combination:
$\Phi_{\bsl K,0}(\zeta)Q_{\bsl K,0}
+\Phi_{\bsl K,1}(\zeta)Q_{\bsl K,1}$,
where $Q_{\bsl K,0}$ and $Q_{\bsl K,1}$ are independent
of $\zeta$, and the two coefficient functions
$\Phi_{\bsl K,0}(\zeta)$ and $\Phi_{\bsl K,1}(\zeta)$
are linearly independent.
Then, the condition of $Q_{\bsl K}(\zeta)\ket{\Psi}=0$ for every
$\zeta$ can be shown to reduce to the condition of
$Q_{\bsl K,0}\ket{\Psi}=Q_{\bsl K,1}\ket{\Psi}=0$, which can be used to directly construct the Hamiltonian.

To do this, we will first prove an identity
\eqa{
&
g_{\bsl k}\left(\zeta+\frac w2\right)
g_{\bsl k'}\left(\zeta-\frac w2\right)
=
\sum_{\mu=0}^{1}
\vartheta
\begin{bmatrix}
\dfrac{k_2+k_2'+\mu}{2}\\[2pt]
-(k_1+k_1')
\end{bmatrix}
\left(\frac{2\zeta}{a_2}\middle|2\tau_0\right)
\vartheta
\begin{bmatrix}
\dfrac{k_2-k_2'+\mu}{2}\\[2pt]
k_1'-k_1
\end{bmatrix}
\left(\frac{w}{a_2}\middle|2\tau_0\right).
\label{eq:seperation_w_zeta}
}
where $\tau_0=a_1/a_2$ is the primitive-cell
modular parameter with $\Im\tau_0>0$, which is distinct from the torus parameter $\tau=L_1/L_2$. 
We use the reciprocal coordinates $k_j$ from
\cref{eq:reciprocal-coordinate-definition}. 
We now prove \cref{eq:seperation_w_zeta}.
To do so, first we note that 
\eqa{
& g_{\bsl k}\left(\zeta+\frac w2\right)
g_{\bsl k'}\left(\zeta-\frac w2\right)\\
&=
\vartheta\genfrac{[}{]}{0pt}{}{k_2}{-k_1}
\left(
\frac{\zeta+w/2}{a_2}
\,\middle|\,
\tau_0
\right)
\vartheta\genfrac{[}{]}{0pt}{}{k_2'}{-k_1'}
\left(
\frac{\zeta-w/2}{a_2}
\,\middle|\,
\tau_0
\right)
\\
&=\sum_{m,n\in\dsZ}\exp\left[\ii\pi\tau_0\bigl[(n+k_2)^2+(m+k_2')^2\bigr]+2\pi \ii(n+k_2)
 \left(\frac{\zeta+w/2}{a_2}-k_1\right)+2\pi \ii(m+k_2')
 \left(\frac{\zeta-w/2}{a_2}-k_1'\right)\right]\\
 &=\sum_{m,n\in\dsZ}\exp\left[\ii\pi\tau_0\bigl[(n+k_2)^2+(m+k_2')^2\bigr]+2\pi \ii(n+k_2+m+k_2')(\frac{\zeta}{a_2} - \frac{k_1+k_1'}{2})+ 2\pi \ii(n+k_2-m-k_2')(\frac{w}{2a_2} - \frac{k_1-k_1'}{2})\right]
}
We can clear see the dependence on $w$ and $\zeta$ starts to separate.
To further do so, we redefine the variable as
\eq{
m+n=2r+\mu\ , -m+n=2s+\mu\ ,
}
where $\mu=0,1$ and $r,s\in\dsZ$.
Then, the expression becomes 
\eqa{
& g_{\bsl k}\left(\zeta+\frac w2\right)
g_{\bsl k'}\left(\zeta-\frac w2\right)\\
&=\sum_{\mu=0,1}\sum_{r,s\in\dsZ}\exp\left[\ii\pi\tau_0\bigl[(r+s+\mu+k_2)^2+(r-s+k_2')^2\bigr]+2\pi \ii(2r+\mu+k_2+k_2')(\frac{\zeta}{a_2} - \frac{k_1+k_1'}{2})\right.\\
& \qquad \qquad \qquad \left. + 2\pi \ii(2s+\mu+k_2-k_2')(\frac{w}{2a_2} - \frac{k_1-k_1'}{2})\right] \\
&=\sum_{\mu=0,1}\sum_{r,s\in\dsZ}\exp\left[\ii\pi\tau_0\bigl[2(r+\frac{\mu+k_2+k_2'}{2})^2+2(s+\frac{\mu+k_2-k_2'}{2})^2\bigr]+2\pi \ii(r+\frac{\mu+k_2+k_2'}{2})(\frac{2\zeta}{a_2} - k_1-k_1')\right.\\
& \qquad \qquad \qquad \left. + 2\pi \ii(s+\frac{\mu+k_2-k_2'}{2})(\frac{w}{a_2} - k_1 + k_1')\right] \\
& =\sum_{\mu=0}^{1}
\vartheta
\begin{bmatrix}
\dfrac{k_2+k_2'+\mu}{2}\\[2pt]
-(k_1+k_1')
\end{bmatrix}
\left(\frac{2\zeta}{a_2}\middle|2\tau_0\right)
\vartheta
\begin{bmatrix}
\dfrac{k_2-k_2'+\mu}{2}\\[2pt]
k_1'-k_1
\end{bmatrix}
\left(\frac{w}{a_2}\middle|2\tau_0\right).
}
Thus \cref{eq:seperation_w_zeta} is proven, which is a clear way to separate the dependence on $\zeta$ and $w$.

Based on \cref{eq:seperation_w_zeta}, we can define
\eq{
\label{eq:Phi_def}
\Phi_{(\bsl{k}+\bsl{k}')_{\mathrm{BZ}},\mu}(\zeta) = \vartheta
\begin{bmatrix}
\dfrac{(k_2+k_2'\,\mod\,1)+\mu}{2}\\[2pt]
-(k_1+k_1'\,\mod\,1)
\end{bmatrix}\left(\frac{2\zeta}{a_2}\middle|2\tau_0\right)\ ,
}
where $(\bsl{k}+\bsl{k}')_{\mathrm{BZ}}$ means the representative in the BZ.
In general, 
\eq{
\bsl{k}+\bsl{k}'=(\bsl{k}+\bsl{k}')_{\mathrm{BZ}} + f_1(\bsl{k}+\bsl{k}') \bsl{b}_1+ f_2(\bsl{k}+\bsl{k}') \bsl{b}_2\ ,
}
where $f_j(\bsl{k}+\bsl{k}')$ is a integer function of $\bsl{k}+\bsl{k}'$.
Thus, 
\eqa{
& \vartheta
\begin{bmatrix}
\dfrac{k_2+k_2'+\mu}{2}\\[2pt]
-(k_1+k_1')
\end{bmatrix}
\left(\frac{2\zeta}{a_2}\middle|2\tau_0\right) = e^{-\pi\ii f_1(\bsl{k}+\bsl{k}')(k_2+k_2'+\mu)}\Phi_{(\bsl{k}+\bsl{k}')_{\mathrm{BZ}},\mu}(\zeta)\ \text{ if $f_2(\bsl{k}+\bsl{k}')$ is even ,}\\
& \vartheta
\begin{bmatrix}
\dfrac{k_2+k_2'+\mu}{2}\\[2pt]
-(k_1+k_1')
\end{bmatrix}
\left(\frac{2\zeta}{a_2}\middle|2\tau_0\right) = e^{-\pi\ii f_1(\bsl{k}+\bsl{k}')(k_2+k_2'+\mu)}\Phi_{(\bsl{k}+\bsl{k}')_{\mathrm{BZ}},\mu+1\,\mod\, 2}(\zeta)\ \text{ if $f_2(\bsl{k}+\bsl{k}')$ is odd}\ ,
}
which leads to 
\eq{
\vartheta
\begin{bmatrix}
\dfrac{k_2+k_2'+\mu}{2}\\[2pt]
-(k_1+k_1')
\end{bmatrix}
\left(\frac{2\zeta}{a_2}\middle|2\tau_0\right) = e^{-\pi\ii f_1(\bsl{k}+\bsl{k}')(k_2+k_2'+\mu)}\Phi_{(\bsl{k}+\bsl{k}')_{\mathrm{BZ}},\mu+f_2(\bsl{k}+\bsl{k}')\,\mod\, 2}(\zeta)\ .
}
As a result, we have 
\eqa{
g_{\bsl k}\left(\zeta+\frac w2\right)
g_{\bsl k'}\left(\zeta-\frac w2\right)
& =
\sum_{\mu=0}^{1}\Phi_{(\bsl{k}+\bsl{k}')_{\mathrm{BZ}},\mu+f_2(\bsl{k}+\bsl{k}')\,\mod\, 2}(\zeta)
 e^{-\pi\ii f_1(\bsl{k}+\bsl{k}')(k_2+k_2'+\mu)}\ 
\vartheta
\begin{bmatrix}
\dfrac{k_2-k_2'+\mu}{2}\\[2pt]
k_1'-k_1
\end{bmatrix}
\left(\frac{w}{a_2}\middle|2\tau_0\right) \\
& =
\sum_{\nu=0}^{1}\Phi_{(\bsl{k}+\bsl{k}')_{\mathrm{BZ}},\nu}(\zeta)
 e^{-\pi\ii f_1(\bsl{k}+\bsl{k}')((k_2+k_2'\,\mod\,1)+\nu)}\ 
\vartheta
\begin{bmatrix}
\dfrac{k_2-k_2'+(\nu-f_2(\bsl{k}+\bsl{k}')\,\mod\,2)}{2}\\[2pt]
k_1'-k_1
\end{bmatrix}
\left(\frac{w}{a_2}\middle|2\tau_0\right) \\
& =
\sum_{\mu=0}^{1}\Phi_{(\bsl{k}+\bsl{k}')_{\mathrm{BZ}},\mu}(\zeta)
 e^{-\pi\ii f_1(\bsl{k}+\bsl{k}')((k_2+k_2'\,\mod\,1)+\mu)}\ 
\vartheta
\begin{bmatrix}
\dfrac{k_2-k_2'+\mu-f_2(\bsl{k}+\bsl{k}')}{2}\\[2pt]
k_1'-k_1
\end{bmatrix}
\left(\frac{w}{a_2}\middle|2\tau_0\right) \\
}

In $Q_{\bsl{K}}(\zeta)$, we care about $J_{\bsl k,\bsl k'}-J_{\bsl k',\bsl k}(\zeta)$, which has the following expression
\eqa{
\frac12
\left[
J_{\bsl k,\bsl k'}(\zeta)
-J_{\bsl k',\bsl k}(\zeta)
\right] & = \frac{1}{\mathcal N_{\bsl k}\mathcal N_{\bsl k'}}
\left.
\partial_w
\left[
g_{\bsl k}\left(\zeta+\frac w2\right)
g_{\bsl k'}\left(\zeta-\frac w2\right)
\right]
\right|_{w=0} \\
& =\sum_{\mu=0}^{1}
\Phi_{(\bsl{k}+\bsl{k}')_{\text{BZ}},\mu}(\zeta)D_{\mu}(\bsl k,\bsl k')\ ,
\label{eq:E-relative-derivative-J}
}
where 
\eq{
D_{\mu}(\bsl k,\bsl k')
\equiv
\frac{1}{\mathcal N_{\bsl k}\mathcal N_{\bsl k'}}e^{-\pi\ii f_1(\bsl{k}+\bsl{k}')((k_2+k_2'\,\mod\,1)+\mu)}\ 
\left.\partial_w\vartheta
\begin{bmatrix}
\dfrac{k_2-k_2'+\mu-f_2(\bsl{k}+\bsl{k}')}{2}\\[2pt]
k_1'-k_1
\end{bmatrix}
\left(\frac{w}{a_2}\middle|2\tau_0\right) 
\right|_{w=0}\ ,
\label{eq:D_def}
}
leading to
\eq{
\frac12
\left[
J_{\bsl k,\bsl k'}(\zeta)
-J_{\bsl k',\bsl k}(\zeta)
\right]
=
\sum_{\mu=0}^{1}
\Phi_{(\bsl{k}+\bsl{k}')_{\text{BZ}},\mu}(\zeta)
D_{\mu}(\bsl k,\bsl k').
\label{eq:E-relative-factorization}
}
Substituting \cref{eq:E-relative-factorization} into the
antisymmetrized expression \cref{eq:Q_K},
\eqa{
Q_{\bsl K}(\zeta)
&=
\sum_{\bsl k,\bsl k'\in B_{\phi_1,\phi_2}^{N_1,N_2}}
\delta^{\mathrm{BZ}}_{\bsl k+\bsl k',\bsl K}
\left[
\sum_{\mu=0}^{1}
\Phi_{(\bsl{k}+\bsl{k}')_{\text{BZ}},\mu}(\zeta)
D_{\mu}(\bsl k,\bsl k')
\right]
\gamma_{\bsl k'}\gamma_{\bsl k}
\\
&\equiv
\sum_{\mu=0}^{1}
\Phi_{\bsl K,\mu}(\zeta)
Q_{\bsl K,\mu},
\label{eq:E-pair-mode-expansion}
}
with
\eq{
Q_{\bsl K,\mu}
\equiv
\sum_{\bsl k,\bsl k'\in B_{\phi_1,\phi_2}^{N_1,N_2}}
\delta^{\mathrm{BZ}}_{\bsl k+\bsl k',\bsl K}
D_{\mu}(\bsl k,\bsl k')
\gamma_{\bsl k'}\gamma_{\bsl k}.
\label{eq:Q_Kmu}
}
Thus, \cref{eq:E-pair-mode-expansion} shows that for each pair momentum $\mathbf{K}$, all dependence of $Q_{\mathbf{K}}(\zeta)$ on the coordinate $\zeta$ is carried by the two functions $\Phi_{\mathbf{K},\mu}(\zeta)$, while $Q_{\mathbf{K},\mu}$ are $\zeta$-independent pair-annihilation operators.

To show that $\Phi_{\bsl K,0}$ and $\Phi_{\bsl K,1}$ are linearly independent, we compare how they transform under the same shift $\zeta\to\zeta+a_2/2$.
Using \cref{eq:theta_property_1,eq:Phi_def}, we obtain
\eq{
\Phi_{\bsl K,\mu}\left(\zeta+\frac{a_2}{2}\right)
=e^{\ii(\bsl{K}\cdot\bsl{a}_2/2+\pi\mu)}\Phi_{\bsl K,\mu}(\zeta),
\qquad \mu=0,1\ .
\label{eq:D-pair-mode-independence}
}
Since $\mu=0,1$, the two functions acquire opposite factors under this shift.
If they were proportional, they would transform by the same factor.
Therefore, the two nonzero functions are linearly independent.
As a result, we have
\eqa{
& Q_{\bsl K}(\zeta)\ket{\Psi}=0
\quad\text{for all }\bsl K\in B_{2\phi_1,2\phi_2}^{N_1,N_2},\ \zeta\in\dsC \\
& \Rightarrow (\Phi_{\bsl K,0}(\zeta_0)
Q_{\bsl K,0} + \Phi_{\bsl K,1}(\zeta_0)
Q_{\bsl K,1})\ket{\Psi}=0\ \&\ (\Phi_{\bsl K,0}(\zeta_0+a_2/2)
Q_{\bsl K,0} + \Phi_{\bsl K,1}(\zeta_0+a_2/2)
Q_{\bsl K,1})\ket{\Psi}=0 \text{ with }\Phi_{\bsl K,\mu}(\zeta_0)\neq 0 \\
& \Rightarrow (\Phi_{\bsl K,0}(\zeta_0)
Q_{\bsl K,0} + \Phi_{\bsl K,1}(\zeta_0)
Q_{\bsl K,1})\ket{\Psi}=0\ \&\ (\Phi_{\bsl K,0}(\zeta_0)
Q_{\bsl K,0} - \Phi_{\bsl K,1}(\zeta_0)
Q_{\bsl K,1})\ket{\Psi}=0 \text{ with }\Phi_{\bsl K,\mu}(\zeta_0)\neq 0 \\
& \Rightarrow Q_{\bsl K,\mu}\ket{\Psi}=0
\quad\text{for all }\bsl K\in B_{2\phi_1,2\phi_2}^{N_1,N_2},\ \mu=0,1\ ,
}
leading to 
\eq{
Q_{\bsl K,\mu}\ket{\Psi}=0
\quad\text{for all }\bsl K\in B_{2\phi_1,2\phi_2}^{N_1,N_2},\ \mu=0,1\quad\Leftrightarrow\quad
Q_{\bsl K}(\zeta)\ket{\Psi}=0
\quad\text{for all }\bsl K\in B_{2\phi_1,2\phi_2}^{N_1,N_2},\ \zeta\in\dsC\ .
}
Note that we do not require $Q_{\bsl K,0}$ and $Q_{\bsl K,1}$ to be independent for the above statement to hold.

We now construct a positive-semidefinite Hamiltonian
whose zero-energy states satisfy
$Q_{\bsl K,0}\ket{\Psi}
=Q_{\bsl K,1}\ket{\Psi}=0$
for every total pair momentum $\bsl K$.
Define the two-component column vector
$\bsl Q_{\bsl K}\equiv
(Q_{\bsl K,0},Q_{\bsl K,1})^{\mathsf T}$.
We choose a Hermitian positive-definite $2\times2$
matrix $M(\bsl K)$ for each
$\bsl K\in B_{2\phi_1,2\phi_2}^{N_1,N_2}$,
so that the energy vanishes only when $Q_{\bsl K,0}\ket{\Psi}=Q_{\bsl K,1}\ket{\Psi}=0$. The projected Hamiltonian is then defined as
\eqa{
H_p
&=g\sum_{\bsl K\in B_{2\phi_1,2\phi_2}^{N_1,N_2}}
\sum_{\mu,\nu=0}^{1}
M_{\mu\nu}(\bsl K)Q_{\bsl K,\mu}^\dagger Q_{\bsl K,\nu}\\
&=g\sum_{\bsl K\in B_{2\phi_1,2\phi_2}^{N_1,N_2}}
\bsl Q_{\bsl K}^\dagger M(\bsl K)\bsl Q_{\bsl K},\quad M(\bsl K)=M(\bsl K)^\dagger>0,
\label{eq:D-general-projected-interaction}
}
where $g>0$ sets the overall interaction strength. 
For any state $\ket{\Psi}$ in the Fock space of the
ideal band,
\eq{
\bra{\Psi}H_p\ket{\Psi}
=g\sum_{\bsl K\in B_{2\phi_1,2\phi_2}^{N_1,N_2}}
\left\|M(\bsl K)^{1/2}\bsl Q_{\bsl K}\ket{\Psi}\right\|^2
\geq0.
\label{eq:D-general-projected-positivity}
}
The squared norm in \cref{eq:D-general-projected-positivity} is the sum of the squared norms of the two components of $M(\bsl K)^{1/2}\bsl Q_{\bsl K}\ket{\Psi}$. Since $M(\bsl K)^{1/2}$ is invertible, the energy vanishes if and only if $Q_{\bsl K,0}\ket{\Psi}=Q_{\bsl K,1}\ket{\Psi}=0$ for every $\bsl K$.
Using \cref{eq:E-pair-mode-expansion,eq:D-pair-mode-independence},
we obtain
\eqa{
H_p\ket{\Psi}=0
&\quad\Leftrightarrow\quad
Q_{\bsl K,\mu}\ket{\Psi}=0
\quad\text{for all }\bsl K\in B_{2\phi_1,2\phi_2}^{N_1,N_2},\ \mu=0,1\\
&\quad\Leftrightarrow\quad
Q_{\bsl K}(\zeta)\ket{\Psi}=0
\quad\text{for all }\bsl K\in B_{2\phi_1,2\phi_2}^{N_1,N_2},\ \zeta\in\dsC.
\label{eq:D-general-zero-condition}
}
Thus, all positive-definite choices of $M(\bsl K)$ yield the same zero-energy states. For a simple example, we can choose $M(\bsl K)=I_2$.
By \cref{eq:D-general-zero-condition} and the discuss in \cref{sec:D-zero-modes-are-FCI,sec:D3}, we know that, for $N_e=N/3$ particles,
\eq{
\ker\left(H_p\right)
=
\operatorname{span}\left\{
\ket{\Psi^{3,1}},\ket{\Psi^{3,2}},\ket{\Psi^{3,3}}
\right\}.
\label{eq:D-exact-ground-state-space}
}
The three states are linearly independent, so the
ground-state degeneracy is exactly three.
For each finite system with $N_e\geq2$, all other
eigenvalues are strictly positive, so there is a positive
gap above the three-dimensional zero-energy space.

\section{Real-Space Lattice Hamiltonian and Its Projection}
\label{app:E-real-space-parent}

In \cref{sec:app_D}, we constructed a class of projected
interactions whose zero-energy space is spanned by
the three fractional Chern states at filling $\nu=1/3$.
Here we construct an explicit real-space unprojected
Hamiltonian on a lattice with two sublattices per unit cell
whose projection belongs to this class.

We use the lattice conventions of \cref{sec:C-setup},
with arbitrary primitive vectors $\bsl a_1,\bsl a_2$
satisfying $\Im(a_1/a_2)>0$ and two inequivalent
sublattice positions $\bsl\delta_A,\bsl\delta_B$.
The unit-cell area is $\Omega=\Im(a_1\bar a_2)>0$,
and we write $\tau_0=a_1/a_2$ and
$\tau_I=\Im\tau_0$. Recall that we impose the twist boundary condition
$c_{\bsl r+\bsl L_j}=e^{\ii\phi_j}c_{\bsl r}$ with $\phi_1=\phi_2=0$ meaning the periodic boundary condition,
where $\bsl L_j=N_j\bsl a_j$, and $N=N_1N_2$ is the number of unit cells. 

The full Hamiltonian is
\eq{
H=H_0+H_{\mathrm{int}}.
\label{eq:E-general-real-space-model}
}
We first specify the single-particle Hamiltonian,
\eq{
H_0=\sum_{\bsl r\in\Lambda}
\sum_{\bsl r'\in\overline\Lambda}
t_{\bsl r\bsl r'}
c_{\bsl r}^{\dagger}c_{\bsl r'},
\label{eq:E-general-hopping-term}
}
where $t_{\bsl r\bsl r'}=t_{\alpha\beta}(\bsl R-\bsl R')$
for $\bsl r=\bsl R+\bsl\delta_\alpha$ and
$\bsl r'=\bsl R'+\bsl\delta_\beta$.
The reference site $\bsl r$ runs over the finite set
$\Lambda$ (\cref{eq:finite-torus-site-set}), while $\bsl r'$ includes all images in
$\overline\Lambda$ (\cref{eq:infinite-lattice-site-set}), with their operators identified
through the twisted boundary conditions.

For $\bsl R=m_1\bsl a_1+m_2\bsl a_2$
with $m_1,m_2\in\dsZ$, the hopping matrix is
\eq{
t(\bsl R)=
\mat{
\mathscr T_{BB}(\bsl R)&-\mathscr T_{AB}(\bsl R)\\
-\mathscr T_{BA}(\bsl R)&\mathscr T_{AA}(\bsl R)
},
\label{eq:E-final-hopping}
}
where
\eq{
\mathscr T_{\alpha\beta}(\bsl R)
=\frac{\eta_\alpha\eta_\beta^*}{\sqrt{2\tau_I}}
(-1)^{m_1m_2}
e^{-\frac{\pi|\bsl R+\bsl\delta_\alpha-\bsl\delta_\beta|^2}{2\Omega}}
\times
e^{\frac{\ii\pi}{\Omega}
\left\{
\bsl R\times(\bsl\delta_\alpha+\bsl\delta_\beta)
+\bsl\delta_\alpha\times\bsl\delta_\beta
\right\}},
\label{eq:E-gaussian-kernel}
}
where
\eq{
\eta_\alpha=
\chi(\bsl\delta_\alpha)
e^{-\pi\mathcal K(\bsl\delta_\alpha)/(2\Omega)},
\qquad \alpha=A,B.
\label{eq:E-site-weights}
}
We define the two pair-annihilation operators
using the hopping amplitudes above:
\eq{
B_{\bsl r}^{(p)}=
\sum_{\substack{
\bsl r'\in\overline\Lambda\\
\bsl r'\ne\bsl r
}}
\frac{t_{\bsl r\bsl r'}}
{(z_{\bsl r'}-z_{\bsl r})^p}
c_{\bsl r'}c_{\bsl r},
\qquad p=1,2,
\label{eq:E-general-pair-operators}
}
where $z_{\bsl r}=r_x-\ii r_y$. The two-body interaction is then constructed from
these operators as
\eq{
H_{\mathrm{int}}=
u\sum_{\bsl r\in\Lambda}\sum_{p=1}^{2}
\Omega^{p-1}
B_{\bsl r}^{(p)\dagger}B_{\bsl r}^{(p)},
\qquad u>0.
\label{eq:E-general-interaction}
}
The Hamiltonian acts on the full Fock space, and the filling such as $\nu=1/3$ can be imposed on the states when determining the ground states of the projected Hamiltonian.

In the following, we will first explain the construction of $H_0$ and derive the hopping amplitudes. 
We then show the ground states of $H$ are  exactly the fractional Chern states in \cref{eq:C-discretized-many-body-state} for filling $1/3$.
We do so by projecting $H$ to the ideal band, because $\ker H = \ker \widetilde{H}$ as discussed in the main text.
In particular, we project $H_{\mathrm{int}}$ onto the ideal band and show that the resulting Hamiltonian $\widetilde{H}_{\mathrm{int}}$ has the form in \cref{eq:D-general-projected-interaction}, with a positive-definite coefficient matrix. Thus, it means that $\widetilde{H}_{\mathrm{int}}$ belongs to the class of projected interactions constructed in \cref{sec:app_D} whose zero-energy space at filling $\nu=1/3$  is exactly spanned by the fractional Chern states in \cref{eq:C-discretized-many-body-state}.

\subsection{Two-Band Single-Particle Hamiltonian}
\label{sec:SP-hamiltonian}

We start from the ideal-band Bloch spinor constructed in \cref{sec:C-band-containment} and choose a two-band Hamiltonian that has this spinor as a zero-energy eigenvector at every momentum.

Using the Bloch spinor in \cref{eq:two-component-characteristic-spinor},
define its unnormalized components by
\eq{
v(\bsl k)
=
\mat{
v_A(\bsl k)\\
v_B(\bsl k)
},
\qquad
v_\alpha(\bsl k)
\equiv
e^{-\ii\bsl k\cdot\bsl\delta_\alpha}
e^{\pi\mathcal B(\bsl\delta_\alpha)/(2\Omega)}
\chi(\bsl\delta_\alpha)
g_{\bsl k}(\delta_\alpha),
\qquad
\alpha=A,B.
\label{eq:E-unnormalized-spinor}
}
It is therefore convenient to define
\eq{S(\bsl k)
\equiv
v^\dagger(\bsl k)v(\bsl k) = |v_A(\bsl k)|^2+|v_B(\bsl k)|^2.
\label{eq:S_def}
}
Comparing with \cref{eq:C-normalization-cell}, we have
$\mathcal N_{\bsl k}^2=N S(\bsl k)$. Then
\eq{
U_{\bsl k}
=
\frac{v(\bsl k)}{\sqrt{S(\bsl k)}}.
\label{eq:E-normalized-spinor}
}
As shown in \cref{app:C-lattice-states}, the two discretized theta functions cannot vanish
simultaneously, so
$S(\bsl k)>0$
for every real $\bsl k$. The projector onto the target band is therefore
\eq{
P_{\bsl k} =
U_{\bsl k}U_{\bsl k}^\dagger =
\frac{v(\bsl k)v^\dagger(\bsl k)}
{S(\bsl k)} =
\frac{1}{S(\bsl k)}
\mat{
|v_A(\bsl{k})|^2 & v_A(\bsl{k})v_B^*(\bsl{k})\\
v_B(\bsl{k})v_A^*(\bsl{k}) & |v_B(\bsl{k})|^2
}.
\label{eq:E-projector}
}
By construction, $P_{\bsl k}U_{\bsl k}=U_{\bsl k}$ and
$(I_2-P_{\bsl k})U_{\bsl k}=0$,
where $I_2$ is the identity in the two-sublattice space.
Thus any positive Hamiltonian proportional to
$I_2-P_{\bsl k}$ has the desired band as a zero-energy eigenband. The simplest projector Hamiltonian would therefore be
$I_2-P_{\bsl k}$. However, from \cref{eq:E-projector} this contains
the denominator $1/S(\bsl k)$. Since our next goal is to Fourier
transform the Bloch Hamiltonian to real space, it is useful to remove
this denominator:
\eq{
h_0(\bsl k)
\equiv
S(\bsl k)
\left(
I_2-P_{\bsl k}
\right).
\label{eq:E-h0-definition}
}
Using \cref{eq:E-projector,eq:S_def}, this becomes
\eq{
h_0(\bsl k)
=
\mat{
|v_B(\bsl k)|^2
&
-v_A(\bsl k)v_B(\bsl k)^*
\\
-v_B(\bsl k)v_A(\bsl k)^*
&
|v_A(\bsl k)|^2
}.
\label{eq:E-bloch-matrix}
}
We next verify the two eigenvalues explicitly.
Since
$v(\bsl k)^\dagger v(\bsl k)=S(\bsl k)$,
\eq{
h_0(\bsl k)v(\bsl k)
=\left[S(\bsl k)I_2-v(\bsl k)v(\bsl k)^\dagger\right]v(\bsl k)=S(\bsl k)v(\bsl k)
-v(\bsl k)\left[v(\bsl k)^\dagger v(\bsl k)\right]=0.
\label{eq:E-target-band-zero}
}
And because this is true for every $\bsl k$, the target ideal band is exactly flat at zero energy. Let $V_{\bsl k}$ be a normalized
vector orthogonal to $v(\bsl k)$, namely
$v(\bsl k)^\dagger V_{\bsl k}=0$. Then
\eq{
h_0(\bsl k)V_{\bsl k}
=\left[S(\bsl k)I_2-v(\bsl k)v(\bsl k)^\dagger\right]V_{\bsl k}=S(\bsl k)V_{\bsl k}
-v(\bsl k)\left[v(\bsl k)^\dagger V_{\bsl k}\right]
=S(\bsl k)V_{\bsl k}.
\label{eq:E-upper-band-action}
}
Therefore, the two band energies are $0$ and $S(\bsl k)>0$.
Since $S(\bsl k)$ is continuous and periodic on the Brillouin
zone, it has a strictly positive minimum. The target band is
therefore isolated from the upper band.

We now derive the hopping amplitudes stated in
\cref{eq:E-gaussian-kernel,eq:E-final-hopping} by Fourier
transforming $h_0(\bsl k)$. Using the Bloch basis defined in \cref{eq:C-orbital-Bloch-basis}, we express the hopping Hamiltonian as
\eq{
 c^\dagger_{\bsl k\alpha}
 =\frac1{\sqrt N}\sum_{\bsl R}
 e^{\ii\bsl k\cdot(\bsl R+\bsl\delta_\alpha)}c^\dagger_{\bsl R+\bsl\delta_\alpha},
 \qquad
 H_0=\sum_{\bsl k\in B_{\phi_1,\phi_2}^{N_1,N_2}}\sum_{\alpha,\beta=A,B}
 c^\dagger_{\bsl k\alpha}[h_0(\bsl k)]_{\alpha\beta}c_{\bsl k\beta}.
 \label{eq:E-fourier-operators}
}
The finite-system momentum sum uses the mesh in \cref{eq:allowed+momenta}.

\subsubsection{Infinite Lattice}

Let us first focus on the infinite lattice, and the finite-torus version is derived later in
\cref{eq:E-h0-finite-site-form}.
On the infinite lattice, the inverse transform is:
\eq{
 t_{\alpha\beta}(\bsl R)
 =\frac1{\Omega^*}\int_{\mathrm{BZ}}d^2k\,
 e^{\ii\bsl k\cdot(\bsl R+\bsl\delta_\alpha-\bsl\delta_\beta)}
 [h_0(\bsl k)]_{\alpha\beta}.
 \label{eq:E-hopping-transform}
}
Here $\Omega^*$ is the area of BZ, and $t_{\alpha\beta}(\bsl R)$ is the hopping amplitude from
sublattice $\beta$ in cell zero to sublattice $\alpha$ in cell
$\bsl R$. 
To distinguish the infinite Bravais lattice from the finite
set of unit-cell representatives, we define
\eq{
\mathcal L_\infty
=
\{n_1\bsl a_1+n_2\bsl a_2\mid n_1,n_2\in\dsZ\},
\qquad
\mathcal L_N
=
\{n_1\bsl a_1+n_2\bsl a_2
\mid 0\leq n_j<N_j,\ n_j\in\dsZ\}.
}
On the infinite lattice, the real-space Hamiltonian is\eq{
H_0=\sum_{\bsl R,\bsl R'\in\mathcal L_\infty}\sum_{\alpha,\beta=A,B}
t_{\alpha\beta}(\bsl R-\bsl R')
c^\dagger_{\bsl R+\bsl\delta_\alpha}
c_{\bsl R'+\bsl\delta_\beta},
\label{eq:E-h0-infinite-site-form}
}
where both cell vectors run over the infinite Bravais lattice. 

Recall the cell modular parameter
$\tau_0=\frac{a_1}{a_2}=\tau_R+\ii\tau_I$
which is distinct from the full-torus modulus $\tau=L_1/L_2$ in \cref{sec:B-continuum-review}.
To evaluate the transform, write
$\bsl R=m_1\bsl a_1+m_2\bsl a_2$, with $m_1,m_2\in\dsZ$,
and use the reciprocal coordinates $k_1,k_2$ from
\cref{eq:reciprocal-coordinate-definition}. Then
we have $d^2k/\Omega^*=dk_1dk_2$.
With $w_\alpha=\delta_\alpha/a_2=x_\alpha+\ii y_\alpha$, the Fourier transform of $v_\alpha v_\beta^*$ gives
\eqa{
 \mathscr T_{\alpha\beta}(\bsl R)
 &\equiv
 \int_0^1dk_1\int_0^1dk_2\,
 e^{\ii\bsl k\cdot(\bsl R+\bsl\delta_\alpha-\bsl\delta_\beta)}
 v_\alpha(\bsl k)v_\beta(\bsl k)^*
 \\
 &=
 \int_0^1dk_1\int_0^1dk_2\,
 e^{\ii\bsl k\cdot(\bsl R+\bsl\delta_\alpha-\bsl\delta_\beta)}
 \left[
 e^{-\ii\bsl k\cdot\bsl\delta_\alpha}
 e^{\pi\mathcal B(\bsl\delta_\alpha)/(2\Omega)}
 \chi(\bsl\delta_\alpha)
 g_{\bsl k}(\delta_\alpha)
 \right]
 \left[
 e^{+\ii\bsl k\cdot\bsl\delta_\beta}
 e^{\pi\mathcal B(\bsl\delta_\beta)^*/(2\Omega)}
 \chi(\bsl\delta_\beta)^*
 g_{\bsl k}(\delta_\beta)^*
 \right]
 \\
 &=
 e^{\frac{\pi}{2\Omega}
 [\mathcal B(\bsl\delta_\alpha)+\mathcal B(\bsl\delta_\beta)^*]}
 \chi(\bsl\delta_\alpha)\chi(\bsl\delta_\beta)^*
 \int_0^1dk_1\int_0^1dk_2\,
 e^{\ii\bsl k\cdot\bsl R}
 g_{\bsl k}(\delta_\alpha)g_{\bsl k}(\delta_\beta)^*
 \\
 &=
 e^{\frac{\pi}{2\Omega}
 [\mathcal B(\bsl\delta_\alpha)+\mathcal B(\bsl\delta_\beta)^*]}
 \chi(\bsl\delta_\alpha)\chi(\bsl\delta_\beta)^*
 \int_0^1dk_2\int_0^1dk_1\,
 e^{2\pi\ii(m_1k_1+m_2k_2)}
 g_{\bsl k}(\delta_\alpha)g_{\bsl k}(\delta_\beta)^*,
 \label{eq:E-kernel-transform}
}
where we substitute \cref{eq:E-unnormalized-spinor}.
Using \cref{eq:A-gk-definition,eq:reciprocal-coordinate-definition}
at $z=\delta_\alpha$ and the theta-function definition in \cref{eq:theta-with-characteristics}, we have
\eq{
g_{\bsl k}(\delta_\alpha) =
\sum_{j\in\dsZ}
\exp\left[
\ii\pi\tau_0(j+k_2)^2
+ 2\pi\ii(j+k_2)(w_\alpha-k_1)
\right]
=
\sum_{j\in\dsZ}
e^{\ii\pi\tau_0(j+k_2)^2}
e^{2\pi\ii(j+k_2)w_\alpha}
e^{-2\pi\ii(j+k_2)k_1}.
\label{eq:E-gk-series}
}
Similarly, taking the complex conjugate
\eqa{
g_{\bsl k}(\delta_\beta)^*
&=
\sum_{l\in\dsZ}
e^{-\ii\pi\bar\tau_0(l+k_2)^2}
e^{-2\pi\ii(l+k_2)\bar w_\beta}
e^{2\pi\ii(l+k_2)k_1}.
\label{eq:E-gk-conjugate-series}
}
Therefore, \cref{eq:E-kernel-transform} becomes
\eqa{
\mathscr T_{\alpha\beta}(\bsl R)
&=
e^{\frac{\pi}{2\Omega}
[\mathcal B(\bsl\delta_\alpha)+\mathcal B(\bsl\delta_\beta)^*]}
\chi(\bsl\delta_\alpha)\chi(\bsl\delta_\beta)^*
\int_0^1dk_2\int_0^1dk_1
e^{2\pi\ii(m_1k_1+m_2k_2)}
\\
&\quad\times
\left[
\sum_{j\in\dsZ}
e^{\ii\pi\tau_0(j+k_2)^2}
e^{2\pi\ii(j+k_2)w_\alpha}
e^{-2\pi\ii(j+k_2)k_1}
\right]
\left[
\sum_{l\in\dsZ}
e^{-\ii\pi\bar\tau_0(l+k_2)^2}
e^{-2\pi\ii(l+k_2)\bar w_\beta}
e^{2\pi\ii(l+k_2)k_1}
\right]
\\
&=
e^{\frac{\pi}{2\Omega}
[\mathcal B(\bsl\delta_\alpha)+\mathcal B(\bsl\delta_\beta)^*]}
\chi(\bsl\delta_\alpha)\chi(\bsl\delta_\beta)^*
\sum_{j,l\in\dsZ}
\int_0^1dk_2\int_0^1dk_1
e^{2\pi\ii(m_1k_1+m_2k_2)}
\\
&\quad\times
e^{\ii\pi\tau_0(j+k_2)^2}
e^{-\ii\pi\bar\tau_0(l+k_2)^2}
e^{2\pi\ii(j+k_2)w_\alpha}
e^{-2\pi\ii(l+k_2)\bar w_\beta}\times
e^{-2\pi\ii(j+k_2)k_1}
e^{2\pi\ii(l+k_2)k_1}
\\
&=
e^{\frac{\pi}{2\Omega}
[\mathcal B(\bsl\delta_\alpha)+\mathcal B(\bsl\delta_\beta)^*]}
\chi(\bsl\delta_\alpha)\chi(\bsl\delta_\beta)^*
\sum_{j,l\in\dsZ}
\int_0^1dk_2\,
e^{2\pi\ii m_2k_2}
\\
&\quad\times
e^{\ii\pi\tau_0(j+k_2)^2-\ii\pi\bar\tau_0(l+k_2)^2}
e^{2\pi\ii(j+k_2)w_\alpha-2\pi\ii(l+k_2)\bar w_\beta}\times
\int_0^1dk_1
e^{2\pi\ii[
m_1-(j+k_2)+(l+k_2)
]k_1}
\\
&=
e^{\frac{\pi}{2\Omega}
[\mathcal B(\bsl\delta_\alpha)+\mathcal B(\bsl\delta_\beta)^*]}
\chi(\bsl\delta_\alpha)\chi(\bsl\delta_\beta)^*
\sum_{j,l\in\dsZ}
\int_0^1dk_2\,
e^{2\pi\ii m_2k_2}
\\
&\quad\times
e^{\ii\pi\tau_0(j+k_2)^2-\ii\pi\bar\tau_0(l+k_2)^2}
e^{2\pi\ii(j+k_2)w_\alpha-2\pi\ii(l+k_2)\bar w_\beta}\times
\int_0^1dk_1
e^{2\pi\ii(m_1-j+l)k_1},
\label{eq:E-expanded-transform}
}
where the last integral is $\delta_{m_1-j+l,0}
=\delta_{l,j-m_1}$.
Thus $l=j-m_1$. Substituting this into
\cref{eq:E-expanded-transform} gives
\eqa{
\mathscr T_{\alpha\beta}(\bsl R)
&=
e^{\frac{\pi}{2\Omega}
[\mathcal B(\bsl\delta_\alpha)+\mathcal B(\bsl\delta_\beta)^*]}
\chi(\bsl\delta_\alpha)\chi(\bsl\delta_\beta)^*
\sum_{j\in\dsZ}\int_0^1dk_2
e^{2\pi\ii m_2k_2}
\\
&\quad\times
e^{\ii\pi\tau_0(j+k_2)^2
-\ii\pi\bar\tau_0(j-m_1+k_2)^2} \times
e^{2\pi\ii(j+k_2)w_\alpha
-2\pi\ii(j-m_1+k_2)\bar w_\beta}.
\label{eq:E-after-k1-integration}
}
Set $q=j+k_2$. Since $m_2,j\in\dsZ$,
$e^{2\pi\ii m_2k_2}=e^{2\pi\ii m_2q}$.
For fixed $j$, the interval $0\leq k_2<1$
corresponds to $j\leq q<j+1$.
Therefore,
$\sum_{j\in\dsZ}\int_0^1dk_2
=\int_{\dsR}dq$, giving
\eq{
\mathscr T_{\alpha\beta}(\bsl R)
=
e^{\frac{\pi}{2\Omega}
[\mathcal B(\bsl\delta_\alpha)+\mathcal B(\bsl\delta_\beta)^*]}
\chi(\bsl\delta_\alpha)\chi(\bsl\delta_\beta)^*
\int_{\dsR}dq
e^{2\pi\ii m_2q}\times
e^{\ii\pi\tau_0q^2-\ii\pi\bar\tau_0(q-m_1)^2}
e^{2\pi\ii q w_\alpha
-2\pi\ii(q-m_1)\bar w_\beta}.
\label{eq:E-q-integral}
}
We center the remaining integral by setting
$x=q-m_1/2$. Using $\tau_0=\tau_R+\ii\tau_I$,
the exponent in \cref{eq:E-q-integral} becomes
\eqa{
&2\pi\ii m_2q
+\ii\pi\tau_0q^2
-\ii\pi\bar\tau_0(q-m_1)^2
+2\pi\ii q w_\alpha
-2\pi\ii(q-m_1)\bar w_\beta\\
&\quad =
-\frac{\pi\tau_I m_1^2}{2}
+\ii\pi m_1(m_2+w_\alpha+\bar w_\beta)
-2\pi\tau_Ix^2
+2\pi\ii
(m_2+\tau_Rm_1+w_\alpha-\bar w_\beta)x.
\label{eq:E-collected-x-exponent}
}
Substituting \cref{eq:E-collected-x-exponent} into
\cref{eq:E-q-integral} gives
\eqa{
 \mathscr T_{\alpha\beta}(\bsl R)
 &=e^{\frac{\pi}{2\Omega}
 [\mathcal B(\bsl\delta_\alpha)+\mathcal B(\bsl\delta_\beta)^*]}
 \chi(\bsl\delta_\alpha)\chi(\bsl\delta_\beta)^*
 e^{-\pi\tau_I m_1^2/2+\ii\pi m_1(m_2+w_\alpha+\bar w_\beta)}\times\int_{\dsR}dx
 e^{-2\pi\tau_I x^2
 +2\pi\ii(m_2+\tau_Rm_1+w_\alpha-\bar w_\beta)x}\\
 &=\frac{
 e^{\frac{\pi}{2\Omega}
 [\mathcal B(\bsl\delta_\alpha)+\mathcal B(\bsl\delta_\beta)^*]}
 \chi(\bsl\delta_\alpha)\chi(\bsl\delta_\beta)^*}
 {\sqrt{2\tau_I}}\times\exp\left[
 -\frac{\pi\tau_I m_1^2}{2}
 +\ii\pi m_1(m_2+w_\alpha+\bar w_\beta)
 -\frac{\pi(m_2+\tau_Rm_1+w_\alpha-\bar w_\beta)^2}
 {2\tau_I}
 \right].
 \label{eq:E-gaussian-integral}
}
We used the convergent Gaussian identity
$\int_{\dsR}dx\,e^{-2\pi\tau_I x^2+2\pi\ii A x}
 =(2\tau_I)^{-1/2}e^{-\pi A^2/(2\tau_I)}$ \cite{Gradshteyn2014},
which also holds for complex $A$ when $\tau_I>0$.
To rewrite the site-dependent prefactor in physical coordinates, we use the $\eta_\alpha$ defined in \cref{eq:E-site-weights}.

Using \cref{eq:C-B-definition},
$\delta_\alpha=a_2w_\alpha$ with
$w_\alpha=x_\alpha+\ii y_\alpha$,
and $\Omega=|a_2|^2\tau_I$, we obtain
\eq{
e^{\pi\mathcal B(\bsl\delta_\alpha)/(2\Omega)}
\chi(\bsl\delta_\alpha)
=
\eta_\alpha
e^{-\pi y_\alpha^2/\tau_I
+\ii\pi x_\alpha y_\alpha/\tau_I}.
\label{eq:E-site-factor-simplification}
}
Substitution into \cref{eq:E-gaussian-integral}, with real and
imaginary parts collected, we get
\eqa{
 \mathscr T_{\alpha\beta}(\bsl R)
 &=
 \frac{\eta_\alpha\eta_\beta^*}{\sqrt{2\tau_I}}
 \exp\left[
 -\frac{\pi(y_\alpha^2+y_\beta^2)}{\tau_I}
 +\frac{\ii\pi(x_\alpha y_\alpha-x_\beta y_\beta)}{\tau_I}
 -\frac{\pi\tau_I m_1^2}{2}
 \right.
 \\
 &\qquad\left.
 +\ii\pi m_1
 \bigl(m_2+x_\alpha+x_\beta+\ii(y_\alpha-y_\beta)\bigr)
 -\frac{\pi}{2\tau_I}
 \bigl(
 m_2+\tau_Rm_1+x_\alpha-x_\beta
 +\ii(y_\alpha+y_\beta)
 \bigr)^2
 \right]
 \\
 &=
 \frac{\eta_\alpha\eta_\beta^*}{\sqrt{2\tau_I}}
 \exp\left[-\frac{\pi}{2\tau_I}
 \left\{(m_2+\tau_Rm_1+x_\alpha-x_\beta)^2
 +(\tau_I m_1+y_\alpha-y_\beta)^2\right\}\right] \\
 &\quad\times
 \exp\left[\ii\pi m_1m_2
 +\ii\pi\left\{
 m_1(x_\alpha+x_\beta)
 -\frac{(m_2+\tau_Rm_1)(y_\alpha+y_\beta)}{\tau_I}
 +\frac{x_\beta y_\alpha-x_\alpha y_\beta}{\tau_I}
 \right\}\right].
 \label{eq:E-collected-exponent}}
Since
\eq{
\frac{R+\delta_\alpha-\delta_\beta}{a_2}
=
m_2+m_1\tau_0+w_\alpha-w_\beta
=
(m_2+\tau_Rm_1+x_\alpha-x_\beta)
+\ii(\tau_Im_1+y_\alpha-y_\beta),
}
we have
\eq{
(m_2+\tau_Rm_1+x_\alpha-x_\beta)^2
+(\tau_Im_1+y_\alpha-y_\beta)^2 =
\left|
\frac{R+\delta_\alpha-\delta_\beta}{a_2}
\right|^2
= \frac{
|\bsl R+\bsl\delta_\alpha-\bsl\delta_\beta|^2
}{|a_2|^2}.
\label{eq:E-distance-identification}
}
where $R=R_x-\ii R_y=m_1a_1+m_2a_2$ is the complex representative of $\bsl R$. Using $\Omega=|a_2|^2\tau_I$, the real exponential becomes
\eq{
\exp\left[
-\frac{\pi}{2\tau_I}
\left\{(m_2+\tau_Rm_1+x_\alpha-x_\beta)^2
+(\tau_Im_1+y_\alpha-y_\beta)^2\right\}
\right]
=
\exp\left[
-\frac{\pi
|\bsl R+\bsl\delta_\alpha-\bsl\delta_\beta|^2}
{2\Omega}
\right].
}
For the phase, using
$\bsl r\times\bsl r'=-\Im(\bar z z')$, we find
\eq{
\frac{\bsl R\times
(\bsl\delta_\alpha+\bsl\delta_\beta)}{\Omega}
=
m_1(x_\alpha+x_\beta)
-\frac{(m_2+\tau_Rm_1)(y_\alpha+y_\beta)}{\tau_I},
\qquad
\frac{\bsl\delta_\alpha\times\bsl\delta_\beta}{\Omega} =
\frac{x_\beta y_\alpha-x_\alpha y_\beta}{\tau_I}.
}
Hence
\eqa{
m_1(x_\alpha+x_\beta)
-\frac{(m_2+\tau_Rm_1)(y_\alpha+y_\beta)}{\tau_I}
+\frac{x_\beta y_\alpha-x_\alpha y_\beta}{\tau_I} =
\frac{
\bsl R\times(\bsl\delta_\alpha+\bsl\delta_\beta)
+\bsl\delta_\alpha\times\bsl\delta_\beta}{\Omega}.
\label{eq:E-phase-identification}
}
Finally, since $m_1,m_2\in\dsZ$,
$e^{\ii\pi m_1m_2}=(-1)^{m_1m_2}$.
Thus \cref{eq:E-collected-exponent} gives the expression in \cref{eq:E-gaussian-kernel}. Applying this result to the four entries of
\cref{eq:E-bloch-matrix} gives the hopping matrix
in \cref{eq:E-final-hopping}. These hopping amplitudes (\cref{eq:E-gaussian-kernel}) decay Gaussianly with physical distance but are not strictly finite range, consistent with the finite-range no-go theorem for isolated Chern-ideal bands established in Ref. \cite{zhao2026idealbandstightbindingmodels}.

We now verify Hermiticity of the infinite-lattice hopping.
For $\bsl R=m_1\bsl a_1+m_2\bsl a_2$, taking the complex
conjugate of \cref{eq:E-gaussian-kernel} after exchanging
$\alpha\leftrightarrow\beta$ and replacing
$\bsl R\to-\bsl R$ gives
\eqa{
 \mathscr T_{\beta\alpha}(-\bsl R)^*
 &=\frac{(\eta_\beta\eta_\alpha^*)^*}{\sqrt{2\tau_I}}
 (-1)^{(-m_1)(-m_2)}
 e^{-\frac{\pi|-\bsl R+\bsl\delta_\beta-\bsl\delta_\alpha|^2}{2\Omega}}
 e^{-\frac{\ii\pi}{\Omega}
 \{(-\bsl R)\times(\bsl\delta_\beta+\bsl\delta_\alpha)
 +\bsl\delta_\beta\times\bsl\delta_\alpha\}}
 \\
 &=\frac{\eta_\alpha\eta_\beta^*}{\sqrt{2\tau_I}}
 (-1)^{m_1m_2}
 e^{-\frac{\pi|\bsl R+\bsl\delta_\alpha-\bsl\delta_\beta|^2}{2\Omega}}
 e^{\frac{\ii\pi}{\Omega}
 \{\bsl R\times(\bsl\delta_\alpha+\bsl\delta_\beta)
 +\bsl\delta_\alpha\times\bsl\delta_\beta\}}
 =\mathscr T_{\alpha\beta}(\bsl R).
 \label{eq:E-kernel-hermiticity-check}
}
Here $\Omega$ and $\tau_I$ are real and positive.
Applying this identity to \cref{eq:E-final-hopping}, we obtain
\eq{
t(-\bsl R)^\dagger =\begin{pmatrix}
\mathscr T_{BB}(-\bsl R)^*&-\mathscr T_{BA}(-\bsl R)^*\\
-\mathscr T_{AB}(-\bsl R)^*&\mathscr T_{AA}(-\bsl R)^*
\end{pmatrix}
=\begin{pmatrix}
\mathscr T_{BB}(\bsl R)&-\mathscr T_{AB}(\bsl R)\\
-\mathscr T_{BA}(\bsl R)&\mathscr T_{AA}(\bsl R)
\end{pmatrix}
=t(\bsl R).
\label{eq:E-infinite-hopping-hermiticity}
}
Equivalently,
$t_{\alpha\beta}(\bsl R)=t_{\beta\alpha}(-\bsl R)^*$
for every Bravais-lattice displacement $\bsl R$.

\subsubsection{Finite Lattice with Periodic/Twist Boundary Condition}

We now move to the finite lattice.
On a finite $N_1\times N_2$ torus with twists
$\phi=(\phi_1,\phi_2)$, the exact hopping is obtained from the
discrete inverse Fourier transform,
\eqa{
t^{(N,\phi)}_{\alpha\beta}(\bsl R)
&=
\frac1N
\sum_{\bsl k\in B^{N_1,N_2}_{\phi_1,\phi_2}}
e^{\ii\bsl k\cdot
(\bsl R+\bsl\delta_\alpha-\bsl\delta_\beta)}
[h_0(\bsl k)]_{\alpha\beta}.
\label{eq:E-torus-hopping-first}
}
To relate this to the infinite-lattice hopping, insert
$[h_0(\bsl k)]_{\alpha\beta}
= \sum_{\bsl R'\in\mathcal L_\infty}
e^{-\ii\bsl k\cdot
(\bsl R'+\bsl\delta_\alpha-\bsl\delta_\beta)}
t_{\alpha\beta}(\bsl R')$.
Then
\eqa{
t^{(N,\phi)}_{\alpha\beta}(\bsl R)
&=
\sum_{\bsl R'\in\mathcal L_\infty}
t_{\alpha\beta}(\bsl R')
\frac1N
\sum_{\bsl k\in B^{N_1,N_2}_{\phi_1,\phi_2}}
e^{\ii\bsl k\cdot(\bsl R-\bsl R')}.
\label{eq:E-torus-mesh-sum}
}
The discrete momentum sum vanishes unless $\bsl R'
= \bsl R+\ell_1\bsl L_1+\ell_2\bsl L_2, \quad
\ell_1,\ell_2\in\dsZ.$
For these surviving terms,
\eq{
e^{\ii\bsl k\cdot(\bsl R-\bsl R')} =
e^{-\ii\ell_1\bsl k\cdot\bsl L_1}
e^{-\ii\ell_2\bsl k\cdot\bsl L_2} =
e^{-\ii[\ell_1(2\pi n_1+\phi_1)
+\ell_2(2\pi n_2+\phi_2)]} =
e^{-\ii(\ell_1\phi_1+\ell_2\phi_2)},
}
where we used the allowed torus momenta in
\cref{eq:allowed+momenta}. Hence
\eqa{
t^{(N,\phi)}_{\alpha\beta}(\bsl R)
&=
\sum_{\ell_1,\ell_2\in\dsZ}
e^{-\ii(\ell_1\phi_1+\ell_2\phi_2)}
t_{\alpha\beta}
(\bsl R+\ell_1\bsl L_1+\ell_2\bsl L_2).
\label{eq:E-torus-hopping}
}
Thus, the finite-torus hopping from sublattice $\beta$ in the reference cell to sublattice $\alpha$ in unit cell $\bsl R$ is obtained by summing the infinite-lattice hopping amplitudes to sublattice $\alpha$ in all cells $\bsl R+\ell_1\bsl L_1+\ell_2\bsl L_2$, weighted by the corresponding twisted-boundary phases.

We first express the finite-torus Hamiltonian in the physical-site basis.
Substituting the Fourier operators in \cref{eq:E-fourier-operators}
into the momentum-space Hamiltonian and using \cref{eq:E-torus-hopping-first} gives
\eqa{
H_0
&=\sum_{\bsl k\in B_{\phi_1,\phi_2}^{N_1,N_2}}
\sum_{\alpha,\beta=A,B}
c^\dagger_{\bsl k\alpha}
[h_0(\bsl k)]_{\alpha\beta}c_{\bsl k\beta}
\\
&=\sum_{\bsl R,\bsl R'\in\mathcal L_N}\sum_{\alpha,\beta=A,B}
\frac1N\sum_{\bsl k\in B_{\phi_1,\phi_2}^{N_1,N_2}}
e^{\ii\bsl k\cdot
(\bsl R-\bsl R'+\bsl\delta_\alpha-\bsl\delta_\beta)}
[h_0(\bsl k)]_{\alpha\beta}
c^\dagger_{\bsl R+\bsl\delta_\alpha}
c_{\bsl R'+\bsl\delta_\beta}
\\
&=\sum_{\bsl R,\bsl R'\in\mathcal L_N}\sum_{\alpha,\beta=A,B}
t_{\alpha\beta}^{(N,\phi)}(\bsl R-\bsl R')
c^\dagger_{\bsl R+\bsl\delta_\alpha}
c_{\bsl R'+\bsl\delta_\beta}.
\label{eq:E-h0-finite-site-form}
}
Both $\bsl R$ and
$\bsl R'$ run over the $N$ independent unit-cell representatives
$n_1\bsl a_1+n_2\bsl a_2$, with $0\leq n_j<N_j$.
We next verify that the finite-torus hopping preserves
Hermiticity. Write
$\bsl T_\ell=\ell_1\bsl L_1+\ell_2\bsl L_2$,
where $\ell=(\ell_1,\ell_2)\in\dsZ^2$.
Using \cref{eq:E-torus-hopping}, we obtain
\eqa{
\left[t_{\beta\alpha}^{(N,\phi)}(-\bsl R)\right]^*
&=\sum_{\ell\in\dsZ^2}
e^{\ii(\ell_1\phi_1+\ell_2\phi_2)}
t_{\beta\alpha}(-\bsl R+\bsl T_\ell)^* =\sum_{\ell\in\dsZ^2}
e^{\ii(\ell_1\phi_1+\ell_2\phi_2)}
t_{\alpha\beta}(\bsl R-\bsl T_\ell)
\\
&=\sum_{\ell\in\dsZ^2}
e^{-\ii(\ell_1\phi_1+\ell_2\phi_2)}
t_{\alpha\beta}(\bsl R+\bsl T_\ell)
=t_{\alpha\beta}^{(N,\phi)}(\bsl R),
\label{eq:E-torus-hopping-hermiticity}
}
here the second equality uses the infinite-lattice Hermiticity in \cref{eq:E-infinite-hopping-hermiticity}, and the third relabels $\ell\to-\ell$.
We next rewrite the finite-site Hamiltonian in
\cref{eq:E-h0-finite-site-form} as an infinite-image sum.
Every site in $\overline\Lambda$ can be written uniquely as
$\bsl s+\bsl T_\ell$, with $\bsl s\in\Lambda$, using
\cref{eq:infinite-lattice-site-set,eq:finite-torus-site-set}.
The twisted boundary conditions imply
\eq{
c_{\bsl s+\bsl T_\ell}
=e^{\ii(\ell_1\phi_1+\ell_2\phi_2)}c_{\bsl s}.
\label{eq:E-hopping-image-boundary}
}
Substituting the image sum in \cref{eq:E-torus-hopping}
into \cref{eq:E-h0-finite-site-form}, we find
\eqa{
H_0
&=\sum_{\bsl R,\bsl R'\in\mathcal L_N}\sum_{\alpha,\beta=A,B}
\sum_{\ell\in\dsZ^2}
e^{-\ii(\ell_1\phi_1+\ell_2\phi_2)}
t_{\alpha\beta}(\bsl R-\bsl R'+\bsl T_\ell)
c^\dagger_{\bsl R+\bsl\delta_\alpha}
c_{\bsl R'+\bsl\delta_\beta}
\\
&=\sum_{\bsl R,\bsl R'\in\mathcal L_N}\sum_{\alpha,\beta=A,B}
\sum_{\ell\in\dsZ^2}
e^{\ii(\ell_1\phi_1+\ell_2\phi_2)}
t_{\alpha\beta}(\bsl R-\bsl R'-\bsl T_\ell)
c^\dagger_{\bsl R+\bsl\delta_\alpha}
c_{\bsl R'+\bsl\delta_\beta}
\\
&=\sum_{\bsl R,\bsl R'\in\mathcal L_N}\sum_{\alpha,\beta=A,B}
\sum_{\ell\in\dsZ^2}
t_{\alpha\beta}(\bsl R-\bsl R'-\bsl T_\ell)
c^\dagger_{\bsl R+\bsl\delta_\alpha}
c_{\bsl R'+\bsl\delta_\beta+\bsl T_\ell}
\\
&=\sum_{\bsl r\in\Lambda}
\sum_{\bsl r'\in\overline\Lambda}
t_{\bsl r\bsl r'}c^\dagger_{\bsl r}c_{\bsl r'}.
\label{eq:E-h0-image-sum-equivalence}
}
The second equality relabels $\ell\to-\ell$,
and the third uses \cref{eq:E-hopping-image-boundary}.
In the last equality,
$\bsl r=\bsl R+\bsl\delta_\alpha\in\Lambda$ and
$\bsl r'=\bsl R'+\bsl\delta_\beta+\bsl T_\ell
\in\overline\Lambda$, with
$t_{\bsl r\bsl r'}=
t_{\alpha\beta}(\bsl R-\bsl R'-\bsl T_\ell)$.

We now verify Hermiticity directly in this infinite-image form.
Write $\ell\cdot\phi=\ell_1\phi_1+\ell_2\phi_2$.
The hopping depends only on the cell displacement and sublattice
indices, so
$t_{\bsl s+\bsl T_\ell,\bsl r}
=t_{\bsl s,\bsl r-\bsl T_\ell}$.
Using this translation identity, the Hermiticity of $t$, and
\cref{eq:E-hopping-image-boundary}, we obtain
\eqa{
H_0^\dagger
&=\sum_{\bsl r,\bsl s\in\Lambda}\sum_{\ell\in\dsZ^2}
t_{\bsl s+\bsl T_\ell,\bsl r}
c_{\bsl s+\bsl T_\ell}^{\dagger}c_{\bsl r}\\
&=\sum_{\bsl r,\bsl s\in\Lambda}\sum_{\ell\in\dsZ^2}
e^{-\ii\ell\cdot\phi}
t_{\bsl s,\bsl r-\bsl T_\ell}
c_{\bsl s}^{\dagger}c_{\bsl r}\\
&=\sum_{\bsl r,\bsl s\in\Lambda}\sum_{\ell\in\dsZ^2}
e^{\ii\ell\cdot\phi}
t_{\bsl r,\bsl s+\bsl T_\ell}
c_{\bsl r}^{\dagger}c_{\bsl s}\\
&=\sum_{\bsl r\in\Lambda}\sum_{\bsl r'\in\overline\Lambda}
t_{\bsl r\bsl r'}c_{\bsl r}^{\dagger}c_{\bsl r'}=H_0.
\label{eq:E-h0-hermiticity}
}
The third line exchanges $\bsl r\leftrightarrow\bsl s$ and
relabels $\ell\to-\ell$, and the last line uses
$c_{\bsl s+\bsl T_\ell}=e^{\ii\ell\cdot\phi}c_{\bsl s}$
and $\bsl r'=\bsl s+\bsl T_\ell$.
So we showed this Hamiltonian is Hermitian.

For comparison, we rewrite the projective coordinate $f(k)$
derived in \cref{eq:explicit-meromorphic-f} to make its relation
to the previous ideal-band construction in
Ref. \cite{zhao2026idealbandstightbindingmodels} explicit. Write the sublattice
displacement as
\eq{
\delta_B-\delta_A=\rho_1a_1+\rho_2a_2,
\qquad \rho_1,\rho_2\in\dsR.
\label{eq:E-displacement}
}
Using \cref{eq:complex-reciprocal-vectors} and
$a_1\bar a_2-a_2\bar a_1=2\ii\Omega$, the
momentum-dependent exponent becomes
\eq{
\frac{\ii}{2}
\left[
\frac{\bar a_1}{a_1}(\delta_B-\delta_A)
-(\bar\delta_B-\bar\delta_A)
\right]k
=
\frac{\rho_2\Omega}{a_1}k
=
-2\pi\ii\rho_2\frac{k}{b_2}.
\label{eq:E-matching-exponential}
}
Therefore, \cref{eq:explicit-meromorphic-f} takes the form
\eq{
f(k)
= \lambda e^{-2\pi\ii\rho_2 k/b_2}
\frac{
\vartheta_1\left(
\frac{k}{b_2}
+\frac{\delta_B-(a_1+a_2)/2}{a_1}
\middle|\frac{b_1}{b_2}
\right)}{
\vartheta_1\left(
\frac{k}{b_2}
+\frac{\delta_A-(a_1+a_2)/2}{a_1}
\middle|\frac{b_1}{b_2}
\right)},
\label{eq:E-matched-f}
}
where the nonzero, momentum-independent prefactor is
\eq{
\lambda
=
e^{\frac{\pi[\mathcal B(\bsl\delta_B)
-\mathcal B(\bsl\delta_A)]}{2\Omega}}
\frac{\chi(\bsl\delta_B)}{\chi(\bsl\delta_A)}
\exp\left[
-\frac{\ii\pi(\delta_B-\delta_A)
(\delta_A+\delta_B-a_2)}{a_1a_2}
\right].
\label{eq:E-matching-prefactor}
}
The numerator and denominator theta-function arguments differ by
$(\delta_B-\delta_A)/a_1=\rho_1-\rho_2b_1/b_2$.
Thus,
\cref{eq:E-matched-f} has the theta-function ratio form of
Eq.(23) in Ref. \cite{zhao2026idealbandstightbindingmodels},
multiplied by the nonzero constant $\lambda$.
This constant leaves the zeros, poles, and reciprocal-lattice
transformation law of $f(k)$ unchanged.
The band obtained in \cref{sec:C-band-containment} therefore
realizes the same type of $|\Ch|=1$ ideal-band construction.

\subsubsection{Sublattice Dependence of the Hopping}

At last, we check whether the infinite-lattice hopping amplitudes
in \cref{eq:E-gaussian-kernel,eq:E-final-hopping} can be expressed
as a single function of the physical displacement, \ie,
\eq{
\text{whether }t_{\alpha\beta}(\bsl R)
=
t_{\mathrm{disp}}\left(
\bsl R+\bsl\delta_\alpha-\bsl\delta_\beta
\right) \text{ can hold for all }
\bsl{R},\alpha,\beta.
\label{eq:E-disp-only-ansatz}
}
We keep the sublattice basis and phase convention of
\cref{sec:SP-hamiltonian} and the inequivalence assumption in
\cref{eq:C-orbital-inequivalence}.
Suppose \cref{eq:E-disp-only-ansatz} is true.
We would have $\frac{t_{AA}(\bsl R)}{t_{BB}(\bsl R)}=1$ for all $\bsl{R}$, which means $\frac{|\eta_B|^2}{|\eta_A|^2}
\exp\left[
\frac{2\pi\ii}{\Omega}
\bsl R\times(\bsl\delta_B-\bsl\delta_A)
\right]=1$.
Then, it would make $\bsl\delta_B-\bsl\delta_A$ a lattice vector leading to contradiction.
Thus, \cref{eq:E-disp-only-ansatz} must be false.

\subsection{Two-Body Interaction and Exact Projected fractional Chern Ground States}
\label{app:E_2}

We now analyze the real-space interaction
$H_{\mathrm{int}}$ defined in
\cref{eq:E-general-interaction}
and show that its lower-band projection
$\widetilde H_{\mathrm{int}}$ belongs to the class
in \cref{eq:D-general-projected-interaction}.
The Hamiltonian is defined for arbitrary electron number and boundary twists. 

\subsubsection{Properties of the Unprojected Interaction}
\label{sec:E-simple-real-space}

We first examine the interaction $H_{\mathrm{int}}$ before projection. We show that it is a positive-semidefinite two-body interaction and explain how it is defined on the finite torus with twisted boundary conditions.

We use the physical-site convention of
\cref{eq:infinite-lattice-site-set,eq:finite-torus-site-set}.
For $\bsl r=\bsl R+\bsl\delta_\alpha$ and
$\bsl r'=\bsl R'+\bsl\delta_\beta$, define
\eq{
t_{\bsl r\bsl r'}
\equiv t_{\alpha\beta}(\bsl R-\bsl R'),
\qquad
t_{\bsl r\bsl r'}=t_{\bsl r'\bsl r}^*.
\label{eq:E-site-hopping}
}
Here $t_{\bsl r\bsl r'}$ is the infinite-lattice hopping
amplitude from $\bsl r'$ to $\bsl r$, given
by \cref{eq:E-gaussian-kernel,eq:E-final-hopping}. Recall the two pair-annihilation
operators at each physical site $\bsl r$:
\eq{
B_{\bsl r}^{(p)}
=
\sum_{\substack{
\bsl r'\in\overline\Lambda\\
\bsl r'\ne\bsl r
}}
\frac{t_{\bsl r\bsl r'}}
{(z_{\bsl r'}-z_{\bsl r})^p}
c_{\bsl r'}c_{\bsl r},
\qquad p=1,2.
\label{eq:E-simple-bond-operators}
}
Here $c_{\bsl r}=c_{\bsl R+\bsl\delta_\alpha}$ annihilates
a fermion at site $\bsl r$, and
$z_{\bsl r}=r_x-\ii r_y$ is its complex coordinate. The sum includes both sublattices and all periodic
images in $\overline\Lambda$.
Let $u>0$ be the real-space interaction strength, where $N=N_1N_2$ is the number of unit cells. The unprojected interaction and full Hamiltonian are
\eqa{
H_{\mathrm{int}}
&=
u
\sum_{\bsl r\in\Lambda}
\sum_{p=1}^{2}
\Omega^{p-1}
B_{\bsl r}^{(p)\dagger}B_{\bsl r}^{(p)},
\qquad u>0,\\
H&=H_0+H_{\mathrm{int}}.
\label{eq:E-simple-full-hamiltonian}
}
The sum over $\bsl r$ runs over the $2N$ independent
physical sites of the finite torus. Each site
contributes two pair operators, with $p=1,2$. Since $u>0$ and $\Omega>0$, the interaction is
Hermitian and positive semidefinite. 

We next specify how the infinite-lattice sums in
\cref{eq:E-simple-bond-operators} are implemented
on the finite torus. For $\ell_1,\ell_2\in\dsZ$,
define the full-period translation
\eq{
\bsl T_\ell
=
\ell_1\bsl L_1+\ell_2\bsl L_2,
\qquad
T_\ell=\ell_1L_1+\ell_2L_2,
\qquad
\ell=(\ell_1,\ell_2)\in\dsZ^2,
}
where $T_\ell$ is the complex representative
of $\bsl T_\ell$. Every site $\bsl r'\in\overline\Lambda$ can be
written uniquely as
$\bsl r'=\bsl s+\bsl T_\ell$, with
$\bsl s\in\Lambda$. Under the twisted boundary
conditions, the annihilation operators satisfy
\eq{
c_{\bsl s+\bsl T_\ell}
=
e^{\ii(\ell_1\phi_1+\ell_2\phi_2)}
c_{\bsl s}.
}
On the finite torus, periodic images are identified
with the same physical site, and their annihilation
operators differ only by the boundary phase. Substituting into \cref{eq:E-simple-bond-operators}, we obtain
\eq{
B_{\bsl r}^{(p)}
=\sum_{\substack{\bsl s\in\Lambda\\\bsl s\ne\bsl r}}
\sum_{\ell\in\dsZ^2}
\frac{t_{\bsl r,\bsl s+\bsl T_\ell}}
{(z_{\bsl s}+T_\ell-z_{\bsl r})^p}
c_{\bsl s+\bsl T_\ell}c_{\bsl r}=\sum_{\substack{\bsl s\in\Lambda\\\bsl s\ne\bsl r}}
\left[
\sum_{\ell\in\dsZ^2}
e^{\ii(\ell_1\phi_1+\ell_2\phi_2)}
\frac{t_{\bsl r,\bsl s+\bsl T_\ell}}
{(z_{\bsl s}+T_\ell-z_{\bsl r})^p}
\right]c_{\bsl s}c_{\bsl r}.
\label{eq:E-simple-periodized-kernel}
}
The restriction excludes the coincident site, and terms with $\bsl s=\bsl r$ and nonzero $\ell$ vanish because $c_{\bsl r}^2=0$. Since translating $B_{\bsl r}^{(p)}$ by a lattice vector $\bsl{R}$ gives $B_{\bsl r + \bsl{R}}^{(p)}$, $H_{\mathrm{int}}$ preserves lattice translation symmetry and conserves
total crystal momentum.

In the following subsection, we project $H_{\mathrm{int}}$ onto the
ideal band and identify the coefficient matrix in the general
form \cref{eq:D-general-projected-interaction}.

\subsubsection{Projection onto the Ideal Band}
\label{sec:E-simple-projection}

We project the physical-site pair operators
$B_{\bsl r}^{(p)}$ onto the ideal band and express them in
terms of the pair modes $Q_{\bsl K,\mu}$ defined in
\cref{eq:Q_Kmu}.
This will put $\widetilde H_{\mathrm{int}}$ in the
general form \cref{eq:D-general-projected-interaction}.

Let $V_{\bsl k}$ be the normalized upper-band Bloch spinor
introduced in \cref{eq:E-upper-band-action}, so that
$U_{\bsl k}^\dagger V_{\bsl k}=0$ and
$V_{\bsl k}^\dagger V_{\bsl k}=1$.
Using the same physical embedding convention as
\cref{eq:C-orbital-Bloch-basis}, define
\eqa{
 \varphi_{\bsl k}(\bsl R+\bsl\delta_\alpha)
 &=\frac{1}{\sqrt N}
 e^{\ii\bsl k\cdot(\bsl R+\bsl\delta_\alpha)}
 (V_{\bsl k})_\alpha,\\
 \gamma^\dagger_{1,\bsl k}
 &=\sum_{\bsl r\in\Lambda}
 \varphi_{\bsl k}(\bsl r)c^\dagger_{\bsl r}.
 \label{eq:E-upper-band-orbitals}
}
All one-particle momenta belong to the mesh in
\cref{eq:allowed+momenta}. These upper-band Bloch states satisfy
\eq{
 \sum_{\bsl r\in\Lambda}
 \varphi_{\bsl k}(\bsl r)^*\varphi_{\bsl k'}(\bsl r)
 =\delta_{\bsl k,\bsl k'},
 \qquad
 \sum_{\bsl r\in\Lambda}
 \psi_{\bsl k}(\bsl r)^*\varphi_{\bsl k'}(\bsl r)=0.
 \label{eq:E-band-basis-orthogonality}
}
Together with the ideal-band Bloch states in
\cref{eq:C-normalized-band-orbital}, they form a complete
orthonormal basis of the physical one-particle space.
Inverting the band-operator definitions therefore gives
\eq{
 c_{\bsl r}=\sum_{\bsl k\in B_{\phi_1,\phi_2}^{N_1,N_2}}
 \left[\psi_{\bsl k}(\bsl r)\gamma_{\bsl k}
 +\varphi_{\bsl k}(\bsl r)\gamma_{1,\bsl k}\right].
 \label{eq:E-two-band-expansion}
}
Projection onto the ideal band means discarding every term
containing $\gamma_{1,\bsl k}$ or
$\gamma_{1,\bsl k}^{\dagger}$.
For the interaction in \cref{eq:E-general-interaction},
this is equivalent to the substitutions
\eq{
 c_{\bsl r}\rightarrow
 \sum_{\bsl k \in B_{\phi_1,\phi_2}^{N_1,N_2}}\psi_{\bsl k}(\bsl r)\gamma_{\bsl k},
 \qquad
 c^\dagger_{\bsl r}\rightarrow
 \sum_{\bsl k\in B_{\phi_1,\phi_2}^{N_1,N_2}}\psi_{\bsl k}(\bsl r)^*
 \gamma^\dagger_{\bsl k},
 \label{eq:E-band-projection-substitution}
}
with the result acting on the ideal-band Fock space. Applying \cref{eq:E-band-projection-substitution} to
the physical pair operators in \cref{eq:E-simple-bond-operators},
we obtain
\eq{
\widetilde B_{\bsl r}^{(p)}\equiv
\sum_{\substack{\bsl r'\in\overline\Lambda\\
\bsl r'\ne\bsl r}}
\frac{t_{\bsl r\bsl r'}}{(z_{\bsl r'}-z_{\bsl r})^p}
\sum_{\bsl k'\in B_{\phi_1,\phi_2}^{N_1,N_2}}\psi_{\bsl k'}(\bsl r')
\gamma_{\bsl k'}\sum_{\bsl k\in B_{\phi_1,\phi_2}^{N_1,N_2}}\psi_{\bsl k}(\bsl r)
\gamma_{\bsl k}
=\sum_{\bsl k,\bsl k'\in B_{\phi_1,\phi_2}^{N_1,N_2}}
\left[
\sum_{\substack{\bsl r'\in\overline\Lambda\\
\bsl r'\ne\bsl r}}
\frac{t_{\bsl r\bsl r'}}
{(z_{\bsl r'}-z_{\bsl r})^p}
\psi_{\bsl k}(\bsl r)
\psi_{\bsl k'}(\bsl r')
\right]
\gamma_{\bsl k'}\gamma_{\bsl k},
\label{eq:E-projected-bond-operators}
}
with $p=1,2$. Then, the projected interaction is
\eq{
 \widetilde H_{\mathrm{int}}
 =\frac{g}{N}\sum_{\bsl r\in\Lambda}\sum_{p=1}^{2}
 \Omega^{p-1}
 \widetilde B_{\bsl r}^{(p)\dagger}
 \widetilde B_{\bsl r}^{(p)}.
 \label{eq:E-projected-interaction-definition}
}
Its spectrum and zero-energy space are considered within the
ideal-band Fock space, with no upper-band particles.
The full interaction remains $H_{\mathrm{int}}$ in \cref{eq:E-simple-full-hamiltonian}.

Write $\bsl r=\bsl R+\bsl\delta_\alpha$ with $\alpha=A,B$.
The key identities, which relate the operators $\widetilde B_{\bsl r}^{(p)}$
to the operators $Q_{\bsl K}(\zeta)$
introduced in \cref{sec:D1}, are
\eqa{
 \widetilde B_{\bsl r}^{(1)}
 &=C_\alpha\sum_{\bsl K\in B_{2\phi_1,2\phi_2}^{N_1,N_2}}
 e^{\ii\bsl K\cdot\bsl R}Q_{\bsl K}(\delta_\alpha),\\
 \widetilde B_{\bsl r}^{(2)}
 &=\frac{C_\alpha}{2}\sum_{\bsl K\in B_{2\phi_1,2\phi_2}^{N_1,N_2}}
 e^{\ii\bsl K\cdot\bsl R}
 \left[Q'_{\bsl K}(\delta_\alpha)
 -\lambda_\alpha Q_{\bsl K}(\delta_\alpha)\right].
 \label{eq:E-simple-bond-projections}
}
Here $Q_{\bsl K}(\zeta)$ is defined in
\cref{eq:Q_K}, with
$Q_{\bsl K}'(\zeta)=\partial_\zeta Q_{\bsl K}(\zeta)$.
The coefficients $C_\alpha$ are defined by
\eq{
C_\alpha=t_{\bsl\delta_\alpha\bsl\delta_\alpha}
e^{\pi\mathcal B(\bsl{\delta}_\alpha)/\Omega}
\chi^2(\bsl{\delta}_\alpha)\neq0,
\label{eq:E-simple-local-constants}
}
where $t_{\bsl\delta_\alpha\bsl\delta_\alpha}
=t_{\alpha\alpha}(\bsl{0})
=|\eta_{\bar\alpha}|^2/\sqrt{2\tau_I}>0$
by \cref{eq:E-gaussian-kernel,eq:E-final-hopping} and $\bar\alpha$ denotes
the other sublattice.
The coefficient $\lambda_\alpha$ is defined as
\eq{
\lambda_\alpha
=\frac{2}{a_2}
\frac{\vartheta_1'\left(
\frac{\delta_\alpha-\delta_{\bar\alpha}}{a_2}
\middle|\tau_0\right)}
{\vartheta_1\left(
\frac{\delta_\alpha-\delta_{\bar\alpha}}{a_2}
\middle|\tau_0\right)}
-\frac{2\pi}{\Omega}
\left(
\frac{\bar a_2}{a_2}\delta_{\bar\alpha}
-\bar\delta_{\bar\alpha}
\right).
\label{eq:E-simple-lambda}
}
The prime on $\vartheta_1$ in \cref{eq:E-simple-lambda}
denotes differentiation with respect to the 
first argument, evaluated at
$(\delta_\alpha-\delta_{\bar\alpha})/a_2$.

We now prove \cref{eq:E-simple-bond-projections}.
Write $\bsl r=\bsl R+\bsl\delta_\alpha$ and define $\bsl s=\bsl r'-\bsl R$, so that $\bsl r'=\bsl R+\bsl s$. Using the hopping translation symmetry and Bloch translation laws, we have $t_{\bsl r\bsl r'}=t_{\bsl\delta_\alpha\bsl s}$, $z_{\bsl r'}-z_{\bsl r}=z_{\bsl s}-\delta_\alpha$, and $\psi_{\bsl k}(\bsl r)\psi_{\bsl k'}(\bsl r')=e^{\ii(\bsl k+\bsl k')\cdot\bsl R}\psi_{\bsl k}(\bsl\delta_\alpha)\psi_{\bsl k'}(\bsl s)$. Substituting into \cref{eq:E-projected-bond-operators},
\eq{
 \widetilde B_{\bsl r}^{(p)}
 =\sum_{\bsl k,\bsl k'\in B_{\phi_1,\phi_2}^{N_1,N_2}}
 e^{\ii(\bsl k+\bsl k')\cdot\bsl R}
 \left[
 \sum_{\substack{\bsl s\in\overline\Lambda\\
                  \bsl s\ne\bsl\delta_\alpha}}
 \frac{t_{\bsl\delta_\alpha\bsl s}}
 {(z_{\bsl s}-\delta_\alpha)^p}
 \psi_{\bsl k}(\bsl\delta_\alpha)\psi_{\bsl k'}(\bsl s)
 \right]\gamma_{\bsl k'}\gamma_{\bsl k}.
 \label{eq:F-bond-band-expansion_2}
}
We now try to use 
\cref{thm:sum_t_r_p}.
To match the notation in \cref{thm:sum_t_r_p}, we define 
\eq{
M_{\boldsymbol\delta_\alpha,\mathbf s}
=e^{\pi\mathcal B(\bsl{\delta}_{\alpha})/(2\Omega)}\chi(\bsl{\delta}_{\alpha}) t_{\boldsymbol\delta_\alpha,\mathbf s}e^{\pi\mathcal B(\mathbf s)/(2\Omega)}\chi(\mathbf s),
}
which leads to $C_\alpha=M_{\boldsymbol\delta_\alpha,\boldsymbol\delta_\alpha}$ and the following rewritten form
\eq{
\widetilde B_{\mathbf r}^{(p)}
=
\sum_{\mathbf k,\mathbf k'\in B_{\phi_1,\phi_2}^{N_1,N_2}}
\frac{e^{\ii(\mathbf k+\mathbf k')\cdot\mathbf R}}
{\mathcal N_{\mathbf k}\mathcal N_{\mathbf k'}}
g_{\mathbf k}(\delta_\alpha)
\left[
\sum_{\mathbf s\ne\boldsymbol\delta_\alpha}
\frac{M_{\boldsymbol\delta_\alpha,\mathbf s}}
{(z_{\mathbf s}-\delta_\alpha)^p}
g_{\mathbf k'}(z_{\mathbf s})
\right]
\gamma_{\mathbf k'}\gamma_{\mathbf k}.
}
As the lower band has zero energy, we have
\eq{
\sum_{\bsl{s}\in\bar{\Lambda}}t_{\boldsymbol\delta_\alpha,\bsl{s}} \psi_{\bsl{k}}(\bsl{s}) = 0 \ \forall \bsl{k}\in\dsR^2\ ,
}
implying that
\eq{
\sum_{\bsl{s}\in\bar{\Lambda}}M_{\boldsymbol\delta_\alpha,\bsl{s}} g_{\bsl{k}}(\bsl{s}) = 0 \ \forall \bsl{k}\in\dsR^2\ .
}
Furthermore, we verify the uniform absolute convergence
required by \cref{thm:sum_t_r_p}.
Every lattice site can be written as
$\bsl s=\bsl T+\bsl\delta_\beta$,
where $\bsl T\in\mathcal L_\infty$ and $\beta=A,B$.
Using the definition of $M_{\bsl\delta_\alpha,\bsl s}$
above and the Bloch wavefunction and translation laws
in \cref{eq:C-normalized-band-orbital,eq:continuous-bloch-translation-a1,eq:continuous-bloch-translation-a2}, we obtain
\eqa{
|M_{\bsl\delta_\alpha,\bsl s}
g_{\bsl k}(z_{\bsl s})|
&=
\left|
e^{\pi\mathcal B(\bsl\delta_\alpha)/(2\Omega)}
\chi(\bsl\delta_\alpha)
\right|
|t_{\bsl\delta_\alpha,\bsl s}|
\mathcal N_{\bsl k}
|\psi_{\bsl k}(\bsl s)|=
\left|
e^{\pi\mathcal B(\bsl\delta_\alpha)/(2\Omega)}
\chi(\bsl\delta_\alpha)
\right|
|t_{\bsl\delta_\alpha,\bsl s}|
\mathcal N_{\bsl k}
|e^{\ii\bsl k\cdot\bsl T}
\psi_{\bsl k}(\bsl\delta_\beta)|\\
&=
\left|
e^{\pi\mathcal B(\bsl\delta_\alpha)/(2\Omega)}
\chi(\bsl\delta_\alpha)
\right|
|t_{\bsl\delta_\alpha,\bsl s}|
\mathcal N_{\bsl k}
|\psi_{\bsl k}(\bsl\delta_\beta)|\le A_\alpha
|t_{\bsl\delta_\alpha,\bsl s}|,
}
where
\eq{
A_\alpha=
\left|
e^{\pi\mathcal B(\bsl\delta_\alpha)/(2\Omega)}
\chi(\bsl\delta_\alpha)
\right|
\max_{\beta=A,B}
\sup_{\bsl k\in\mathrm{BZ}}
\left[
\mathcal N_{\bsl k}
|\psi_{\bsl k}(\bsl\delta_\beta)|
\right]
<\infty.
}
Here $A_\alpha$ is finite because
$\mathcal N_{\bsl k}\psi_{\bsl k}(\bsl\delta_\beta)$
is continuous on the compact BZ, and is
independent of $\bsl s$ and $\bsl k$.
Furthermore, the Gaussian hopping amplitudes
in \cref{eq:E-gaussian-kernel,eq:E-final-hopping}
satisfy
\eq{
\sum_{\bsl s\in\bar\Lambda}
|t_{\bsl\delta_\alpha,\bsl s}|
\le
D_\alpha
\sum_{\beta=A,B}
\sum_{\bsl T\in\mathcal L_\infty}
e^{-\frac{\pi}{2\Omega}
|\bsl T+\bsl\delta_\beta-\bsl\delta_\alpha|^2}
<\infty,
}
for some constant $D_\alpha>0$.
Combining these results, we obtain
\eq{
\sum_{\bsl s\in\bar\Lambda}
|M_{\bsl\delta_\alpha,\bsl s}
g_{\bsl k}(z_{\bsl s})|
\le
A_\alpha
\sum_{\bsl s\in\bar\Lambda}
|t_{\bsl\delta_\alpha,\bsl s}|
<\infty.
}
Since $A_\alpha$ is independent of $\bsl k$
and the hopping amplitudes are absolutely
summable, the Weierstrass M-test guarantees
uniform absolute convergence for
$\bsl k\in\mathrm{BZ}$, as required by
\cref{thm:sum_t_r_p}.
Then, \cref{thm:sum_t_r_p} gives
\eqa{
& \widetilde B_{\mathbf r}^{(1)}
=
C_\alpha\sum_{\mathbf k,\mathbf k'\in B_{\phi_1,\phi_2}^{N_1,N_2}}
\frac{e^{\ii(\mathbf k+\mathbf k')\cdot\mathbf R}}
{\mathcal N_{\mathbf k}\mathcal N_{\mathbf k'}}
g_{\mathbf k}'(\delta_\alpha)g_{\mathbf k'}(\delta_\alpha)
\gamma_{\mathbf k'}\gamma_{\mathbf k} \\
& \widetilde B_{\mathbf r}^{(2)}
=
\frac{C_\alpha}{2}
\sum_{\mathbf k,\mathbf k'\in B_{\phi_1,\phi_2}^{N_1,N_2}}
\frac{e^{\ii(\mathbf k+\mathbf k')\cdot\mathbf R}}
{\mathcal N_{\mathbf k}\mathcal N_{\mathbf k'}}
\left[
g_{\mathbf k}''(\delta_\alpha)g_{\mathbf k'}(\delta_\alpha)
-\lambda_\alpha
g_{\mathbf k}'(\delta_\alpha)g_{\mathbf k'}(\delta_\alpha)
\right]
\gamma_{\mathbf k'}\gamma_{\mathbf k} \ ,
}
where the $j=0$ term in \cref{thm:sum_t_r_p} is proportional to
$g_{\mathbf k}(\delta_\alpha)g_{\mathbf k'}(\delta_\alpha)$ and thus get canceled by the fermionic commutation relation, and 
\eq{
\lambda_\alpha
=\frac{2}{C_\alpha}
\sum_{\mathbf s\ne\boldsymbol\delta_\alpha}
\frac{M_{\boldsymbol\delta_\alpha,\mathbf s}}
{z_{\mathbf s}-\delta_\alpha}\ .
}
Then, using the definition of $Q_{\bsl K}(\zeta)$
in \cref{eq:discrete-V1-pair-annihilator}
and $Q'_{\bsl K}(\zeta)=\partial_\zeta Q_{\bsl K}(\zeta)$,
we obtain \cref{eq:E-simple-bond-projections}.
It remains to prove \cref{eq:E-simple-lambda}.
Now we do so by considering a special momentum for the $p=1$ case of \cref{thm:sum_t_r_p}.
Let us define 
$
\delta_{\bar\alpha}=u_1a_1+u_2a_2,
\qquad u_1,u_2\in\dsR,
$
and choose $\mathbf k_*$ with
$
k_{*,1}=u_2-\frac12,
$
and 
$
k_{*,2}=\frac12-u_1.
$
Then $g_{\mathbf k_*}(\delta_{\bar\alpha})=0$ and thus $\psi_{\bsl{k}_*}(\delta_{\bar\alpha}+\bsl{R})=0$, while
$g_{\mathbf k_*}(\delta_\alpha)\ne0$ because the sublattices are inequivalent. 
Then, we have
\eqa{
\sum_{\bsl{s}\in\bar{\Lambda}\backslash \{\bsl{\delta}_{\alpha}\}} \frac{M_{\bsl{\delta}_{\alpha},\bsl{s}} }{(z_{\bsl{s}}-z_{\bsl{\delta}_{\alpha}})}g_{\bsl{k}_*}(\bsl{s}) & = 
\N_{\bsl{k}_*}
e^{\pi\mathcal B(\bsl{\delta}_{\alpha})/(2\Omega)}
\chi(\bsl{\delta}_{\alpha})
\sum_{\bsl{s}\in\bar{\Lambda}\backslash \{\bsl{\delta}_{\alpha}\}} \frac{t_{\bsl{\delta}_{\alpha},\bsl{s}} }{(z_{\bsl{s}}-z_{\bsl{\delta}_{\alpha}})}\psi_{\bsl{k}_*}(\bsl{s}) \\
& = 
\N_{\bsl{k}_*}
e^{\pi\mathcal B(\bsl{\delta}_{\alpha})/(2\Omega)}
\chi(\bsl{\delta}_{\alpha})
\sum_{(m_1,m_2)\in\dsZ^2\backslash\{(0,0)\}} \frac{t_{\bsl{\delta}_{\alpha},\bsl{\delta}_{\alpha}+m_1\bsl{a}_1+m_2\bsl{a}_2} }{z_{m_1\bsl{a}_1+m_2\bsl{a}_2}}\psi_{\bsl{k}_*}(\bsl{\delta}_{\alpha}) e^{\ii \bsl{k}_*\cdot (m_1\bsl{a}_1+m_2\bsl{a}_2)} \\
& = 
\N_{\bsl{k}_*}
e^{\pi\mathcal B(\bsl{\delta}_{\alpha})/(2\Omega)}
\chi(\bsl{\delta}_{\alpha})
t_{\bsl{\delta}_{\alpha},\bsl{\delta}_{\alpha}}\psi_{\bsl{k}_*}(\bsl{\delta}_{\alpha}) \sum_{(m_1,m_2)\in\dsZ^2\backslash\{(0,0)\}} \frac{(-1)^{m_1m_2+m_1+m_2}
e^{-\pi|m_1\bsl{a}_1+m_2\bsl{a}_2|^2/(2\Omega)}}{z_{m_1\bsl{a}_1+m_2\bsl{a}_2}} \\
& = 0\ .
}
As a result, we have
\eq{
0=
-C_\alpha g_{\mathbf k_*}'(\delta_\alpha)
+\frac{C_\alpha\lambda_\alpha}{2}
g_{\mathbf k_*}(\delta_\alpha)
\Rightarrow \lambda_\alpha
=
2\frac{g_{\mathbf k_*}'(\delta_\alpha)}
{g_{\mathbf k_*}(\delta_\alpha)}\ .
}
Combined with $
u_1=\frac{\delta_{\bar\alpha}\bar a_2
-\bar\delta_{\bar\alpha}a_2}{2i\Omega}
$, we proved \cref{eq:E-simple-lambda}.

\subsubsection{Projected Interaction Matrix}
\label{sec:E-projected-matrix}

We now express the projected interaction in terms of
the two pair operators $Q_{\bsl K,0}$ and $Q_{\bsl K,1}$ using \cref{eq:E-pair-mode-expansion}.
This allows us to write the interaction in the matrix form introduced in \cref{sec:app_D}.

Using the decomposition in \cref{eq:E-pair-mode-expansion}, with $\Phi_{\bsl K,\mu}$ and $Q_{\bsl K,\mu}$ defined in \cref{eq:Phi_def,eq:Q_Kmu}:
\eqa{
Q_{\bsl K}'(\zeta)
=\frac{\partial}{\partial\zeta}
Q_{\bsl K}(\zeta)=\frac{\partial}{\partial\zeta}
\left[
\sum_{\mu=0}^{1}
\Phi_{\bsl K,\mu}(\zeta)
Q_{\bsl K,\mu}
\right]=\sum_{\mu=0}^{1}
\Phi_{\bsl K,\mu}'(\zeta)
Q_{\bsl K,\mu},
}
since $Q_{\bsl K,\mu}$ is independent of
$\zeta$. We first substitute
\cref{eq:E-pair-mode-expansion}
into the first identity in
\cref{eq:E-simple-bond-projections}:
\eq{
\widetilde B_{\bsl r}^{(1)}=C_\alpha\sum_{\bsl K\in B_{2\phi_1,2\phi_2}^{N_1,N_2}}
e^{\ii\bsl K\cdot\bsl R}
\left[
\sum_{\mu=0}^{1}
\Phi_{\bsl K,\mu}(\delta_\alpha)
Q_{\bsl K,\mu}
\right]=\sum_{\bsl K\in B_{2\phi_1,2\phi_2}^{N_1,N_2}}\sum_{\mu=0}^{1}
e^{\ii\bsl K\cdot\bsl R}
C_\alpha\Phi_{\bsl K,\mu}(\delta_\alpha)
Q_{\bsl K,\mu}.
\label{eq:p=1result}
}
For the second identity in \cref{eq:E-simple-bond-projections}, we substitute $Q_{\bsl K}$ and its
derivative:
\eqa{
\widetilde B_{\bsl r}^{(2)}
&=\frac{C_\alpha}{2}
\sum_{\bsl K\in B_{2\phi_1,2\phi_2}^{N_1,N_2}}e^{\ii\bsl K\cdot\bsl R}
\left[
\sum_{\mu=0}^{1}
\Phi_{\bsl K,\mu}'(\delta_\alpha)
Q_{\bsl K,\mu}
-\lambda_\alpha
\sum_{\mu=0}^{1}
\Phi_{\bsl K,\mu}(\delta_\alpha)
Q_{\bsl K,\mu}
\right]\\
&=\sum_{\bsl K\in B_{2\phi_1,2\phi_2}^{N_1,N_2}}\sum_{\mu=0}^{1}
e^{\ii\bsl K\cdot\bsl R}
\frac{C_\alpha}{2}
\left[
\Phi_{\bsl K,\mu}'(\delta_\alpha)
-\lambda_\alpha
\Phi_{\bsl K,\mu}(\delta_\alpha)
\right]Q_{\bsl K,\mu}.
\label{eq:p=2result}
}
Recall that the projected interaction in
\cref{eq:E-projected-interaction-definition}
contains the weight $\Omega^{p-1}$.
Since $\Omega>0$, we may write
\eq{
\Omega^{p-1}
\widetilde B_{\bsl r}^{(p)\dagger}
\widetilde B_{\bsl r}^{(p)}=
\left(
\Omega^{(p-1)/2}\widetilde B_{\bsl r}^{(p)}
\right)^\dagger
\left(
\Omega^{(p-1)/2}\widetilde B_{\bsl r}^{(p)}
\right).
}
Therefore, it is convenient to include
the factor $\Omega^{(p-1)/2}$ in our
operator expressions. For $p=1$, this factor is $1$.
For $p=2$, it is $\sqrt{\Omega}$.
Multiplying \cref{eq:p=2result} by
$\sqrt{\Omega}$:
\eq{
\sqrt{\Omega}\widetilde B_{\bsl r}^{(2)} =\sum_{\bsl K\in B_{2\phi_1,2\phi_2}^{N_1,N_2}}\sum_{\mu=0}^{1}
e^{\ii\bsl K\cdot\bsl R}
\frac{\sqrt{\Omega}C_\alpha}{2}
\left[
\Phi_{\bsl K,\mu}'(\delta_\alpha)
-\lambda_\alpha
\Phi_{\bsl K,\mu}(\delta_\alpha)
\right]Q_{\bsl K,\mu}.
\label{eq:p=2result_final}
}
We have now expressed both weighted pair
operators as linear combinations of the
same two operators
$Q_{\bsl K,0}$ and $Q_{\bsl K,1}$ in \cref{eq:p=1result} and \cref{eq:p=2result_final}.
We collect their coefficients into a
$4\times2$ matrix $E(\bsl K)$ which is defined as
\eqa{
E_{\alpha1,\mu}(\bsl K)
&\equiv C_\alpha
\Phi_{\bsl K,\mu}(\delta_\alpha),\\
E_{\alpha2,\mu}(\bsl K)
&\equiv\frac{\sqrt{\Omega}C_\alpha}{2}
\left[
\Phi_{\bsl K,\mu}'(\delta_\alpha)
-\lambda_\alpha
\Phi_{\bsl K,\mu}(\delta_\alpha)
\right].
\label{eq:E-simple-E-matrix}
}
Explicitly, its matrix structure is
\eq{
E(\bsl K)=
\begin{pmatrix}
E_{A1,0}(\bsl K) & E_{A1,1}(\bsl K)\\
E_{A2,0}(\bsl K) & E_{A2,1}(\bsl K)\\
E_{B1,0}(\bsl K) & E_{B1,1}(\bsl K)\\
E_{B2,0}(\bsl K) & E_{B2,1}(\bsl K)
\end{pmatrix}.
}
The four rows correspond to $(\alpha, p)$,
while the two columns correspond to the
original pair modes $\mu=0,1$. Using $E_{\alpha p,\mu}$ defined
above, \cref{eq:p=1result} and \cref{eq:p=2result_final} become:
\eqa{
\widetilde B_{\bsl r}^{(1)}
&=\sum_{\bsl K\in B_{2\phi_1,2\phi_2}^{N_1,N_2}}\sum_{\mu=0}^{1}
e^{\ii\bsl K\cdot\bsl R}
E_{\alpha1,\mu}(\bsl K)
Q_{\bsl K,\mu},\\
\sqrt{\Omega}\widetilde B_{\bsl r}^{(2)}
&=\sum_{\bsl K\in B_{2\phi_1,2\phi_2}^{N_1,N_2}}\sum_{\mu=0}^{1}
e^{\ii\bsl K\cdot\bsl R}
E_{\alpha2,\mu}(\bsl K)
Q_{\bsl K,\mu}.
}
They can be combined into
the single identity
\eq{
\Omega^{(p-1)/2}
\widetilde B_{\bsl r}^{(p)}
=
\sum_{\bsl K\in B_{2\phi_1,2\phi_2}^{N_1,N_2}}\sum_{\mu=0}^{1}
e^{\ii\bsl K\cdot\bsl R}
E_{\alpha p,\mu}(\bsl K)
Q_{\bsl K,\mu},
\qquad p=1,2.
\label{eq:E-simple-bond-projections-matrix}
}
The two-component column vector $\bsl Q_{\bsl K}$ was introduced in \cref{sec:D1}.
Substituting \cref{eq:E-simple-bond-projections-matrix}
into \cref{eq:E-projected-interaction-definition} then gives
\eqa{\widetilde H_{\mathrm{int}}
 &=u\sum_{\bsl r\in\Lambda}\sum_{p=1}^{2}\Omega^{p-1}
 (\widetilde B_{\bsl r}^{(p)})^\dagger(\widetilde B_{\bsl r}^{(p)})\\
 &=u\sum_{\bsl K,\bsl K'\in B_{2\phi_1,2\phi_2}^{N_1,N_2}}\sum_{\mu,\nu=0}^{1}
 \left[\sum_{\bsl{R}}e^{\ii(\bsl{K}'-\bsl{K})\cdot\bsl{R}}\right]
 \left[\sum_{\alpha=A,B}\sum_{p=1}^{2}
 E_{\alpha p,\mu}(\bsl{K})^*E_{\alpha p,\nu}(\bsl{K}')\right] Q_{\bsl{K},\mu}^\dagger Q_{\bsl{K}',\nu}\\
 &=uN\sum_{\bsl K\in B_{2\phi_1,2\phi_2}^{N_1,N_2}}\sum_{\mu,\nu=0}^{1}
 M_{\mu\nu}(\bsl{K}) Q_{\bsl{K},\mu}^\dagger Q_{\bsl{K},\nu}\\
 &=uN\sum_{\bsl K\in B_{2\phi_1,2\phi_2}^{N_1,N_2}}\bsl{Q}_{\bsl{K}}^\dagger
 M(\bsl{K})\bsl{Q}_{\bsl{K}},
 \label{eq:E-simple-projected-interaction}
}
where $\sum_{\bsl R}e^{\ii(\bsl K'-\bsl K)\cdot\bsl R}=N\delta^{\mathrm{BZ}}_{\bsl K,\bsl K'}$. We define
\eq{
 M_{\mu\nu}(\bsl{K})\equiv \sum_{\alpha=A,B}\sum_{p=1}^{2}E_{\alpha p,\mu}(\bsl{K})^*
 E_{\alpha p,\nu}(\bsl{K}),\qquad
 M(\bsl{K})=E(\bsl{K})^\dagger E(\bsl{K}).
 \label{eq:E-simple-M-definition}
} 

Thus $\widetilde H_{\mathrm{int}}$ has the general form
\cref{eq:D-general-projected-interaction}, with the specific
coefficient matrix $M(\bsl K)=E(\bsl K)^\dagger E(\bsl K)$. To apply the ground-space result of \cref{sec:app_D}, it remains
to prove that this matrix is positive definite.

We first prove that $M(\bsl K)>0$ for every allowed pair momentum $\bsl K$.

Since $M(\bsl K)=E(\bsl K)^\dagger E(\bsl K)$,
for any $\bsl v=(v_0,v_1)^{\mathsf T}\in\dsC^2$,
we have $\bsl v^\dagger M(\bsl K)\bsl v =\bsl v^\dagger E(\bsl K)^\dagger
E(\bsl K)\bsl v =\norm{E(\bsl K)\bsl v}^2\geq0.$
Thus, to prove $M(\bsl K)>0$, it suffices to show
\eq{
E(\bsl K)\bsl v=0
\quad\rightarrow\quad
\bsl v=0,
\qquad \bsl v\in\dsC^2.
\label{eq:Ev=0impliesv=0}
}
Fix an allowed $\bsl K$ and suppose
$E(\bsl K)\bsl v=0$, where
$\bsl v=(v_0,v_1)^{\mathsf T}$.

Define the holomorphic function
\eq{
F(z)
=\sum_{\mu=0}^{1}v_\mu\Phi_{\bsl K,\mu}(z).
\label{eq:E-simple-test-section}
}
Here $F'(z)=\partial_z F(z)$, with
$v_0$ and $v_1$ independent of $z$.

Using the definition of $E(\bsl K)$ in
\cref{eq:E-simple-E-matrix},
for $p=1$,
\eqa{
\sum_{\mu=0}^{1}E_{\alpha1,\mu}(\bsl K)v_\mu
&=\sum_{\mu=0}^{1}
C_\alpha\Phi_{\bsl K,\mu}(\delta_\alpha)v_\mu
=C_\alpha\sum_{\mu=0}^{1}
v_\mu\Phi_{\bsl K,\mu}(\delta_\alpha)=C_\alpha F(\delta_\alpha).
}
Similarly, for $p=2$,
\eqa{
\sum_{\mu=0}^{1}E_{\alpha2,\mu}(\bsl K)v_\mu
&=\frac{\sqrt{\Omega}C_\alpha}{2}
\left[
\sum_{\mu=0}^{1}
v_\mu\Phi'_{\bsl K,\mu}(\delta_\alpha)
-\lambda_\alpha
\sum_{\mu=0}^{1}
v_\mu\Phi_{\bsl K,\mu}(\delta_\alpha)
\right]
=\frac{\sqrt{\Omega}C_\alpha}{2}
\left[
F'(\delta_\alpha)
-\lambda_\alpha F(\delta_\alpha)
\right].
}
Combining the results for $p=1,2$ and
$\alpha=A,B$, we obtain
\eq{
E(\bsl K)\bsl v =
\begin{pmatrix}
\sum_{\mu=0}^1 E_{A1,\mu}(\bsl K)v_\mu\\
\sum_{\mu=0}^1 E_{A2,\mu}(\bsl K)v_\mu\\
\sum_{\mu=0}^1 E_{B1,\mu}(\bsl K)v_\mu\\
\sum_{\mu=0}^1 E_{B2,\mu}(\bsl K)v_\mu
\end{pmatrix}=
\begin{pmatrix}
C_A F(\delta_A)\\
\frac{\sqrt{\Omega}C_A}{2}
\left[F'(\delta_A)-\lambda_A F(\delta_A)\right]\\
C_B F(\delta_B)\\
\frac{\sqrt{\Omega}C_B}{2}
\left[F'(\delta_B)-\lambda_B F(\delta_B)\right]
\end{pmatrix}
=0.
}
Since $C_A,C_B\neq0$ and $\Omega>0$,
the first and third components give $F(\delta_A)=0$ and $F(\delta_B)=0$, substituting these conditions into the second
and fourth components gives $F'(\delta_A)=0$ and $F'(\delta_B)=0$.

For a holomorphic function that is not identically zero, the conditions
$F(z_0)=F'(z_0)=0$ imply that $z_0$
is a zero of multiplicity at least two \cite{conway1978functions}.
Therefore, $F$ has a zero of multiplicity
at least two at each of $\delta_A$ and $\delta_B$.
Since these positions are inequivalent modulo
the primitive lattice, a nonzero $F$ would
have at least four zeros per primitive cell,
counting multiplicity.

However, a nonzero $F$ has exactly two zeros per primitive
cell, counting multiplicity, as we now show.
Applying \cref{eq:theta_property_1,eq:theta_property_2} to
\cref{eq:Phi_def} gives translation multipliers independent
of $\mu$. Their linear combination $F$ therefore satisfies
\eqa{
F(z+a_2)
&=e^{\ii\bsl K\cdot\bsl a_2}F(z),\\
F(z+a_1)
&=e^{\ii\bsl K\cdot\bsl a_1}
e^{-2\pi\ii\tau_0-4\pi\ii z/a_2}F(z).
\label{eq:E-simple-test-translations}
}
Suppose $F$ is not identically zero.
Taking logarithmic derivatives gives $\frac{F'(z+a_2)}{F(z+a_2)} =\frac{F'(z)}{F(z)}$ and $\frac{F'(z+a_1)}{F(z+a_1)} =\frac{F'(z)}{F(z)}
-\frac{4\pi\ii}{a_2}$.
Let $\mathcal C$ be a positively oriented
primitive parallelogram whose boundary
contains no zeros of $F$.
Since $F$ is holomorphic, the argument principle
counts its zeros, with multiplicity:
\eq{
N_{\mathrm{zeros}} =\frac{1}{2\pi\ii}
\oint_{\partial\mathcal C}
\frac{F'(z)}{F(z)} dz =\frac{1}{2\pi\ii}
\int_{z_0}^{z_0+a_2}
\left[
\frac{F'(z)}{F(z)}
-\frac{F'(z+a_1)}{F(z+a_1)}
\right]dz =\frac{1}{2\pi\ii}
\int_{z_0}^{z_0+a_2}
\frac{4\pi\ii}{a_2}dz =2.
\label{eq:E-simple-two-zeros}
}
Thus, any $F$ that is not identifically zero and that satisfies these
translation laws has exactly two zeros
per primitive cell, counting multiplicity. This contradicts the four zeros required
by $E(\bsl K)\bsl v=0$.

Therefore, due to this contradiction, we have $F(z)\equiv0$. Since
$\Phi_{\bsl K,0}$ and $\Phi_{\bsl K,1}$
are linearly independent by \cref{eq:D-pair-mode-independence},
the identity
\eq{
v_0\Phi_{\bsl K,0}(z)
+v_1\Phi_{\bsl K,1}(z)\equiv0
}
implies $v_0=v_1=0$. We have therefore proved
\cref{eq:Ev=0impliesv=0}. Consequently, for every nonzero $\bsl v$,
\eq{
\bsl v^\dagger M(\bsl K)\bsl v
=\norm{E(\bsl K)\bsl v}^2>0,
\qquad
M(\bsl K)>0.
\label{eq:E-simple-M-positive}
}
This proves the strict positive definiteness
of $M(\bsl K)$ for every allowed $\bsl K$.

By \cref{eq:E-simple-projected-interaction,eq:E-simple-M-positive},
$\widetilde H_{\mathrm{int}}$ is a member of the class
\cref{eq:D-general-projected-interaction}, with
$M(\bsl K)=E(\bsl K)^\dagger E(\bsl K)>0$. Moreover, the ideal band of $H_0$ has zero energy, so the
projection of $H=H_0+H_{\mathrm{int}}$ onto this band is precisely
$\widetilde H_{\mathrm{int}}$. Applying \cref{eq:D-exact-ground-state-space} therefore gives
\eq{
\ker\left(\widetilde H_{\mathrm{int}}\right)
=\operatorname{span}\left\{
\ket{\Psi^{3,1}},\ket{\Psi^{3,2}},\ket{\Psi^{3,3}}
\right\},
\qquad N=3N_e,\quad N_e\geq2.
\label{eq:E-projected-FC-ground-space}
}
Thus, the three constructed fractional Chern states form a basis of the complete zero-energy ground-state space of $\widetilde H_{\mathrm{int}}$.

\subsection{Exact-Diagonalization Checks}
\label{sec:E-numerical-checks}
In this section, we provide numerical values of the gap of the FCI states.
For the ED calculation of the full unprojected Hamiltonian, our ED code takes in a 2-band Hamiltonian that is given by the Fourier transform of $H$ to momentum space.

For the ED calculation of the projected Hamiltonain, our ED code takes in a generic single-band ED Hamiltonian, which reads:
\begin{align}
    \mathcal{H}^{\mathrm{ED}}_0 &= \sum_{\bsl{k}} \epsilon_{\bsl{k}}\, \gamma_{\bsl{k}}^\dagger \gamma_{\bsl{k}},
    \label{eq:code-H0} \\
    \mathcal{H}^{\mathrm{ED}}_{\text{int}} &= \sum_{\bsl{k}_1 \bsl{k}_2 \bsl{k}_3 \bsl{k}_4}
    V_{\mathrm{raw}}(\bsl{k}_1, \bsl{k}_2, \bsl{k}_3, \bsl{k}_4)\,\gamma_{\bsl{k}_1}^\dagger \gamma_{\bsl{k}_2}^\dagger \gamma_{\bsl{k}_3} \gamma_{\bsl{k}_4},
    \label{eq:code-Hint}\\
    \mathcal{H}^{\mathrm{ED}} &= \mathcal{H}^{\mathrm{ED}}_0 + \mathcal{H}^{\mathrm{ED}}_{\text{int}},
\end{align}
where $\epsilon\in\mathbb{R}^N$ and $V_{\mathrm{raw}}\in\mathbb{C}^{N\times N\times N \times N}$. We need to construct $\epsilon$ and $V_{\mathrm{raw}}$ appropriately such that $\mathcal{H}^{\mathrm{ED}} = \widetilde H_{\mathrm{int}}$.
To do so, we first (numerically) Fourier-transform $H_0$ in \cref{eq:main-real-space-h0} from real to momentum space. The result is a $2\times 2$ matrix per momentum, $\ie$, the numerical version of $h_0(\bsl{k})\in\mathbb{C}^{2\times 2}$ in \cref{eq:main-h0}.
Next, we diagonalize this Hermitian matrix numerically at each $\bsl{k}$ to obtain real eigenvalues $E_{-}(\bsl{k})< E_{+}(\bsl{k})$, and their corresponding eigenvectors $\bsl{\xi}_{\bsl{k}}^{\pm}\in\mathbb{C}^{2}$, which should differ from $U_{\bsl{k}}$ in \cref{eq:E-normalized-spinor} by a phase factor.
For our purposes, we eventually want to project to the lowest band, so we are mainly interested in the lowest eigenpair $(E_{-}(\bsl{k}), \bsl{\xi}_{\bsl{k}}^{-})$. For ease of notation, we write $\bsl{\xi}_{\bsl{k}} = \bsl{\xi}_{\bsl{k}}^{-}$ henceforth. The next task at hand is to convert the whole Hamiltonian to the band basis via \cref{eq:E-band-projection-substitution}. 
We intentionally do not use the analytical eigenfunctions $\psi_k$ of $h_0(\bsl{k})$ from \cref{eq:C-normalized-band-orbital} directly, in order to perform a numerical cross-check. 
Instead, we use a Bloch function $\tilde\psi_{\bsl k}$ which is normalized across all momenta like $\tilde\psi_{\bsl k}(\bsl R+\bsl\delta_\alpha)=\frac{1}{\sqrt N}\,e^{\ii\bsl k\cdot(\bsl R+\bsl\delta_\alpha)}\,\xi_{\bsl k\alpha}$. For the single-particle part, the projection to the lower band is quite simple, obtaining $\widetilde H_{0} = \sum_{\bsl{k}}E_{-}(\bsl{k})\,\gamma_{\bsl{k}}^{\dagger}\gamma_{\bsl{k}}$,
where we have used the notation $\widetilde H_{0}$ for the projected version of the single-particle Hamiltonian. Note that analytically, $\widetilde H_{0}$ is supposed to be identically zero, which is why we omitted it earlier and contented ourselves with $\widetilde H_{\mathrm{int}}$. However, for a faithful benchmark, we include it for the purposes of our ED calculation. Numerically, we still find that $E_{-}(\bsl{k}) = 0$ to machine precision for all $\bsl{k}$. The interaction term is slightly more involved. Define
\begin{equation}\label{eq:D-lattice-sum}
    \mathcal{D}_{\alpha\beta}^{(p)}(\bsl{k}) = \sum_{\bsl{U}}^{\prime}e^{\ii\bsl{k}\cdot\bsl{U}}\frac{t_{\alpha\beta}(-\bsl{U})}{(U+\delta_\beta - \delta_\alpha)^p},
\end{equation}
where the sum runs over every Bravais lattice vector $\bsl{U}$ on the infinite lattice, excluding $\bsl U=0$ when $\beta=\alpha$, which is implied by the prime on top of the summation. (Here, $U$, $\delta_\beta$, $\delta_\alpha$ are complex coordinates.) In practice, we sum over the $\bsl U$ with $\abs{U}\le6\sqrt{\Omega}+\abs{\delta_B-\delta_A}$ ($\abs{U + \delta_{\beta} - \delta_{\alpha}}\le6\sqrt{\Omega}$ for the unprojected case). Doing some algebra to obtain the projected interaction in the form of \cref{eq:code-Hint}, we finally obtain
\begin{equation}
    V_{\mathrm{raw}}(\bsl{k}_1,\bsl{k}_2,\bsl{k}_3,\bsl{k}_4) = \delta_{\bsl{k}_1+\bsl{k}_2, \bsl{k}_3+\bsl{k}_4}\, g\sum_{\alpha=A,B}\sum_{p=1}^{2}\Omega^{p-1}\sum_{\beta,\beta'=A,B}[\mathcal{D}_{\alpha\beta}^{(p)}(\bsl{k}_2)^*\tilde{\psi}_{\bsl{k}_2}^*(\bsl{\delta}_{\beta})][\mathcal{D}_{\alpha\beta'}^{(p)}(\bsl{k}_3)\tilde{\psi}_{\bsl{k}_3}(\bsl{\delta}_{\beta'})]\,\tilde{\psi}_{\bsl{k}_1}^*(\bsl{\delta}_{\alpha})\tilde{\psi}_{\bsl{k}_4}(\bsl{\delta}_{\alpha}).
    \label{eq:V-raw-unprojected}
\end{equation}
(Here $\delta_{\bsl{k}_1+\bsl{k}_2,\bsl{k}_3+\bsl{k}_4}$ is a Kronecker delta imposed modulo a reciprocal lattice vector, and should not be confused with the sublattice positions $\bsl{\delta}_\alpha$, $\bsl{\delta}_\beta$.) Thus, with $\epsilon_{\bsl{k}}=E_{-}(\bsl{k})$ and $V_{\mathrm{raw}}$ as in \cref{eq:V-raw-unprojected}, we have $\mathcal{H}^{\mathrm{ED}} $ set for the ED calculation.

\subsubsection{Numerical Results}\label{subsubsec:numerical_results}
\begin{figure}[hbtp]
\centering
\includegraphics[width=\linewidth]{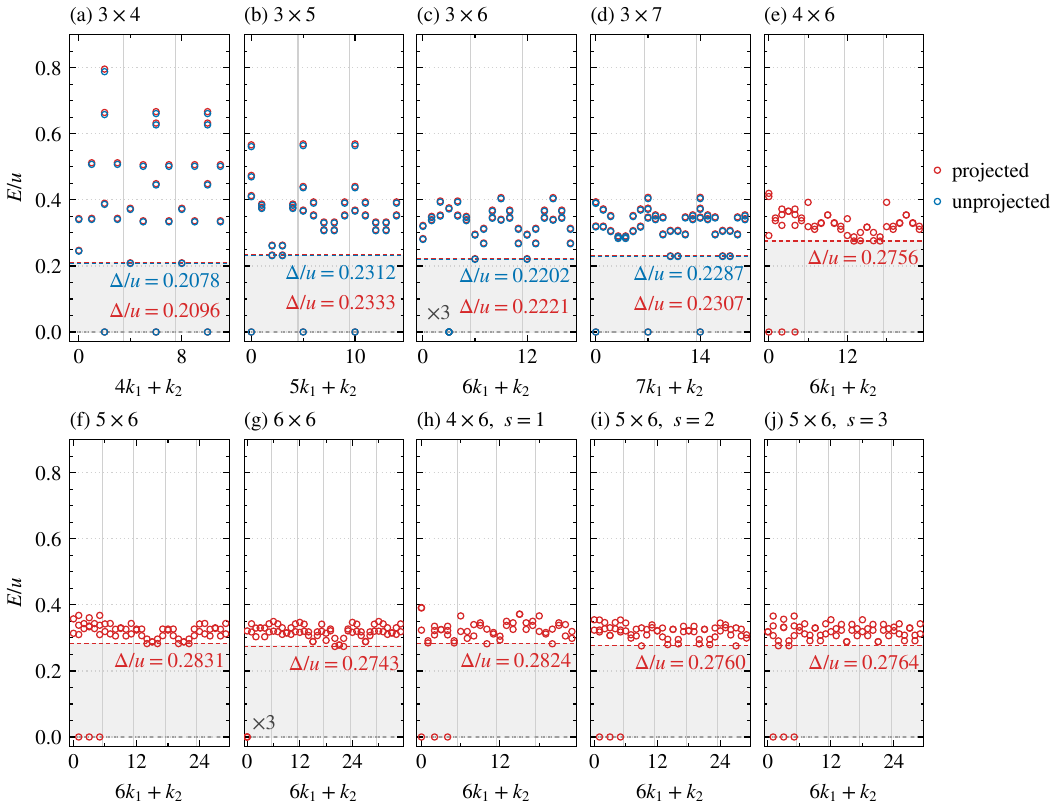}
\caption{ED spectra of the unprojected Hamiltonian $H$ (blue markers) and the numerically-projected Hamiltonian $\widetilde{H}$ (red markers) at filling $1/3$, $N_e=N/3$, on $N_1\times N_2$ with the number of unit cells ranging from $N=N_1 N_2=12$ to $36$.
The numerical parameters used in our calculation can be seen in \cref{tab:ed-inputs}.
Each panel shows the four lowest eigenvalues in the three ground-state momentum sectors, and two eigenvalues each for every other total-momentum sector $(k_1,k_2)$, $k_j=0,\ldots,N_j-1$, in units of $u$ and plotted against $N_2k_1+k_2$. The three zero-energy ground states lie in the sectors predicted for the three $\nu=1/3$ states by the counting rule of Ref.~\cite{regnault2011}; when all three are in the same sector, as in (c) and (g), it is indicated by the ``$\times 3$''. Here $\Delta$ represents the lowest nonzero eigenvalue. }
\label{fig:ed-spectra}
\end{figure}

We ran ED calculations across a range of mesh sizes for both the unprojected Hamiltonian $H$ (with an interaction strength $u$ that is $1/500$ of the single-particle band gap) and the projected Hamiltonian $\widetilde{H}$. 
The numerical parameters used in our calculation can be seen in \cref{tab:ed-inputs}. 
In \cref{fig:ed-spectra} (a)-(d) we show a comparison between the ED calculation on $H$ and $\widetilde{H}$ for $3\times 4$, $3\times 5$, $3\times 6$, and $3\times 7$ for $s=0$. As can be seen, the eigenvalue spectra of $H$ and $\widetilde{H}$ agree closely when they share the same mesh size (as expected due to the weak interaction strength), so it stands to reason that the spectrum of $\widetilde{H}$ can be used to approximate that of $H$ at larger sizes. As such, in \cref{fig:ed-spectra} (e)-(j), we show the ED results for $\widetilde{H}$ on $4\times 6$, $5\times 6$, and $6\times 6$ meshes.
For the eigensolver, we used PRIMME~\cite{Stathopoulos2010PRIMME}. Recall that for projected ED, we use a lattice cutoff $\le6\sqrt{\Omega}+\abs{\delta_B-\delta_A}$ in order to approximate the infinite summations.
To test the effect of changing this cutoff, we also tried using a cutoff of $2\sqrt{N \Omega}+\abs{\delta_B-\delta_A}$ for ED on the $4\times 6$ and $5\times 6$ meshes for the projected ED, and the results were unchanged to machine precision for the chosen parameter values.
\begin{table}[hbtp]
\centering
\renewcommand{\arraystretch}{1.25}
\setlength{\tabcolsep}{9pt}
\begin{tabular}{l|l}
\hline\hline
Input & Value \\
\hline
$\bsl a_1,\ \bsl a_2$ & $(-s, -1),\ (1,0)$ \\
$\Omega$ & $1$ \\
$\bsl\delta_A,\ \bsl\delta_B$ & $(0,0),\ (1/2,0)$ \\
$\eta_A,\ \eta_B$ & $1,\ 1$ \\
$\phi_1,\ \phi_2$ & $0,\ 0$ \\
Real-space cutoff ($H$) & $6\sqrt\Omega$ \\
Real-space cutoff ($\widetilde{H}$) & $6\sqrt\Omega+\abs{\delta_B-\delta_A}$ \\
$\bsl L_1,\ \bsl L_2$ & $N_1\bsl a_1,\ N_2\bsl a_2$ \\
\hline\hline
\end{tabular}
\caption{Numerical parameters for the ED calculations.}
\label{tab:ed-inputs}
\end{table}

Finally, we would like to compute the overlap with the analytical expression for the ground-state eigenvectors from \cref{eq:C-discretized-many-body-state}. Note that since $\eta_\alpha=\chi(\bsl\delta_\alpha)e^{-\pi\mathcal K(\bsl\delta_\alpha)/(2\Omega)}$ for $\alpha\in\{A, B\}$, any value of $\chi(\bsl{\delta}_A), \chi(\bsl{\delta}_B)$, and the K\"ahler potential $\mathcal K(\bsl r)$ that is consistent with our choice of $\eta_A=\eta_B=1$ from \cref{tab:ed-inputs} will yield the same many-body wavefunction according to \cref{eq:C-discretized-amplitude}. We computed the overlap as follows. 
When the three states have different many-body momenta, let $\bsl{K}_{MB}^{i}$ be the many-body momentum of the ED ground state $\ket{\Psi^{\mathrm{ED}}_i}$, for $i\in\{1,2, 3\}$. From the three analytical eigenvectors $\ket{\Psi^{3,\sigma}}$ -- where $\sigma\in\{1,2,3\}$ -- we form a linear combination which gives us the momentum eigenstate $\ket{\Psi_{\bsl{K}_{MB}^{i}}}$. Finally, we normalize $\ket{\Psi_{\bsl{K}_{MB}^{i}}}$, take the inner product with the ED eigenvector $\ket{\Psi^{\mathrm{ED}}_i}$, and find
\begin{equation}
    \abs{\braket{\Psi^{\mathrm{ED}}_i}{\Psi_{\bsl{K}_{MB}^{i}}}} = 1
\end{equation}
to machine precision for the $4\times 6$ and $5\times 6$ meshes. When the three states have the same many-body momenta, we also check that the span of the three analytical eigenvectors is equal to the span of the numerically-obtained ground states, which serves as an additional check in the case of degenerate eigenstates in the ground-state momentum sectors, such as for the $6\times6$ untilted calculation in \cref{fig:ed-spectra}.

\section{Anyonic Excitations: Quasi-holes}
\label{sec:qh-der-construction}

In this section, we will discuss the construction of one-quasihole wavefunctions on lattice and show that they have exactly flat dispersion under the projected Hamiltonian.
Our quasihole wavefunctions on lattice is motivated by the continuum
construction of Ref. \cite{iyer2026dispersionanyonblochbands},
but with electron coordinates restricted to discrete lattice
sites and the continuum spinor replaced by a scalar lattice
factor $\chi(\bsl r)$.
For the one-quasihole construction, we consider 
$N=N_1N_2=3N_e+1$ unit cells with $N_e\geq2$ electrons on a torus. Based on the lattice fractional Chern wavefunctions in
\cref{eq:C-discretized-amplitude,eq:ideal-fci-holomorphic-factor} with $m=3$, we construct the one-quasihole wavefunctions as
\eq{
 \Psi_{\qh}^{3,\sigma}(\{\bsl r_i\};\wh)
 =F_{\qh}^{\sigma}(\{z_i\};\wh)
 \mathcal S(\{\bsl r_i\}),
 \qquad \sigma=1,2,3,
 \label{eq:qh-der-Psi}
}
\eq{
 F_{\qh}^{\sigma}(\{z_i\};\wh)
 ={}e^{\ii K_\sigma(Z+\wh/3)}
 \mathcal P_\sigma(Z+\wh/3)\prod_{a<b}^{N_e}
 \vartheta_1\left(\frac{z_a-z_b}{L_2}\middle|\tau\right)^3
 \prod_{a=1}^{N_e}
 \vartheta_1\left(\frac{z_a-\wh}{L_2}\middle|\tau\right) \ ,
 \label{eq:qh-der-F}
}
where $\bsl r_i\in\Lambda$ are the physical lattice
sites defined in \cref{eq:finite-torus-site-set} with $z_i=r_{i,x}-\ii r_{i,y}$, and $\wh=w_x-\ii w_y$ denotes the quasihole
position which can be anywhere on the torus and does not need to be on the lattice site. 
We use the lattice conventions of
\cref{sec:C-setup} and the center-of-mass
parameters $K_\sigma$ and $Z_\sigma^\gamma$
from \cref{eq:K-Z-sector-definition} with $m=3$, and $\mathcal B$ and $\chi$ are defined in \cref{eq:C-B-definition,eq:C-chi-translation}.

To satisfy the twisted boundary conditions
$\Psi_{\qh}^{3,\sigma}
(\ldots,\bsl r_i+\bsl L_j,\ldots;\wh)
=e^{\ii\phi_j}
\Psi_{\qh}^{3,\sigma}(\{\bsl r_i\};\wh)$
for $j=1,2$, using the quasiperiodicity of the theta functions in \cref{eq:qh-der-F}, the corresponding factors are $(-1)^N e^{\ii K_1L_2}$ along $\bsl L_2$
and $(-1)^N e^{\ii K_1L_1+2\pi\ii Z_1}$ along $\bsl L_1$. Setting them equal to the boundary phases gives
\eq{
(-1)^N e^{\ii K_1L_2}=e^{\ii\phi_2},
\qquad
(-1)^N e^{\ii K_1L_1+2\pi\ii Z_1}
=e^{\ii\phi_1},
\qquad N=3N_e+1.
\label{eq:qh-der-constraints}
}
Here $K_1$ and $Z_1$ are the center-of-mass
parameters of the $\sigma=1$ sector defined in
\cref{eq:K-Z-sector-definition}.
They are chosen to satisfy \cref{eq:qh-der-constraints} for the one-quasihole state with $N=3N_e+1$.
where $\mathcal S$ is defined in  \cref{eq:commonlatticefactor} and $P_{\sigma}$ is defined in \cref{eq:com-theta-product}.

We now check that the quasihole wavefunctions
satisfy the twisted boundary conditions in \cref{eq:C-twisted-boundary}.
First, recall that the theta-function translation identities are given in \cref{eq:direct-theta1-shift-one,eq:direct-theta1-shift-tau}. 
\cref{eq:C-site-factor-full-periods} shows that the combination of $\mathcal B$ and $\chi$, as the lattice factor $\mathcal S$ in  \cref{eq:commonlatticefactor}, transforms as
\eqa{
\mathcal S(\ldots,\bsl r_i+\bsl L_2,\ldots)
&=\mathcal S(\{\bsl r_i\}),\\
\mathcal S(\ldots,\bsl r_i+\bsl L_1,\ldots)
&=e^{2\pi\ii Nz_i/L_2+\ii\pi N\tau}
\mathcal S(\{\bsl r_i\}).
\label{eq:qh-der-S-full}
}
Then,  under $z_i\mapsto z_i+L_2$, with $\wh$ fixed,
the exponential, center-of-mass theta product,
and cubic Jastrow factor acquire
$e^{\ii K_\sigma L_2}$, $(-1)^3$, and
$(-1)^{3(N_e-1)}$, respectively, as discussed in
\cref{sec:TBC} with $m=3$. 
The additional quasihole factor contributes $-1$ owing to \cref{eq:direct-theta1-shift-one}, while the lattice factor is unchanged according to
\cref{eq:qh-der-S-full}. 
Combining these factors gives
\eqa{
\Psi_{\qh}^{3,\sigma}
(\ldots,\bsl r_i+\bsl L_2,\ldots;\wh)
&=e^{\ii K_\sigma L_2}
(-1)^3(-1)^{3(N_e-1)}(-1)
\Psi_{\qh}^{3,\sigma}(\{\bsl r_i\};\wh)\\
&=(-1)^N e^{\ii K_1L_2}
\Psi_{\qh}^{3,\sigma}(\{\bsl r_i\};\wh)\\
&=e^{\ii\phi_2}
\Psi_{\qh}^{3,\sigma}(\{\bsl r_i\};\wh),
\label{eq:qh-der-L2-total}
}
where we used $N=3N_e+1$ and
\cref{eq:K-Z-sector-definition,eq:qh-der-constraints}.
Under $z_i\mapsto z_i+L_1$, with $\wh$ fixed,
the exponential factor in \cref{eq:qh-der-F}
gets $e^{\ii K_\sigma L_1}$.
Using \cref{eq:direct-theta1-shift-tau},
the center-of-mass theta product $\mathcal P_\sigma$ transforms as
\eqa{
\mathcal P_\sigma(Z+\wh/3+L_1)
&=(-1)^3 e^{-3\ii\pi\tau}
e^{-6\pi\ii Z/L_2-2\pi\ii\wh/L_2}
e^{\frac{2\pi\ii}{L_2}\sum_{\gamma=1}^{3}
Z_\sigma^\gamma}
\mathcal P_\sigma(Z+\wh/3).
\label{eq:qh-der-L1-COM}
}
As derived in \cref{sec:TBC} for $m=3$,
the cubic Jastrow factor gets
$(-1)^{3(N_e-1)}
e^{-3\ii\pi\tau(N_e-1)}
e^{-6\pi\ii(N_ez_i-Z)/L_2}$. The additional quasihole factor contributes
$-e^{-\ii\pi\tau-2\pi\ii(z_i-\wh)/L_2}$,
while the lattice factor contributes
$e^{\ii\pi N\tau+2\pi\ii Nz_i/L_2}$
by \cref{eq:qh-der-S-full}. Multiplying these factors, the exponents
can be collected into
\eq{
\Psi_{\qh}^{3,\sigma}
(\ldots,\bsl r_i+\bsl L_1,\ldots;\wh)=(-1)^{3N_e+1}
e^{\ii K_\sigma L_1+
\frac{2\pi\ii}{L_2}\sum_{\gamma=1}^{3}
Z_\sigma^\gamma}
e^{\ii\pi\tau(N-3N_e-1)}
e^{\frac{2\pi\ii}{L_2}
(N-3N_e-1)z_i}
\Psi_{\qh}^{3,\sigma}(\{\bsl r_i\};\wh).
}
Using \cref{eq:K-Z-sector-definition,eq:sum_Z}
with $m=3$, we also have
$\sum_{\gamma=1}^{3}Z_\sigma^\gamma
=Z_1L_2+L_2-(\sigma-1)L_1$, which gives
$e^{\ii K_\sigma L_1+
\frac{2\pi\ii}{L_2}\sum_\gamma Z_\sigma^\gamma}
=e^{\ii K_1L_1+2\pi\ii Z_1}$, and using \cref{eq:qh-der-constraints},
\eqa{
\Psi_{\qh}^{3,\sigma}
(\ldots,\bsl r_i+\bsl L_1,\ldots;\wh)
&=(-1)^N e^{\ii K_1L_1+2\pi\ii Z_1}
\Psi_{\qh}^{3,\sigma}(\{\bsl r_i\};\wh)\\
&=e^{\ii\phi_1}
\Psi_{\qh}^{3,\sigma}(\{\bsl r_i\};\wh).
\label{eq:qh-der-L1-final}
}
Therefore, \cref{eq:qh-der-L2-total,eq:qh-der-L1-final} show that both twisted boundary conditions are satisfied.

We now show that the quasihole wavefunction still lives in the Fock space fo the lower energy band.
We have showed that the full quasihole wavefunction gets $e^{\ii\phi_j}$ under an electronic translation by $L_j$ (\cref{eq:qh-der-L2-total,eq:qh-der-L1-final}). Since the full wavefunction is $F_{\qh}^\sigma \mathcal S$ by
\cref{eq:qh-der-Psi}, dividing out the site-factor transformations in \cref{eq:qh-der-S-full} gives
\eqa{
 F_{\qh}^{\sigma}(\ldots,z_i+L_2,\ldots;\wh)
 &=e^{\ii\phi_2}F_{\qh}^{\sigma}(\{z_i\};\wh),\\
 F_{\qh}^{\sigma}(\ldots,z_i+L_1,\ldots;\wh)
 &=\frac{e^{\ii\phi_1}}{e^{\ii\pi N\tau+2\pi\ii Nz_i/L_2}}
 F_{\qh}^{\sigma}(\{z_i\};\wh)=e^{\ii\phi_1}e^{-\ii\pi N\tau-2\pi\ii Nz_i/L_2}
 F_{\qh}^{\sigma}(\{z_i\};\wh).
 \label{eq:qh-energy-clean-holomorphic-BC}
}
Note that these boundary conditions match those of the single-particle functions $g_{\bsl k}(z)$ in \cref{thm:f-gk-completeness}. Since $F_{\qh}^\sigma$ is holomorphic and satisfies the required twisted boundary conditions in each $z_i$, \cref{thm:f-gk-completeness} allows us to expand $F_{\qh}^\sigma$ in products of the single-particle basis functions $g_{\bsl k}(z_i)$:
\eqa{
F_{\qh}^\sigma(\{z_i\};\wh)
&=\sum_{\bsl k_1,\ldots,\bsl k_{N_e}
\in B_{\phi_1,\phi_2}^{N_1,N_2}}
A_{\bsl k_1,\ldots,\bsl k_{N_e}}^\sigma(\wh)
\prod_{i=1}^{N_e}g_{\bsl k_i}(z_i).
\label{eq:F_qh_holomorphic_expansion}
}
Recall from \cref{eq:C-normalized-band-orbital} that
$e^{\pi\mathcal B(\bsl r)/(2\Omega)}
\chi(\bsl r)g_{\bsl k}(z)
=\mathcal N_{\bsl k}\psi_{\bsl k}(\bsl r).$
Substituting this identity into \cref{eq:qh-der-Psi},
\eqa{
\Psi_{\qh}^{3,\sigma}(\{\bsl r_i\};\wh)
&=\sum_{\bsl k_1,\ldots,\bsl k_{N_e}}
A_{\bsl k_1,\ldots,\bsl k_{N_e}}^\sigma(\wh)
\prod_{i=1}^{N_e}
\left[
e^{\pi\mathcal B(\bsl r_i)/(2\Omega)}
\chi(\bsl r_i)g_{\bsl k_i}(z_i)
\right]\\
&=\sum_{\bsl k_1,\ldots,\bsl k_{N_e}}
C_{\bsl k_1,\ldots,\bsl k_{N_e}}^\sigma(\wh)
\prod_{i=1}^{N_e}\psi_{\bsl k_i}(\bsl r_i),
}
where
$C_{\bsl k_1,\ldots,\bsl k_{N_e}}^\sigma(\wh)
=A_{\bsl k_1,\ldots,\bsl k_{N_e}}^\sigma(\wh)
\prod_i\mathcal N_{\bsl k_i}$,
and all momentum sums are over
$B_{\phi_1,\phi_2}^{N_1,N_2}$ in
\cref{eq:allowed+momenta}. Thus, the quasihole wavefunction is expressed entirely
in terms of the lower-band Bloch wavefunctions $\psi_{\bsl k}$.

To obtain the corresponding Fock-space state, we now
substitute the above expansion into the many-body state definition in \cref{eq:C-discretized-many-body-state}. Exchanging the finite sums over lattice sites and momenta, and using the lower-band creation operators $\gamma_{\bsl k}^\dagger
=\sum_{\bsl r\in\Lambda}
\psi_{\bsl k}(\bsl r)c_{\bsl r}^\dagger$ from \cref{eq:creation_operator}, we get
\eqa{
\ket{\Psi_{\qh}^{3,\sigma}}
&=\frac{1}{\sqrt{N_e!}}
\sum_{\bsl r_1,\ldots,\bsl r_{N_e}\in\Lambda}
\Psi_{\qh}^{3,\sigma}(\{\bsl r_i\};\wh)
c_{\bsl r_1}^\dagger\cdots
c_{\bsl r_{N_e}}^\dagger\ket{0}\\
&=\frac{1}{\sqrt{N_e!}}
\sum_{\bsl k_1,\ldots,\bsl k_{N_e}}
C_{\bsl k_1,\ldots,\bsl k_{N_e}}^\sigma(\wh)
\prod_{i=1}^{N_e}
\left[
\sum_{\bsl r_i\in\Lambda}
\psi_{\bsl k_i}(\bsl r_i)
c_{\bsl r_i}^\dagger
\right]\ket{0}\\
&=\frac{1}{\sqrt{N_e!}}
\sum_{\bsl k_1,\ldots,\bsl k_{N_e}}
C_{\bsl k_1,\ldots,\bsl k_{N_e}}^\sigma(\wh)
\gamma_{\bsl k_1}^\dagger\cdots
\gamma_{\bsl k_{N_e}}^\dagger\ket{0}.
\label{eq:quasihole_state_expansion}
}
Notice that this expansion contains only lower-band
creation operators $\gamma_{\bsl k}^\dagger$, and thus it is contained in the Fock space of the lower band.

The one-quasihole states are FCI states with a charge-$e/3$ excitation.
To see this, we note that the FCI states satisfy lattice translation symmetry and thus have a charge density of $-e/3$ per unit cell when they are grouped into states with definite many-body momentum, where $-e$ is the electron charge.
Treating this as the background for the one-quasihole state, we find that the additional charge in a one-quasihole state is $-e N_e - (-e)(3N_e+1)/3 = e/3$.
Therefore, the excitation has a charge of $e/3$.

\begin{figure}[hbtp]
\centering
\includegraphics[width=0.5\linewidth]{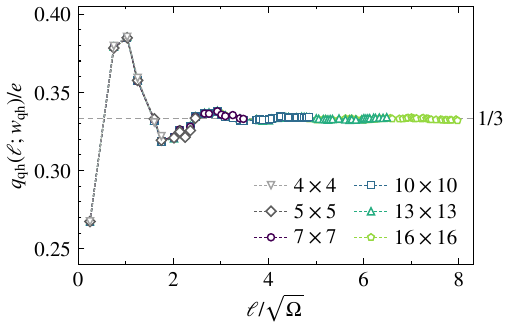}
\caption{We show the average excess charge $q_{\qh}$ (in units of $e$), as function of $\ell/\sqrt{\Omega}$. 
We choose the same model parameters \cref{tab:ed-inputs} with $s=0$, and perform the calculation for a variety of meshes, namely $4\times 4$, $5\times 5$, $7\times 7$, $10\times10$, $13\times 13$, and $16\times 16$. In each case, we take $\wh$ to be one of the $A$ sublattice sites closest to the center unit cell. 
Specifically, in complex form we have $\wh=2+2\ii$ for $4\times 4$ and $5\times 5$ meshes, $\wh=3+3\ii$ for $7\times 7$, $\wh=5+5\ii$ for $10\times 10$, $\wh=6+6\ii$ for $13\times 13$, and $\wh=8+8\ii$ for $16\times 16$. The calculation for $4\times 4$ and $5\times 5$ was done via exact implementation of \cref{eq:qh-charge-enclosed} and \cref{eq:qh-density-cell}, whereas the calculation for larger sizes was done via Monte-Carlo approximation of the expectation value of the number operator in \cref{eq:qh-density-cell}.}
\label{fig:quasihole_charge}
\end{figure}

Now we still need to show that the charge is localized near the quasihole position $\wh$. For each unit cell $\bsl R$, the average electron number is
\eq{\rho_{\qh}^{\sigma}(\bsl R;\wh) = \sum_{\alpha=A,B}
\frac{\bra{\Psi_{\qh}^{3,\sigma}(\wh)}c_{\bsl R+\bsl\delta_\alpha}^{\dagger}c_{\bsl R+\bsl\delta_\alpha}
\ket{\Psi_{\qh}^{3,\sigma}(\wh)}}{\braket{\Psi_{\qh}^{3,\sigma}(\wh)}
{\Psi_{\qh}^{3,\sigma}(\wh)}}.\label{eq:qh-density-cell}}
To measure the charge near $\wh$, we choose a region $\mathcal D_\ell$ centered at $\wh$ that contains lattice vectors within the distance $\ell$.
The excess charge in this region is calculated as
\eq{q_{\qh}(\ell;\wh) = e\frac{1}{3}\sum_{\sigma}\sum_{\bsl R\in\mathcal D_\ell} \left[\frac13-\rho_{\qh}^{\sigma}(\bsl R;\wh)\right],\label{eq:qh-charge-enclosed}}
and we expect $q_{\qh}(\ell;\wh)\approx e/3$ while $\ell$ is large enough to contain the charge distribution but still much smaller than the torus.
We numerically checked it: up to a $16\times 16$ lattice, the $e/3$ charge of the excitation is mainly ($\sim 1$\% error) localized within $3|\bsl{a}_1|$ away from the $w_{qh}$, as seen in \cref{fig:quasihole_charge}.

\subsection{Zero Energy of the one-quasihole States}
\label{app:zero_energy_quasihole}

We now show that the energy of the one-quasihole states is zero
under the total Hamiltonian $H=H_0+H_{\mathrm{int}}$ from
\cref{eq:E-simple-full-hamiltonian}.
Again, we do so by projecting $H$ to the lower band to obtain $\widetilde H$, since $\ker H = \ker \widetilde H$, where $\widetilde H=\widetilde H_0+\widetilde H_{\mathrm{int}}$,
and $\widetilde H_0$ and
$\widetilde H_{\mathrm{int}}$ are the corresponding
lower-band projections of $H_0$ and $ H_{\mathrm{int}}$, respectively.
In particular, $\widetilde H_0=0$ because the
lower band has zero single-particle energy
(\cref{eq:E-target-band-zero}).

We will show that $\ket{\Psi_{\qh}^{3,\sigma}}$ is \cref{eq:E-projected-interaction-definition} is a zero eigenstate of $\widetilde{H}$ for $3N_e+1 = N$, regardless of the anyon position $\wh$.
In particular, we will show 
\eq{
\widetilde B_{\bsl r}^{(p)}
\ket{\Psi_{\qh}^{3,\sigma}}=0,
\qquad \forall \bsl r\in\Lambda,\quad p=1,2\ ,
\label{eq:qh-energy-clean-targets}
}
where $\widetilde B_{\bsl r}^{(p)}$ is derived in \cref{eq:E-simple-bond-projections}.

To do so, we first note that for $\bsl r=\bsl R+\bsl\delta_\alpha$, \cref{eq:E-simple-bond-projections} gives
\eqa{
\widetilde B_{\bsl r}^{(1)}\ket{\Psi_{\qh}^{3,\sigma}} &=C_\alpha\sum_{\bsl K}e^{\ii\bsl K\cdot\bsl R}
Q_{\bsl K}(\delta_\alpha)\ket{\Psi_{\qh}^{3,\sigma}},\\
\widetilde B_{\bsl r}^{(2)}\ket{\Psi_{\qh}^{3,\sigma}}&=\frac{C_\alpha}{2}\sum_{\bsl K}e^{\ii\bsl K\cdot\bsl R} \left[Q_{\bsl K}'(\delta_\alpha) -\lambda_\alpha Q_{\bsl K}(\delta_\alpha)\right] \ket{\Psi_{\qh}^{3,\sigma}}.
\label{eq:qh-energy-clean-B-action}
}
Here $\bsl K$ is the total electron-pair momentum that takes value on the mesh in \cref{eq:D-allowed-pair-momenta}.
Thus, to show that both projected pair operators
annihilate the quasihole state, it enough to prove
$Q_{\bsl K}(\zeta)\ket{\Psi_{\qh}^{3,\sigma}}=0$
for every $\bsl K$ and $\zeta$, which also implies
$Q'_{\bsl K}(\zeta)\ket{\Psi_{\qh}^{3,\sigma}}=0$ since $Q'_{\bsl K}(\zeta)\ket{\Psi_{\qh}^{3,\sigma}}=\partial_{\zeta}[Q_{\bsl K}(\zeta)\ket{\Psi_{\qh}^{3,\sigma}}]$.

The key property is that the quasihole wavefunction
still has cubic zeros when two electrons coincide.
Using the relative coordinates in
\cref{eq:D-relative-coordinate-convention},
we write $z_1=\zeta+w/2$ and $z_2=\zeta-w/2$,
where $w=z_1-z_2$. From \cref{eq:qh-der-F}, the factor involving $w$ is $\vartheta_1(w/L_2|\tau)^3$. Since $\vartheta_1$ is odd and vanishes at the origin, its Taylor expansion around $w=0$ is $\vartheta_1(w/L_2|\tau) =\vartheta_1'(0|\tau)w/L_2+O(w^3)=O(w)$, and consequently, $\vartheta_1(w/L_2|\tau)^3=O(w^3)$. Since all other factors in $F_{\qh}^{\sigma}$
are holomorphic at $w=0$,
\eq{
F_{\qh}^{\sigma}
\left(\zeta+\frac w2,\zeta-\frac w2,
z_3,\ldots,z_{N_e};\wh\right)
=O(w^3).
}
This means that the wavefunction has no constant
or linear term in its Taylor expansion around $w=0$, so its first derivative vanishes there:
\eq{
\left.\partial_w F_{\qh}^{\sigma}
\left(\zeta+\frac w2,\zeta-\frac w2,
z_3,\ldots,z_{N_e};\wh\right)
\right|_{w=0}=0.
\label{eq:qh-energy-clean-cubic}
}
We now use \cref{eq:qh-energy-clean-cubic} to show that
$Q(\zeta)\ket{\Psi_{\qh}^{3,\sigma}}=0$.
Recall that $Q(\zeta)=\sum_{\bsl K}Q_{\bsl K}(\zeta)$
from \cref{eq:D-total-pair-annihilator}.
Using the definitions in
\cref{eq:discrete-V1-pair-annihilator,eq:relative-one-jet-wronskian}, we obtain
$Q(\zeta)=
\sum_{\bsl k,\bsl k'}
\frac{g'_{\bsl k}(\zeta)g_{\bsl k'}(\zeta)}
{\mathcal N_{\bsl k}\mathcal N_{\bsl k'}}
\gamma_{\bsl k'}\gamma_{\bsl k}$. We apply this operator to the quasihole Fock-space
expansion in \cref{eq:quasihole_state_expansion}.
Using $\{\gamma_{\bsl k},\gamma_{\bsl k'}^\dagger\}
=\delta_{\bsl k,\bsl k'}$ and the antisymmetry of the coefficients $C^\sigma$,
the same calculation as in
\cref{eq:D-Q-action-generic_k} gives
\eq{
Q(\zeta)\ket{\Psi_{\qh}^{3,\sigma}} =\frac{N_e(N_e-1)}{\sqrt{N_e!}} \sum_{\bsl k_1,\ldots,\bsl k_{N_e}}
C_{\bsl k_1,\ldots,\bsl k_{N_e}}^\sigma(\wh)
\frac{g'_{\bsl k_1}(\zeta)g_{\bsl k_2}(\zeta)}
{\mathcal N_{\bsl k_1}\mathcal N_{\bsl k_2}}
\gamma_{\bsl k_3}^\dagger\cdots
\gamma_{\bsl k_{N_e}}^\dagger\ket{0}.
}
Next, using $C_{\bsl k_1,\ldots,\bsl k_{N_e}}^\sigma(\wh)
=A_{\bsl k_1,\ldots,\bsl k_{N_e}}^\sigma(\wh)
\prod_{i=1}^{N_e}\mathcal N_{\bsl k_i}$
from the earlier expansion,
\eq{
Q(\zeta)\ket{\Psi_{\qh}^{3,\sigma}} =\frac{N_e(N_e-1)}{\sqrt{N_e!}}
\sum_{\bsl k_1,\ldots,\bsl k_{N_e}}
A_{\bsl k_1,\ldots,\bsl k_{N_e}}^\sigma(\wh)
g'_{\bsl k_1}(\zeta)g_{\bsl k_2}(\zeta)
\prod_{j=3}^{N_e}
\left(\mathcal N_{\bsl k_j}
\gamma_{\bsl k_j}^\dagger\right)\ket{0}.
}
We use \cref{eq:C-normalized-band-orbital,eq:creation_operator},
which give
$\mathcal N_{\bsl k}\gamma_{\bsl k}^\dagger
=\sum_{\bsl r\in\Lambda}
e^{\pi\mathcal B(\bsl r)/(2\Omega)}
\chi(\bsl r)g_{\bsl k}(z)c_{\bsl r}^\dagger$.
Substituting this identity for every
$j=3,\ldots,N_e$,
\eq{
Q(\zeta)\ket{\Psi_{\qh}^{3,\sigma}}=\frac{N_e(N_e-1)}{\sqrt{N_e!}}\sum_{\bsl r_3,\ldots,\bsl r_{N_e}\in\Lambda}
\prod_{j=3}^{N_e}
e^{\pi\mathcal B(\bsl r_j)/(2\Omega)}
\chi(\bsl r_j)
\left[
\sum_{\bsl k_1,\ldots,\bsl k_{N_e}}
A_{\bsl k_1,\ldots,\bsl k_{N_e}}^\sigma(\wh)
g'_{\bsl k_1}(\zeta)g_{\bsl k_2}(\zeta)
\prod_{j=3}^{N_e}g_{\bsl k_j}(z_j)
\right]
c_{\bsl r_3}^\dagger\cdots
c_{\bsl r_{N_e}}^\dagger\ket{0}.
\label{eq:Q_zeta_on_state}
}
We now show that the expression in the square brackets is $\left.\partial_w F_{\qh}^{\sigma}
\left(\zeta+\frac w2,\zeta-\frac w2,
z_3,\ldots,z_{N_e};\wh\right)\right|_{w=0}$. Using the expansion in \cref{eq:F_qh_holomorphic_expansion},
Substituting $z_1=\zeta+w/2$ and
$z_2=\zeta-w/2$, and differentiating with
respect to $w$ using
\cref{eq:D-relative-derivative-pair}, gives
\eqa{
\left.\partial_w F_{\qh}^{\sigma}
\left(\zeta+\frac w2,\zeta-\frac w2,
z_3,\ldots,z_{N_e};\wh\right)\right|_{w=0}
&=\frac12
\sum_{\bsl k_1,\ldots,\bsl k_{N_e}}
A_{\bsl k_1,\ldots,\bsl k_{N_e}}^\sigma(\wh)
\left[
g'_{\bsl k_1}(\zeta)g_{\bsl k_2}(\zeta)
-g'_{\bsl k_2}(\zeta)g_{\bsl k_1}(\zeta)
\right]
\prod_{j=3}^{N_e}g_{\bsl k_j}(z_j)\\
&=\sum_{\bsl k_1,\ldots,\bsl k_{N_e}}
A_{\bsl k_1,\ldots,\bsl k_{N_e}}^\sigma(\wh)
g'_{\bsl k_1}(\zeta)g_{\bsl k_2}(\zeta)
\prod_{j=3}^{N_e}g_{\bsl k_j}(z_j).
\label{eq:derivative_relation_for_F_qh}
}
The last equality follows by exchanging
$\bsl k_1\leftrightarrow\bsl k_2$ in the
second term and using
$A_{\bsl k_2,\bsl k_1,\ldots}^\sigma
=-A_{\bsl k_1,\bsl k_2,\ldots}^\sigma$. Substituting \cref{eq:derivative_relation_for_F_qh} into \cref{eq:Q_zeta_on_state},
and by \cref{eq:qh-energy-clean-cubic},
\eqa{
Q(\zeta)\ket{\Psi_{\qh}^{3,\sigma}}
&=\frac{N_e(N_e-1)}{\sqrt{N_e!}}
\sum_{\bsl r_3,\ldots,\bsl r_{N_e}\in\Lambda}
\prod_{j=3}^{N_e}
e^{\pi\mathcal B(\bsl r_j)/(2\Omega)}
\chi(\bsl r_j)\\
&\qquad\times
\left.\partial_w
F_{\qh}^{\sigma}
\left(\zeta+\frac w2,\zeta-\frac w2,
z_3,\ldots,z_{N_e};\wh\right)
\right|_{w=0}
c_{\bsl r_3}^{\dagger}\cdots
c_{\bsl r_{N_e}}^{\dagger}\ket{0}\\
&=0,
\label{eq:qh-energy-clean-Q-zero}
}
for every $\zeta\in\dsC$.

So far, \cref{eq:qh-energy-clean-Q-zero} shows that
$\sum_{\bsl K}Q_{\bsl K}(\zeta) \ket{\Psi_{\qh}^{3,\sigma}}=0$. However, to satisfy \cref{eq:qh-energy-clean-B-action}, we need to prove that $Q_{\bsl K}(\zeta)\ket{\Psi_{\qh}^{3,\sigma}}=0$
for every $\bsl K$ and $\zeta$. Now we show it.
From \cref{eq:E-pair-mode-expansion},
$Q_{\bsl K}(\zeta)=\sum_{\mu=0}^{1}\Phi_{\bsl K,\mu}(\zeta)Q_{\bsl K,\mu}$,
with all $\zeta$ dependence contained in $\Phi_{\bsl K,\mu}$.
We already derived their translation factors in
\cref{eq:E-simple-test-translations}. Because these factors are the same for $\mu=0,1$, they also give the translation laws for
$Q_{\bsl K}(\zeta)$:
\eqa{
Q_{\bsl K}(\zeta+a_2)
&=\sum_{\mu=0}^{1}\Phi_{\bsl K,\mu}(\zeta+a_2)Q_{\bsl K,\mu} =e^{\ii\bsl K\cdot\bsl a_2}Q_{\bsl K}(\zeta),\\
Q_{\bsl K}(\zeta+a_1)
&=\sum_{\mu=0}^{1}\Phi_{\bsl K,\mu}(\zeta+a_1)Q_{\bsl K,\mu}
=e^{\ii\bsl K\cdot\bsl a_1}
e^{-2\pi\ii\tau_0-4\pi\ii\zeta/a_2}Q_{\bsl K}(\zeta), \label{eq:qh-energy-clean-QK-primitive}
}
where $\tau_0=a_1/a_2$ as before.
For a unit-cell translation $\bsl R=n_1\bsl a_1+n_2\bsl a_2$,
write $R=n_1a_1+n_2a_2$, with $0\leq n_j<N_j$. Applying the $a_1$ shift $n_1$ times and the $a_2$ shift $n_2$ times, with $s=0,\ldots,n_1-1$ labeling the successive $a_1$ shifts, gives
\eqa{Q_{\bsl K}(\zeta+R)
&=e^{\ii\bsl K\cdot\bsl R}
\left[\prod_{s=0}^{n_1-1}
e^{-2\pi\ii\tau_0-4\pi\ii(\zeta+sa_1)/a_2}\right]
Q_{\bsl K}(\zeta)\\
&=e^{\ii\bsl K\cdot\bsl R}e^{-2\pi\ii n_1\tau_0-4\pi\ii n_1\zeta/a_2-4\pi\ii\tau_0\sum_{s=0}^{n_1-1}s}Q_{\bsl K}(\zeta)\\
&=e^{\ii\bsl K\cdot\bsl R}e^{-2\pi\ii n_1^2\tau_0-4\pi\ii n_1\zeta/a_2}Q_{\bsl K}(\zeta),\label{eq:qh-energy-clean-QK-R}}
In the last step, we used $\sum_{s=0}^{n_1-1}s=n_1(n_1-1)/2$. Notice that in \cref{eq:qh-energy-clean-QK-R}, the second exponential is the same
for every $\bsl K$, and only $e^{\ii\bsl K\cdot\bsl R}$ depends on $\bsl K$.
Since \cref{eq:qh-energy-clean-Q-zero} holds for every $\zeta$, we have
$Q(\zeta+R)\ket{\Psi_{\qh}^{3,\sigma}}=0$.
We then use the definition $Q(\zeta+R)=\sum_{\bsl K'}Q_{\bsl K'}(\zeta+R)$
and substitute \cref{eq:qh-energy-clean-QK-R} for each $\bsl K'$:
\eq{
0 =Q(\zeta+R)\ket{\Psi_{\qh}^{3,\sigma}}=\sum_{\bsl K'}Q_{\bsl K'}(\zeta+R)
\ket{\Psi_{\qh}^{3,\sigma}}=e^{-2\pi\ii n_1^2\tau_0-4\pi\ii n_1\zeta/a_2} \sum_{\bsl K'}e^{\ii\bsl K'\cdot\bsl R} Q_{\bsl K'}(\zeta)\ket{\Psi_{\qh}^{3,\sigma}}.
}
Since the exponential prefactor is nonzero and
independent of $\bsl K'$,
\eq{0=\sum_{\bsl K'}
e^{\ii\bsl K'\cdot\bsl R} Q_{\bsl K'}(\zeta)\ket{\Psi_{\qh}^{3,\sigma}}.
\label{eq:qh-energy-clean-translated-zero}
}
To isolate one $\bsl K$, multiply by $e^{-\ii\bsl K\cdot\bsl R}/N$
and sum over the $N$ unit cells:
\eqa{
 0
 &=\frac1N\sum_{\bsl R}e^{-\ii\bsl K\cdot\bsl R}
 \sum_{\bsl K'}e^{\ii\bsl K'\cdot\bsl R}
 Q_{\bsl K'}(\zeta)\ket{\Psi_{\qh}^{3,\sigma}}\\
 &=\sum_{\bsl K'}\left[\frac1N\sum_{\bsl R}
 e^{\ii(\bsl K'-\bsl K)\cdot\bsl R}\right]
 Q_{\bsl K'}(\zeta)\ket{\Psi_{\qh}^{3,\sigma}}\\
 &=Q_{\bsl K}(\zeta)\ket{\Psi_{\qh}^{3,\sigma}},
 \label{eq:qh-energy-clean-QK-zero}
}
where we used the orthogonality used in \cref{eq:E-simple-projected-interaction}. Thus every $Q_{\bsl K}(\zeta)$ annihilates the quasihole state. By \cref{eq:qh-energy-clean-QK-zero} we also get
\eq{
 Q_{\bsl K}'(\zeta)\ket{\Psi_{\qh}^{3,\sigma}}
 =\partial_\zeta\left[Q_{\bsl K}(\zeta)\ket{\Psi_{\qh}^{3,\sigma}}\right]
 =0.
 \label{eq:qh-energy-clean-QK-prime-zero}
}
We can now return to the pair operators in
\cref{eq:qh-energy-clean-B-action}, as
\cref{eq:qh-energy-clean-QK-zero,eq:qh-energy-clean-QK-prime-zero}
make
\eq{
\widetilde B_{\bsl r}^{(1)}\ket{\Psi_{\qh}^{3,\sigma}}
=\widetilde B_{\bsl r}^{(2)}\ket{\Psi_{\qh}^{3,\sigma}}=0,
\label{eq:qh-energy-clean-B-zero}
}
with $\bsl r\in\Lambda$. 

We have now proved \cref{eq:qh-energy-clean-targets}. 
Thus, the quasihole states are exact zero-energy
eigenstates of the band-projected Hamiltonian:
\eq{
\widetilde H\ket{\Psi_{\qh}^{3,\sigma}}
= (\widetilde H_0+\widetilde H_{\mathrm{int}})
\ket{\Psi_{\qh}^{3,\sigma}} =0, \qquad \sigma=1,2,3,
\label{eq:qh-energy-clean-eigenstate}}
for every quasihole position $\wh$. 
Notice that these zero-energy quasihole states contain $N_e$ electrons on a torus with $N=3N_e+1$ unit cells, which is different from the $N=3N_e$ case where we proved exactly three ground states in \cref{sec:app_D}.

\end{document}